%% file: PhD_thesis.tex
\documentclass[twoside, 
	openright,
	titlepage, 
	numbers=noenddot, 
	headinclude,
	footinclude=true, 
	cleardoublepage=empty, 
	BCOR=10mm, 
	paper=a4,
	fontsize=10pt,
	american,
]{scrbook} 

\usepackage[T1]{fontenc}    
\usepackage[utf8]{inputenc} 
\usepackage{amsthm}
\usepackage{amsfonts}
\usepackage{mathtools}
\usepackage{booktabs,siunitx}		
\input{classicthesis-config}

\usepackage{makeidx} 
\makeindex

\DeclareMathOperator{\e}{e}
\newcommand{\D}{{\rm d}}
\newcommand{\I}{{\rm i}} 
\newcommand{\upleft}[2]{\prescript{(#1)}{}{\! #2}} 

\begin{document}
\frenchspacing
\raggedbottom
\selectlanguage{american} % american ngerman
%\renewcommand*{\bibname}{new name}
%\setbibpreamble{}
\pagenumbering{roman}
\pagestyle{plain}
%********************************************************************
% Frontmatter
%*******************************************************
%\include{FrontBackmatter/DirtyTitlepage}
\include{FrontBackmatter/Titlepage}
\cleardoublepage\include{FrontBackmatter/Dedication}
%\cleardoublepage\include{FrontBackmatter/Foreword}
\cleardoublepage\include{FrontBackmatter/Abstract}
\cleardoublepage\include{FrontBackmatter/Publication}

\cleardoublepage\include{FrontBackmatter/Acknowledgments}

\pagestyle{scrheadings}
\cleardoublepage\include{FrontBackmatter/Contents}
%********************************************************************
% Mainmatter
%*******************************************************
\pagenumbering{arabic}
%\setcounter{page}{90}
% use \cleardoublepage here to avoid problems with pdfbookmark
\cleardoublepage

\part{Introduction and basic tools}

\include{chapters/1-01-introduction}

\include{chapters/1-02-chapter}

\include{chapters/1-03-chapter}
\cleardoublepage
\ctparttext{Effective theories of a scalar $\phi$ invariant under internal \textit{Galileon symmetry} transformations, $\phi\to\phi+b_\mu x^\mu+c$, have been extensively studied due to their special theoretical and phenomenological properties.
In this part, we introduce the notion of \textit{weakly broken Galileon invariance}, which characterizes the unique class of couplings of such theories to gravity that maximally retain their defining symmetry. The curved-space remnant of the Galileon's quantum properties allows to construct (quasi) de Sitter backgrounds largely insensitive to loop corrections. We exploit this fact to build novel cosmological models with interesting phenomenology, relevant for both inflation and late-time acceleration of the Universe.}
\cleardoublepage
\part{WBG Symmetry}
\include{chapters/2-01-chapter}
\include{chapters/2-02-chapter}
\include{chapters/2-03-chapter}
\include{chapters/2-04-chapter}

\include{chapters/2-05-chapter}

\cleardoublepage
\ctparttext{We propose a class of scalar models that, once coupled to gravity, lead to cosmologies that smoothly and stably connect an inflationary quasi-de Sitter Universe to a low, or even zero-curvature, maximally symmetric space-time in the asymptotic past, strongly violating the \acf{NEC} ($\dot H\gg H^2$) at intermediate times. The models are deformations of the conformal Galileon Lagrangian and are therefore based on symmetries, both exact and approximate, that give rise to the quantum robustness of the whole picture. The resulting cosmological backgrounds can be viewed as regularized extensions of the \emph{Galilean Genesis} scenario, or, equivalently, as ``early-time complete'' realizations of inflation.
The late-time inflationary dynamics possesses phenomenologically interesting properties: it can produce a large tensor-to-scalar ratio within the regime of validity of the effective field theory and can lead to sizeable equilateral non-Gaussianity.}
\cleardoublepage	
\part{An alternative to the standard cosmological evolution}
\include{chapters/3-01-chapter}

%\addtocontents{toc}{\protect\clearpage} % <--- just debug stuff, ignore
%\include{multiToC} % <--- just debug stuff, ignore for your documents
% ********************************************************************
% Backmatter
%*******************************************************
\appendix
\cleardoublepage
\part{Appendices}
\include{chapters/appendix-ADM}
\include{chapters/appendix-bispectrum}
\include{chapters/appendix-WBG}
\include{chapters/appendix-Eij}
\include{chapters/appendix-Extended_Genesis}
%********************************************************************
% Other Stuff in the Back
%*******************************************************
\cleardoublepage\include{FrontBackmatter/Bibliography}

\cleardoublepage\include{FrontBackmatter/Colophon}
\cleardoublepage\include{FrontBackmatter/Index}
% ********************************************************************
% Game Over: Restore, Restart, or Quit?
%*******************************************************
\end{document}

%% file: classicthesis-config.tex
\PassOptionsToPackage{eulerchapternumbers, %eulerchapternumbers: usa Euler come font per i numeri dei capitoli
	listings, %listings: you can add non-formatted text as you would do with \begin{verbatim} but its main aim is to include the source code of any programming language within your document.
	pdfspacing,
	floatperchapter, %floatperchapter: numera le figure in base al capitolo 1.1, 1.2,... 2.1, 2.2,..., 3.1, 3.2,...
	linedheaders, %linedheaders: mette una linea orizzontale sopra e sotto il titolo
	subfig,
	beramono,
	parts, %parts: serve per dividere in parti il book
	dottedtoc,%dottedtoc: nell'indice generale allinea a destra i numeri di pagina
	manychapters%manychapters: nell'indice generale aumenta lo spazio tra il numero e il titolo del capitolo
}{classicthesis}				
\usepackage{ifthen}
\newboolean{enable-backrefs} % enable backrefs in the bibliography
\setboolean{enable-backrefs}{false} % true false
\newcommand{\myTitle}{Weakly Broken Galileon Symmetry in Cosmology\xspace}
\newcommand{\mySubtitle}{Ph.D. Thesis\xspace}

\newcommand{\myName}{Luca Santoni\xspace}
\newcommand{\myProf}{Prof. Enrico Trincherini\xspace}

\newcommand{\myFaculty}{Classe di Scienze Matematiche e Naturali\xspace}

\newcommand{\myUni}{Put data here\xspace}

\newcommand{\myTime}{October 2016\xspace}
\newcommand{\myVersion}{version\xspace}

\newcounter{dummy} % necessary for correct hyperlinks (to index, bib, etc.)
\providecommand{\mLyX}{L\kern-.1667em\lower.25em\hbox{Y}\kern-.125emX\@}

\PassOptionsToPackage{latin9}{inputenc}	% latin9 (ISO-8859-9) = latin1+"Euro sign"
 \usepackage{inputenc}				

 \usepackage{babel}					

\PassOptionsToPackage{square,numbers}{natbib}
 \usepackage{natbib}				

 \usepackage{amssymb,amsmath}
 \usepackage{adjustbox} 

\PassOptionsToPackage{T1}{fontenc} % T2A for cyrillics
	\usepackage{fontenc}     
\usepackage{textcomp} % fix warning with missing font shapes
\usepackage{scrhack} % fix warnings when using KOMA with listings package          
\usepackage{xspace} % to get the spacing after macros right  
\usepackage{mparhack} % get marginpar right
\usepackage{fixltx2e} % fixes some LaTeX stuff 
\PassOptionsToPackage{printonlyused,smaller}{acronym}
	\usepackage{acronym} % nice macros for handling all acronyms in the thesis
\usepackage{tabularx} % better tables
\usepackage{caption}
\usepackage{subfig}  
\usepackage{listings} 
\PassOptionsToPackage{pdftex,hyperfootnotes=false,pdfpagelabels}{hyperref}
	\usepackage{hyperref}  % backref linktocpage pagebackref
\PassOptionsToPackage{pdftex}{graphicx}
	\usepackage{graphicx} 

\newcommand{\backrefnotcitedstring}{\relax}%(Not cited.)
\newcommand{\backrefcitedsinglestring}[1]{(Cited on page~#1.)}
\newcommand{\backrefcitedmultistring}[1]{(Cited on pages~#1.)}
\ifthenelse{\boolean{enable-backrefs}}%
{%
		\PassOptionsToPackage{hyperpageref}{backref}
		\usepackage{backref} % to be loaded after hyperref package 
		   \renewcommand*{\backref}[1]{}  % disable standard
		   \renewcommand*{\backrefalt}[4]{% detailed backref
		      \ifcase #1 %
		         \backrefnotcitedstring%
		      \or%
		         \backrefcitedsinglestring{#2}%
		      \else%
		         \backrefcitedmultistring{#2}%
		      \fi}%
}{\relax}    

\hypersetup{%
    colorlinks=true, linktocpage=true, pdfstartpage=3, pdfstartview=FitV,%
    breaklinks=true, pdfpagemode=UseNone, pageanchor=true, pdfpagemode=UseOutlines,%
    plainpages=false, bookmarksnumbered, bookmarksopen=true, bookmarksopenlevel=1,%
    hypertexnames=true, pdfhighlight=/O,%nesting=true,%frenchlinks,%
    urlcolor=webbrown, linkcolor=RoyalBlue, citecolor=webgreen, %pagecolor=RoyalBlue,%
    pdftitle={\myTitle},%
    pdfauthor={\textcopyright\ \myName, \myUni, \myFaculty},%
    pdfsubject={},%
    pdfkeywords={},%
    pdfcreator={pdfLaTeX},%
    pdfproducer={LaTeX with hyperref and classicthesis}%
}   

\makeatletter
\@ifpackageloaded{babel}%
    {%
       \addto\extrasamerican{%
				}%
       \addto\extrasngerman{% 
				}%	
    }{\relax}
\makeatother

\usepackage{classicthesis} 

%% file: FrontBackmatter/Titlepage.tex
%*******************************************************
% Titlepage
%*******************************************************
\begin{titlepage}
	% if you want the titlepage to be centered, uncomment and fine-tune the line below (KOMA classes environment)
	\begin{addmargin}[-1cm]{-3cm}
    \begin{center}
    		\includegraphics[width=6cm]{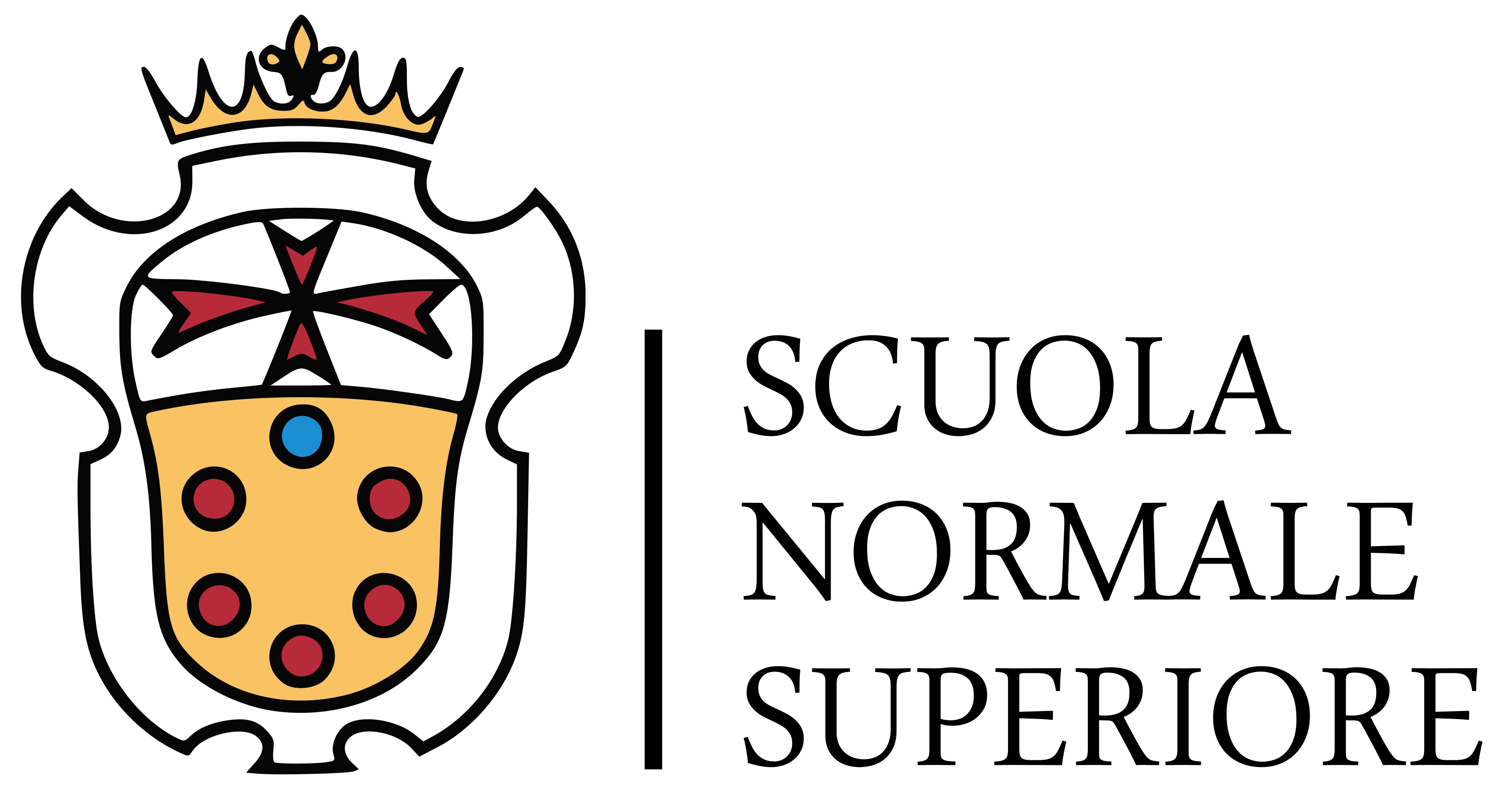} \\ 
    		
    		\rule{15cm}{0.01cm}
    		
    		%\vspace{0.3cm}
    		\medskip     		
    		{\large \spacedallcaps{\myFaculty}} \\
    		\bigskip
    		{\large Corso di Perfezionamento in \spacedallcaps{Fisica}}\\
    		    		
    		%\bigskip
    		\vfill
    		%\hfill
        
        \spacedallcaps{\mySubtitle}
        
        \large

        \vfill

        \begingroup
            {\LARGE\color{Maroon} \spacedallcaps{\myTitle} \\ \hspace{1 mm} } \\
            \bigskip
        \endgroup

        \vfill

        %\mySubtitle \\ \medskip   
        %\myDegree \\
        %\myDepartment \\                            
        %\myFaculty \\
        %\myUni \\ \bigskip

        %\myTime\ -- \myVersion
        
        %%%%%%%%%%	Sezione di candidato e relatore:
        \noindent Candidate \hfill Supervisor \par
        \smallskip
        \noindent \spacedlowsmallcaps{\myName} \hfill \spacedlowsmallcaps{\myProf} \par
        %%%%%%%%%%

        \vfill        
        
        \myTime              

    \end{center}  
  \end{addmargin}       
\end{titlepage}   

%% file: FrontBackmatter/Dedication.tex
%*******************************************************
% Dedication
%*******************************************************
\thispagestyle{empty}
%\phantomsection 
\refstepcounter{dummy}
\pdfbookmark[1]{Dedication}{Dedication}

\vspace*{3cm}

\begin{center}
    \textit{To Martina, my family and in memory of my father.}   
\end{center}

%% file: FrontBackmatter/Abstract.tex
%*******************************************************
% Abstract
%*******************************************************
%\renewcommand{\abstractname}{Abstract}
\pdfbookmark[1]{Abstract}{Abstract}
\begingroup
\let\clearpage\relax
\let\cleardoublepage\relax
\let\cleardoublepage\relax

\chapter*{Abstract}

One of the main challenges of modern cosmology consists of confirming, or even disproving, inflation. In the simplest models, a single scalar field drives the acceleration of the early Universe thanks to a flat potential or derivative self-interactions. In the latter case, however, in order to avoid possible instabilities, only single derivatives acting on the field are usually considered in the Lagrangians.

In the present work, using an effective field theory point of view, we explore theories of single-field inflation where higher derivative operators become relevant, affecting in a novel way the dynamics and therefore the observations. For instance, concerning the scalar spectrum, they allow for measurable equilateral non-Gaussianity, whose amplitude can differ significantly from the predictions of other existing models.

Moreover, we show that the stability and the consistency of such theories are ensured by an approximate Galileon symmetry. Indeed, being generically possible to build an invariant theory under Galileon transformations in flat space-time, it is instead well known that such a symmetry is unavoidably broken by gravity. In principle, this might ruin the nice and interesting properties of the Galileons in flat backgrounds, such as the non-renormalization theorem. However, we find that this does not happen if the Galileon invariance is broken only \textit{weakly}, in a well defined sense, by a suitable coupling to gravity, providing therefore an extension of the quantum non-renormalization properties in curved space-times.

Hence, besides discussing the phenomenological consequences and the observational predictions for inflation, we apply such Galileon theories to the context of the late-time acceleration of the Universe.

In the last part, in order to probe non-standard primordial scenarios, they are also employed in a cosmology where the Big Bang singularity is smoothed down and the Universe emerges from a Minkowski space-time, in a well defined extension at all times of the Galilean Genesis scenario.

\vfill

%\pdfbookmark[1]{Sommario}{Sommario} %Zusammenfassung
%\chapter*{Sommario}

\endgroup			

\vfill

%% file: FrontBackmatter/Publication.tex
%*******************************************************
% Publications
%*******************************************************
\pdfbookmark[1]{Publications}{publications}
\chapter*{Publications}
%Some ideas and figures have appeared previously in the following publications:

\bigskip

%\noindent Put your publications from the thesis here. The packages \texttt{multibib} or \texttt{bibtopic} etc. can be used to handle multiple different bibliographies in your document.

David Pirtskhalava, Luca Santoni, Enrico Trincherini and Patipan Uttayarat. \textit{Inflation from Minkowski Space}. JHEP 12 (2014) 151, \href{http://arxiv.org/pdf/1410.0882v1.pdf}{arXiv:1410.0882 [hep-th]}.
\\ 

\noindent
David Pirtskhalava, Luca Santoni, Enrico Trincherini and Filippo Vernizzi. \textit{Weakly Broken Galileon Symmetry}. JCAP 09 (2015) 007, \href{http://arxiv.org/pdf/1505.00007v1.pdf}{arXiv:1505.00007 [hep-th]}.
\\ 

\noindent
David Pirtskhalava, Luca Santoni, Enrico Trincherini and Filippo Vernizzi. \textit{Large Non-Gaussianity in Slow-Roll Inflation}. JHEP 04 (2016) 117, \href{http://arxiv.org/pdf/1506.06750v1.pdf}{arXiv:1506.0675 [hep-th]}.
\\ 

\noindent
David Pirtskhalava, Luca Santoni and Enrico Trincherini. \textit{Constraints on Single-Field Inflation}. JCAP 06 (2016) 051, \href{http://arxiv.org/pdf/1511.01817.pdf}{arXiv:1511.01817 [hep-th]}.
\\

\noindent
Paolo Creminelli, David Pirtskhalava, Luca Santoni and Enrico Trincherini. \textit{Stability of Geodesically Complete Cosmologies}. JCAP 11 (2016) 047, \href{https://arxiv.org/pdf/1610.04207v2.pdf}{arXiv:1610.04207 [hep-th]}.

\vspace{2cm}
\noindent
Other publications during the Ph.D., not related to the present work:
\\ 

%\noindent
%Mihail Mintchev, Luca Santoni and Paul Sorba. \textit{Thermoelectric efficiency of critical quantum junctions}. \href{http://arxiv.org/pdf/1310.2392v2.pdf}{arXiv:1310.2392 [cond-mat.stat-mech]}.
%\\ 

\noindent
Mihail Mintchev, Luca Santoni and Paul Sorba. \textit{Energy transmutation in non-equilibrium quantum systems}. J. Phys. A: Math. Theor. 48:5 (2015) 055003, \href{http://arxiv.org/pdf/1409.2994v2.pdf}{arXiv: 1409.2994 [cond-mat.stat-mech]}.
\\ 

\noindent
Mihail Mintchev, Luca Santoni and Paul Sorba. \textit{Non-linear quantum noise effects in scale invariant junctions}. J. Phys. A: Math. Theor, 48:28 (2015) 285002, \href{http://arxiv.org/pdf/1502.05234v1.pdf}{arXiv:1502.05234 [cond-mat.stat-mech]}.
\\ 

\noindent
Mihail Mintchev, Luca Santoni and Paul Sorba. \textit{Non-equilibrium current cumulants and moments with a point-like defect}. J. Phys. A: Math. Theor. 49 (2016) 265002, \href{http://arxiv.org/pdf/1601.01819.pdf}{arXiv:1601.01819 [cond-mat.stat-mech]}.
\\

\noindent
Mihail Mintchev, Luca Santoni and Paul Sorba. \textit{Quantum Transport in Presence of Bound States -- Noise Power}. \href{https://arxiv.org/pdf/1609.05427.pdf}{arXiv:1609.05427 [cond-mat.stat-mech]}.

%% file: FrontBackmatter/Acknowledgments.tex
%*******************************************************
% Acknowledgments
%*******************************************************
\pdfbookmark[1]{Acknowledgments}{acknowledgments}

\begin{flushright}{\slshape    
	ESTRAGON: We always find something, eh Didi, to give us the impression we exist?} \\ \medskip
	--- Samuel Beckett, \textit{Waiting for Godot}, Act II.
%	--- \defcitealias{knuth:1974}{Donald E. Knuth}\citetalias{knuth:1974} \citep{knuth:1974}
\end{flushright}

\bigskip

\begingroup
\let\clearpage\relax
\let\cleardoublepage\relax
\let\cleardoublepage\relax
\chapter*{Acknowledgments}

I am irretrievably in debt to Enrico Trincherini not only for having taught me most of what I have learnt during my Ph.D. but especially for his guidance, for always inspiring me, and for constantly suggesting new perspectives and precious ways to look at the physical problems. I always find our discussions extremely illuminating. Moreover, I would like to thank David Pirtskhalava, whom I had the opportunity to meet and work with, for all his suggestions and teachings. It has been a great pleasure to work also with Filippo Vernizzi: I really benefited from the discussions with him during the preparation of our works. I want to thank Mihail Mintchev for all I have learnt from him, for all our discussions and for having so fruitfully inspired me in many circumstances. Furthermore, I immensely benefited also from having met Riccardo Barbieri, from his lessons on particle physics and our group meetings. Finally, I would like to thank Lorenzo Anselmetti, Alberto Biella, Marco Cè, Christopher Murphy, Fabrizio Senia for fruitful discussions and all I met during these years, not less strongly inspiring me.

\endgroup

%% file: FrontBackmatter/Contents.tex
%*******************************************************
% Table of Contents
%*******************************************************
%\phantomsection
\refstepcounter{dummy}
\pdfbookmark[1]{\contentsname}{tableofcontents}
\setcounter{tocdepth}{2} % <-- 2 includes up to subsections in the ToC
\setcounter{secnumdepth}{3} % <-- 3 numbers up to subsubsections
\manualmark
\markboth{\spacedlowsmallcaps{\contentsname}}{\spacedlowsmallcaps{\contentsname}}
\tableofcontents 
\automark[section]{chapter}
\renewcommand{\chaptermark}[1]{\markboth{\spacedlowsmallcaps{#1}}{\spacedlowsmallcaps{#1}}}
\renewcommand{\sectionmark}[1]{\markright{\thesection\enspace\spacedlowsmallcaps{#1}}}
%*******************************************************
% List of Figures and of the Tables
%*******************************************************
\clearpage

\begingroup 
    \let\clearpage\relax
    \let\cleardoublepage\relax
    \let\cleardoublepage\relax
    %*******************************************************
    % List of Figures
    %*******************************************************    
    %\phantomsection 
    \refstepcounter{dummy}
    %\addcontentsline{toc}{chapter}{\listfigurename}
    \pdfbookmark[1]{\listfigurename}{lof}
    \listoffigures

    \vspace*{8ex}

    %*******************************************************
    % List of Tables
    %*******************************************************
    %\phantomsection 
    \refstepcounter{dummy}
    %\addcontentsline{toc}{chapter}{\listtablename}
    \pdfbookmark[1]{\listtablename}{lot}
    \listoftables
        
    \vspace*{8ex}
%   \newpage
    
    %*******************************************************
    % List of Listings
    %*******************************************************      
	  %\phantomsection 
%    \refstepcounter{dummy}
    %\addcontentsline{toc}{chapter}{\lstlistlistingname}
%    \pdfbookmark[1]{\lstlistlistingname}{lol}
%    \lstlistoflistings 

%    \vspace*{8ex}
       
    %*******************************************************
    % Acronyms
    %*******************************************************
    %\phantomsection 
    \refstepcounter{dummy}
    \pdfbookmark[1]{Acronyms}{acronyms}
    \markboth{\spacedlowsmallcaps{Acronyms}}{\spacedlowsmallcaps{Acronyms}}
    \chapter*{Acronyms}
    \begin{acronym}[UML]
        \acro{ADM}{Arnowitt-Deser-Misner}
        \acro{CMB}{Cosmic Microwave Background}
        \acro{COBE}{Cosmic Background Explorer}
        \acro{DBI}{Dirac-Born-Infeld}
        \acro{DGP}{Dvali-Gabadadze-Porrati}
        \acro{dRGT}{de Rham-Gabadadze-Tolley}
        \acro{EFT}{Effective Field Theory}
        \acro{EFTI}{Effective Field Theory of Inflation}
        \acro{FRW}{Friedmann-Robertson-Walker}
        \acro{GG}{Galilean Genesis}
        \acro{KSW}{Komatsu-Spergel-Wandelt}
        \acro{KWBG}{Kinetically driven inflation with Weakly Broken Galileon symmetry}
        \acro{LSS}{Large Scale Structure}
        \acro{NEC}{Null Energy Condition}
        \acro{SEC}{Strong Energy Condition}
        \acro{SMICA}{Spectral Matching of foregrounds implementing Independent Components Analysis}
        \acro{SRWBG}{Slow-Roll inflation with Weakly Broken Galileon symmetry}
        \acro{WBG}{Weakly Broken Galileon}
        \acro{1PI}{One-particle irreducible}
    \end{acronym}                     
\endgroup

\cleardoublepage

%% file: chapters/1-01-introduction.tex
\chapter{Introduction}
\label{introduction}

\begin{flushright}{\slshape    
	Sed quoniam docui solidisssima materiai\\
	corpora perpetuo volitare invicta per aevom,\\
	nunc age, summai quaedam sit finis eorum\\
	necne sit, evolvamus; item quod inane repertumst\\
	seu locus ac spatium, res in quo quaeque gerantur,\\
	pervideamus utrum finitum funditus omne\\
	constet, an immenum pateat vasteque profundum.}
	\\ \medskip
    --- Titus Lucretius Carus, \textit{De Rerum Natura} (951-957, Liber I).
\end{flushright}

Physics is the highest human effort to get a faithful description and knowledge at fundamental level of Nature. The complementarity between observations on one side and theoretical work on the other defines our current scientific paradigm. Evading the concept of reproducibility in the common sense in physical experiments, cosmology crucially relies almost uniquely on an accurate measurement and understanding of light signals from the sky. For this reason, many important discoveries have been attained only recently, thanks to significant technological improvements.

In this context, an example is provided by the \ac{CMB}\index{Cosmic Microwave Background (\ac{CMB})}. Since its discovery in 1964 by Arno Penzias and Robert Wilson, it has been playing a key role in the understanding of the features of the early Universe. Consisting in black body radiation at a temperature of about $2.73$°K, homogeneously and isotropically distributed in the sky, it represents a fundamental snapshot of the Universe at redshift $z\simeq 1100$ ($\sim 10^5$ yrs).

\begin{figure}[t]
  \caption{The Planck 2015 temperature power spectrum \cite{planck:2015}. The best-fit base $\Lambda$CDM theoretical spectrum fitted to the \textit{Planck} TT+lowP likelihood is plotted in the upper panel. Residuals with respect to this model are shown in the lower panel. The error bars show $\pm 1 \sigma$ uncertainties.}
  \centering
    \includegraphics[width=1.0\textwidth]{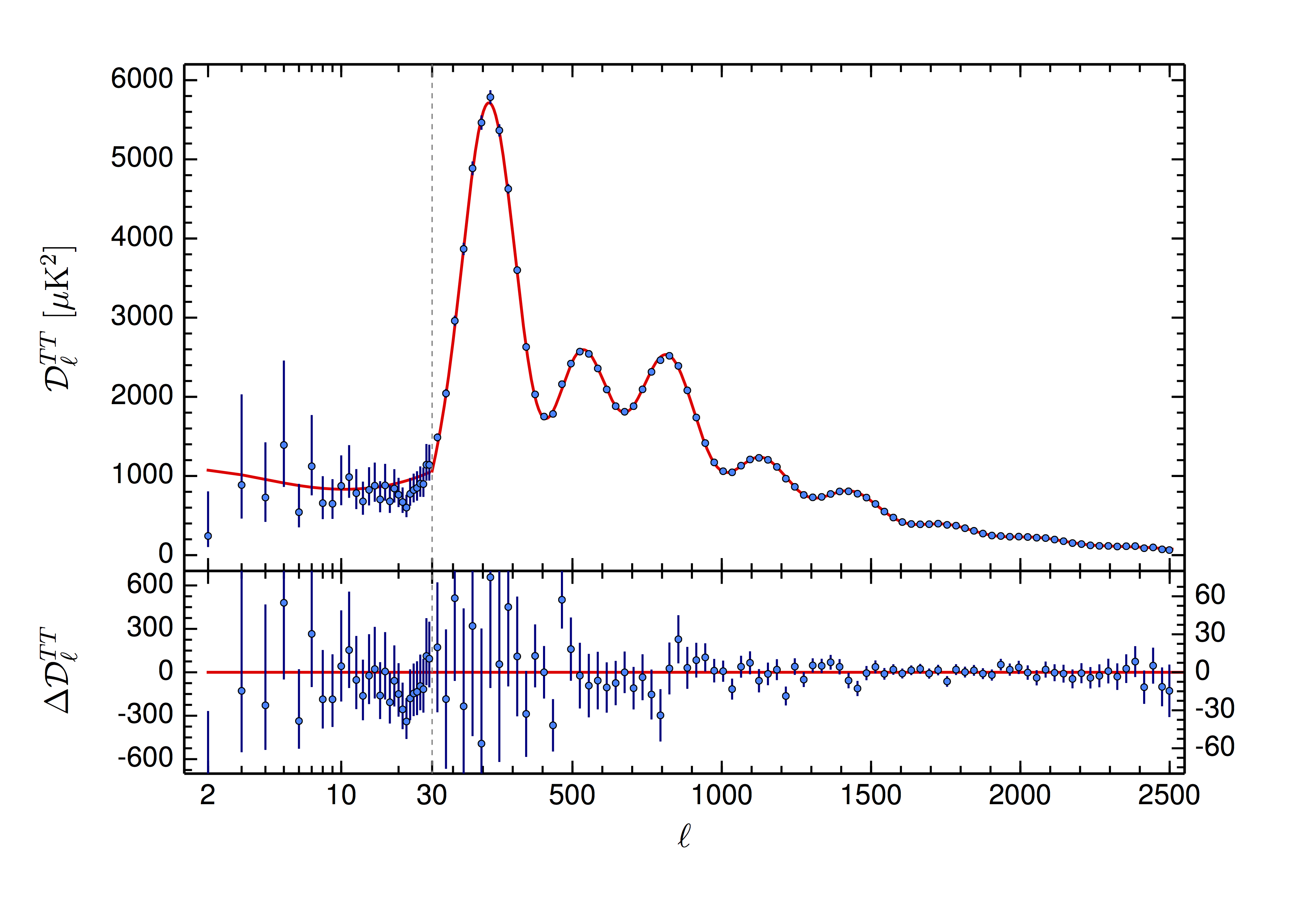}
\label{fig-planck}
\end{figure}

More recently, precision measurements on the \acs{CMB} map, that started with the \acs{COBE} experiment\index{experiments!COBE}, launched in 1989, were able to reveal small temperature fluctuations.
%, whose detection earned jointly John C. Mather and George F. Smoot the Nobel Prize in 2006.
These anisotropies\index{Cosmic Microwave Background (\ac{CMB})!anisotropies}, that appear in the temperature distribution as a result of an intrinsic anisotropy on the last-scattering surface\index{Cosmic Microwave Background (\ac{CMB})!last-scattering} and combined effects acquired by photons during their travel, are surprisingly very small, being of order $\delta T/T \sim 10^{-5}$.
%representing an aspect of extraordinary importance for cosmology.
Fig. \ref{fig-planck} shows the measurement of the \ac{CMB}\index{Cosmic Microwave Background (\ac{CMB})} angular power spectrum\index{power spectrum},
%that is the size of the temperature anisotropy\index{Cosmic Microwave Background (\ac{CMB})!anisotropies},
obtained by Planck Collaboration\index{experiments!Planck} \cite{planck:2015}.

The \ac{CMB}\index{Cosmic Microwave Background (\ac{CMB})} represents the overall dominant component in the photon content of the Universe, but it is not the only radiation source. Indeed, valuable information can be gained from the spectroscopy of the photons coming from galaxies and stars. The first example of this, which contributed to the definition of modern cosmology, is due to Edwin P. Hubble\index{Hubble!law}, who in 1929 was able to relate the recession velocity of galaxies to their relative distance, studying from Mount Wilson Observatory the brightness of Cepheid stars\index{Cepheid stars} and the redshift in the spectrum of radiation.
Soon after Hubble's discovery, supernovae became the most promising candidates for probing the large-scale geometry of the Universe and measuring the cosmic expansion. Indeed, thanks to the higher luminosity, they can be observed from more than $10^3$ Mpc away, a large number in comparison with the maximal distance of a visible Cepheid star, being of order of $10$ Mpc. Moreover, supernovae of type Ia\index{type Ia supernovae} turn out to have remarkably similar peak luminosities, leading to the possibility of estimating the relative distances just by looking at the light curves. The use of supernovae as reference candles allowed to improve the estimation of the Hubble parameter and led to the first evidence of the current cosmic acceleration in the late 1990s \cite{Riess:1998,Perlmutter:1998}. 

Finally, combining the analysis of the \ac{CMB}\index{Cosmic Microwave Background (\ac{CMB})} power spectrum and the study of the light signals coming from the other sources of radiation in the Universe offers the following summarizing picture, that describes the current status of knowledge:
\begin{itemize}
  \item the Universe is expanding and accelerating\index{cosmic acceleration};
  \item it is very homogeneous and isotropic;
  \item it is extremely flat;
  \item temperature fluctuations in the \ac{CMB}\index{Cosmic Microwave Background (\ac{CMB})} are tiny and are the tracks of the density perturbations on the last-scattering surface\index{Cosmic Microwave Background (\ac{CMB})!last-scattering}, which are in turn intimately related to spatial gradients in the gravitational potential;
  \item primordial scalar perturbations are almost scale-invariant\index{scale-invariance} and adiabatic\index{adiabatic mode};
  \item within the limits of the experimental sensitivity the distribution of the primordial perturbations is Gaussian;
  \item there is not any observation of primordial tensor modes, whose amplitude is currently constrained in terms of an upper bound.
\end{itemize}

Given these experimental evidences, the global cosmic history can be reconstructed according to our understanding of particle physics and fundamental interactions. The current picture suggests that the Universe should have started in a very hot era, gradually reducing its temperature during the expansion and passing through different phases, where composite particles are generated and decoupling phenomena occur depending on the typical energy scales (Hot Big Bang model\index{Big Bang model}).

However, in a standard decelerated expansion the current spatial flatness requires a tuning of the initial value in the past and the observed sky turns out to be composed by many sectors, which would correspond to causally disconnected patches at early times, characterized by an almost equally tuned value of the mean temperature.
Therefore, the high isotropy in the \ac{CMB}\index{Cosmic Microwave Background (\ac{CMB})} temperature map and the accurate degree of flatness\index{flatness problem} observed in the current Universe would be explainable only in terms of very specific initial conditions\footnote{They are usually called horizon and flatness problems, respectively.}  without a further justification\index{horizon problem}.

In modern cosmology the standard accepted paradigm, providing a simple explanation of these features, is called \textit{inflation}\index{inflation}. Believed to have occurred
%about $10^{-34}s$
soon after the Big Bang, it consists in an epoch of exponential expansion of the early Universe, able in principle to explain the high level of flatness,
the observed large-scale homogeneity and the isotropy in the \ac{CMB}\index{Cosmic Microwave Background (\ac{CMB})} temperature map.
Moreover, inflation\index{inflation} provides a fair explanation of the origin of the primordial density perturbations, which have been the seeds for the large-scale structure formation later in the Universe evolution, being directly related to the quantum fluctuations of the \textit{inflaton}\index{inflaton}, the quantum field responsible for the exponential expansion of the early stage.

Nevertheless it is fair to say that inflation\index{inflation} is not the only possibility. In fact, in accordance with observations, there exist different models which seem to be good candidates to shed light on the early Universe evolution. It is worth emphasizing that, in order to either disprove or confirm inflation, a thorough analysis of these alternatives is deserved and necessary.

After this first part, where notations and formal tools are presented, in Part II the standard inflationary paradigm will be investigated thoroughly from the point of view of a particular kind of symmetry transformations, and a new class of theories will be introduced and applied to the early Universe evolution, briefly commenting also on the current dark energy driven phase.

Furthermore, in order to investigate the possibility of deviations from the standard Big Bang scenario, Part III will be devoted to the introduction of an alternative cosmological evolution, which we will show to be degenerate with inflation in the observational constraints fixed by recent experiments.

Finally, more technical comments are collected in the appendices.

%% file: chapters/1-02-chapter.tex
\chapter{The inflationary epoch}
\label{chap-infepoch}

\begin{flushright}{\slshape    
    We are left in a situation which would be untenable with the old mechanics. If the universe were simply the motion which follows from a given scheme of equations of motion with trivial initial conditions, it could not contain the complexity we observe. Quantum mechanics provides an escape from the difficulty.} \\ \medskip
    --- Paul Adrien Maurice Dirac, \textit{The Relation between Mathematics and Physics}, Proceedings of the Royal Society (Edinburgh) Vol. 59, 1938-39, Part II.
%    --- \defcitealias{bentley:1999}{BBBBB}\citetalias{bentley:1999} \citep{bentley:1999}
\end{flushright}

In the introductory chapter, we outlined the relevant observational features of the Universe and we gave a hint of the commonly accepted theoretical paradigm. As one could have inferred, at the very fundamental level, our understanding of the Universe relies on General Relativity and Quantum Mechanics. The former clearly provides an explanation of the large scale physics, where gravity becomes the most relevant interaction. On the other hand, because of the primordial accelerated expansion, also the latter affects the macroscopic scales. Indeed, during inflation the quantum fluctuations of the inflaton are excited and stretched over large distances, exceed the Hubble radius\index{Hubble!radius} and their amplitudes remain frozen until they re-enter the horizon after the end of the accelerated phase, leaving their imprints in the \acs{CMB}\index{Cosmic Microwave Background (\ac{CMB})} anisotropies and being responsible for the growth of the structures we observe today.

In this chapter, we provide some notations and terminology, which will be useful in the subsequent parts, and we define the standard inflationary paradigm, emphasizing the role of symmetries in the understanding of the general features.

\section{The Friedmann-Robertson-Walker Universe}

Because of the observed isotropy and homogeneity of the Universe, it is natural to adopt the metric describing a maximally spatially symmetric \ac{FRW}\index{Friedmann-Robertson-Walker (\ac{FRW})!metric} space, given by
\begin{equation}
{\rm d}s^2 = -c^2{\rm d}t^2 + a(t)^2\left[\frac{{\rm d}r^2}{1-kr^2}
	+ r^2\left({\rm d}\theta^2+\sin^2\theta \, {\rm d}\varphi^2\right)\right] \, ,
\label{FRW-metric}
\end{equation}
where $a(t)$ is the cosmic-scale factor\index{scale factor}, $k$ is the curvature signature and $(r,\theta,\varphi)$ are the comoving coordinates\index{comoving!coordinates}. The curvature radius is given by the combination $R_{\rm curv} = a(t)\vert k\vert^{-1/2}$.

Moreover, because the Universe appears spatially flat, it is enough to consider the case of null curvature, that is $k=0$:
\begin{equation}
{\rm d}s^2 = -c^2{\rm d}t^2 + a(t)^2\left[{\rm d}r^2
	+ r^2\left({\rm d}\theta^2+\sin^2\theta \, {\rm d}\varphi^2\right)\right] \, .
\label{FRW-flatmetric}
\end{equation}

The conformal time\index{conformal!time} $\tau$ is usually defined through the differential equation
\begin{equation}
{\rm d}\tau = \frac{{\rm d}t}{a(t)} 
\label{ct}
\end{equation}
and the Hubble function\index{Hubble!constant} is introduced as
\begin{equation}
H(t) \equiv \frac{\dot{a}(t)}{a(t)} \, .
\label{H}
\end{equation}

In order to determine the cosmic evolution, the \acs{FRW} metric \eqref{FRW-metric} is employed in the Einstein equations\index{Einstein equations},
\begin{equation}
R_{\mu\nu} - \frac{1}{2}Rg_{\mu\nu} + \Lambda g_{\mu\nu} = \frac{8\pi G}{c^4}T_{\mu\nu} \, ,
\label{Eeqs}
\end{equation}
where $T_{\mu\nu}$ is the stress-energy tensor and $\Lambda$ the cosmological constant\index{cosmological constant}, obtaining the Friedmann equations
\begin{align}
H^2 
	= \left(\frac{\dot{a}}{a}\right)^2 
&	= \frac{8\pi G}{3}\rho - \frac{kc^2}{a^2} + \frac{\Lambda c^2}{3} \, ,
\label{Fe1}\\
\dot{H}+H^2
	= \frac{\ddot{a}}{a}
&	= -\frac{4\pi G}{3}\left(\rho+\frac{3p}{c^2}\right) + \frac{\Lambda c^2}{3} \, .
\label{Fe2}
\end{align}
\textit{Inflation}\index{inflation} is defined to be an evolutionary stage such that
\begin{equation}
\ddot{a} > 0
\qquad \Leftrightarrow \qquad
\frac{\D}{\D t}\left(aH\right)^{-1} < 0 \, ,
\label{inflation}
\end{equation}
that is an accelerating expansion or equivalently a phase with a decreasing Hubble radius\index{Hubble!radius}.
In principle, such an evolution could seem a quite exotic way to explain the observational data and solve the apparent primordial inconsistencies of the Hot Big Bang model, but actually it is not so peculiar given that our current Universe is undergoing an accelerated expansion as well.
Even though in principle $\Lambda$ satisfies the requirement \eqref{inflation}, the physics of the primordial acceleration and the succeeding evolution can not be explained in terms of a mere cosmological constant, therefore we will neglect it for the moment. Inflation can be achieved violating the \ac{SEC}\index{Strong Energy Condition (\acs{SEC})}, $\rho c^2 + 3p \geq 0$. In particular, a phase characterized by an equation of state of the form $p=-\rho c^2$, saturating the \ac{NEC}\index{Null Energy Condition (\acs{NEC})}, $\rho c^2 + p \geq 0$, is called \textit{de Sitter}\index{de Sitter!space} and has clearly $H = \text{constant}$.

Geometrically thought as an hyperboloid embedded in a Minkowski space of one higher dimension, a de Sitter space-time is a maximally symmetric space-time described by the metric
\begin{equation}
\D s^2 = \frac{1}{H^2\tau^2}\left(-c^2\D\tau^2+\D\vec{x}^2\right)
	= -c^2\D t^2+\e^{2Ht}\D\vec{x}^2 \, ,
\label{intro-dS-gmn}
\end{equation}
whose isometry group is $SO(4,1)$\index{de Sitter!$SO(4,1)$ group}. The corresponding algebra $so(4,1)$ is generated by $10$ isometries\index{de Sitter!isometries}, defining its dimension. These are the spatial translations and rotations, the dilations\index{dilations}
\begin{equation}
\tau \rightarrow \lambda\tau \, ,
\qquad
\vec{x} \rightarrow \lambda\vec{x} \,  ,
\qquad
\lambda\in\mathbb{R} \, ,
\label{intro-dS-dil}
\end{equation}
and the additional transformations\index{conformal!transformations}
\begin{equation}
\tau \rightarrow \tau - 2\tau \, \vec{b}\cdot\vec{x} \, ,
\qquad
x^i \rightarrow x^i + b^i \left( -c^2\tau^2+\vec{x}^2\right) -2x^i \, \vec{b}\cdot\vec{x} \, ,
\label{intro-dS-special}
\end{equation}
parametrized by an infinitesimal vector $\vec{b}$.

Assuming that a (quasi) de Sitter space-time is a valid description of the early Universe accelerated expansion, the previous isometries can be used to infer general and model independent predictions for the observables \cite{Creminelli:2010ba,Creminelli:2012ci,Creminelli:2012ccrsfi}. This makes the analysis of exact and approximate symmetries extremely useful.
For instance, the time-evolving background does not spoil the homogeneity and the isotropy, making spatial translations and rotations exact symmetries of the system. This constrains in an clear way the two-point function of the spectrum \cite{Creminelli:2010ba,Creminelli:2012ci} that we are going to define in \eqref{dS-ps0}. Moreover, because of the time-dependence, the background spontaneously breaks the other isometries\index{de Sitter!isometries}, that are therefore non-linearly realized\index{symmetry!non-linearly realized} and will not show up as an invariance at the level of the expectation values\footnote{They will show up instead as non-trivial relations between different correlation functions\index{consistency relations}. Indeed, it is worth emphasising that, while unbroken symmetries yield invariant correlators, spontaneously broken symmetries are related to the presence of consistency relations.}.
This allows to interpret inflation in terms of a spontaneous breaking of \textit{global} symmetries and the scalar perturbations as the Goldstone mode\footnote{As stressed in \cite{Hinterbichler:2012csam, baumann:2015book}, because the breaking pattern \eqref{symmbreakpattdS} involves space-time symmetries, the number of Goldstone bosons\index{Goldstone field} does not necessarily match the number of broken generators \cite{Volkov:1973,Ogievetsky:1974,Low:2002gt}.} $\zeta$ associated with the symmetry breaking pattern\footnote{For further details, we refer also to \cite{Hinterbichler:2012csam,Hinterbichler:2014is,Creminelli:2013crcm,Assassi:2012sl}. In particular in \cite{Hinterbichler:2012csam}, using the isomorphism between $SO(4,1)$ and the conformal group on $\mathbb{R}^3$ due to a one-to-one correspondence between the ten solutions of the Killing equations for the maximally symmetric de Sitter group and those of the conformal Killing equations\index{Killing equations} in $\mathbb{R}^3$ \cite{Antoniadis:2009ci}, the existence of certain non-linear realizations\index{symmetry!non-linearly realized} of the conformal symmetries\index{symmetry!conformal} for the curvature perturbation $\zeta$ -- that we will define in Eq. \eqref{ug-zeta} -- is studied in generality in any \acs{FRW} background. Obviously, the spontaneously broken symmetries are restored, and therefore linearly realized on the fields, in an exact de Sitter space-time.}
\begin{equation}
SO(4,1) \rightarrow \text{spatial rotations} + \text{translations} \, .
\label{symmbreakpattdS}
\end{equation}
Actually, the approximate scale-invariance\index{scale-invariance} in the observed power spectrum can be recovered imposing an additional internal shift-symmetry\index{symmetry!shift} on the background field. This automatically makes the two-point function approximately invariant under scale transformations \cite{Creminelli:2010ba,Creminelli:2012ci} and justifies why in the great majority of inflationary models the shift-symmetry\index{symmetry!shift} plays a key role. It is worth noticing \cite{Creminelli:2012ccrsfi} that one could in principle try to do the same with the transformations \eqref{intro-dS-special}, looking for an extra symmetry to require for the scalar field. The result would be an additional Galilean invariance, that we will define later, being at the core of the present work. However, as we will widely discuss, it can only be defined unambiguously in a Minkowski space-time, being explicitly broken by the coupling to the metric in a curved background. Therefore, the correlation functions, probing the theory at a scale comparable with the curvature, can not be invariant under the isometries \eqref{intro-dS-special} \cite{Creminelli:2012ccrsfi}.

We emphasise again that sooner or later the accelerated expansion has to be connected to a standard hot \acs{FRW} cosmology: in other words, the de Sitter phase is equipped with a physical ``clock'' signalling the end of inflation and defining a preferred time slicing of the space-time. This is provided by the Hubble function $H(t)$, which measures the changes during the background evolution.
As we will review in Chap. \ref{chap-EFTinflation}, being the \textit{leitmotiv} of the present work, this represents a guide principle in building a different approach to study inflation, based on an effective field theory\index{inflation!EFT of} \cite{Creminelli:2006xe,Cheung:2007st,Weinberg:2008}. Here, the breaking of the time-diffeomorphisms induces the presence of a Goldstone\index{Goldstone field} boson\footnote{The relation between the field $\pi$, associated with the spontaneous breaking of the time-diffeomorphisms\index{time-diffeomorphisms!breaking of}, and the curvature perturbation $\zeta$ in the comoving gauge is given in \eqref{zetapi}.} $\pi$, interpreted as the adiabatic mode\index{adiabatic mode} during inflation.

\section{Quantum field theory in de Sitter}

From now on we will set the units to be $\hbar=c=1$. We will introduce the Planck mass\index{Planck mass} $M_{\rm Pl}\equiv(8\pi G)^{-1/2}$.

\subsection{Background evolution}

An inflationary epoch can be achieved just by means of a single scalar field. The simplest action for the inflaton\index{inflaton} can be written as
\begin{equation}
S = \int{\rm d}^4x \, \sqrt{-g} \left[ 
	\frac{1}{2}M_{\rm Pl}^2R - \frac{1}{2}\partial_\mu\phi\partial^\mu\phi - V(\phi) \right] \, .
\label{dS-action0}
\end{equation}
Splitting the scalar field in
\begin{equation}
\phi(t,\vec{x}) = \phi_0(t) + \delta\phi(t,\vec{x}) \, ,
\label{ssplit}
\end{equation}
where $\phi_0$ is the \textit{classical} field, which governs the homogeneous and isotropic evolution, and $\delta\phi$ represents the fluctuation, the background equations of motion are
\begin{equation}
\ddot{\phi}_0 + 3 H \dot{\phi}_0 + V'(\phi_0) = 0 
\label{backEoM}
\end{equation}
and the stress-energy tensor is
\begin{align}
\rho = T_{00} & = \frac{1}{2}\dot{\phi}^2_0 + V(\phi_0) \, ,
\label{T00}\\
p = a^{-2}T_{ii} & = \frac{1}{2}\dot{\phi}^2_0 - V(\phi_0) \, .
\label{Tii}
\end{align}

In general, the requirement \eqref{inflation} of accelerating expansion can be also expressed through the following inequality:
\begin{equation}
\varepsilon \equiv - \frac{\dot{H}}{H^2} < 1 \, .
\label{eps}
\end{equation}
Moreover, a de Sitter configuration is reached for $\varepsilon\ll 1$.
In the particular case of the theory \eqref{dS-action0}, this reduces to $\dot{\phi}_0^2\ll V(\phi_0)$, namely a potentially dominated expansion\index{potentially dominated evolution}. Inflation can be sustained for a sufficient period of time only if such a \textit{rolling down} evolution is \textit{slow}\index{slow-roll conditions}, that is if $\vert\ddot{\phi}_0\vert \ll \vert H\dot{\phi}_0\vert , \, V'(\phi_0)$.
In the simplest case of a harmonic oscillator, $V(\phi)=m^2\phi^2/2$, the slow-roll conditions yield $\phi\gg M_{\text{Pl}}$. In principle, this does not ruin the reliability of the model, provided that $\rho\ll M_{\text{Pl}}^4$.

It is worth emphasising that this standard basic example is not the only way to achieve a de Sitter evolution: as we will discuss, different configurations can yield a wider and more interesting phenomenology. A more model-independent approach will be introduced in the next chapters.

\subsection{Quantum fluctuations during inflation}
\label{sub-chap1-qf}

As emphasised before, the quantum nature of the field during inflation is crucial for the explanation of several phenomena during the cosmic evolution, yielding effects also at macroscopic scales.
The fluctuations $\delta\phi$ of the scalar field represent small deviations from the absolute homogeneity and isotropy, and play a fundamental role in the construction of the current Universe. Being intimately related to the fluctuations of the metric through the energy-momentum tensor $T_{\mu\nu}$ and the Einstein equations, they represent the seeds for the large-scale structure formation, which occurs essentially via gravitational instability. In other words, the quantum field fluctuations, which populate the accelerating expanding phase, are transferred to both scalar density perturbations and gravitational waves.

First of all, during the shrinking of the comoving Hubble radius, once they exit the horizon, the inflaton fluctuations are stretched, freeze and become ``squeezed'', giving us a direct access to the physics governing the primordial evolution, without knowing the details of the \textit{reheating}\index{reheating} phase which links the final steps of inflation and the beginning of a standard evolution dominated by radiation and matter.

Moreover, once inflation ends and the comoving Hubble radius starts growing up restoring a decelerating expansion, these classical frozen fluctuations eventually re-enter the horizon during the radiation or matter dominated phase and imprint deviations to the homogeneous Universe, giving rise via the Poisson equation to \textit{matter} (and consequently to \textit{temperature}) \textit{perturbations} $\delta\rho$ ($\delta T$), which grow and become non-linear. For example, they manifest and contribute in the map of the \ac{CMB} anisotropies\index{Cosmic Microwave Background (\ac{CMB})!anisotropies}, as measured today.

Therefore, the quantization of $\delta\phi(t,\vec{x})$ and the analysis of its dynamics is necessary in order to understand the structure and the features of the Universe. In order to work in full generality, we consider a generic scalar field $\zeta(t,\vec{x})$ and quantize it in a (quasi) de Sitter background.

Given an action of the form\index{curvature perturbation!quadratic action}
\begin{equation}
S^{(2)}_\zeta = \int{\rm d}^4x \, a^3 \mathcal{N} \left[ \dot{\zeta}^2 - c_s^2 \frac{(\partial_i\zeta)^2}{a^2} \right] \, ,
\label{dS-action}
\end{equation}
where the field $\zeta(t,\vec{x})$ has to be canonically normalized as $\zeta_c \equiv \zeta\sqrt{2\mathcal{N}}$ and the \textit{speed} $c_s$ is now restored and generically set to be different from $1$ for future convenience\footnote{As anticipation, we assert that in presence of a time-dependent background and a privileged time-slicing of the space-time, that breaks Lorentz symmetry, the presence of a non-unitary speed of sound, $c_s\neq 1$, should not surprise. We refer to Chap. \ref{chap-EFTinflation} for further details.}, the solution in de Sitter space is
\begin{equation}
\zeta(\tau,\vec{x}) = \int\frac{{\rm d}^3\vec{k}}{(2\pi)^3}
	\left[f^*_{\vec{k}}a_{\vec{k}}^\dagger\e^{-\I\vec{k}\cdot\vec{x}}
	+ f_{\vec{k}}a_{\vec{k}}\e^{\I\vec{k}\cdot\vec{x}}\right] \, ,
\label{dS-sol}
\end{equation}
with
\begin{equation}
f^*_{\vec{k}} \equiv \frac{{\rm i}H}{2\sqrt{\mathcal{N}(kc_s)^3}}(1-\I kc_s\tau){\rm e}^{\I kc_s\tau} \, .
\label{dS-sol-f}
\end{equation}

Under the hypothesis of homogeneity and isotropy the field correlator $\langle\zeta(\tau,\vec{x}_1)\zeta(\tau,\vec{x}_2)\rangle$ is expected to depend only on $\vert\vec{x}_1-\vec{x}_2\vert$. Hence, the \textit{power spectrum}\index{power spectrum} is defined as the Fourier transform of the two-point function of the field,
\begin{equation}
\langle\zeta(\tau,\vec{x}_1)\zeta(\tau,\vec{x}_2)\rangle 
	= \int\frac{\D^3\vec{k}}{(2\pi)^3}
	P_\zeta(k)\e^{\I\vec{k}\cdot(\vec{x}_1-\vec{x}_2)} \, ,
\label{dS-ps0}
\end{equation}
and is equal to
\begin{equation}
P_\zeta(k) = \vert f_{\vec{k}}\vert^2
	\simeq \frac{H^2}{4\mathcal{N}k^3c_s^3}
\label{dS-ps}
\end{equation}
on super-horizon scales, \textit{i.e.} for $kc_s\tau\ll 1$. Often, the notation
\begin{equation}
\Delta_\zeta^2(k) \equiv \frac{k^3}{2\pi^2}P_\zeta(k)
\label{dS-ps-delta}
\end{equation}
is preferred. Observations indicate a nearly flat power spectrum, namely a power-law parametrization
\begin{equation}
\Delta_\zeta^2(k) = \mathcal{A}_s\left(\frac{k}{k_*}\right)^{n_s-1} \, ,
\label{dS-ps-pl}
\end{equation}
where $\mathcal{A}_s$ is the amplitude and $k_*$ a fiducial moment, seems to be the most natural. The parameter $n_s$ is known as \textit{scalar spectral index} (or \textit{scalar tilt}). An exact flat spectrum has $n_s=1$, while spectra with $n_s<1$ and $n_s>1$ are identified as ``red'' and ``blue'', respectively. To account for more complex behaviours, a \textit{running index} is introduced by
\begin{equation}
\Delta_\zeta^2(k) = \mathcal{A}_s\left(\frac{k}{k_*}\right)^{n_s-1+\frac{\D n_s}{\D\ln k}\ln\frac{k}{k_*}} \, ,
\label{dS-ps-run}
\end{equation}
representing the rate of the spectral index in proximity of $k_*$.

We stress that the field $\zeta$ represents the comoving curvature perturbation\index{curvature perturbation}, that we will define later in \eqref{ug-zeta}.
The confidence limits on some cosmological parameters are displayed\index{experiments!Planck} in Tab. \ref{planck2015parinf} \cite{planckXX:2015}. The amplitude of the scalar spectrum is found to be of order $\mathcal{A}_s\sim 10^{-9}$, while an unambiguous deviation from the scale invariant point $n_s=1$ has been confirmed. This is clearly visible in Fig. \ref{planck2015parinf} \cite{planckXX:2015}.

\begin{table}[t]
\begin{center}
\begin{tabular}{
 l
 S[table-format=-2.4,
   table-figures-uncertainty=1]
 S[table-format=-2.4,
   table-figures-uncertainty=1]
 S[table-format=-2.4,
   table-figures-uncertainty=1]
}
\toprule
 & {TT+lowP} & {TT+lowP+lensing} & {TT,TE,EE+lowP} \\
\midrule
$\ln(10^{10} A_s)$		& 3.089 \pm 0.036 & 3.062 \pm 0.029  & 3.094 \pm 0.034 \\
$n_s$					& 0.9655  \pm 0.0062 & 0.9677 \pm 0.0060 & 0.9645 \pm 0.0049 \\
$\frac{\D n_s}{\D\ln k}$	& -0.0084  \pm 0.0082 & -0.0033 \pm 0.0074 & -0.0057 \pm 0.0071 \\
$H_0$\index{Hubble!constant}					& 67.31 \pm 0.96 & 67.81 \pm 0.92 & 67.27 \pm 0.66 \\
\bottomrule
\end{tabular}
\caption{Confidence limits on some parameters of the base $\Lambda$CDM model, for various combinations of \textit{Planck} 2015 data \cite{planckXX:2015} at $68$\% confidence level, are shown. The results are taken from \cite{planckXX:2015}. In the analysis the pivot scale is $k_*=0.05 \, \text{Mpc}^{-1}$.}
\label{planck2015parinf}
\end{center}
\end{table}

\begin{figure}[t]
  \caption{The figure is taken from \cite{planckXX:2015}. It shows the marginalized joint confidence contours for $(n_s, \text{d}n_s/\text{d}\ln k)$ at the $68$\% and $95$\% CL (without considering the tensor contribution), using \textit{Planck} TT+lowP and \textit{Planck} TT,TE,EE+lowP. For comparison, also the \textit{Planck} 2013 results are shown.} 
  \centering
    \includegraphics[width=1.0\textwidth]{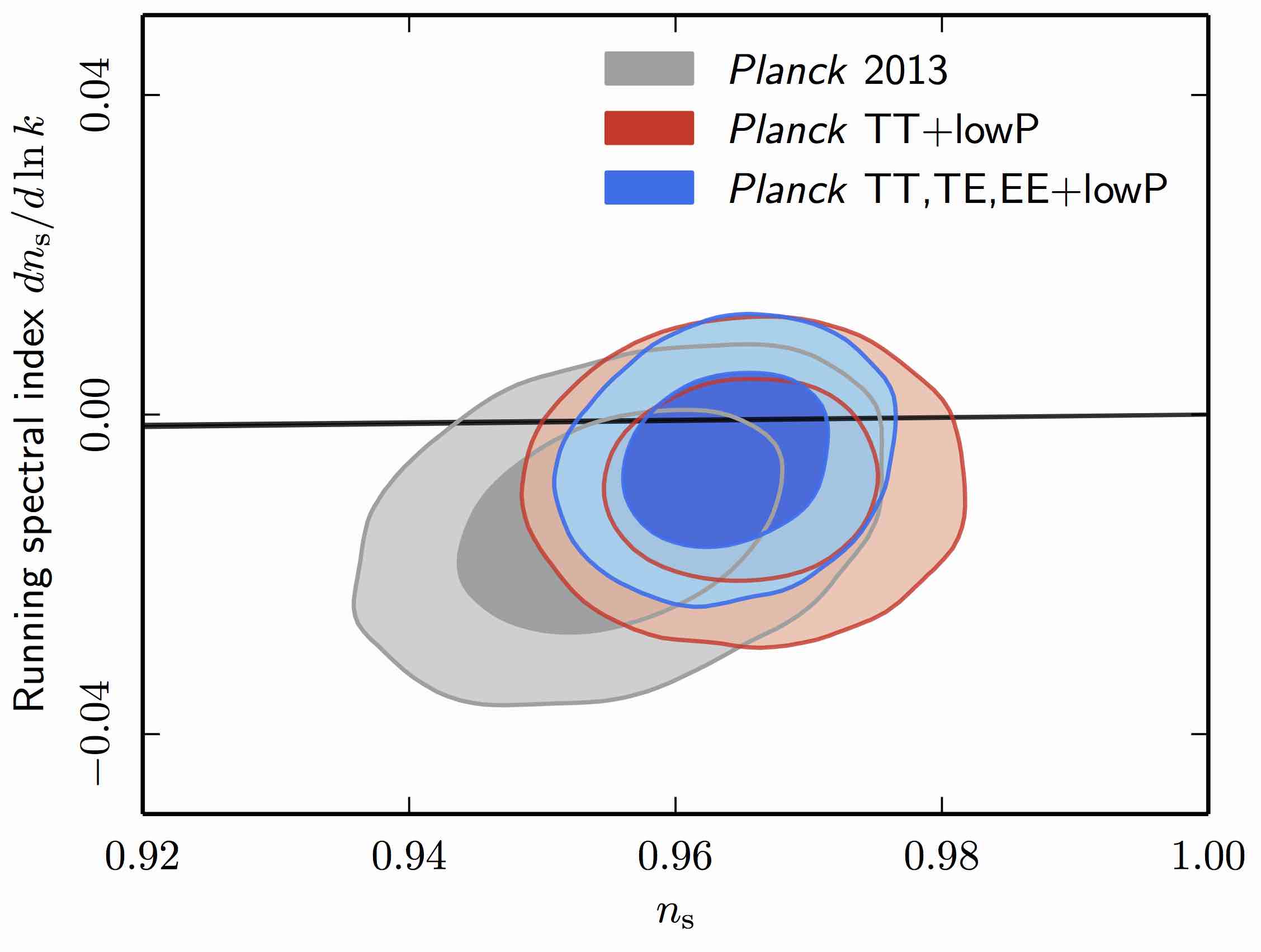}
\label{fig-planck2015parinf}
\end{figure}

Some comments about the power spectrum are now in order. Even though it has played an important role in the understanding of the Universe evolution, on the theoretical side it does not offer stringent constraints in the model building and therefore is not able to shed light on the plethora of inflationary theories. Indeed, as it is clear from \eqref{dS-ps}, being sensitive to many details, it has a lot of freedom in its flexible parameter dependence: for instance, one can play with the shape of the potential (observe that in the simplest models $\mathcal{N}\propto\varepsilon/c_s^2$ \cite{Cheung:2007st,Maldacena:2002vr}), the speed of propagation $c_s$ of the scalar perturbations and the number of fields that might have been present during inflation\index{inflation!multi-field}. In other words, it is not difficult to adjust the free parameters in \eqref{dS-ps} to accommodate to the measured values. In this sense, the analysis of possible departures from free-theories and a Gaussian spectrum is potentially a discriminant among interacting models. In Sec. \ref{N-Gsec} we will provide some terminology and tools concerning theories with non-trivial self-interactions.

\begin{figure}[t]
  \caption{The Planck 2015 \cite{planckXX:2015} marginalized joint contours for $(n_s,r)$, at the $68$\% and $95$\% CL, in the presence of running of the spectral indices, using \textit{Planck} TT+lowP and \textit{Planck} TT,TE,EE+lowP. Comparisons from Planck 2013 data release are shown for comparison.}
  \centering
    \includegraphics[width=1.0\textwidth]{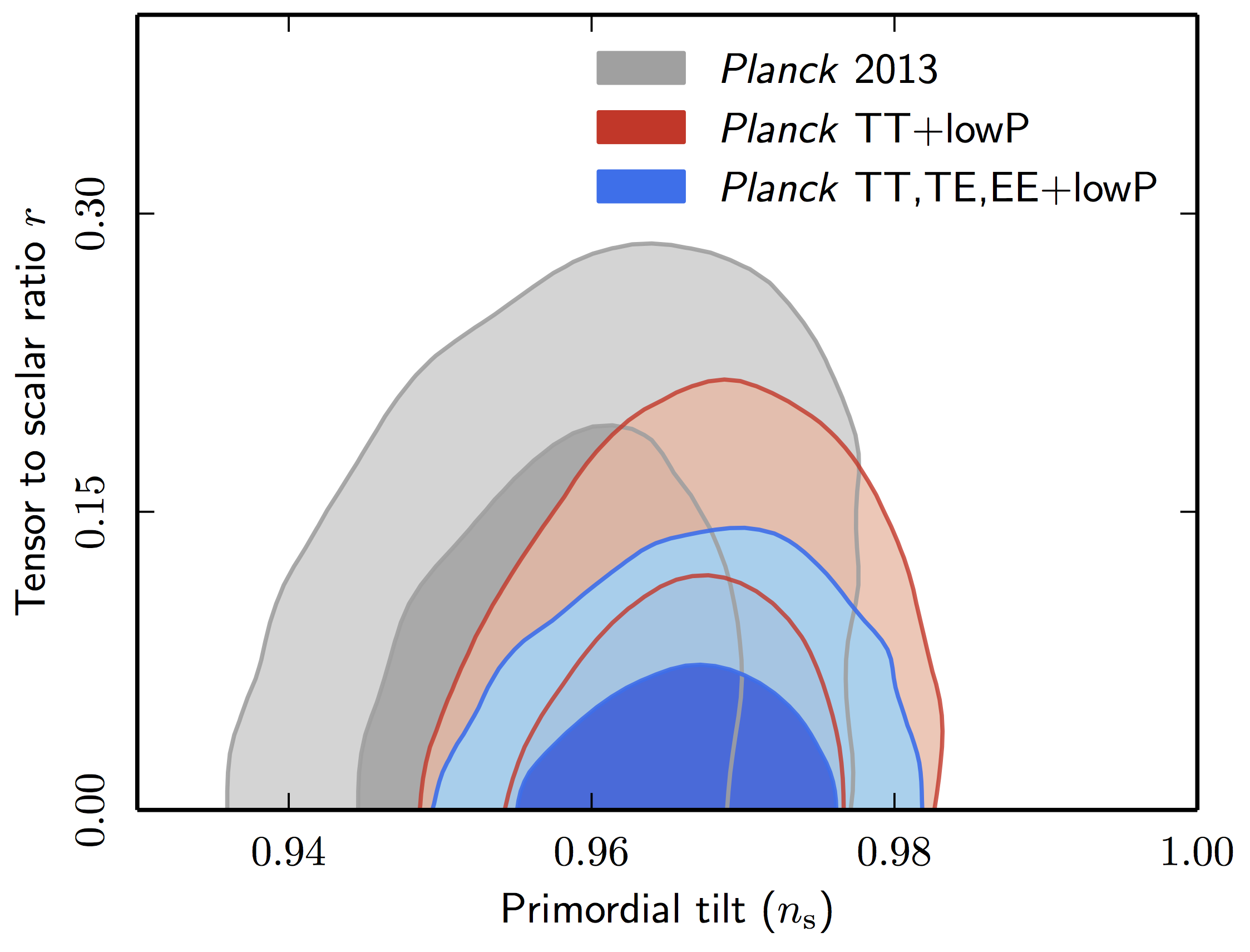}
\label{fig-planck-r-n_s}
\end{figure}

The previous analysis is valid also for the quadratic action of tensor perturbations obtained as leading order expansion of the Einstein-Hilbert action\index{Einstein-Hilbert action},
\begin{equation}
S^{(2)}_h = \frac{M_{\rm Pl}^2}{8}
	\int{\rm d}^4x \, a^3 \left[ (\dot{h}_{ij})^2 - \frac{(\partial_kh_{ij})^2}{a^2} \right] \, ,
\label{dS-action-h}
\end{equation}
where $h_{\mu\nu}$ represents the deviation from the \ac{FRW} background metric $\bar{g}_{\mu\nu}={\rm diag}(-1,a^2,a^2,a^2)$, namely $g_{\mu\nu}=\bar{g}_{\mu\nu}+h_{\mu\nu}$. In the case of \eqref{dS-action-h}, the power spectrum for tensor modes\index{tensor modes} is
\begin{equation}
P_T(k) = 2P_h(k) \simeq \frac{4H^2}{M_{\rm Pl}^2k^3} \, ,
\label{dS-ps-h}
\end{equation}
where the factor $2$ stands for the two polarizations and, as before, $H$ is evaluated at horizon crossing. In analogy with the scalar sector, the following parametrization is introduced, 
\begin{equation}
\Delta_T^2(k) \equiv \frac{k^3}{2\pi^2}P_T(k) = \mathcal{A}_T\left(\frac{k}{k_*}\right)^{n_T} \, ,
\label{dS-ps-h-pl}
\end{equation}
where $n_T$ is the \textit{tensor tilt}. The relative magnitude of the scalar and tensor perturbations is measured by the ratio
\begin{equation}
r \equiv \frac{\mathcal{A}_T}{\mathcal{A}_s} \, ,
\label{te-to-s}
\end{equation}
known as \textit{tensor-to-scalar ratio}\index{tensor-to-scalar ratio}, which is experimentally constrained to be $r_{0.002}<0.11$ ($95$\% CL) at the pivot scale $k_*=0.002 \, \text{Mpc}^{-1}$ \cite{planckXX:2015}.
Fig. \ref{fig-planck-r-n_s} shows the confidence regions for $(n_s,r)$, obtained by Planck Collaboration\index{experiments!Planck} \cite{planckXX:2015}. For a more recent analysis see also \cite{Bicepnew}.

In standard slow-roll inflation with minimal coupling to Einstein gravity, the parameters $r$ and $n_T$ are not mutually independent but are related by the \textit{consistency condition} \cite{Lidsey:1995}
\begin{equation}
r = -8n_T \, .
\label{consistency-cond}
\end{equation}

It is fair to say that, contrary to what we have done in the action \eqref{dS-action} for the scalar field, we have not restored the speed of propagation $c_T$ of tensor modes in \eqref{dS-action-h}. This would have led to a modification of the tensor spectrum by a factor $1/c_T$. Now one might wonder about the reason of this discrepancy with respect to the scalar case. The motivation relies on the fact that one can not change the tensor quadratic and cubic action at the leading order in derivatives \cite{Creminelli:2014wna}. Indeed, thanks to a suitable field re-parametrization of the form
\begin{equation}
g_{\mu\nu} \rightarrow g_{\mu\nu}+\left(1-c_T^2(t)\right)n_\mu n_\nu \, ,
\label{disf-cT}
\end{equation}
called \textit{disformal transformation}\index{disformal transformations} \cite{Bekenstein:1993}, where $n_\mu$ is the unit vector perpendicular to the surfaces of constant time\footnote{We refer to App. \ref{Appendix-ADM} for further details.}, the speed of propagation of tensor modes can always set to be $c_T=1$. Moreover, an additional conformal transformation of the metric,
\begin{equation}
g_{\mu\nu} \rightarrow c_T^{-1}(t)g_{\mu\nu} \, ,
\label{conf-cT}
\end{equation}
can remove any time dependence in the overall factor, restoring the standard $M_{\text{Pl}}$ coefficient and bringing the action back to the Einstein frame. In other words, at the leading order in the derivative expansion the gravity sector is not modified \cite{Creminelli:2014wna}. Nevertheless, some remnant of the non-standard dynamics of the tensor modes in the previous frame remains, for instance, in the violation of the consistency condition \eqref{consistency-cond} \cite{Baumann:2015hs}, which is as well as all physical observables frame-independent.

The robustness of the tensor spectrum would make a possible measurement of primordial gravitational waves extremely interesting, because these fix the energy scale of inflation, given by the ratio $H/M_{\text{Pl}}$ in \eqref{dS-ps-h}. Obviously, experiments with this purpose have to face several practical difficulties, essentially related to the purification of the signal from contamination effects due to dust and non-primordial sources in the \acs{CMB} polarization map.

Finally, we comment on the fact that sometimes any eventual evidence of primordial tensor modes and a large tensor-to-scalar ratio is considered a smoking-gun for the \textit{standard} inflationary paradigm, ruling out alternative cosmologies. Actually, this statement is not completely true and without loopholes. In Part III of the present work, we will discuss about this point, providing an example of an alternative cosmic evolution that could in principle accommodate sizeable values of $r$.

\subsection{Non-Gaussianity}
\label{N-Gsec}

If the probability distribution function for the scalar perturbations is Gaussian, the two-point correlator is enough to encode the whole information. This happens if the self-couplings of the scalar field driving the accelerated expansion of the Universe are negligible: indeed, in this case the Fourier modes can be treated independently. This occurs for instance in the case of slow-roll inflation\index{inflation!slow-roll} where the scalar potential is extremely flat. Otherwise, in interacting theories deviations from a Gaussian statistics are expected.

The main reason of the interest in primordial non-Gaussianity\index{non-Gaussianity} lies in the fact that it can probe the physics at the origin of the Universe, which involves regimes of extreme high energy, inaccessible to laboratory experiments. A possible measure of sizeable deviations from exact Gaussianity would discriminate among competing models of evolution. Indeed, a plethora of consistent mechanisms regarding the early Universe stage characterizes the current literature, predicting different amplitudes and shapes. As already emphasized, the \acs{CMB}\index{Cosmic Microwave Background (\ac{CMB})} map contains the fingerprints of possible primordial non-Gaussianity, which is transferred during the cosmological evolution from inflaton fluctuations to temperature anisotropies\index{Cosmic Microwave Background (\ac{CMB})!anisotropies}.

However, this analysis has to deal with the presence of several mechanisms of generation of non-Gaussianity. Indeed, there are various potential non-primordial sources, which contaminate the original signal one is actually interested in. They can be classified into four broad categories \cite{planckXXIV:2013}: instrumental systematic effects, residual foregrounds and point sources, secondary \acs{CMB} anisotropies\index{Cosmic Microwave Background (\ac{CMB})!anisotropies} such as the gravitation lensing, and non-linear or second order perturbative phenomena. For an exhaustive introduction to all these effects we refer to \cite{planckXXIV:2013} and references therein\index{experiments!Planck}.

Now we briefly summarize some notation and terminology\footnote{See for instance \cite{baumann:2015book} for a clear compendium on this topic and more in general for a nice introduction to inflation and related models.}.
We decide to adopt the notation of \cite{planckXVII:2015} for the analysis of non-Gaussianity\footnote{See also \cite{planckXXIV:2013,gorbunov:2011,Komatsu:2000,bartolo:2004,Creminelli:2005hu,Senatore:2009ng} and the references therein.}, which involves the gravitational potential $\Phi$, that enters the Newtonian gauge\index{Newtonian gauge} metric parametrizing the scalar perturbations in the following way:
\begin{equation}
\D s^2 = a(\tau)^2\left[-(1+2\Phi)\D\tau^2+(1-2\Psi)\D\vec{x}^2\right] \, ,
\label{Ngm}
\end{equation}
where the condition $\Psi=\Phi$ arises from Einstein equations in absence of anisotropic stress.

If the probability distribution of the primordial density perturbations turned out to be non-Gaussian, in general it would exhibit non-trivial multiple correlators. Therefore, the first quantity to analyse is given by the three-point function:
\begin{equation}
\langle\Phi(\vec{k}_1)\Phi(\vec{k}_2)\Phi(\vec{k}_3)\rangle
	= (2\pi)^3\delta(\vec{k}_1+\vec{k}_2+\vec{k}_3)B_\Phi(k_1,k_2,k_3) \, ,
\label{NG-3bisdef}
\end{equation}
where $B_\Phi$ is called \textit{bispectrum}\index{bispectrum} and the delta-function is a consequence of the spatial homogeneity. However, experimental results are not given in terms of $B_\Phi$ but in terms of the combination
\begin{equation}
f_{\rm NL} = \frac{B_\Phi(k,k,k)}{6P_{\Phi}(k)^2} \, ,
\label{NG-3biseq2}
\end{equation}
\textit{i.e.} normalizing with the power spectrum $P_\Phi$ in the equilateral configuration, $\vert\vec{k}_1\vert=\vert\vec{k}_2\vert=\vert\vec{k}_3\vert=k$. 
As we have already emphasised, the moments of the distribution function could signal the presence of non-trivial interactions in the inflaton Lagrangian. Thus, crucial information can be inferred from the bispectrum in order to clarify the physics of inflation. Indeed, different models can produce very different templates in the momentum dependence of $B_\Phi(k_1,k_2,k_3)$. We remind now the three common shapes.

\subsubsection{Local shape}

A simple parametrization of non-Gaussianity preserving locality in space is achieved by considering a non-linear function in the Gaussian field $\Phi_{\rm L}(\vec{x})$:
\begin{equation}
\Phi(\vec{x}) = \Phi_{\rm L}(\vec{x}) + f_{\rm NL}^{\text{local}}\left[\Phi_{\rm L}(\vec{x})^2
	- \langle\Phi_{\rm L}(\vec{x})^2\rangle\right] \, ,
\label{NG-1}
\end{equation}
where $f_{\rm NL}^{\text{local}}$ is dimensionless. Since the amplitude of scalar perturbations is small, $\Phi\sim10^{-5}$, the second term in Eq. \eqref{NG-1} represents a subleading contribution if $f_{\rm NL}^{\text{local}}$ is not too big. Using the definition of power spectrum $\langle\Phi_{\rm L}(\vec{k}_1)\Phi_{\rm L}(\vec{k}_2)\rangle=(2\pi)^3P_{\Phi}(k_1)\delta(\vec{k}_1+\vec{k}_2)$ and the fact that the field $\Phi_{\rm L}$ follows a Gaussian distribution, at the leading order the cubic correlator is
\begin{multline}
\langle\Phi(\vec{k}_1)\Phi(\vec{k}_2)\Phi(\vec{k}_3)\rangle
	= 2(2\pi)^3\delta(\vec{k}_1+\vec{k}_2+\vec{k}_3) \times
\\
	\times f_{\rm NL}^{\text{local}}
	\left[ P_{\Phi}(k_1)P_{\Phi}(k_2) + P_{\Phi}(k_1)P_{\Phi}(k_3) + P_{\Phi}(k_2)P_{\Phi}(k_3) 
	\right] .
\label{NG-3bis}
\end{multline}
From Eq. \eqref{NG-3bis} and the definition \eqref{NG-3bisdef} we find
\begin{equation}
B_\Phi^{\text{local}}(k_1,k_2,k_3) = 2f_{\rm NL}^{\text{local}}\left[ P_{\Phi}(k_1)P_{\Phi}(k_2) + P_{\Phi}(k_1)P_{\Phi}(k_3) + P_{\Phi}(k_2)P_{\Phi}(k_3) \right] \, ,
\label{NG-3bis2}
\end{equation}
which defines the \textit{local} template\index{non-Gaussianity!local}.
In the equilateral configuration, 
\begin{equation}
\langle\Phi(\vec{k}_1)\Phi(\vec{k}_2)\Phi(\vec{k}_3)\rangle
	= 6(2\pi)^3\delta(\vec{k}_1+\vec{k}_2+\vec{k}_3)f_{\rm NL}^{\text{local}}P_{\Phi}(k)^2 \, ,
\label{NG-3biseq}
\end{equation}
in accordance with the definition \eqref{NG-3biseq2}. It is clear that this kind of shape peaks in the region where one of the three momenta vanishes and the others become equal. This configuration is called \textit{squeezed limit}\index{squeezed limit} and usually strongly characterizes multi-field models\index{inflation!multi-field} of inflation.

\subsubsection{Equilateral shape}

The \textit{equilateral} shape\index{non-Gaussianity!equilateral} is well described by the template \cite{planckXVII:2015,baumann:2015book,Creminelli:2006lNG}
\begin{equation}
B_\Phi^{\text{equil}}(k_1,k_2,k_3) = f_{\rm NL}^{\text{equil}}\left[
	6(P_1^3P_2^2P_3)^{1/3}
	- 3P_1P_2 - 2(P_1P_2P_3)^{2/3}
	+ \text{perms.}
	\right] \, ,
\label{NG-Bequil}
\end{equation}
where $P_i\equiv P_\Phi(k_i)$. In this case, the momentum equilateral configuration maximizes the shape. This template, as well as the orthogonal one, typically arises in higher derivative interacting theories, as we will see later on.

\subsubsection{Orthogonal shape}

The \textit{orthogonal} shape\index{non-Gaussianity!orthogonal} is parametrized by \cite{planckXVII:2015,baumann:2015book,Senatore:2009ng}
\begin{equation}
B_\Phi^{\text{ortho}}(k_1,k_2,k_3) = f_{\rm NL}^{\text{ortho}}\left[
	18(P_1^3P_2^2P_3)^{1/3}
	- 9P_1P_2 - 8(P_1P_2P_3)^{2/3}
	+ \text{perms.}
	\right] \, .
\label{NG-Bortho}
\end{equation}
The introduction of this new template with respect to the equilateral one was inspired \cite{Senatore:2009ng} by the observation that the combination \eqref{NG-Bequil} does not cover alone all the possible shapes in the most general class of single-field models\index{inflation!single-field}. In other words, the template \eqref{NG-Bequil} arises for instance in \acs{DBI}\index{inflation!DBI} \cite{Silverstein:2003hf,Alishahiha:2004eh} and ghost inflation\index{inflation!ghost} \cite{ArkaniHamed:2003uz,Senatore:2004rj} but it is not enough to parametrize all the cases that are encoded in the effective action of inflation (see Chap. \ref{chap-EFTinflation} for further details), namely the parameter space of non-Gaussian shapes, generated by the different Lagrangian operators, is larger than that one characterized simply by $f_{\rm NL}^{\text{equil}}$. \\

It is worth noticing that, beyond the huge amount of inflationary scenarios, some model independent statements and consistency relations\index{consistency relations} can be proved \cite{Creminelli:2004consrel,Cheung:2008cr}. In particular, under the very general assumption of single-field evolution, without any slow-roll approximation, one finds that the bispectrum is proportional to the scalar tilt in the squeezed limit \cite{Creminelli:2004consrel}. In symbols,
\begin{equation}
\lim_{k_1\rightarrow 0} \langle\Phi(\vec{k}_1)\Phi(\vec{k}_2)\Phi(\vec{k}_3)\rangle
\propto (1-n_s)P_\Phi(k_1)P_\Phi(k_2) \, .
\label{NG-consrel}
\end{equation}
This has been verified in \cite{Cheung:2008cr} using the effective theory of inflation \cite{Creminelli:2006xe,Cheung:2007st}, that encodes in full generality all single-field theories\index{inflation!single-field}. The main interests in the consistency condition \eqref{NG-consrel} are based on the fact that any experimental violation could support multi-field models\index{inflation!multi-field}, whose signals peak in the squeezed configuration. The most recent bounds on  $f_{\rm NL}$, strongly constraining the possibility of non-Gaussian deviations in the scalar spectrum, are displayed in Tab. \ref{planck2015fNL} \cite{planckXVII:2015}\index{experiments!Planck}. Concerning the \acs{CMB} data analysis, the Planck sensitivity seems to be almost saturated, due to the fact that not all scales are accessible. Indeed, as it can be inferred from Fig. \ref{fig-planck}, Silk effects consisting in damped oscillations fix a sort of upper limit on the multiple numbers at $\ell_{\text{max}}\sim 2000$. However, some improvements can be reached in \ac{LSS} surveys\index{large-scale structure surveys}, increasing $k_{\text{max}}$ (see \cite{baumann:2015book} and references therein).

\begin{table}[t]
\begin{center}
\begin{tabular}{
 l
 S[table-format=-2.1,
   table-figures-uncertainty=1]
 S[table-format=-2.1,
   table-figures-uncertainty=1]
}
\toprule
 & {\acs{SMICA}($T$)} & {\acs{SMICA}($T$+$E$)} \\
\midrule
$f_{\rm NL}^{\text{local}}$	& 2.5 \pm 5.7 & 0.8 \pm 5.0  \\
$f_{\rm NL}^{\text{equil}}$	& -16  \pm 70 & -4 \pm 43 \\
$f_{\rm NL}^{\text{ortho}}$	& -34  \pm 33 & -26 \pm 21 \\
\bottomrule
\end{tabular}
\caption{This table is taken from Planck Collaboration \cite{planckXVII:2015} and summarizes the resulting values of $f_{\rm NL}$ for different shapes, determined by the \acs{KSW} estimator from the \acs{SMICA} foreground-cleaned map, for temperature alone ($T$) and for the combined temperature - polarization analysis ($T$+$E$). The integrated Sachs-Wolfe lensing is subtracted. Error bars are $68$\% CL. For notations and details about data analysis we refer to \cite{planckXVII:2015}, where the comparison among different estimators and validation tests can be found.}
\label{planck2015fNL}
\end{center}
\end{table} 

In practise, given the importance of the three-point function in clarifying the fundamental theory of inflation, a thorough understanding of contamination mechanisms and secondary effects and the construction of multiple statistical bispectrum estimators\footnote{See \cite{planckXVII:2015}, \cite{planckXXIV:2013} and references therein for details.} are crucial in order to isolate the real primordial signal. Moreover, such independent techniques in providing foreground-cleaned maps for \acs{CMB} data are used in order to improve the confidence of the final result, summarized in Tab. \ref{planck2015fNL}\index{experiments!Planck}.

In single-field models of slow-roll inflation\index{inflation!slow-roll} a tiny level of non-Gaussianity is predicted \cite{Maldacena:2002vr,Acquaviva:2003}, namely $f_{\rm NL}\sim\mathcal{O}(\varepsilon)$. In order to have a long accelerated expansion, intuitively this follows from the requirement of a very flat inflaton potential: in this sense, every non-derivative self-interaction of the form $\phi^n/\Lambda^{n-4}$ can only distort the Gaussian distribution of a tiny factor, at least proportional to the slow-roll parameter, as shown by \cite{Maldacena:2002vr,Acquaviva:2003}. The minimal extensions of single-field inflationary models, able to allow for possible detections of non-Gaussianity, can be roughly summarized in some points:
\begin{itemize}
  \item \textit{Multi-field models.}\index{inflation!multi-field} Being not constrained as the inflaton by slow-roll conditions\index{slow-roll conditions} for the background evolution, the fluctuations of these new degrees of freedom can in principle possess a less Gaussian distribution, which can be imprinted into curvature perturbations.
  \item \textit{Theories with non-canonical kinetic term and higher-derivative operators.} In principle, derivative operators of the form $(\partial\phi)^{2n}/\Lambda^{4(n-1)}$, which do not affect the background evolution, can be considered in the Lagrangian. However, some care is required. In fact, for instance in the case $n=2$, the cubic Lagrangian for the fluctuation $\varphi\equiv\delta\phi$ is
 \begin{equation}
 \mathcal{L}^{(3)} \sim \frac{\dot{\phi_0}}{\Lambda^4}\dot{\varphi}(\partial\varphi)^2
 \label{L3-hd}
 \end{equation}
and a rough estimation of the size of non-Gaussianity yields \cite{Creminelli:2003iq}
 \begin{equation}
 f_{\rm NL} \sim \frac{\dot{\phi_0}^2}{\Lambda^4} \, .
 \label{L3-hdnG}
 \end{equation}
Eq. \eqref{L3-hdnG} leads to sizeable effects if $\vert\dot{\phi_0}\vert\gtrsim\Lambda^2$, but in this regime the derivative expansion can not be trusted any longer\footnote{See the next chapter for further details regarding effective field theories.}. The only loophole seems to be the presence of symmetries: only if the infinite series of derivative operators can be re-summed, the condition $f_{\rm NL} \gtrsim 1$ is consistent. Such an example is provided by the \ac{DBI} model\index{inflation!DBI} \cite{Silverstein:2003hf,Alishahiha:2004eh}, which predicts equilateral non-Gaussianity \cite{Creminelli:2006lNG} with the amplitude $f_{\rm NL}\sim 1/c_s^2$ \cite{Alishahiha:2004eh,Chen:2006nt}. We will come back to this point in Chap. \ref{chap-EFTinflation}. In Part II we will propose a different slow-roll scenario with enhanced non-Gaussianity ($f_{\rm NL}\sim 1/c_s^4$) in a well defined low energy regime for an effective field theory.
Another example is provided by ghost inflation\index{inflation!ghost} \cite{ArkaniHamed:2003uz,Senatore:2004rj}, based on a derivatively coupled ghost scalar field, which condenses in the background with a non-zero velocity, predicting a sizeable magnitude for non-Gaussianity with respect to the conventional inflation.
  \item \textit{Non-Bunch-Davies vacua.} Standard cosmological models on a curved space-time are based on the assumption that at early times and short distances the inflaton quantum fluctuations behave as in a flat space. This hypothesis can be relaxed: possible deviations from the standard Bunch-Davies vacuum\index{Bunch-Davies vacuum} in the cosmic evolution would be imprinted into the \acs{CMB} map with a non-trivial bispectrum.
\end{itemize}

Obviously, more exotic scenarios than these minimal extensions to the commonly accepted inflationary regime exist and can provide explanations to possible detections of non-Gaussianity.

%% file: chapters/1-03-chapter.tex
\chapter{The Effective Field Theory of Inflation}
\label{chap-EFTinflation}

\begin{flushright}{\slshape    
	Namque non potest aedis ulla sine symmetria atque proportione rationem habere compositionis [...].} \\ \medskip
    --- Marcus Vitruvius Pollio, \textit{De Architectura}, [III, 1].
%    --- \defcitealias{bentley:1999}{BBBBB}\citetalias{bentley:1999} \citep{bentley:1999}
\end{flushright}

% For an example of a full page figure, see Fig.~\ref{fig:myFullPageFigure}.

Being the main theme of the present work, effective theories are briefly introduced in the following section. Then, their application to inflation will be discussed. Hence, this chapter will conclude the presentation of the key ingredients that will be employed in Part II and III.

\section{An overview}

In most of systems in physics, a simple recipe seems to hold, that is the possibility of organizing the physical features at different scales. This means that, up to some precision and at some energy or distance, only a limited number of ingredients is required in order to have a sufficiently accurate description of the related phenomena. In practice, because in most cases the knowledge of the details of the physics at short distances does not affect qualitatively the description at large length scales, the only understanding of the relevant physical degrees of freedom and symmetries that govern the low-energy dynamics is enough. This allows to parametrize the system independently of the details of the microscopic theory and represents a powerful approach whenever the UV completion\index{UV completion} is not at hand.

This crucial property, which usually arises as an empirical evidence without requiring any justification, is at the basis of the \acf{EFT}\index{Effective Field Theory (\acs{EFT})} method. Formally, the heavy degrees of freedom ($\Phi$) are integrated out as
\begin{equation}
\e^{\text{i}S_{\text{EFT}}(\phi)} \equiv \int\mathcal{D}\Phi \e^{\text{i}S(\phi,\Phi)}
\label{intoutEFT}
\end{equation}
and one is left with an effective Lagrangian for the light fields ($\phi$), which contains a finite number of renormalizable terms of dimension four or less and an infinite tower of non-renormalizable operators of larger dimension, allowed by the symmetries of the system and suppressed by some scale $\Lambda$:
\begin{equation}
\mathcal{L}_{\rm EFT}(\phi) = \sum_i c_i \frac{\mathcal{O}_i(\phi)}{\Lambda^{\Delta_i-4}} \, ,
\label{EFT-lag}
\end{equation}
where $\Delta_i$ are the dimensions of the operators $\mathcal{O}_i$, which depend only on the light fields $\phi$, and the couplings $c_i$ are the dimensionless \textit{Wilson coefficients}\index{Wilson coefficients} that contain the information about the heavy degrees of freedom.
For evident reasons, depending on the dimensions $\Delta_i$, we distinguish three types of operators: \textit{relevant}\index{relevant operator} ($\Delta_i<4$), \textit{marginal}\index{marginal operator} ($\Delta_i=4$) and \textit{irrelevant}\index{irrelevant operator} ($\Delta_i>4$).
The effective Lagrangian \eqref{EFT-lag} provides a faithful description up to a certain region of energies where new degrees of freedom are supposed to start contributing significantly and the theory breaks down. The exact domain of validity is known if one has access to the full theory or to the characteristic energy scales of the heavy fields. Otherwise, a fair estimation of the upper limit for the effective description can be reasonably given in terms of the \textit{strong coupling scale}\index{strong coupling scale}, which identifies the breaking of the loop expansion or the violation of the perturbative unitarity\index{unitarity bound}.

Furthermore, even though infinite non-renormalizable operators are introduced in the Lagrangian \eqref{EFT-lag}, the predictive power and the consistency of the theory below the cutoff are not ruined. In practise, in order to reproduce the experimental results up to some finite accuracy, only a finite sub-set of the sum \eqref{EFT-lag} has to be considered. This makes the effective Lagrangian as useful as the standard renormalizable theories.

The main strategy to build up explicitly an effective theory of the form \eqref{EFT-lag} essentially consists in defining the relevant low-energy degrees of freedom, identifying the symmetries of the system and writing down all the admitted combinations of the field operators, organized in an appropriate derivative expansion\footnote{Some of the many existing introductory references to effective field theories are \cite{Kaplan:1995,Manohar:1996,Pich:1998,Rothstein:2003, Luty:2004tasi,Kaplan:2005,Skiba:2010}.}. We stress that the role of symmetries, exact and approximate, is crucial both to identify the significant effective Wilson coefficients\index{Wilson coefficients} and for their technical\index{technical naturalness} naturalness\footnote{For example, chiral symmetry\index{symmetry!chiral}, although inexact, is the central reason behind the technical naturalness\index{technical naturalness} of fermion masses in the standard model of particle physics.}.

Two classic examples of effective theories are provided by the Fermi theory\index{Fermi theory} of weak interactions and the QCD chiral\index{chiral perturbation theory} Lagrangian\footnote{For further details we refer to the reviews \cite{Leutwyler:1994,Gerhard:1995,Pich:1995cpt,Scherer:2002}.}. In the former case, the theory contains four-fermion vertices, which can be obtained integrating out the massive gauge boson fields in the full Standard Model Lagrangian. In the latter case, one tries to describe the physics of strong interactions at low energies, identifying the relevant ingredients as the Goldstone bosons associated with the symmetry breaking pattern $SU(3)_\text{L}\otimes SU(3)_\text{R}\rightarrow SU(3)_\text{V}$. The important point is that, even if the fields of the resulting pseudo-scalar octet do not represent the microscopic degrees of freedom of the full theory, which are known to be quarks and gluons, one is able to capture the low energy physics of the strong interactions just in terms of symmetry considerations.

Some crucial information about the microscopic structure, underlying the low-energy description \eqref{EFT-lag}, is encoded in the couplings $c_i$. For instance, the locality and the Lorentz invariance of the effective Lagrangian are not enough to guarantee the existence of a UV completion\index{UV completion} in the form of a Lorentz-invariant, local quantum field theory or a perturbative string theory \cite{Adams:2006sv}. At low energies, this obstruction translates into particular constraints on the couplings of the higher dimensional operators of the effective action. Indeed, in \cite{Adams:2006sv} it is argued that some apparently consistent low-energy effective field theories, defined by local and Lorentz invariant Lagrangians, are secretly non-local and can not be embedded in any UV theory whose $S$-matrix satisfies the usual analyticity conditions, losing therefore any Lorentz-invariant notion of causality at the microscopic level. One IR manifestation of such a failure is the superluminal propagation\index{superluminal propagation} of the fluctuations around non-trivial backgrounds for some choices of the effective coefficients.
The simplest example is provided by the following Lagrangian for a massless scalar field $\pi$ with a shift symmetry:
\begin{equation}
\mathcal{L} = -(\partial\pi)^2 + \frac{c_4}{\Lambda^4}(\partial\pi)^4 + \ldots
\label{IRUVob-1}
\end{equation}
At first sight, from the point of view of the mere construction of an effective Lagrangian in accordance with the standard rules of a quantum field theory, the coefficient $c_4$ is completely arbitrary. Nevertheless, the theory \eqref{IRUVob-1} can be embedded in a UV completion respecting the standard axioms of the $S$-matrix only if $c_4>0$ \cite{Adams:2006sv}. As an example, this is exactly the constraint that one finds integrating out the heavy Higgs field $h$ in the Lagrangian
\begin{equation}
\mathcal{L} = -\vert\partial\Phi\vert^2 - \lambda\left(\vert\Phi\vert^2-v^2\right)^2 \, ,
\label{IRUVob-2}
\end{equation}
where
\begin{equation}
\Phi = (v+h)\e^{\text{i}\pi/v} \, .
\label{IRUVob-3}
\end{equation}
It is fair to say that the low-energy vacuum of the effective theory \eqref{IRUVob-1} is well defined even if $c_4\leq 0$, provided that the kinetic terms have the correct sign. Problems arise once fluctuations around non-trivial backgrounds are considered: indeed, expanding the Goldstone model \eqref{IRUVob-1} around the solution\footnote{We will assume $C_\mu$ to be a constant vector.} $C_\mu\equiv\partial_\mu\pi_0$ and introducing the fluctuation $\varphi\equiv \pi-\pi_0$, the linearized equations of motion are \cite{Adams:2006sv}
\begin{equation}
\left(-\eta^{\mu\nu}+\frac{4c_4}{\Lambda^4}C^\mu C^\nu+\ldots\right) \partial_\mu\partial_\nu
\varphi =0 \, ,
\label{IRUVob-4}
\end{equation}
which in momentum space read
\begin{equation}
-k^\mu k_\mu+\frac{4c_4}{\Lambda^4}(C\cdot k)^2 =0 \, .
\label{IRUVob-5}
\end{equation}
The absence of superluminal excitations requires $c_4>0$. There are analogous results for more familiar effective theories in particle physics. For instance, in the case of the chiral Lagrangian the coefficients of some of the four-derivative operators are forced to be positive as well. Indeed, we know that the microscopic theory is QCD, which is a local quantum field theory.

\section{An effective theory for inflation}
\label{sec-efti}

As we emphasized before, basing the structure of an effective Lagrangian on spontaneously broken symmetries makes it largely independent of the specific physical realization. This justifies why effective field theories are ubiquitous in physics. 
%Therefore, giving the possibility to control the low-energy dynamics of systems and parametrize in full generality all microscopic completions that share the same macroscopic behaviour, effective field theories are ubiquitous in physics.
In particular, in the cosmological context we are interested in, this approach can offer new insights for instance in inflation \cite{Creminelli:2006xe,Cheung:2007st,Weinberg:2008}, dark energy\index{dark energy} \cite{Creminelli:2008wc,Gubitosi:2012hu,Bloomfield:2012ff} and the large-scale structure evolution \cite{Carrasco:2012cv}\index{Effective Field Theory (\acs{EFT})!for large-scale structures}. In the rest of the chapter we concentrate on inflation, following closely the procedure introduced by \cite{Creminelli:2006xe,Cheung:2007st}.

In order to build an effective theory of inflation, we have to capture its most essential features, discussed in Chap. \ref{chap-infepoch}.
By definition, inflation is a quasi-de Sitter Universe satisfying the condition $\vert\dot{H}\vert\ll H^2$, that is an approximate time-translation invariance is assumed. Indeed, this period of accelerated expansion has to end and has to be connected to a standard decelerated evolution. Therefore a physical ``clock'', measuring and scanning the status of the Universe, has to exist.
This is a very general statement and transcends the microscopic details of the physics governing the evolution. As in the case of the pion chiral Lagrangian, where the relevant low-energy dynamics follows solely from symmetry breaking patterns and ignores the fundamental constituents of the macroscopic degrees of freedom that enter the effective theory, here the nature of such a clock is completely irrelevant. Moreover, no matter what it is, performing a particular coordinate transformation, one can always choose a suitable setting for the clock.
To be more concrete, let us assume that inflation is caused by a scalar field $\phi(t,\vec{x})=\phi_0(t)+\delta\phi(t,\vec{x})$, where $\phi_0(t)$ drives the homogeneous background evolution and defines a privileged time-slicing, while $\delta\phi(t,\vec{x})$ identifies the field fluctuation.
As a general fact, one can choose the \textit{unitary gauge}\index{unitary gauge} defined by the condition $\delta\phi\equiv0$, fixing the particular slicing in which the perturbations disappear from the scalar field and are eaten by the metric.
Now, in order to have a description of the scalar perturbations around the inflationary solution $\phi_0(t)$ in terms of an \acs{EFT}, one has to write the most general Lagrangian containing all the operators compatible with the symmetries, organized in a polynomial expansion in the number of derivatives: this guarantees that only few terms will be relevant at low energies. 
In the unitary gauge, this is attained considering all the operators invariant under the unbroken time-dependent spatial diffeomorphisms $x^i\rightarrow x^i+\xi^i(t,\vec{x})$, while any non-invariant term under the time diffeomorphisms\index{time-diffeomorphisms!breaking of} $t\rightarrow t+\xi^0(t,\vec{x})$, which are conversely broken, is allowed\footnote{It is fair to say that there exist also other interesting symmetry breaking patterns realizing inflation, see \textit{e.g.} \cite{Endlich:2012si}.}.
For example, it is easy to realize \cite{Cheung:2007st} that $g^{00}$ is a scalar under spatial diffeomorphisms, therefore arbitrary polynomials of $g^{00}$ can appear, determining the only part of the effective action without derivatives. Moreover, opting for the more convenient \acs{ADM} formalism\footnote{We refer to Appendix \ref{Appendix-ADM} for notations and more details.}\index{Arnowitt-Deser-Misner (\acs{ADM}) formalism},
\begin{equation}
{\rm d}s^2=-N^2 {\rm d}t^2+\gamma_{ij}(N^i {\rm d}t+{\rm d}x^i)(N^j {\rm d}t+{\rm d}x^j) \; ,
\end{equation}
one can show that the most generic Lagrangian in unitary gauge is a function of the form $F(R_{\mu\nu\rho\sigma},N,K_{\mu\nu},\nabla_\mu, t)$, where $N=(-g^{00})^{-1/2}$ and $K_{\mu\nu}$ is the extrinsic curvature\index{extrinsic curvature}, given in Eq. \eqref{ex-curvij}. Explicitly, it can be written at the leading order in derivatives as\index{inflation!EFT of}
%\begin{equation}
%\begin{split}
%S = & \int\D^4x \, \sqrt{-g} \bigg[
%	\frac{1}{2}M_{\rm Pl}^2R - c(t)g^{00} - \Lambda(t)
%\\
%&	+ \frac{1}{2!}M_2(t)^4(g^{00}+1)^2 + \frac{1}{3!}M_3(t)^4(g^{00}+1)^3 + \ldots
%\\
%&	- \frac{1}{2}\hat{M}_1(t)^3(g^{00}+1)\delta {K^\mu}_\mu
%	- \frac{1}{2}\hat{M}_2(t)^3(g^{00}+1)^2\delta {K^\mu}_\mu + \ldots
%\\
%&	- \frac{1}{2}\bar{M}_1(t)^2(g^{00}+1)(\delta {K^\mu}_\mu)^2
%	- \frac{1}{2}\bar{M}_2(t)^2(g^{00}+1)\delta {K^\mu}_\nu\delta {K^\nu}_\mu + \ldots
%\bigg]
%\label{EFTI-action}
%\end{split}
%\end{equation}
\begin{align}
S = & \int\D^4x \, N\sqrt{\gamma} \Bigg[
	\frac{M_{\rm Pl}^2}{2}\left(\upleft{3}{R} + K_{\mu\nu}K^{\mu\nu} - K^2 \right)
	+ \frac{c(t)}{N^2} - \Lambda(t)
\notag \\
&	+ \frac{1}{2}M_2(t)^4\delta N^2 + M_3(t)^4\delta N^3 + \ldots
\notag \\
&	- \hat{M}_1(t)^3\delta N\delta K
	+ \hat{M}_2(t)^3\delta N^2\delta K + \ldots
\notag \\
&	- \frac{1}{2}\bar{M}_1(t)^2\delta K^2
	- \frac{1}{2}\bar{M}_2(t)^2\delta {K^\mu}_\nu\delta {K^\nu}_\mu 
	+ \bar{M}_3(t)^2 \upleft{3}{R} \delta N + \ldots
\Bigg] \, ,
\label{EFTI-action-2}
\end{align}
where $\delta N\equiv N-1$ and $\delta K_{\mu\nu}  \equiv K_{\mu\nu} - K^{(0)}_{\mu\nu}$, denoting $K^{(0)}_{\mu\nu}$ the background value of the extrinsic curvature\index{extrinsic curvature}.
The first line in \eqref{EFTI-action-2} is fixed by the \acs{FRW} background evolution,
\begin{align}
c(t) & = -M_{\rm Pl}^2\dot{H} \, ,
\label{eft-be1}\\
\Lambda(t) & = M_{\rm Pl}^2(3H^2+\dot{H})  \, ,
\label{eft-be2}
\end{align}
while the other operators, with some arbitrary time-dependent coefficients to be constrained experimentally, parametrize all the possible different theories of perturbations with the same unperturbed solution \cite{Cheung:2007st}.
In the case of the particular single field theory \eqref{dS-action0}, at the background level one gets
\begin{align}
\dot{\phi}_0^2 & = -2M_{\text{Pl}}^2 \dot{H} \, ,
\label{eft-exslowroll1}\\
V(\phi_0) & = M_{\text{Pl}}^2 (3H^2 + \dot{H})  \, ,
\label{eft-exslowroll2}
\end{align}
which agree with Eqs. \eqref{Fe1}-\eqref{Fe2} and \eqref{T00}-\eqref{Tii}.
In other words, the standard slow-roll scenario\index{inflation!slow-roll} corresponds to the case in which only the first line survives while all the effective coefficients are set to zero.
On the other hand, the presence of higher order operators in Eq. \eqref{EFTI-action-2} yields possible deviations from it. Hence, these operators can provide in principle an interesting phenomenology, imputable eventually to more complex evolutions than the simple slow-roll inflation. For instance, they are responsible for a non-unitary speed of sound $c_s$ -- as a possible consequence of the spontaneous breaking of Lorentz symmetry\index{symmetry!Lorentz} -- and sizeable amplitudes or new shapes for non-Gaussianity in the scalar spectrum. All these possible effects are captured by the single action \eqref{EFTI-action-2}, irrespective of the knowledge of any UV completion\index{UV completion}. In other words, all single field models\index{inflation!single-field} are unified in a single framework, allowing to infer very general conclusions and encoding all deviations from the standard slow-roll in the size of the higher order operators \cite{Creminelli:2006xe,Cheung:2007st}. In terms of these, one can classify all the ``microscopic'' theories of inflation (see the discussion in Chap. \ref{chap-WBG-constraints} and Tab. \ref{tabsummodels}).

It is worth stressing that, as in every \acs{EFT}, one should take care also of loop corrections. This means that the choice of setting to zero some of the tree-level coefficients in \eqref{EFTI-action-2} might not be exact, because they could be generated quantum mechanically, unless symmetry arguments intervene.
We will further comment on these topics later on in the next paragraphs.

For future practical convenience, it is useful to introduce dimensionless combinations of the effective parameters entering \eqref{EFTI-action-2}. We define\footnote{Normalizing as in Eqs. \eqref{coeffs}-\eqref{coeffs-bis} will be useful because we will explore the case in which the parameters $\alpha$, $\beta$, $\gamma$ and $\delta$ take the values roughly in the interval $(0,1)$, as shown in \eqref{def-ps}. The apparently complicated definitions \eqref{coeffs}-\eqref{coeffs-bis} have been chosen purely for matter of convenience. This choice turns out to be more transparent with the different parametrization of the \acs{EFTI} of App. \ref{Appendix-Eij}, where the relations \eqref{massr-1}-\eqref{massr-4} clearly justify it.}
\begin{gather}
\alpha =\frac{\hat M_1^3}{2M_\text{Pl}^2 H}~, \quad \beta =\frac{M_2^4+6H\hat{M}_1^3}{2M_\text{Pl}^2 H^2}~,
\label{coeffs}\\
\gamma=\frac{\hat M_1^3 + \hat M_2^3}{M_\text{Pl}^2 H}~, \quad \delta = \frac{M_3^4-3H(\hat M_1^3 + \hat M_2^3)}{M_\text{Pl}^2 H^2}~,
\label{coeffs-bis}
\end{gather}
that we will use several times in the rest of the work.

In Eq. \eqref{EFTI-action-2} the full diffeomorphism invariance can be restored using the St\"{u}ckelberg trick\index{St\"{u}ckelberg trick}, namely performing a broken transformation of the form $t\rightarrow t+\xi^0(t,\vec{x})$ and promoting the parameter $\xi^0(t,\vec{x})$ to a new field $\pi(t,\vec{x})$, which has to transform as $\pi(t,\vec{x})\rightarrow \pi(t,\vec{x})-\xi^0(t,\vec{x})$ under time diffeomorphisms, in order to achieve a full gauge invariance.
The new action obtained in such a way is dynamically equivalent to the previous one, as can be easily understood counting the number of degrees of freedom: indeed, the extra degree of freedom contained in the action \eqref{EFTI-action-2} and due to the lack of time diffeomorphism invariance disappears from the metric and is simply made explicit in $\pi(t,\vec{x})$ once the full invariance under the combined transformations
\begin{equation}
 \left\{ \begin{aligned}
        x^\mu & \rightarrow x^\mu+\xi^\mu(t,\vec{x}) \\
        \pi   & \rightarrow \pi-\xi^0
       \end{aligned}
 \right.
\label{IFTI-gt}
\end{equation}
is restored. This will be explicitly done in Sec. \ref{sec-stucktrick-dc}. The field $\pi$, introduced as the space-time transformation along the non-invariant direction, is nothing but the Goldstone mode\index{Goldstone field} which non-linearly realizes the spontaneously broken time diffeomorphism invariance and parametrizes the scalar perturbations.
This makes the analysis of inflation very general, independently of what is really driving the accelerated expansion. In other words, $\pi(t,\vec{x})$ represents the perturbation corresponding to a common local transformation in time for all matter fields, that are present during the evolution. In this sense, it is legitimate to interpret it as the so-called \textit{adiabatic mode}\index{adiabatic mode} \cite{Creminelli:2006xe}. 
The convenience in the introduction of $\pi$ is based on the fact that there could exist a sufficiently high energy regime in which the mixing with gravity can be neglected\footnote{In particle physics, this is known as \textit{equivalence theorem}\index{equivalence theorem} for massive gauge bosons. Analogously, if the vector particle carries a mass, the lack of gauge symmetry is responsible for the appearance of extra longitudinal degrees of freedom. For a boson at rest, the different polarizations are equivalent, being related simply by spatial rotations. Conversely, for a highly boosted particle, the longitudinal components become more relevant in such a way that for enough high energies they determine the dynamics and the amplitudes in scattering processes \cite{Cornwall:1974,Chanowitz:1985}.}: in this case Eq. \eqref{EFTI-action-2} reduces to a simple action of $\pi$ only, which now carries all the information about the dynamics. Otherwise, if gravity is relevant at all scales a more involved computation has to be performed: this will be widely discussed in the rest of the work.

It is worth noticing that in the de Sitter limit, without any explicit dependence on time in the action \eqref{EFTI-action-2}, \textit{i.e.} if the effective coefficients are assumed to be constant, the Goldstone field\index{Goldstone field} $\pi$ appears with at least one derivative, therefore the theory becomes automatically invariant under shifts,
\begin{equation}
\pi(t,\vec{x}) \rightarrow \pi(t,\vec{x}) + \text{constant} \, .
\label{pishifts}
\end{equation}

%We stress that the scalar excitation $\pi(t,\vec{x})$ is the Goldstone mode\index{Goldstone field} associated with a spontaneous symmetry breaking, being introduced as the space-time transformation along the non-invariant direction.

A common choice for the metric fluctuations in the unitary gauge\index{unitary gauge} is called \textit{comoving gauge}\index{comoving!gauge} and is given by
\begin{equation}
g_{ij} = a(t)^2 \e^{2\zeta(t,\vec{x})}\delta_{ij} \, ,
\label{ug-zeta}
\end{equation}
where $\zeta(t,\vec{x})$ is the comoving curvature perturbation\index{curvature perturbation}. The relation that links $\zeta(t,\vec{x})$ to $\pi(t,\vec{x})$ at the quadratic level is
\begin{equation}
\zeta = -H\pi + H\pi\dot{\pi} + \frac{1}{2}\dot{H}\pi^2 + \ldots
\label{zetapi}
\end{equation}
Moreover, we remind that on super-horizon scales, in the matter dominated era, the relation between the comoving gauge and the Newtonian gauge\index{Newtonian gauge} \eqref{Ngm} is $\zeta=-\frac{5}{3}\Phi$.

From the definition \eqref{ug-zeta} we expect $\zeta=\text{constant}$ to be a legitimate solution, since $g_{ij}$ in this case is obtained from the unperturbed \acs{FRW} metric\index{Friedmann-Robertson-Walker (\ac{FRW})!metric} by a mere constant rescaling of the spatial coordinates. Therefore, any operator that would generate non-invariant terms under constant shifts in the equations of motion, is forbidden in the action. In particular, this means that the field $\zeta$ is massless.

\section{St\"{u}ckelberg trick and decoupling limit}
\label{sec-stucktrick-dc}

This section is devoted to the restoration of the full diffeomorphism invariance in the action for perturbations. To this purpose the following transformations
\begin{equation}
\left\{
	\begin{aligned}
		t & \rightarrow \tilde{t} = t+\pi(t,\vec{x}) \\
		x^i & \rightarrow \tilde{x}^i = x^i 
	\end{aligned}
\right.
\label{stuck-ts}
\end{equation}
have to be performed. Moreover, recalling that under a general coordinate transformation $x^\mu\rightarrow\tilde{x}^\mu(x)$ the metric varies as
\begin{equation}
\tilde{g}_{\mu\nu}(\tilde{x}) = \frac{\partial x^\alpha}{\partial\tilde{x}^\mu}
\frac{\partial x^\beta}{\partial\tilde{x}^\nu} g_{\alpha\beta}(x(\tilde{x})) \, ,
\label{gcmetric}
\end{equation}
the particular St\"{u}ckelberg transformations\index{St\"{u}ckelberg trick} \eqref{stuck-ts} up to second order in $\pi$ lead to
\begin{multline}
g^{\mu\nu} \rightarrow \tilde{g}^{\mu\nu} =
\\
\begin{pmatrix}
	g^{00}(1+\dot{\pi})^2
		+ 2g^{i0}(1 + \dot{\pi})\partial_i\pi
		+ g^{ij}\partial_i\pi\partial_j\pi
		& 	(1 + \dot{\pi})g^{0i} + g^{ki}\partial_k\pi
	\\
	(1 + \dot{\pi})g^{0i} + g^{ki}\partial_k\pi
		& 	g^{ij}
	\end{pmatrix}
\label{StuckForm-g^munu}
\end{multline}
\begin{multline}
g_{\mu\nu} \rightarrow \tilde{g}_{\mu\nu} =
\\
\begin{pmatrix}
	g_{00}(1-2\dot{\pi} + 3\dot{\pi}^2)
		& 	-g_{00}(1 - 2\dot{\pi})\partial_i\pi
		+ g_{0i}(1-\dot{\pi})
	\\
	-g_{00}(1-2\dot{\pi})\partial_i\pi
		+ g_{0i}(1-\dot{\pi})
		& 	g_{00}\partial_i\pi\partial_j\pi	- g_{0j}\partial_i\pi - g_{0i}\partial_j\pi	+ g_{ij}
	\end{pmatrix}
\label{StuckForm-g_munu}
\end{multline}
where the inverse transformation $t=\tilde{t}-\pi+\dot{\pi}\pi+\ldots$ has been used.

Assuming a negligible mixing with gravity\index{decoupling limit} and a background metric of the form $g_{\mu\nu}=\text{diag}(-1,a^2,a^2,a^2)$, the previous equations simplify and yield the following relations for the \acs{ADM} variables in the action \eqref{EFTI-action-2}:
\begin{align}
\delta N & \rightarrow
	\delta N - \dot{\pi} + \dot{\pi}^2 + \dfrac{\partial_i\pi\partial^i\pi}{2a^2} + \mathcal{O}(\pi^3) \, ,
\label{StuckDeltaN}
\\
\delta K_{ij} & \rightarrow
	\delta K_{ij} -(1-\dot{\pi})\partial_i\partial_j\pi
	+ \partial_i\dot{\pi}\partial_j\pi + \partial_i\pi\partial_j\dot{\pi}
\notag \\
	& \qquad + H \left(	- 2 \partial_i\pi\partial_j\pi
	+ \dfrac{1}{2}\delta_{ij}\partial_k\pi\partial^k\pi\right)
	+ \mathcal{O}(\pi^3) \, ,
\label{StuckDeltaK}
\\
\delta K & \rightarrow
	\delta K + a^{-2} \left[ -(1-\dot{\pi})\partial^2\pi
		+ 2\partial_k\dot{\pi}\partial^k\pi 
		+ \dfrac{H}{2}\partial_k\pi\partial^k\pi
		\right] 	
		+ \mathcal{O}(\pi^3) \, .
\label{StuckDeltaTraceK}
\end{align}
The results \eqref{StuckDeltaN}-\eqref{StuckDeltaTraceK} are obtained by simply neglecting all the terms which contain the metric perturbation $\delta g_{\mu\nu}$ after a St\"{u}ckelberg transformation. It is worth stressing that this applies only if a suitable energy regime, in which the mixed terms are subleading and the evolution is driven only by the St\"{u}ckelberg field $\pi$, exists.

\begin{figure}[t]
  \caption{The figure, taken from \cite{planckXVII:2015}, shows the $68$\%, $95$\% and $99.7$\% confidence regions. \textit{Left}: the bounds on the parameter space $(f_{\text{NL}}^{\text{equil}},f_{\text{NL}}^{\text{ortho}})$ are obtained from the $T+E$ constraints for a $\chi^2$ statistic with two degrees of freedom and thresholds $\chi^2\leq 2.28$, $5.59$ and $11.62$ respectively. \textit{Right}: the confidence regions on the parameter space $(c_s,\tilde{c}_3)$ are shown; marginalizing, \cite{planckXVII:2015} finds $c_s\geq 0.024$ at $95$\% CL ($T+E$).} 
  \centering
    \includegraphics[width=0.49\textwidth]{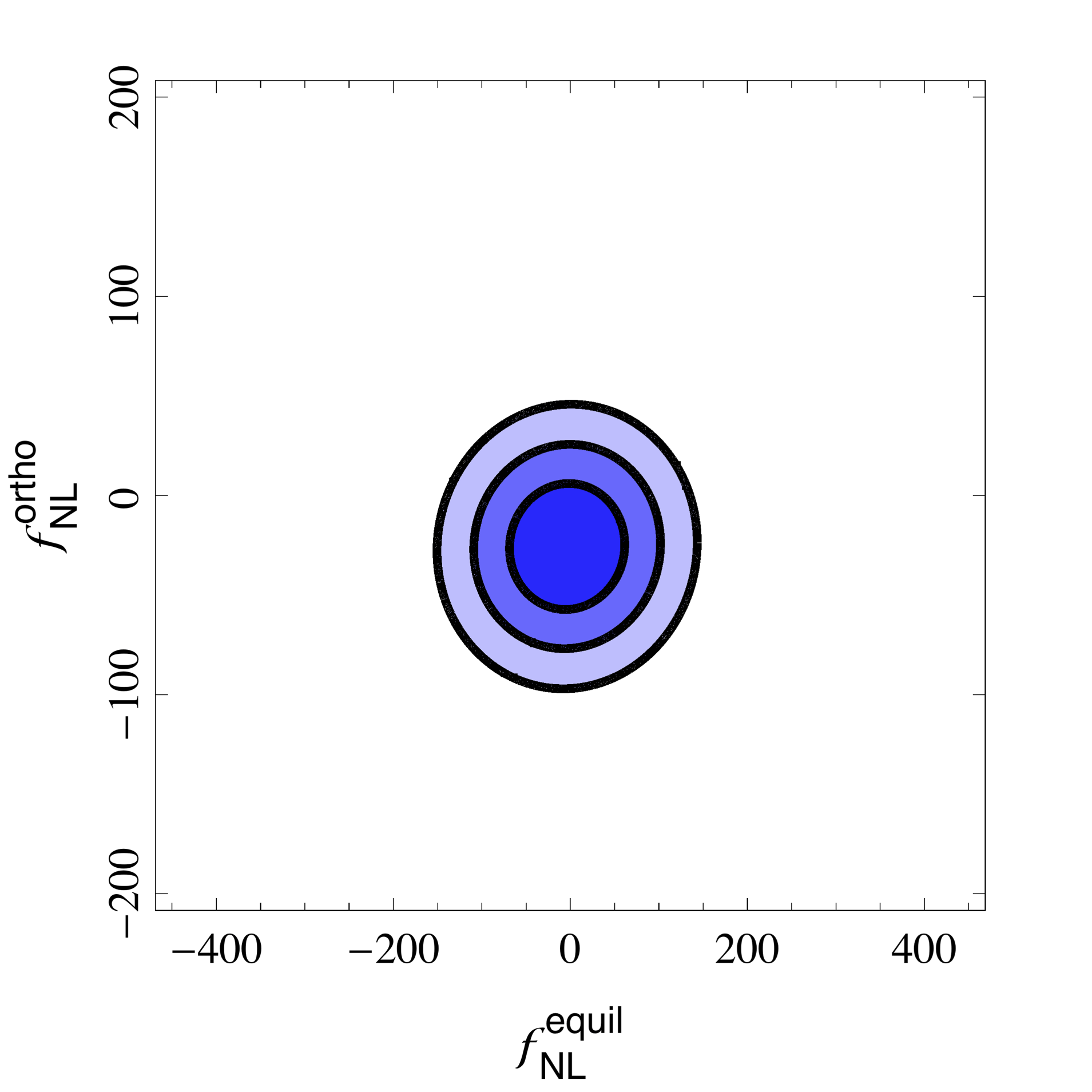}
    \includegraphics[width=0.49\textwidth]{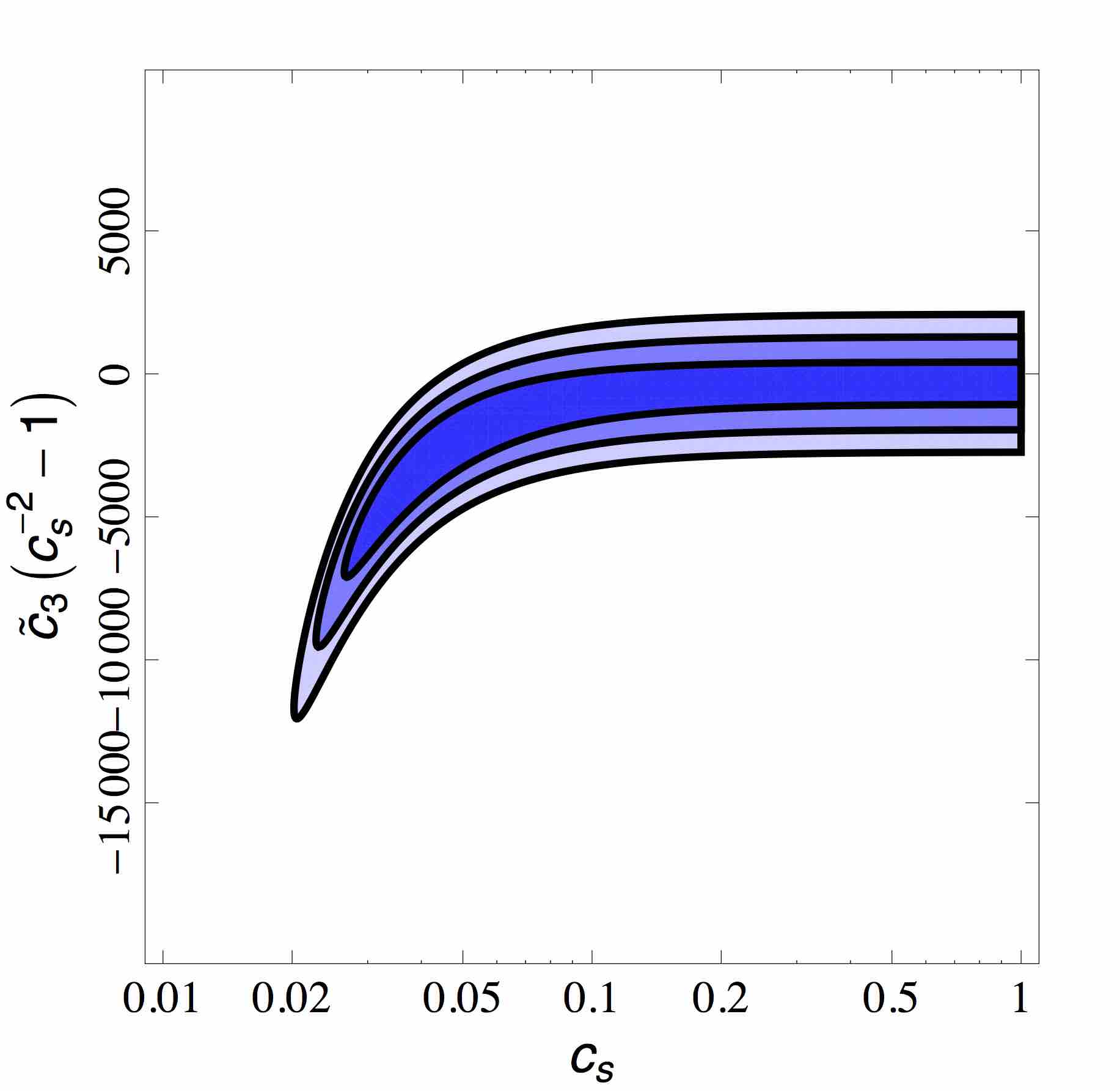}
\label{fig-planck2015-EFTbounds}
\end{figure}

At this point, an illustrative example might be appropriate \cite{Cheung:2007st}. For the moment, we focus only on the lowest order in derivatives, which is expected to contain all the most relevant operators in the low energy limit. Hence, assuming that the dominant mixing terms in the effective action \eqref{EFTI-action-2} are of the form $\sim M_2^4\dot{\pi}\delta N$, the decoupling limit\index{decoupling limit} is attained at energies $E\gg E_{\text{mix}}\sim M_2^2/M_{\text{Pl}}$, once the fields are canonically normalized. The resulting Goldstone action is
\begin{multline}
S_\pi = \int\D^4 x \sqrt{-g}\bigg[
	-\frac{M_{\text{Pl}}^2\dot{H}}{c_s^2}\left(\dot{\pi}^2-c_s^2\frac{(\partial_i\pi)^2}{a^2}\right)
\\
	+M_{\text{Pl}}^2\dot{H}\left(1-\frac{1}{c_s^2}\right)\left(\dot{\pi}^3-\dot{\pi}\frac{(\partial_i\pi)^2}{a^2}\right)
	- \frac{4}{3}M_3^4\dot{\pi}^3
	+ \ldots
\bigg] \, , 
\label{ex-Gaction-dc}
\end{multline}
where
\begin{equation}
c_s^2 = \frac{M_{\text{Pl}}^2\dot{H}}{M_{\text{Pl}}^2\dot{H}-2M_2^4}
\label{ex-Gaction-dc-cs}
\end{equation}
is the speed of propagation of the scalar perturbations. In a Lorentz invariant\index{symmetry!Lorentz} theory, the speed of massless modes is fixed by the covariant structure of the kinetic term $\partial_\mu\pi\partial^\mu\pi$ to be $c_s=1$. In the present case, as already anticipated, the circumstance $c_s\neq 1$ should not be surprising, because the background spontaneously breaks the Lorentz invariance\index{symmetry!Lorentz}. Phenomenologically, the configuration $M_2^4\gg \varepsilon M_{\text{Pl}}^2H^2$ is very attractive. Indeed, this corresponds to small values of the speed of sound \eqref{ex-Gaction-dc-cs}, $c_s^2\ll 1$, and leads to potentially observable non-Gaussianity. A rough estimation of the dominant contribution yields
\begin{equation}
\frac{\mathcal{L}_{\dot{\pi}(\partial_i\pi)^2}}{\mathcal{L}_2} \sim \frac{H\pi}{c_s^2} \sim 
\frac{\zeta}{c_s^2} \, ,
\label{ex-Gaction-dc-fNL}
\end{equation}
where we used the fact that $\omega\sim H$ and $k\sim H/c_s$ around freezing. Taking $\zeta\sim 10^{-5}$, that one infers from Tab. \ref{planck2015parinf}, Eq. \eqref{ex-Gaction-dc-fNL} provides the size of the non-linear corrections. The estimator of non-Gaussianity scales as
\begin{equation}
f_{\text{NL}} \sim \frac{1}{\zeta}\frac{\mathcal{L}_{\dot{\zeta}(\partial_i\zeta)^2}}{\mathcal{L}_2} \sim \frac{1}{c_s^2} \, ,
\label{ex-Gaction-dc-fNL2}
\end{equation}
leading to possible measurable effects.
The Planck Collaboration\index{experiments!Planck} \cite{planckXVII:2015} is able to constrain the values of the free parameters in the effective action \eqref{ex-Gaction-dc} from the bounds on $f_{\text{NL}}$. Introducing the parameter $\tilde{c}_3$ as $\tilde{c}_3(c_s^{-2}-1)=2M_3^4c_s^2/(M_{\text{Pl}}^2\dot{H})$ \cite{Senatore:2009ng}, the experimental constraints on the effective coefficients are reported in Fig. \ref{fig-planck2015-EFTbounds}.

We will come back to this example later in this chapter.

\section{Complete solution with mixing with gravity}
\label{sec-cswmwg}

The effective action \eqref{ex-Gaction-dc} has been derived under two assumptions: first we focused only on the operator proportional to $M_2^4$ and then we neglected mixing terms between the Goldstone field\index{Goldstone field} and the gravity sector.
The decoupling limit\index{decoupling limit} procedure provides a very straightforward tool in the study of the evolutionary dynamics because in this regime the action strikingly simplifies and contains only the field $\pi$, yielding a plain description of scalar perturbations. However, there might exist situations in which it does not apply. In these cases, also the metric perturbations have to be taken into account and they will appear in the action as well.
In the present section, we analyse this possibility, extending the results of \cite{Maldacena:2002vr,Chen:2006nt} by including higher derivative operators. Other useful references might be \cite{Seery:2011pc,Kobayashi:2011pc}. Most of the discussion is technical and the uninterested reader can skip it and move to the next section. 

The complete form of the transformation laws \eqref{StuckDeltaN}-\eqref{StuckDeltaTraceK} of the previous section is
\begin{align}
\gamma_{ij} 
	& \rightarrow  \gamma_{ij} - N_j\partial_i\pi - N_i\partial_j\pi + \ldots
\label{StcuckSumm1-lin2} \\
N_i 
	& \rightarrow  N_i(1-\dot{\pi}) +(N^2-N^k N_k)\partial_i\pi + \ldots
\label{StcuckSumm2-lin2} \\
N
	& \rightarrow  N\left(1 - \dot{\pi} + N^i\partial_i\pi \right)
		+ \ldots
\label{StcuckSumm3-lin2} \\
K_{ij}
& \rightarrow K_{ij}
	- N \partial_i\partial_j\pi
	- 2\left(\partial_i N \partial_j\pi + \partial_j N \partial_i\pi \right)
\notag \\
	& \qquad
	+ \dfrac{1}{2a^2} \partial^k\pi \left( \partial_i \gamma_{jk}
			+ \partial_j \gamma_{ik} - \partial_k \gamma_{ji}
		 \right) + \ldots 
\label{Mixing-K_ij}
\end{align}
up to second order in perturbations and in the de Sitter limit.

Nevertheless, for what we are going to deal with in the next part, the unitary gauge with the parametrization \eqref{ug-zeta} for scalar perturbations results to be a more convenient choice \cite{Maldacena:2002vr}. The action for $\pi$ can be recovered straightforwardly using the non-linear relation \eqref{zetapi}. In the next paragraph, we scan the procedure leading to the action for the scalar perturbations, highlighting the key steps of the computation.

\subsection{Hamiltonian constraints in the effective theory of inflation}

Ignoring tensor modes, the choice \eqref{ug-zeta} fixes completely the gauge freedom, with the field $\zeta$ describing the scalar fluctuations. One should notice that the \acs{ADM} variables $N$ and $N^i$ in \eqref{EFTI-action-2} appear without time derivatives: this means that they are not dynamical but act instead as Langrange multipliers, whose equations of motions are simply Hamiltonian constraints\index{Hamiltonian constraints}. Therefore, in order to find the theory for the scalar degree of freedom, one has to solve the constraint equations, obtained by varying the action with respect to $N$ and $N^i$, and plug the solutions back into the action.

Explicitly, the equations of motion of $N$ and $N^i$ are
\begin{multline}
\frac{M_\text{Pl}^2}{2} \left[ \upleft{3}{R} - K^{ij}K_{ij}+K^2  
		+\frac{2}{N^2}\dot{H}
		- 2(3H^2+\dot{H})\right] 
\\
	+ M_2^4\delta N 
	-\hat{M}_1^3 \left(\delta K - 3H\delta N\right)
	+ (3 \bar{M}_1^2 + \bar{M}_2^2)H\delta K
		+ \bar{M}_3^2\upleft{3}{R} = 0 
\label{constraint-N}
\end{multline}
and
\begin{equation}
\hat{\nabla}_i \left[ M_\text{Pl}^2(K^i_j - K \delta^i_j)
	- \hat{M}^3_1 \delta^i_j \delta N
	- \bar{M}_1^2 \delta^i_j \delta K
	- \bar{M}_2^2 \delta K^i_j \right] = 0 \, ,
\label{constraint-Ni}
\end{equation}
respectively.
It is worth noticing that, in order to obtain the action up to the third order in $\zeta$, it is enough to solve the equations \eqref{constraint-N}-\eqref{constraint-Ni} at the linear level in perturbation theory\footnote{For the proof we refer to \cite{Chen:2006nt}.}. The solutions are then used to remove the Lagrange multipliers and to obtain an action for $\zeta$ only, containing the free part and the interactions.

\subsection{Quadratic and cubic actions}
\label{sec-cswmwg-sub2}

At this point, it is instructive to provide an example of solution of the equations \eqref{constraint-N}-\eqref{constraint-Ni} in the case $\bar{M}_i^2=0$. This choice will be conveniently justified and the result will be employed in Chap. \ref{WBG-kin}. With the ansatz $N_i=\partial_i\psi$, one finds
\begin{align}
\delta N & = \frac{2M_\text{Pl}^2}{2M_\text{Pl}^2H-\hat{M}^3_1}\dot{\zeta} \, ,
\label{chap3-ce-1}\\
\psi & = - \frac{2M_\text{Pl}^2}{2M_\text{Pl}^2H-\hat{M}^3_1}\zeta
	+ \frac{-4M_\text{Pl}^4\dot{H}+2M_\text{Pl}^2M_2^4+3\hat{M}^6_1}{(2M_\text{Pl}^2H-\hat{M}^3_1)^2} 
	\left(\frac{\partial}{a}\right)^{-2}\dot{\zeta} 
\notag \\
	& \equiv C\zeta + \chi	 \, .
\label{chap3-ce-2}
\end{align}
Plugging these solutions back into \eqref{EFTI-action-2}, all the tadpole terms cancel and the quadratic action\index{curvature perturbation!quadratic action} for the scalar perturbation reduces to the form \eqref{dS-action} with
\begin{align}
\mathcal{N} & = M_\text{Pl}^2 \frac{-4M_\text{Pl}^4\dot{H}+2M_\text{Pl}^2M_2^4+3\hat{M}_1^6}{(2M_\text{Pl}^2H-\hat{M}^3_1)^2} \, ,
\label{mg-N}\\
c_s^2 & = \frac{4M_\text{Pl}^4\dot{H}-2M_\text{Pl}^2H\hat{M}^3_1+\hat{M}_1^6-2M_\text{Pl}^2\partial_t\hat{M}^3_1}{4M_\text{Pl}^4\dot{H}-2M_\text{Pl}^2M_2^4-3\hat{M}_1^6} \, .
\label{mg-cs}
\end{align}
Obtaining the cubic-order action requires more work, but the procedure is straightforward. Using the solutions \eqref{chap3-ce-1} and \eqref{chap3-ce-2} for the lapse and shift perturbations yields the following result 
\begin{equation}
\begin{split}
S^{(3)}_\zeta &= \int\D^4x \, \bigg\{
	-a\mathcal{N}c_s^2\zeta(\partial_i\zeta)^2  
	+ a^3
	\left[C \mathcal{N}+\frac{C^3}{M_\text{Pl}^4}\lambda\right]\dot{\zeta}^3
	+ 3a^3\mathcal{N}\zeta\dot{\zeta}^2  +
\\
&	+ \frac{M_\text{Pl}^2}{2 a}\left(3\zeta+C\dot{\zeta}\right)
	\left[(\partial_i\partial_j\psi)^2-(\partial^2\psi)^2\right]
	- \frac{2M_\text{Pl}^2}{a}\partial_i\zeta\partial_i\psi\partial^2\psi +
\\
&	+ aC^2(\hat{M}_1^3-\hat{M}_2^3) \dot{\zeta}^2\partial^2\psi
\bigg\}~,
\label{mg-a3}
\end{split}
\end{equation}
where
\begin{equation}
\lambda \equiv - \frac{M_\text{Pl}^2}{2}\left[2M_\text{Pl}^2(M_2^4+M_3^4)
	+3\hat{M}_1^3(\hat{M}_1^3-\hat{M}_2^3)\right] .
\label{mg-lambda}
\end{equation}
It will be convenient to recast the action \eqref{mg-a3} into a slightly different form. In doing so, we omit a number tedious but straightforward manipulations. At the end, one finds the following expression, equivalent to \eqref{mg-a3} up to a total derivative,\index{curvature perturbation!cubic action}
\begin{equation}
\begin{split}
S^{(3)}_\zeta
	&= \int\D^4x \,  \bigg\{
	a^3
	\left[\mathcal{N}C\left(1+\frac{HC}{c_s^2}\right)
	- \lambda'
	\right]\dot{\zeta}^3 +
 \frac{a^3\mathcal{N}(HC)^2}{c_s^2}\left(\varrho-3+\frac{3c_s^2}{(HC)^2}\right) \zeta\dot{\zeta}^2 
\\
&	+ a\mathcal{N}(HC)^2\left(\varrho-2s+1-\frac{c_s^2}{(HC)^2}\right) \zeta(\partial_i\zeta)^2
	+ 2a\mathcal{N}HC\dot{\zeta}\partial_i\zeta\partial_i\chi 
\\
&	+ \frac{a^3\mathcal{N}}{2}\frac{\D}{\D t}\left(\frac{n(HC)^2}{c_s^2}\right)\zeta^2\dot{\zeta}
	+ \frac{\mathcal{N}c_s^2}{2a}\partial_i\zeta\partial_i\chi\partial^2\chi
	+ \frac{\mathcal{N}c_s^2+2\hat{M}_1^3C}{4a}\partial^2\zeta(\partial_i\chi)^2 
\\
&	+ \frac{\hat{M}_1^3C^2}{a}\partial^2\zeta\partial_i\zeta\partial_i\chi
	+ \frac{\hat{M}_1^3C^3}{2a}(\partial_i\zeta)^2\partial^2\zeta
	+ aC^3 (\hat{M}_1^3-\hat{M}_2^3)\dot{\zeta}^2\partial^2\zeta 
	- g(\zeta)\frac{\delta L}{\delta\zeta}
\bigg\}~.
\label{our-14}
\end{split}
\end{equation}
In the last expression, we have defined a number of quantities 
\begin{equation}
n=\frac{1}{H}\frac{\D}{\D t}\ln(\mathcal{N} c_s^2)\, , 
\quad
\varrho =\frac{1}{H}\frac{\D}{\D t}\ln(H C^2) \, ,
\quad
s = \frac{1}{H}\frac{\D}{\D t}\ln( c_s)~,
\end{equation}
as well as 
\begin{equation}
\lambda' \equiv C^3(M_2^4+M_3^4)
	+ C^2 (\hat{M}_1^3-\hat{M}_2^3)\left(3+3HC-\frac{\mathcal{N}}{M_\text{Pl}^2}\right) \, .
\label{our-15}
\end{equation}
Furthermore, $\delta L/\delta \zeta$ denotes the variation of the quadratic Lagrangian with respect to $\zeta$,
\begin{equation}
\frac{\delta L}{\delta \zeta} = -2 M_\text{Pl}^2 \partial_t (a\partial^2\chi) +2 a\mathcal{N} c_s^2 \partial^2 \zeta~,
\end{equation}
and the coefficient of this term in the action \eqref{our-14} is
\begin{multline}
g(\zeta) 
	= \frac{n(HC)^2}{4c_s^2}\zeta^2+\frac{HC^2}{c_s^2}\zeta\dot{\zeta}
	+ \frac{C^2}{4a^2}\left[-(\partial_i\zeta)^2+\partial^{-2}(\partial_i\partial_j(\partial_i\zeta\partial_j\zeta))\right] +
\\
	- \frac{C}{2a^2}\left[\partial_i\zeta\partial_i\chi - \partial^{-2}(\partial_i\partial_j(\partial_i\zeta\partial_j\chi))\right] \, .
\label{our-g}
\end{multline}

As a quick consistency check, we note that the action \eqref{our-14} reduces to the particular case of \cite{Chen:2006nt}, with the following substitutions: $\hat{M}_1^3=\hat{M}_2^3=0$, $C\rightarrow-H^{-1}$, $\mathcal{N}c_s^2\rightarrow\varepsilon$, $n\rightarrow\eta\equiv (\D \ln\varepsilon/\D t)/H$ and $\varrho\rightarrow\varepsilon$. In that case,
a small speed of sound implies large (equilateral) non-Gaussianity, $f_\text{NL}\sim 1/c_s^2$. 
Our result generalizes the cubic action of \cite{Chen:2006nt} to the case of non-zero $\hat{M}_1^3$ and $\hat{M}_2^3$, which opens up qualitatively novel ways of generating large non-Gaussianity, as we will explain in Chap. \ref{WBG-kin}.
For analogous computations, we suggest also Refs. \cite{Seery:2011pc,Kobayashi:2011pc}.

The scalar power spectrum can be directly read off Eq. \eqref{dS-ps}, where the normalization factor $\mathcal{N}$ is given in \eqref{mg-N}. Moreover, the cubic action \eqref{our-14} can be employed in computing non-Gaussianity. One finds that it contains eight operators non-trivially contributing to the bispectrum. Using the standard \textit{in-in} formalism\index{in-in formalism} \cite{Maldacena:2002vr,Chen:2010png,Wang:2013ng} and the definition \eqref{NG-3bisdef} for the bispectrum in terms of the gravitational potential $\Phi$, one finds
\begin{equation}
B_{\Phi}(k_1,k_2,k_3)= 2 \left(\frac{3}{20}\right)^3 \frac{H^{4}}{\mathcal{N}^3 c_s^6}\sum_{i=1}^8 c_i S_i(k_1,k_2,k_3) + \text{cyclic}~,
\label{mgbispr}
\end{equation}
where the sum of $S_i(k_1,k_2,k_3)$ encodes the relevant contributions. The explicit expressions are collected in App. \ref{Appendix-bispectrum}.

We have left out the operator $\zeta^2\dot\zeta$ in \eqref{our-14}, since its coefficient is at most of order $\sim\varepsilon^2$, and it is not expected to affect the analysis in any significant way. Moreover, the last term in Eq. \eqref{our-14}, being proportional to the lower-order $\zeta$ equations of motion, can be removed by a field redefinition. The latter also contributes to the three-point function of the conserved scalar mode through the function $g(\zeta)$ in \eqref{our-g}. The contributions of the terms in this function that include derivatives are suppressed at superhorizon distances. Moreover, the first term contributes to $f_\text{NL}$ by an amount that scales as $\eta/c_s^2$ (see \textit{e.g.} \cite{Chen:2006nt}). We neglect this piece in the analysis, since it is always expected to be sub-dominant: whenever non-trivial constraints from bispectrum arise, there are leading contributions, enhanced by at least a factor of $1/\eta$ compared to it.

\section{The role of symmetries in the EFT of inflation}
\label{subsec-stucktrick-dc}

Undoubtedly, the use of symmetries is fundamental in physics, allowing to have a plain and smart description of the analysed system. In the case of an \acs{EFT}, as already emphasized, the notion of breaking pattern is one of the main ingredients in the construction of the low-energy Lagrangian. Moreover, symmetry considerations are able for instance to justify why a small Wilson coefficient\index{Wilson coefficients} can be technically natural. Conversely, the experimental bounds on the size of the effective couplings can provide in principle some indications of a possible underlying symmetry and the features of the microscopic theory.

This section is devoted to highlight the role of the symmetries in the \acs{EFT} of inflation that are useful for our purposes.
To this end, we start with a straightforward fact, but of great relevance. Being an expansion in the ratio ``derivatives/cutoff'', the leading operators in the action \eqref{EFTI-action-2}, mainly affecting the observables, are expected to be those with the least number of derivatives. In other words, for example, it seems reasonable to disregard the Lagrangian operators proportional to $\hat{M}_i^3$ compared to those with the coefficients $M_i^4$. Namely, using the effective action \eqref{EFTI-action-2} one expects to have control only over the region $\hat{M}_i^3H\ll M_i^4$ of the parameter space. The rest would be out of the regime of validity of the \acs{EFT}. Actually, one might wonder whether there exist consistent and well defined effective theories of inflation where for instance the operators $\delta N^2$ and $\delta N\delta K$ are equally dominant. This would be possible only because of suitable symmetries of the system. One of the main original points of the present work is to assert that this is indeed possible: we will provide an example in terms of a weakly broken symmetry during inflation in Part II and roughly discuss the experimental bounds on the effective parameters, in light of the recent observational results.

Furthermore, let us make some more quantitative remarks.
To this end, we come back to the simple example \eqref{ex-Gaction-dc} and estimate first of all the unitary cutoff scales\index{strong coupling scale} $\Lambda_{\star}$ corresponding to the interacting operators $\dot{\pi}^3$ and $\dot{\pi}(\partial_i\pi)^2$ in the effective action. One finds \cite{Cheung:2007st,Senatore:2009ng}
\begin{align}
\Lambda_{\star,\dot{\pi}(\partial_i\pi)^2}^4 &
\simeq 16\pi^2M_{\text{Pl}}^2\vert\dot{H}\vert\frac{c_s^5}{1-c_s^2} \, ,
\label{cutoffscalesEFT-1}\\
\Lambda_{\star,\dot{\pi}^3}^4 &
\simeq \frac{\Lambda_{\star,\dot{\pi}(\partial_i\pi)^2}^4}{\left(c_s^2+2\tilde{c}_3/3\right)^2} \, ,
\label{cutoffscalesEFT-1bis}
\end{align}
which are of the same order if $\tilde{c}_3\sim 1$, while the theory becomes more and more strongly coupled as $c_s\rightarrow 0$. One can prove that the choice $\tilde{c}_3\sim 1$ and $c_s\ll 1$, or equivalently $M_3^4\sim M_2^4/c_s^2$, is technically\index{technical naturalness} natural\footnote{In particular, this scaling for the coefficients in the effective Lagrangian enhances the relevance of the operator $\dot{\pi}^3$ to the operator $\dot{\pi}(\partial_i\pi)^2$, equally contributing in the estimation \eqref{ex-Gaction-dc-fNL}.} \cite{Senatore:2009ng,Baumann:2011su,Baumann:2014cja}. In other words, if there are not symmetry considerations that protect the Lagrangian coefficients against large quantum corrections, it does not seem natural to have small $M_3^4$ and large $M_2^4$: loop contributions would spoil such a hierarchy.
In our model of Part II, we will show that a weakly broken symmetry during inflation conversely allows to have effective parameters of the same order, namely the choice $M_3^4\sim M_2^4$ will turn out to be technically natural\index{technical naturalness}.
%This fact will be of crucial importance and should be kept in mind.

For the moment, let us just provide an example of microscopic theory which yields the scaling \eqref{ex-Gaction-dc-fNL2} and the hierarchy $M_3^4\sim M_2^4/c_s^2$. It is \acf{DBI} inflation\index{inflation!DBI} \cite{Silverstein:2003hf,Alishahiha:2004eh}, whose action reads
\begin{equation}
S_{\text{DBI}} = \int \D^4 x \sqrt{-g} \left[ - f(\phi)^{-1} \sqrt{1+ f(\phi)g^{\mu\nu}(\partial_\mu\phi)(\partial_\nu\phi)} - V(\phi) \right]  \, ,
\label{4_0-action2}
\end{equation}
where $f(\phi)$ is a function of the scalar field $\phi$.
The Friedmann equations \eqref{Fe1}-\eqref{Fe2} yield
\begin{align}
H^2 
	& = \dfrac{1}{3M_P^2}(f^{-1}\gamma + V),
\label{4_1-Fr21} \\
\dot{H} 
	& = \dfrac{1}{2M_P^2}\dfrac{1-\gamma^2}{f\gamma}
	= -\dfrac{\gamma \dot{\phi}^2}{2M_P^2} ,
\label{4_1-Fr23}
\end{align}
where $\gamma^{-1}\equiv\sqrt{1-f\dot{\phi}^2}$.
Expanding the action in the \acs{ADM} perturbation $\delta N$,
\begin{multline}
S_{\text{DBI}} = \int \D^4 x \sqrt{-g} \bigg[-\frac{M_{\rm Pl}^2\dot{H}}{N^2}-M_{\rm Pl}^2(3H^2+\dot{H})
\\
	+\frac{1}{2}\gamma^3f\dot{\phi}^4(\delta N)^2 + \frac{1}{2}\gamma^5f\dot{\phi}^4(2f\dot{\phi}^2-3)(\delta N)^3 
	+ \ldots\bigg] \, .
\label{4_0-action2-exp}
\end{multline}
The effective coefficients in \eqref{EFTI-action-2} are
\begin{align}
M_2^4 & = \gamma^3f\dot{\phi}^4 \, ,
\label{coeff-1}\\
M_3^4 & = \frac{1}{2}\gamma^5f\dot{\phi}^4(2f\dot{\phi}^2-3) \, .
\label{coeff-2}
\end{align}
It is worth defining
\begin{equation}
\frac{\beta}{\varepsilon} \equiv -\frac{M_2^4}{2M_{\rm Pl}^2\dot{H}}
	= \frac{f\dot{\phi}^2}{1-f\dot{\phi}^2} \, ,
\label{betacoeff}
\end{equation}
\begin{equation}
\frac{\delta}{\varepsilon} \equiv -\frac{M_3^4}{M_{\rm Pl}^2\dot{H}}
	= \frac{f\dot{\phi}^2(2f\dot{\phi}^2-3)}{(1-f\dot{\phi}^2)^2} \, ,
\label{nucoeff}
\end{equation}
where the parameter \eqref{eps} is
\begin{equation}
\varepsilon = \frac{3}{2}\frac{\gamma fZ^2}{\gamma+fV} \, .
\label{epDBI}
\end{equation}
The interesting regime is reached when $f\dot{\phi}^2\rightarrow 1^-$, or equivalently $fV\gg\gamma\gg 1$.
In this case, $\gamma\sim 1/c_s$ and
\begin{align}
\varepsilon & \sim \frac{3}{2}\frac{\gamma}{fV} \ll 1 \, ,
\label{dbi-r1}\\
\beta & \sim \frac{\varepsilon}{c_s^2} \, ,
\label{dbi-r2}\\
\delta & \sim - \frac{\varepsilon}{c_s^4} \, .
\label{dbi-r3}
\end{align}
In this configuration, one has $\vert\delta\vert\gg\beta\gg\varepsilon$. Finally, we only report the result of the exact computation\footnote{We will further comment on this point later on. For the full procedure including the gravity sector, see Sec. \ref{sec-cswmwg}.} of non-Gaussianity in \acs{DBI} inflation,
\begin{equation}
f_{\text{NL}} = \frac{5}{27} - \frac{35}{108}\frac{1}{c_s^2} + \frac{10}{27}\beta + \mathcal{O}(c_s^2) \, ,
\label{fnldbifull}
\end{equation}
where one recognizes the leading contribution $\sim 1/c_s^2$.
The constraint on the speed of sound obtained by\index{experiments!Planck} \cite{planckXVII:2015} for \acs{DBI} inflation is $c_s^{\text{DBI}}\geq 0.087$ at $95$\% CL ($T+E$).

Now, it is worth emphasising again what we have anticipated in Sec. \ref{N-Gsec}: if one started from a generic effective Lagrangian
\begin{equation}
\mathcal{L} \sim \Lambda^4 P(X) - V(\phi) + \ldots \, ,
\label{eftdbi}
\end{equation}
where $\Lambda$ is an arbitrary scale and $P(X)$ a generic polynomial in the dimensionless variable $X=(\partial\phi)^2/\Lambda^4$,
\begin{equation}
P(X)\sim\sum_{n=1}^\infty \frac{(\partial\phi)^{2n}}{\Lambda^{4n}} \, ,
\label{polyPX}
\end{equation}
the regime $c_s\ll 1$ would require equally dominant operators, $(\partial\phi)^2\sim\Lambda^4$, invalidating the derivative expansion. On the other hand, the presence of symmetries could save the calculability of the theory. This is what happens in \acs{DBI} inflation (as well as in generalizations thereof \cite{Hinterbichler:2012fr,Hinterbichler:2012yn}), where the series \eqref{eftdbi} can be re-summed as \eqref{4_0-action2} taking $f^{-1}\equiv\Lambda^4$. Moreover, one might worry about the role of quantum corrections in the theory. Actually, the \acs{DBI} action benefits from non-renormalization properties, being also protected against higher derivative corrections in the ultra-relativistic limit \cite{baumann:2015book,deRham:2010eu,Goon:2011qf,Maldacena:1998Nl,Tseytlin:1999bi}. All these properties make the \acs{DBI} action a theoretically well defined model of inflation, admitting consistently large non-Gaussianity. In Part II, motivated by the requirement of controlling quantum corrections, we will demonstrate how a different symmetry of the scalar field for a certain class of theories makes them valid also in the region $f_{\text{NL}}\gtrsim 1$.

Starting from the Lagrangian \eqref{eftdbi}, one can also compute the values of the effective coefficients in Eq. \eqref{EFTI-action-2}.
%Before turning to the case of models with Galileon symmetry, it is instructive to recall what happens in more standard scalar effective theories. There, one starts with an action organized in derivative expansion containing a single scale $\Lambda$ -- the \acs{EFT} cutoff,
%\begin{equation}
%\label{simplescalartheory}
%S = S_{EH} + \int {\rm d}^4 x \sqrt{-g}\bigg[  \Lambda_c^4 G_2\left(X\right) -V(\phi) +     \frac{(\Box\phi)^2}{\Lambda_c^2}+\dots\bigg ]~,
%\end{equation}
%where, as before, $G_2(X)$ is a dimensionless function of the variable $X=-(\partial\phi)^2/\Lambda_c^4$, parametrizing the leading derivative effects. 
For a flat enough potential, one can show\footnote{See also the discussion in Chap. \ref{WBG-secfour}.} that the scalar background profile can be approximated by a linear function of time $\phi_0(t) =c(t)\Lambda^2 t$, where $c(t)$ is a slowly varying function. This is true both for slow-roll inflation\index{inflation!slow-roll} and for models such as the ghost condensate\index{ghost condensate}, where the field ``velocity'' is explicitly constant, \textit{i.e.} $c =$  const. at the leading order \cite{ArkaniHamed:2003uz}. For definiteness, let us  concentrate on the latter class of models and assume that $c\sim 1$ solves $P'(c^2)=0$. On such solutions, the Hubble rate can be estimated as $M_{\rm Pl}^2 H^2\sim \Lambda^4$ and is completely determined by the leading $P(X)$ piece, the higher-derivative operators being unimportant for the backgrounds at hand. Indeed, in principle the effective theory \eqref{eftdbi} contains also operators like $(\Box\phi)^2/\Lambda^2$, but on the background it can be estimated as $(\Box\phi_0)^2/\Lambda^2\sim \Lambda^4 (H/M_{\rm Pl})$ and thus it is negligible for $H\ll M_{\rm Pl}$, as required for a consistent classical description. 

Restricting to the unitary gauge, $\delta\phi(t,x)=0$, one straightforwardly finds that all \acs{EFT} coefficients are determined by the cutoff of the theory
\begin{equation}
M^4_2\sim \Lambda^4 \;,\qquad \hat{M}^3_1\sim \Lambda^3 \; ,\qquad \bar{M}^2_1\sim\Lambda^2 \;,  \qquad \text{etc.} 
\label{ordeftparest}
\end{equation}
In this case, the dynamics of small perturbations is fully dominated by the only zero-derivative quadratic operator in the effective theory -- $\delta N^2$ -- and most of the phenomenology is thus determined by the single coefficient $M^4_2$.
%This leads to interesting characteristic features, such as the possibility of small speed of sound $c_s^2$ of scalar perturbations (if $|M^4|\gg |M_{\rm Pl}^2 \dot H|$) and, associated to it, large non-Gaussianity $f_{\rm NL}\sim 1/c_s^2$~\cite{Chen:2006nt,Cheung:2007st}.
Higher-order terms contribute only to slightly correct the leading results. For example, one can show that the correction to the speed of sound from the operator $\bar{M}^2_1$ (which arises entirely due to mixing with gravity) is of order $\delta c_s^2 \sim \bar{M}_1^2 / M_{\rm Pl}^2\sim H/M_{\rm Pl}$ and can be safely ignored. Consequently, one can consistently consider the \text{perturbations} of the inflaton\index{inflaton} field as weakly coupled over a sufficiently broad range of distances encompassing the Hubble scale and straightforwardly apply the derivative expansion. 
This would correspond to the case of \textit{k}-inflation \cite{ArmendarizPicon:1999rj}, ghost inflation\index{inflation!ghost} \cite{ArkaniHamed:2003uz} and other related models.

\section{Energy scales in the EFT of inflation}
\label{sec-energyscales}

The identification of the energy scales in a physical system is a useful practice, in order to understand the general dynamical structure. A first scale is obviously fixed by the typical energies involved in the experiments, which in the case of cosmological observables are of order $\sim H$. Then, in the context of an effective field theory, we have already discussed about the importance of the strong coupling scale\index{strong coupling scale}, defining the upper limit of validity of the effective description. Moreover, in dealing with spontaneously broken symmetries, another relevant quantity can be introduced: the \textit{symmetry breaking scale}\index{symmetry!breaking scale} \cite{Baumann:2011su,Baumann:2014cja}. It essentially identifies the energy scale at which the spontaneous symmetry breaking occurs, defining the regime of validity of the description in terms of Goldstone bosons\index{Goldstone field}. Indeed, let $J^\mu$ be the conserved Noether current associated with the symmetry of the Lagrangian and
\begin{equation}
Q_R(t) = \int_{\vert\vec{x}\vert\leq R}\text{d}^3\vec{x} \, J^0(t,\vec{x}) 
\label{QR}
\end{equation}
the corresponding charge. As an example, we consider the case
\begin{equation}
J^\mu = -f_\pi \partial^\mu\pi + \ldots
\label{Jmu}
\end{equation}
characterizing for instance a non-linear sigma model, having $f_\pi$ the dimension of a mass. From the canonical commutation relations, one finds that
\begin{equation}
\text{i}\lim_{R\rightarrow\infty} \left[Q_R,\pi(x)\right] = f_\pi + \ldots
\label{Qpi}
\end{equation}
In other words, $f_\pi$ represents the order parameter of the symmetry breaking. Moreover, one can show that below such scale the correlation functions of the current are IR divergent and the charge can not be defined, as well known. For these reasons, one is persuaded to interpret $f_\pi$ as the symmetry breaking scale $\Lambda_\text{b}$ of the theory.

In the context of the \acs{EFT} of inflation, in the same spirit one would want to identify such $\Lambda_\text{b}$, at which the spontaneous symmetry breaking of time-translations during the quasi de Sitter phase occurs. For the effective action \eqref{ex-Gaction-dc}, in \cite{Baumann:2011su} it is estimated to be
\begin{equation}
\Lambda_\text{b}^4 = 2M_{\rm Pl}^2\vert\dot{H}\vert c_s \, .
\label{baum-sbs}
\end{equation}
This has to be compared with the strong coupling scale $\Lambda_\star$, estimated before.
%As already discussed, $\Lambda_\star$ identifies the unitary cutoff of the theory, namely the regions of energies where additional modes, integrated out of the \acs{EFT}, are expected to enter and cure the unitarity violation\index{unitarity bound}.
The relative magnitude between $\Lambda_\star$ and $\Lambda_\text{b}$ is relevant, signalling two different qualitative behaviours of the theory. In other words, if $\Lambda_\star>\Lambda_\text{b}$ the theory is essentially weakly coupled, resembling a standard slow-roll evolution with small non-Gaussianity, while if $\Lambda_\star<\Lambda_\text{b}$ in its whole range of applicability the Goldstone action exhibits some strongly coupled dynamics, showing up in possibly measurable interactions in the spectrum. Indeed, in terms of these quantities the rough estimation of non-Gaussianity and $f_{\text{NL}}$ yields
\begin{equation}
\frac{\mathcal{L}_{\dot{\pi}(\partial_i\pi)^2}}{\mathcal{L}_2} \sim \left(\frac{H}{\Lambda_\star}\right)^2 \, ,
\qquad
f_{\text{NL}} \sim \left(\frac{\Lambda_\text{b}}{\Lambda_\star}\right)^2 \, .
\label{baum-ng}
\end{equation}
Therefore, in order to have a valid perturbative description, one has to require $H\lesssim \Lambda_\star$. Rephrasing in these notations what we have stressed several times, a hierarchy of the form $H\lesssim\Lambda_\star\lesssim\Lambda_\text{b}$ could provide measurable non-Gaussian signals.

%% file: chapters/2-01-chapter.tex
\chapter{Weakly Broken Galileon Symmetry}
\label{chap-WBG}

\begin{flushright}{\slshape    
    SALVIATI: [...] Shut yourself up with some friend in the largest room below decks of some large ship and there procure gnats, flies, and such other small winged creatures. Also get a great tub full of water and within it put certain fishes; [...]. 
    You shall not be able to discern the least alteration in all the forenamed effects, nor can you gather by any of them whether the ship moves or stands still.} \\ \medskip
    --- Galileo Galilei, \textit{Dialogue Concerning the Two Chief World Systems}.
%    --- \defcitealias{bentley:1999}{BBBBB}\citetalias{bentley:1999} \citep{bentley:1999}
\end{flushright}

% For an example of a full page figure, see Fig.~\ref{fig:myFullPageFigure}.

In Part I, we have essentially learnt that symmetries are the crucial and fundamental ingredients in the definition of an effective field theory and in establishing the relative weights of the Lagrangian operators.
For instance, as we have already discussed, the \ac{EFTI}\index{inflation!EFT of} \eqref{EFTI-action-2} is based on the observation that the dynamics of the most general theory of ``single-clock'' inflation can be universally captured by an \ac{EFT}\index{Effective Field Theory (\acs{EFT})} non-linearly realizing\index{symmetry!non-linearly realized} time diffeomorphisms\index{time-diffeomorphisms!breaking of} $t\to t+\xi^0(t,\vec{x})$, with spatial diffeomorphisms $x^i\to x^i+\xi^i (t,\vec{x})$ realized linearly. Therefore, the spectrum of perturbations consists of the two polarizations of the graviton plus the Goldstone boson\index{Goldstone field} of time diffeomorphism symmetry breaking.
%That the latter mode has to be present in the spectrum is a direct consequence of the symmetry breaking pattern and has little to do with the exact UV details of the microscopic theory of inflation. This means that any UV theory of inflation that does not lead to extra degrees of freedom around the Hubble energies is equivalent to single-clock inflation and is thus captured by the action \eqref{EFTI-action-2}.
We have also mentioned that symmetries are at the basis of a possible re-summation of the series \eqref{eftdbi}, constraining the particular form of the \ac{DBI} action \eqref{4_0-action2} and justifying a particular phenomenology which otherwise would have been unreliable uniquely from the point of view of the \acs{EFT}.
Moreover, we have emphasised the role of an approximate shift symmetry\index{symmetry!shift} for the scalar field in inflationary models, being related to a restoration of the quasi scale-invariance\index{scale-invariance} of the power spectrum in an approximate de Sitter background, whose symmetries highly constrain the form of the correlation functions.

In this part, we study yet another possible -- and, as we argue below, necessarily approximate -- symmetry of cosmological scalar fields: the invariance under \textit{internal} Galileon transformations\index{Galileon!transformations}\index{symmetry!Galileon}
\begin{equation}
\phi \to \phi + b_\mu x^\mu + c~.
\label{g_inv}
\end{equation}

Theories invariant under \eqref{g_inv} have appeared in various contexts before. To start with, Eq.~\eqref{g_inv} is a symmetry (up to a total derivative) of the simplest possible quantum field theory: that one of a free scalar field.
The most general scalar theory with the symmetry \eqref{g_inv} has been proposed for the first time in \cite{Nicolis:2008in}, where the analogy with the Galilean symmetry $\dot{x}\rightarrow\dot{x}+v$ in non-relativistic mechanics suggested the appellative \textit{Galileon}\index{Galileon!theories}. Then, this has been found \cite{deRham:2010ik} to describe the scalar polarization of the ghost-free \acs{dRGT} massive graviton \cite{deRham:2010kj}. Moreover, it has appeared in \cite{Luty:2003vm} in the context of the \acs{DGP} model\index{DGP model} \cite{Dvali:2000hr}.

In the next section, we review the flat-space Galileon\index{Galileon!theories!in flat space} of \cite{Nicolis:2008in} and its main properties. Then, having in mind the curved-space generalization where the invariance \eqref{g_inv} can not be exact any more, we introduce a \textit{small} breaking in the theory and discuss the consequences. Despite its simplicity, this flat-space example with weakly broken Galileon symmetry encloses the crucial points and the general philosophy, that will be directly exported to the case with gravity.

\section{Flat space Galileons and non-renormalization theorem}
\label{sec-wbg-ngt}

Consider a trivial, free theory of a scalar $\phi$. As emphasized above, in addition to more familiar symmetries (such as the ones under constant shifts or conformal transformations), this theory possesses an extra invariance under internal Galileon transformations \eqref{g_inv}. The latter leave the action invariant only up to a boundary term. Apart from this and even a more trivial tadpole term, there exist in principle infinitely many operators that satisfy the symmetry \eqref{g_inv}, namely those with at least two derivatives per field. However, at the lowest order in derivatives, as shown in \cite{Nicolis:2008in}, apart from the kinetic operator, there are exactly three more interaction terms in four space-time dimensions that share invariance under internal Galileon transformations in a non-trivial way. The corresponding theory can be written as
\begin{equation}
\label{gallag}
\mathcal{L}=(\partial\phi)^2 +\sum_{I=3}^5\frac{c_I}{\Lambda_3^{3(I-2)}}\mathcal{L}_I ~,
\end{equation}
where those three interaction terms are 
\begin{align}
\label{gal1}
\mathcal{L}_3 &= (\partial\phi)^2~[\Phi]~, \\
\label{gal2}
\mathcal{L}_4 &= (\partial\phi)^2~ \big([\Phi]^2-
[\Phi^2]\big )~,\\
\label{gal3}
\mathcal{L}_5 &= (\partial\phi)^2~\big([\Phi]^3-3[\Phi][\Phi^2]+ 2[\Phi^3]\big )~,
\end{align}
where we denote $[\Phi]\equiv \Box\phi,~[\Phi^2]\equiv \partial^\mu\partial_\nu\phi\partial^\nu\partial_\mu\phi$, etc.
In addition to being invariant under \eqref{g_inv}, the Galileon theory\index{Galileon!theories} \eqref{gallag} shares other special properties.
First of all, despite the higher-derivative interactions in Eqs. \eqref{gal1}-\eqref{gal3}, the associated scalar equations of motion are second order, both in time and in space. This guarantees that there are no extra propagating degrees of freedom hidden in $\phi$.
Furthermore, the particular structure of the Lagrangian \eqref{gallag} results in the so-called \textit{non-renormalization theorem}\index{non-renormalization theorem} \cite{Luty:2003vm}, that allows to radiatively generate only terms trivially invariant under \eqref{g_inv}, \textit{i.e.} terms with at least two derivatives acting on $\phi$. In particular, it states that the coefficients $c_I$ of the Galileon interactions are not renormalized by quantum loops in the presence of exact Galileon invariance. The proof of the theorem follows simply by integrations by parts \cite{Luty:2003vm}. For instance, in the case of the cubic operator \eqref{gal1}, one can show that every \acs{1PI} diagram is forced to have only external fields derived twice. Indeed, if one started from \acs{1PI} diagram with two derivatives acting on an internal field as
\begin{equation}
\partial^\mu\phi_\text{ext}\partial_\mu\phi_\text{int}\square\phi_\text{int}
	= \partial^\mu\phi_\text{ext}\partial^\nu \left[
	\partial_\mu\phi_\text{int}\partial_\nu\phi_\text{int}
	-\frac{1}{2}\eta_{\mu\nu}\partial^\rho\phi_\text{int}\partial_\rho\phi_\text{int}\right] \, ,
\label{non-ren-proof}
\end{equation}
a simple turn of the derivative would prevent loop corrections to $c_3$, leading to a diagram with only $\partial^2\phi_\text{ext}$ as external legs. Concerning this topic, we refer also to \cite{Hinterbichler:2010nrt,deRham:2010sky,Heisenberg:2014qcml}.

Such non-renormalization properties make Galileon theories extremely attractive because, for instance, classical solutions to the equations of motion are automatically protected against quantum corrections and fine-tuning\index{fine-tuning} problems do not arise. However, the symmetry \eqref{g_inv} is known to be broken in presence of gravity and therefore the non-renormalization theorem is expected to fail. Nevertheless, we will prove that there exists some sort of generalization assuring that loop corrections are actually tiny if the symmetry is broken by a small coupling to the gravity sector.

%Moreover, it is worth recalling that in theories of modified gravity with matter, Galileons couple to the rest of the degrees of freedom via symmetry-breaking Planck-suppressed operators\footnote{In massive gravity\index{massive gravity}, where $\phi$ describes the helicity-$0$ polarization of the graviton, this simply follows from the equivalence principle.}, usually making the renormalization of $c_I$ parametrically weak.

%Before going into the details, some further explanatory comments are in order.
Apart from guaranteeing the avoidance of fine-tuning problems, the resilience of non-renormalization properties is particularly relevant whenever ``large'' background solutions are involved, as happens in cosmology. Indeed, in such situations, the higher order operators in the effective Lagrangian are enhanced if some of the fields are evaluated on the large classical background, invalidating the derivative expansion and in some cases making the physical predictions unreliable, unless one of the following situations occurs: \textit{i)} all but only a finite number of operators, that are the leading ones on the background, can be set to zero because of non-renormalization properties protecting against quantum corrections, or \textit{ii)} one is able to re-sum all terms of the series.
The latter case, apart from the already discussed \acs{DBI} theory\footnote{See Sec. \ref{subsec-stucktrick-dc}.}, is what happens also in General Relativity.
%and, more in general, in its effective formulation\index{Effective Field Theory (\acs{EFT})!of General Relativity}.
Indeed, denoting with $h_{\mu\nu}$ the canonically normalized tensor fluctuation over the flat metric, the expansion of the Einstein-Hilbert Lagrangian contains terms of the form
\begin{equation}
\frac{\partial^2h^{n+2}}{M_\text{Pl}^{n}} \, ,
\label{GRexpEH}
\end{equation}
where the higher order contributions are negligible if $h<M_\text{Pl}$. However, whenever one considers for example non-trivial sources of mass $M_*$ and approaches distances of order of the Schwarzschild radius\index{Schwarzschild radius} $r_S\sim M_*/M_\text{Pl}^2$, the classical formulation remains valid but non-linearities become of the same order of the kinetic term and therefore can not be neglected any longer. Nevertheless, in the case at hand, even though there is an infinite number of operators equally contributing in the perturbative expansion, their coefficients are not arbitrary, but they are fixed by General Relativity, crucially allowing a re-summation of the series.
Conversely, in the rest of the work we will focus on case \textit{i)}: here, the reliability of the physical predictions in regimes of large backgrounds will strongly depend on the presence of non-renormalization properties, due to a weakly broken symmetry.

In this context, it is fair to remind also that, as in every effective field theory, the Lagrangian is expected to include also higher derivatives operators, like powers of $\square\phi$. They are generically associated with higher-order equations of motion and ghost-like degrees of freedom, but these are by no means a problem if the typical energy scale of the instability is of order of (or above) the cutoff \cite{Creminelli:2005gmg}. However, in presence of sizeable backgrounds, the ghost mass might be dramatically lowered, unless some controlling criterion is found.

All these aspects will be discussed in details, leading to the final result of Eq. \eqref{stabilitybiss}. 

\section{Weak breaking in flat space-time}

Before turning to the case of a curved space-time, it is instructive to discuss the notion of weak symmetry breaking in the flat-space version \eqref{gallag} of the theory. This is the content of the present section.
As already said, despite the interesting properties, the invariance under Galileon transformations cannot be exact in Nature because the couplings of the scalar to gravity break it explicitly. Therefore, we are now interested in characterizing the theories that preserve as much as possible the attractive quantum non-renormalization properties of the exactly invariant case. This naturally leads to the notion of \textit{weakly broken} invariance\index{Weakly Broken Galileon (\acs{WBG})!invariance} under \eqref{g_inv}, which we will define in a precise way in what follows. 
The simplest sufficiently non-trivial theory with \textit{weakly broken Galileon invariance} (and exact shift symmetry\index{symmetry!shift}) in a flat space-time is of the following form
\begin{equation}
\label{simpleaction}
\mathcal{L}=-\frac{1}{2}(\partial\phi)^2 +\frac{1}{\Lambda_3^3}(\partial\phi)^2\Box\phi+\frac{1}{\Lambda_2^4}(\partial\phi)^4~.
\end{equation}
While the first two operators in \eqref{simpleaction} are exactly invariant under \eqref{g_inv} (up to boundary terms), the quartic operator is a small breaking as far as $\Lambda_2 \gg \Lambda_3$. 
In general we would expect other symmetry breaking operators of the form $(\partial \phi)^{2n}$ to be generated by quantum corrections at the scale $\Lambda_2$. However, 
in this case, a stronger result, the remnant of the non-renormalization properties\index{non-renormalization theorem} of the invariant action, holds: all the symmetry breaking operators are generated at a scale that is parametrically higher than $\Lambda_2$. This means in particular that the operator $(\partial \phi)^4$ gets only small corrections through loop effects. Indeed, in principle one could expect loop diagrams obtained with one insertion of the quartic vertex and two insertions of the cubic one, but actually these are forbidden\footnote{In fact, as we have shown before in \eqref{non-ren-proof}, the non-renormalization properties of the Galileon Lagrangian forbid loop diagrams (involving Galileon vertices) with external legs all derived only once.}. Therefore, the only possibility is the loop diagram built by means of two quartic vertices: as stated above, this kind of quantum correction is suppressed by a tiny factor $(\Lambda_3/\Lambda_2)^4$ with respect to the lagrangian operator $(\partial \phi)^4$ in \eqref{simpleaction}.

In the presence of gravity, as we will show in the next section, the weak breaking gives rise to even more non-trivial consequences.

\section{Weak breaking in curved space-time}
\label{WBG-sectwo}

In this section we study the fate of Galileon invariance upon inclusion of coupling to gravity\index{Galileon!theories!in curved space}, showing that it unavoidably breaks the symmetry and, at the same time, defines the sense in which the breaking can be considered weak. 
Indeed, as we will see below, out of the infinite number of possible ways one can couple the Galileon to gravity, there is a unique set of non-minimal couplings\index{non-minimal coupling} that can be qualified as being \textit{more invariant} under Galileon transformations than all the rest. The resulting theory turns out to be the \textit{covariant Galileon}\index{Galileon!covariant} of \cite{Deffayet:2009wt}, representing a particular curved-space extension of the theory that keeps the property of second-order equations of motion intact. On the other hand, since the Galileon symmetry has to be weakly broken by the couplings to gravity, one should in principle include in the effective theory symmetry-breaking operators of the form $(\partial\phi)^{2n}$. The latter will be generated by quantum loops with Wilson coefficients\index{Wilson coefficients} bounded from above, so as to be consistent with the approximate symmetries at hand. As we will see, requiring  invariance under \eqref{g_inv} to be only \textit{weakly} broken will naturally lead to a particular sub-class of Horndeski theories\footnote{Horndeski theories\index{Horndeski theories} \cite{Horndeski:1974wa} represent the most general class of scalar, Lorentz-invariant\index{symmetry!Lorentz} theories in a curved space-time with second-order equations of motion both for the scalar field and the metric. Since, however, a generic Horndeski theory does not have much to do with the invariance under \eqref{g_inv}, we prefer to refer to the sub-class that we study here as ``theories with \ac{WBG} invariance''.}\index{Horndeski theories}, which is however much more general than just the covariant Galileon\index{Galileon!covariant}. 
The purpose of this part is to introduce this particular sub-class of theories and to set the stage for their detailed phenomenological studies.

Being generally impossible to couple a Galileon-invariant theory to gravity\footnote{Unless gravity is massive, see \cite{Gabadadze:2012tr}.}, the symmetry is expected to be broken even in the purely scalar sector -- \textit{i.e.} if one sets the metric perturbation to zero ``by hand'' -- due to  loop-generated operators not invariant under \eqref{g_inv} (these operators of course have to be suppressed by at least one power of the Planck mass). For the sake of concreteness, let us concentrate on the operators of the form $(\partial\phi)^{2n}$. After all, it is the absence, or smallness, of these operators that makes the Galileon special compared to a generic shift-symmetric theory. These operators can be generated from loops with the appropriate number of insertions of symmetry-breaking vertices, that include at least one graviton line. 
Interpreting $\Lambda_3$ -- the smallest scale by which interactions are suppressed in the effective theory\footnote{We always assume in this paper that $\Lambda_3\ll M_{\rm Pl}$.} -- as the genuine cutoff, any loop-generated operator can be written as
\begin{equation}
\label{oneder}
\frac{(\partial\phi)^{2n}}{\Lambda_{k,n}^{4(n-1)}}~, \qquad \Lambda_{k,n}\equiv \left[ M_{\rm Pl}^k \Lambda_3^{4n-k-4}\right]^{\frac{1}{4n-4}}~,
\end{equation}
where $k$ denotes some positive integer and we have omitted  factors of $16\pi^2$ for simplicity. For  fixed $k$ and $n$ sufficiently large, the scale $\Lambda_{k,n} $ suppressing a given single-derivative operator can in principle be arbitrarily close to $\Lambda_3$. If this were true, the resulting theory would by no means be considered as a theory with \acs{WBG} invariance, since symmetry-breaking operators would be order-one in the units of the cutoff. 

We wish to show here that: \textit{i)} this in fact does not happen for the Galileon, even if minimally coupled to gravity; \textit{ii)} for a very particular, non-minimal coupling\index{non-minimal coupling}, the scale suppressing the symmetry-breaking operators can be made parametrically higher than the analogous scale characterizing all the other ways of introducing gravity into the system. For the latter theories, there is a well-defined separation between the scale suppressing the invariant Galileon interactions and the quantum-mechanically generated single-derivative operators: while the former are suppressed by powers of $\Lambda_3$, the scale suppressing the latter is parametrically higher, asymptotically reaching 
\begin{equation}
\Lambda_2\equiv (M_{\rm Pl} \Lambda_3^3)^{1/4} 
\end{equation}
for large $n$ in \eqref{oneder}. Perhaps not surprisingly, we will find that the special type of couplings to gravity required to suppress symmetry-breaking operators is of the Horndeski class -- the unique curved-space extension of the (generalized) Galileons that leads to second-order equations of motion both for the scalar and the metric. 

\begin{figure}
\includegraphics[width=0.9\textwidth, natwidth=610,natheight=642]{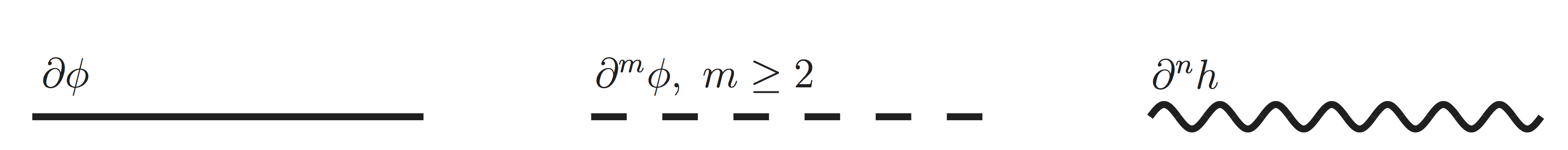}\centering
\caption{For every given vertex, a solid straight line corresponds to a possible external leg resulting in a scalar with at most one derivative acting on it. A dashed line corresponds to an external scalar with at least two derivatives, while a wiggly line denotes an external graviton.}
\label{fig1}
\end{figure}

We now prove all of the above-described statements, starting with a few definitions. Consider an arbitrary vertex of the form $(\partial\phi)^{k}(\partial^m\phi)^n\partial^l h^p$, with $m\geq 2$. In terms of Feynman diagrams, we will adopt the convention according to which a solid straight line corresponds to an external leg coming out of this vertex, which results in a scalar with one derivative acting on it. A dashed line corresponds to an external scalar with at least two derivatives, while a wiggly line denotes an external graviton, see Fig.~\ref{fig1}. For example, a $k=2,~n=1,~p=1$ vertex would correspond to the first diagram on the upper line of Fig.~\ref{fig2}. In certain cases, the number of solid lines can be less than the number of factors of $\partial\phi$ in a Lagrangian interaction term. For example, the special structure of the pure Galileon interactions makes them equivalent to vertices with only dashed lines, even though on average these terms contain less than two derivatives per field. This is because a Galileon vertex can only lead to external states with at least two space-time derivatives, as discussed in Sec. \ref{sec-wbg-ngt}. We now wish to show that there exist more redundant vertices of this kind in the suitable non-minimal extension of the theory to curved space.

\subsection{Minimal coupling to gravity}

Now we have a look at all possible vertices with a single graviton line in the minimally\index{minimal coupling} coupled Galileon theory, obtained from \eqref{gallag} by replacing all derivatives with covariant ones ($\partial_\mu\to\nabla_\mu$). A straightforward inspection yields that there are six of such vertices, shown in Fig.~\ref{fig2}. Vertices with three solid lines can in principle arise from picking up a factor of $\partial h$ from covariant derivatives\footnote{Moving the derivative on $h$ to the rest of the fields in the vertex generically does not reduce the number of solid lines for the quartic and quintic Galileons, since this extra derivative can act on the factor $\partial^2\phi$, representing the dashed line.}, \textit{i.e.} $\nabla^2\phi \sim \partial^2\phi+\partial h\partial\phi$, where we denote by $\nabla^2$  any product of two covariant derivatives applied on $\phi$. 
This means that, in the minimally coupled Galileon theory, the smallest scale by which the operators of the form \eqref{oneder} with $n=3 a, ~k=2 a$ are suppressed, is\footnote{This can be seen by inserting enough number of vertices with one graviton and three solid lines into a generic \acs{1PI} loop diagram.}
\begin{equation}
\Lambda_\text{mc} \equiv (M_{\rm Pl}\Lambda_3^{5})^{1/6} \, .
\label{minscaleWB}
\end{equation}
The latter suppression scale is still too small to be consistent with the generic definition of \acs{WBG} invariance, which we will introduce and discuss extensively in what follows.

\subsection{Non-minimal coupling to gravity}

We will now show that it is possible to modify the theory by non-minimal couplings\index{non-minimal coupling} to gravity that result in the elimination of vertices with three solid lines, leaving just a factor of  $(\partial\phi)^2$ per graviton line. Insertion of these vertices into a generic loop diagram can only lead to symmetry-breaking operators of the form \eqref{oneder} suppressed by at least the scale $\Lambda_2 = (M_{\rm Pl}\Lambda_3^3)^{1/4}$, parametrically higher than $ (M_{\rm Pl}\Lambda_3^{5})^{1/6}$. Therefore, there is a well-defined sense in which the resulting theory is ``more invariant'' under internal Galileon transformations than a generic curved-space extension of the Galileon. This defines what we mean by ``theories with \acs{WBG} invariance'' throughout the present work and we will sharpen this definition in the next chapter, where non-trivial vacua of the resultant theories are considered.

\begin{figure}
\includegraphics[width=0.7\textwidth, natwidth=610,natheight=642]{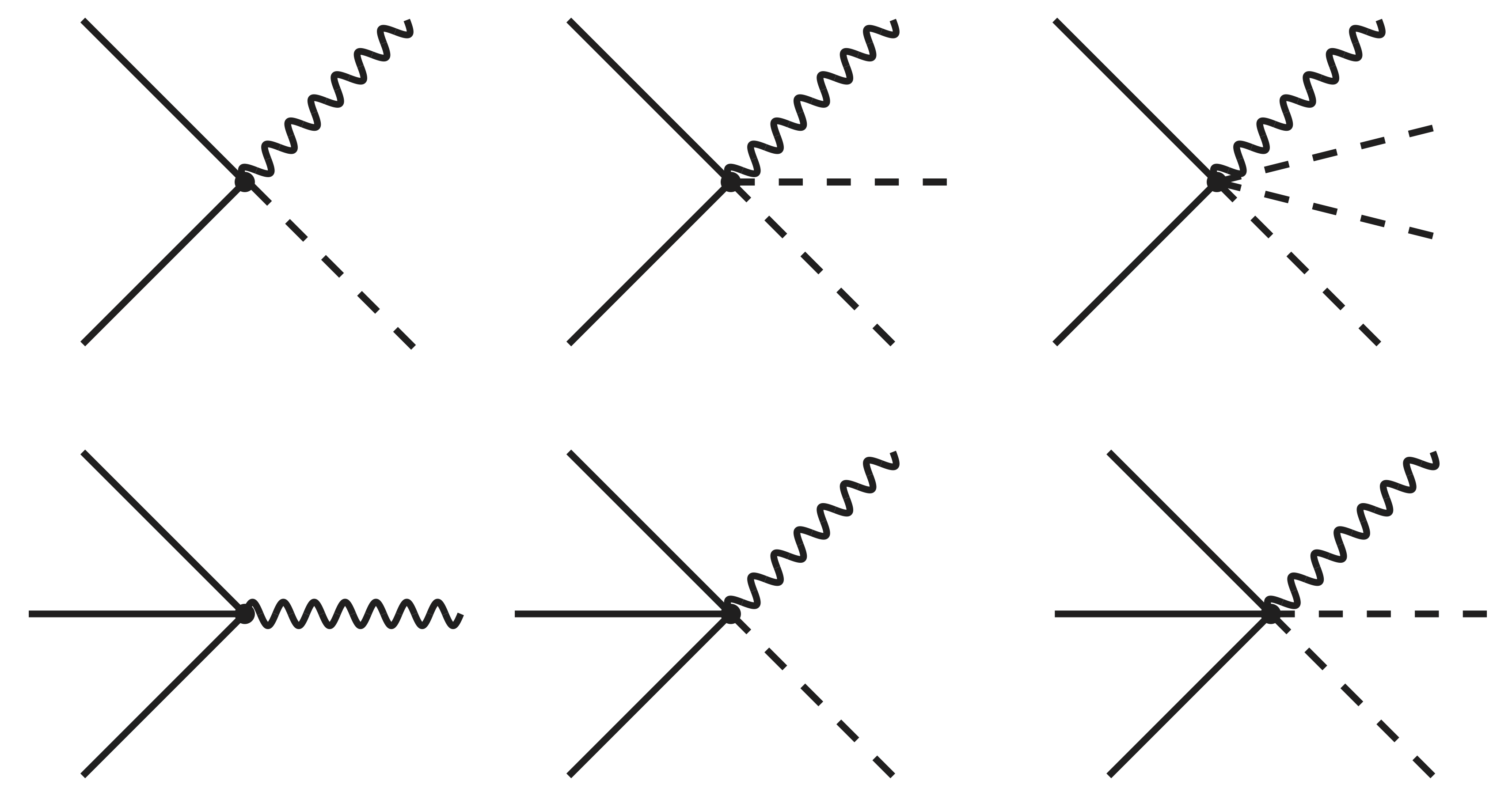}\centering
\caption{Single-graviton vertices, which can contribute to a \acs{1PI} vertex made of two or three external scalars with one derivative ($(\partial\phi)^2$ or $(\partial\phi)^3$). The vertices from the second line exactly cancel for the unique (non-minimal) curved-space extension of the Galileon that retains second-order equations of motion.}
\label{fig2}
\end{figure}

For the cubic covariant Galileon, it is easy to show that the first vertex from the lower line of Fig.~\ref{fig2} is absent: the only way to generate it is by picking up a metric perturbation from the covariant laplacian. This yields a term of the form $(\partial\phi)^3\partial h$. Putting the derivative from $h$ on the rest of the fields via partial integration however makes it evident that the corresponding vertex can only have two solid lines, but not three.  The case of quartic and quintic Galileons is more non-trivial, but straightforward; we show in App. \ref{Appendix-WBG} that vertices with three solid lines and one graviton, as well as five solid lines and two gravitons can be removed by suitably adding non-minimal couplings\index{non-minimal coupling} to gravity.
The resultant theory is the covariant Galileon of \cite{Deffayet:2009wt} -- characterized, as a bonus, by second-order equations of motion both for the scalar and the metric. 

In summary, the ``most symmetric'' generalization of the Galileon coupled to gravity consists of the following operators 
\begin{align}
\label{g1}
\frac{1}{\Lambda_3^3}\mathcal{L}_3 &\to \sqrt{-g} 
~\frac{1}{\Lambda_3^3}
~\mathcal{L}_3^{\text{min}}, \\
\label{g2}
\frac{1}{\Lambda_3^6} \mathcal{L}_4 &\to  \sqrt{-g} 
~ \frac{1}{\Lambda_3^6}
~\bigg [(\partial\phi)^4~ R-4 ~\mathcal{L}_4^{\text{min}}\bigg ]~, \\
\label{g3}
\frac{1}{\Lambda_3^9} \mathcal{L}_5 &\to \sqrt{-g}
~\frac{1}{\Lambda_3^9}
~\bigg [(\partial\phi)^4 ~G^{\mu\nu}\nabla_\mu\nabla_\nu\phi+\frac{2}{3} ~\mathcal{L}_5^{\text{min}}\bigg ]~,
\end{align}
where $\mathcal{L}_I^{\text{min}}$ denote the Galileons \eqref{gal1}-\eqref{gal3} minimally coupled to gravity.
The structure of the full effective theory is such that every pair of external scalars with no more than one derivative on each unavoidably comes with a suppression of at least one power of the Planck scale. With this in mind, quantum-mechanically generated operators of the form $(\partial\phi)^{2n}$ can be estimated simply on dimensional grounds to be at most of the following magnitude
\begin{equation}
\label{qcorr}
\frac{(\partial\phi)^{2n}}{M_{\rm Pl}^n\Lambda_3^{3n-4}}~.
\end{equation}
The scales $(M_{\rm Pl}^n\Lambda_3^{3n-4})^{1/(4n-4)}$ suppressing such operators approach $\Lambda_2$ only for asymptotically large $n$, otherwise being parametrically larger. 

In the rest of the work, we will argue that the statement of approximate Galileon invariance can be viewed as a statement about \textit{non-trivial} classical vacua, generically present in the shift-invariant theories at hand. If these vacua are to be insensitive to UV physics, the symmetry-breaking operators in the effective theory can be \textit{at most} of order \eqref{qcorr} in magnitude. In such cases we will say that the Galileon invariance is broken only weakly by couplings to gravity.

\subsection{Theories with WBG invariance}
\label{WBG-secthree}
Perhaps, the most important lesson that one can draw from the discussion of  the previous section is that the covariant Galileon is in fact \textit{not} the most general theory enjoying the above-described quantum properties. Consider the effective theory with the leading operators given by the following Horndeski sub-class 
\begin{align}
\label{hor1}
\mathcal{L}^{\rm WBG}_2&=\Lambda_2^4 ~G_2(X) ~,\\
\mathcal{L}^{\rm WBG}_3&=\frac{\Lambda_2^4}{\Lambda_3^3} ~G_3(X)[\Phi] ~,\\
\label{planck}
\mathcal{L}^{\rm WBG}_4&=\frac{\Lambda_2^8}{\Lambda_3^6}  ~ G_{4}(X) R + 2 \frac{\Lambda_2^4}{\Lambda_3^6}  ~G_{4X}(X)\left( [\Phi]^2-[\Phi^2] \right)~,\\
\label{hor4} 
\mathcal{L}^{\rm WBG}_5&=\frac{\Lambda_2^8}{\Lambda_3^9} ~G_{5}(X)G_{\mu\nu}\Phi^{\mu\nu}-\frac{\Lambda_2^4}{3 \Lambda_3^9} ~ G_{5X}(X)\left([\Phi]^3-3[\Phi][\Phi^2]+2[\Phi^3] \right) ~,
\end{align}
where we now extend the notation employed in Eqs.~\eqref{gal1}-\eqref{gal3} by replacing the partial derivatives with covariant ones, \textit{i.e.} $[\Phi]\equiv g^{\mu \nu} \nabla_\mu \nabla_\nu\phi,~[\Phi^2]\equiv \nabla^\mu\nabla_\nu\phi\nabla^\nu\nabla_\mu\phi$, etc. 
Moreover, $G_I$ are arbitrary dimensionless functions of the dimensionless variable 
\begin{equation}
X\equiv - \frac{1}{\Lambda^4_2} g^{\mu\nu}\partial_\mu\phi\partial_\nu\phi ~,
\label{X_def}
\end{equation}
and the subscript ``$X$'' means differentiation with respect to this variable\footnote{The particular form of the interactions \eqref{hor1}-\eqref{hor4} (the relative coefficients between two operators in $\mathcal{L}^{\rm WBG}_{4}$ or  $\mathcal{L}^{\rm WBG}_{5}$, for example) may appear tuned, and this is sometimes presented as an unfortunate feature in the literature. We stress that this ``tuning'', motivated by restoring unitarity in the theory, is in fact natural and stable under quantum corrections. This is very similar to why we work with a gauge-invariant kinetic term $- \text{Tr}~ F^{\mu \nu} F_{\mu \nu}$ in theories with massive spin-$1$ particles, despite of there being no gauge invariance in the presence of the mass term.}. 
We will then assume that the functions $G_I$ are defined through the Taylor expansion
\begin{equation}
\label{gfunct}
G_I = c^{(0)}_{I}-c^{(1)}_{I} X+c^{(2)}_{I} X^2+\dots=c^{(0)}_{I}+c^{(1)}_{I} \frac{(\partial\phi)^2}{\Lambda_2^4}+c^{(2)}_{I} \frac{(\partial\phi)^4}{\Lambda_2^8}+\dots~,
\end{equation}
where $c^{(n)}_{I}$ are dimensionless, order-one coefficients. 
Note that by setting all the coefficients $c^{(n)}_{I}$ to zero except  $c^{(1)}_2 $, $c^{(1)}_{3}$, $c^{(2)}_{4}$ and $c^{(2)}_{5}$ we recover the covariant Galileon of Eqs.~\eqref{g1}-\eqref{g3}. 

As in the example of the introductory section, the theory at hand is characterized by \textit{two} scales\footnote{The Planck scale arises here from ${\cal L}^{\rm WBG}_4$ in Eq.~\eqref{planck} as $M_{\rm Pl}^2=2 c^{(0)}_{4}\Lambda_2^8/\Lambda_3^6$.}: $\Lambda_2$ and $\Lambda_3\ll \Lambda_2$. From the \acs{EFT} standpoint that we are adopting here, all Wilson coefficients\index{Wilson coefficients} $g^{(n)}_{I}$ are measured in the units of the cutoff $\Lambda_3$. The theory \eqref{hor1}-\eqref{hor4} is the one for which a well-defined class of (canonically normalized) operators -- those that respect the Galileon symmetry $\phi\to\phi+b_\mu x^\mu+c$ -- have order-one coefficients, 
\begin{equation}
g^{(n)}_{I} \sim {\cal O}(1)~, 
\end{equation} 
while all the others are parametrically suppressed by positive powers of the small ratio $\Lambda_3/\Lambda_2$ (for example, the Wilson coefficient corresponding the the symmetry-breaking operator $(\partial\phi)^4\Box\phi$ is of order $g_3^{(2)}\sim c_3^{(2)}\left(\Lambda_3/\Lambda_2\right)^4$). 
As a consequence, one can define a scaling limit
\begin{equation}
\label{declim}
M_{\rm Pl}\to \infty,\qquad \Lambda_2\to \infty, \qquad  \Lambda_3=\text{finite}~,
\end{equation}
in which the system recovers the exact internal Galileon symmetry. 

A comment about the terms $c^{(1)}_{4}$ and $c^{(1)}_{5}$ is in order here\footnote{The operator $c^{(0)}_{2}$ is just the cosmological constant, which for simplicity we tune to zero, $c^{(0)}_{2}=0$. Moreover, the operators $c^{(1)}_{2}$ and $c^{(0)}_{4}$ set the normalization of the scalar and graviton kinetic terms respectively, while $c^{(0)}_{3}$ and $c^{(0)}_{5}$ give inconsequential total derivatives and can be disregarded.}. 
These terms are absent in the covariant Galileon, which starts with $G_{4}, G_{5} \sim  (\partial\phi)^4$. Thus, one can wonder whether they modify our conclusions regarding quantum corrections. By expanding ${\cal L}_4$ and ${\cal L}_5$ at linear order in metric perturbations, it is straightforward to verify that $c^{(1)}_{4}$ and $c^{(1)}_{5}$ do lead to new operators at the scale $\Lambda_3$. However, these operators are just a scalar-tensor generalization of the Galileon familiar from the decoupling limit\index{decoupling limit} of \acs{dRGT} massive gravity\index{massive gravity} \cite{deRham:2010ik},
\begin{equation}
\begin{split}
\label{massgravityterms}
&\mathcal{L}^{\rm WBG}_4\sim h_{\mu\nu}\varepsilon_{\mu\alpha\rho\lambda}\varepsilon_{\nu\beta\sigma\lambda}
\partial_\alpha\partial_\beta\phi\partial_\rho\partial_\sigma\phi+\dots~, \\ &\mathcal{L}^{\rm WBG}_5\sim h_{\mu\nu}\varepsilon_{\mu\alpha\rho\lambda}\varepsilon_{\nu\beta\sigma\delta}
\partial_\alpha\partial_\beta\phi\partial_\rho\partial_\sigma\phi\partial_\lambda
\partial_\delta\phi+\dots \;.
\end{split}
\end{equation}
The latter interactions obey the same non-renormalization theorem as the Galileon \cite{deRham:2012ew}, forbidding asymptotic states with less than two derivatives acting on them. In particular, this means that the counting of $M_{\rm Pl}^{-1}$ factors in loop diagrams of the previous section goes through unaltered: each factor of $(\partial\phi)^2$ arising from a generic \acs{1PI} vertex still comes with a suppression of at least a factor of $M_{\rm Pl}^{-1}$, and our analysis of quantum corrections performed for the covariant Galileon also applies for the generalized theory \eqref{hor1}-\eqref{hor4}.\footnote{That the terms \eqref{massgravityterms} can be re-written as a certain limit of  non-minimally coupled scalar-tensor theory has been noticed in \cite{Chkareuli:2011te} and the cosmology of that theory has been studied in \cite{deRham:2011by}.} Therefore, for simplicity we will disregard these terms and assume that $G_{4}$ and $G_5$ start at least quadratic in $X$. Indeed, our discussion of the previous section guarantees that, once tuned to zero, $G_{4X}$ and $G_{5X}$ do not receive any appreciable quantum corrections as a result of the approximate Galileon symmetry.

Notice that the theories that we propose do \textit{not} reduce to the Galileon for general functions $G_I$, even if one switches off gravity. Rather, they describe a certain generalization thereof, which still has no more than second-order equations of motion. The symmetry \eqref{g_inv}  is thus broken even when one sets $h_{\mu \nu}=0$ -- note that this is different from taking the decoupling limit\index{decoupling limit} \eqref{declim}. Nevertheless, the corresponding breaking of invariance under \eqref{g_inv} is in a well-defined sense \textit{not stronger} than the one already present once gravity is turned on. In particular, it follows from the very construction of the operators in \eqref{g1}-\eqref{g3} that each pair of factors of $\partial\phi$ still comes with a suppression of at least a factor of $M_{\rm Pl}$ in the full effective theory. Thus, 
the lagrangians \eqref{hor1}-\eqref{hor4}  define the most general scalar-tensor theory consistent with our definition of \acs{WBG} invariance. 

The peculiar quantum properties characteristic of the theories at hand put loop corrections under complete control for a broad class of physical backgrounds. This can be established by noting that loop-generated operators always have at least one extra factor of $M_{\rm Pl}^{-1}$, compared to their ``tree-level'' counterparts. Indeed, working in the units in which $\Lambda_3=1$, each factor of $\Lambda_2^4$ becomes equivalent to $M_{\rm Pl}$, and the non-minimal terms in \eqref{hor1}-\eqref{hor4} can be schematically written as 
\begin{equation}
M_{\rm Pl}  R (\partial\phi)^2 \left(1+\frac{(\partial\phi)^2}{M_{\rm Pl}}+\dots\right) (\nabla^2\phi)^n~,
\end{equation}
while the similar loop-generated terms can never have the $M_{\rm Pl}$-enhancement. Likewise, the symmetry-breaking scalar operators, quantum-mechanically generated, are bound to have the following schematic form 
\begin{equation}
\label{loops}
\frac{\left(\partial\phi\right)^{2n}}{M_{\rm Pl}^n}\left(\nabla^2\phi\right)^m~,
\end{equation}
and the analogous terms in the original action are all enhanced by at least a factor of $M_{\rm Pl}$ compared to this expression. 
Restoring the factors of $\Lambda_3$, we conclude that if the field expectation values are such that
\begin{equation}
\label{stabilitybiss}
|X|\lesssim 1, \qquad |Y|\lesssim 1 ~, 
\end{equation}
where $X$ is defined in Eq.~\eqref{X_def} and 
\begin{equation}
Y\equiv \frac{\nabla^2 \phi}{ \Lambda_3^3} \;,
\label{Z_def}  
\end{equation}
the loop corrections are negligible: they can never compete with the operators from the classical action \eqref{hor1}-\eqref{hor4} and all predictions based on the latter can be trusted in the full quantum theory. 
Moreover, not only the magnitudes of the various operators are technically natural\index{technical naturalness}, but any possible tuning of these can only be spoiled by corrections of higher order in $\Lambda_3/M_{\rm Pl}$. This generalizes the Galileon non-renormalization theorem\index{non-renormalization theorem} in the presence of gravity. 

On the other hand, if the conditions \eqref{stabilitybiss} are not satisfied, the background becomes too strong to be trusted: the magnitude of loop corrections becomes pumped up beyond that of the leading operators and Galileon invariance alone is useless in organizing the low-energy expansion. One could only trust the classical theory beyond that point if there is an extra structure, leading to a re-summation of the series \eqref{gfunct}. As mentioned in Sec. \ref{subsec-stucktrick-dc}, an example is provided by \acs{DBI} theories, where this happens due to non-linearly realized\index{symmetry!non-linearly realized} higher-dimensional space-time symmetries. However, here we want to remain as general as possible and do not assume any extra symmetry beyond the (weakly broken) Galileon one. 
In the latter case, the requirement of quantum stability expressed by Eq. \eqref{stabilitybiss} places a strong upper bound on how large the predictions for various physical quantities can be, but at the same time lays the foundation for some novel technically natural\index{technical naturalness} predictions. Indeed, in Sec. \ref{subsec-stucktrick-dc}, we learnt how some natural hierarchies among the Wilson coefficients\index{Wilson coefficients} in the \acs{EFT} of inflation can arise because of quantum corrections to the Lagrangian operators and how symmetries can alter this appraisal, protecting the tree-level operators. In this spirit, the rest of Part II will be devoted to the understanding of the physical consequences of an underlying \acs{WBG} symmetry in cosmology, showing that it can provide unexpected predictions, which would have been unreliable in the standard view.

%% file: chapters/2-02-chapter.tex
\chapter{WBG theories in Cosmology: generalities}
\label{WBG-secfour}

\begin{flushright}{\slshape    
Quem etiam, quo grandior 
sit et quodam modo excelsior, ne physicorum quidem esse ignarum volo. Omnia 
profecto, cum se a caelestibus rebus referet ad humanas, excelsius magnificentiusque et dicet et sentiet.} \\ \medskip
	--- Marcus Tullius Cicero, \textit{Orator}, [\textbf{34},119].
%    --- \defcitealias{bentley:1999}{BBBBB}\citetalias{bentley:1999} \citep{bentley:1999}
\end{flushright}

% For an example of a full page figure, see Fig.~\ref{fig:myFullPageFigure}.

In the previous chapter we have proved the existence of a class of theories which maximally retain in a curved space-time the remnants of the Galileon non-renormalization theorem\index{non-renormalization theorem}. Now, the present chapter is devoted to the application of these results to cosmology.

Before proceeding, in order to appreciate the novel physical implications and the differences with the known literature, it is worth recalling that the study of Galileon-invariant theories in the inflationary context have been initiated in \cite{Burrage:2010cu}. The model, referred to as \textit{Galileon inflation}\index{inflation!Galileon}, is based on the covariant Galileon of \cite{Deffayet:2009wt}. It has been pointed out in \cite{Burrage:2010cu} that such theories enjoy more freedom compared to ghost/\acs{DBI}\index{inflation!DBI}\index{inflation!ghost} inflation-like models as far as phenomenology is concerned. In particular, they can lead to the possibility of lifting the ``large speed of sound/small non-gaussianities'' correspondence, characteristic of the latter models. 
Our results extend the findings of \cite{Burrage:2010cu} in several directions. While it has been assumed in \cite{Burrage:2010cu} that the covariant Galileon is the unique model capable of extending the phenomenological virtues of shift-symmetric theories in a radiatively stable way, we have found that there is in fact a wider class of models that can achieve this. Unlike the covariant Galileon, the theories we are interested in do not generically reduce to the standard Galileon once gravity is turned off. Nevertheless, neither  non-renormalization nor the second-order equations of motion of the Galileon need to be given up, leading to the possibility of strong -- and quantum-mechanically robust -- phenomenological differences from slow-roll inflation\index{inflation!slow-roll}. This generalizes the well-studied case of \acs{DBI} models in an interesting way. 

The best way to illustrate the physical consequences of the theories of Chap. \ref{chap-WBG} is to resort to a concrete example. To this end, we wish to consider the following theory
\begin{equation}
\label{full}
S=\int {\rm d}^4x \sqrt{-g} ~\bigg[\frac{1}{2}M_{\rm Pl}^2 R -\frac{1}{2}(\partial\phi)^2 -V(\phi)+\sum_{I=2}^5 \mathcal{L}^{\rm WBG}_I +\dots\bigg]~,
\end{equation}
where $\mathcal{L}^{\rm WBG}_I$ are the  operators defined in Eqs.~\eqref{hor1}--\eqref{hor4} of the previous chapter. We have extracted and canonically normalized the scalar and graviton kinetic terms, so that $G_{2}$, $G_{4}$ and $G_{5}$ are assumed to start \textit{at least quadratic}\footnote{See the discussion below Eq. \eqref{massgravityterms}.} in $X$, while $G_3$ can have a linear piece. In general, one can allow for a small potential, 
\begin{equation}
V(\phi)=-\lambda^3\phi +\frac{1}{2}m^2_\phi \phi^2+\dots~,
\end{equation}
where the parameters $\lambda$, $m_\phi$, etc. can be (naturally) arbitrarily small, since they are the only ones that break the scalar shift symmetry, which otherwise is exact even on curved space.
In the next chapters, we will discuss the implications of a weakly broken Galileon symmetry for inflation. We will thus be interested in background solutions given by an approximate de Sitter space with the Hubble parameter $H$.
%Now we present a preliminary study of such solutions. 

For the theory \eqref{full}, the scalar equation of motion on a flat \acs{FRW} background\index{Friedmann-Robertson-Walker (\ac{FRW})!equations} reads
\begin{equation}
\label{phieqcosm}
\frac{1}{a^3}\frac{{\rm d}}{{\rm d}t} \big[2 a^3 \dot\phi F(X,Z)\big]=-\frac{{\rm d}V}{{\rm d}\phi} ~,
\end{equation}
where the function $F$ is given by the following expression\footnote{Note that both $G_{4X}/X$ and $G_{5X}/X$ in this equation are finite even in the $X\to 0$ limit due to our assumption that $G_4$ and $G_5$ start at least quadratic in $X$; see the discussion below Eq.~\eqref{massgravityterms}.}
\begin{multline}
\label{f}
F(X,Z) =  \frac{1}{2}+G_{2X}-3 Z G_{3X}+6 Z^2 \left(\frac{G_{4X}}{X}+2 G_{4XX}\right)
\\
	+ Z^3\left(3\frac{G_{5X}}{X}+2 G_{5XX}\right)~,
\end{multline}
with the variable $Z$ defined so as to roughly coincide in the limit $\ddot\phi\ll H\dot\phi$ with the background value of $Y$ defined in Eq.~\eqref{Z_def}, 
\begin{equation}
\label{Z_def1}
 Z \equiv \frac{ H \dot\phi}{\Lambda_3^3}~. 
\end{equation}
Moreover, the two Friedmann equations\index{Friedmann-Robertson-Walker (\ac{FRW})!equations} can be written in the following way
\begin{align}
3M_{\rm Pl}^2H^2= 
&	\ V +\Lambda_2^4X\bigg[\frac{1}{2}-\frac{G_2}{X}+2G_{2X}-6ZG_{3X}
\nonumber \\
&	-6Z^2\left(\frac{G_{4}}{X^2}-4\frac{G_{4X}}{X} -4G_{4XX} \right)
	+2Z^3\left(5\frac{G_{5X}}{X}+2G_{5XX}\right)\bigg ]
\label{fried1}\\
M_{\rm Pl}^2\dot H = &-[\Lambda_2^4 X F +M_{\rm Pl}\ddot \phi(X G_{3X}-4 ZG_{4X}-8ZXG_{4XX}-3 Z^2 G_{5X}
\nonumber \\
&	-2Z^2XG_{5XX})]\left[1+2 G_{4}-4X G_{4X}-2ZXG_{5X}\right]^{-1}~. 
\label{fried2}
\end{align}
By choosing appropriate combinations of the coefficients $c_I^{(n)}$ and in the absence of the potential, one can check that there exist \textit{exact} linear solutions $\phi_0\propto t$ to \eqref{phieqcosm} with $F(X_0,Z_0)=0$, sourcing an exact de Sitter space. Upon turning on a small enough potential for $\phi$, these backgrounds can slightly deviate from de Sitter, $\dot H \ll H^2$. Moreover, if both $X_0$ and $Z_0$ are of order one on these solutions, the background curvature is of the order $H^2\sim \Lambda_2^4/M_{\rm Pl}^2$ and all terms involving $\phi$ in \eqref{full} contribute equally to the energy density. 

We observe that for $X\ll 1$ and $Z\ll 1$ one reproduces the standard potentially dominated\index{potentially dominated evolution} slow-roll\index{inflation!slow-roll} regime. In the opposite limit $X\sim 1$ and for small enough $V(\phi)$, the acceleration can be supported by the kinetic part of the action. We will comment more on this points later. At the same time, we have argued below Eq. \eqref{stabilitybiss} that the two parameters $X$ and $Z$ control the magnitude of quantum corrections. 
Indeed, in the $X \lesssim 1$ and $Z\lesssim 1$ regime loop corrections are fully under control, as can be seen by estimating typical loop-generated terms, \textit{e.g.} of the form
\begin{equation}
\label{estimates}
\frac{(\nabla^2 \phi)^n}{\Lambda_3^{3n-4}}\sim Z^n\Lambda_3^4~,\qquad \frac{(\partial\phi)^{2n}}{M_{\rm Pl}^n \Lambda_3^{3n-4}} \sim X^n \Lambda_3^4 ~.
\end{equation}
For $X$ and $Z$ less than or of order one, these are parametrically suppressed with respect to the background energy density contributed by the classical Lagrangian, which, as follows from Eq.~\eqref{fried1}, generically satisfies $3M_{\rm Pl}^2 H^2 \gtrsim X\Lambda_2^4$. 
Saturating either one or both bounds in Eq. \eqref{stabilitybiss} leads to a non-linear regime where various next-to-leading order terms in the derivative expansion become important, while quantum corrections are still under control. This case will be of our prime interest in the rest of the work.
%In particular, in the next section we will explore the properties of the corresponding backgrounds in the context of inflationary physics. 

\section{Inflation in WBG theories}
\label{WBG-secfive}

\subsection{The effective field theory for perturbations}

The consequences of an approximate Galileon symmetry\index{inflation!WBG} for the physics of the early Universe can be conveniently studied in the \ac{EFTI}\index{inflation!EFT of} framework discussed in Chap. \ref{chap-EFTinflation}. 
In the rest of the present chapter, we will concentrate only on the quadratic part of the action \eqref{EFTI-action-2}: it is enough to capture the relevant physical implications of the \acs{WBG} symmetry\footnote{The only effect of the cubic part of \eqref{EFTI-action-2} is to modify the coefficients of the operators describing the self-interactions in the Lagrangian for the adiabatic mode. We refer to Sec. \ref{sec-cswmwg} and to the next chapters for further details.}. Moreover, for convenience we will drop some indices, \textit{i.e.} we will call $\hat{M}^3\equiv\hat{M}^3_1$ and $M^4\equiv M_2^4$. Moreover, we anticipate that the theory at hand \eqref{full} satisfies $\bar{M}_1^2=-\bar{M}_2^2=2\bar{M}_3^2\equiv\bar{M}^2$, as we will discuss in Eq. \eqref{comb}.

Therefore, given such a generic model of inflation, defined via the set of \acs{EFT} coefficients $M^4(t), \hat{M}^3(t), \bar{M}^2(t), \dots$, it is important to understand what are, if any, the possible hierarchies among these. Indeed, the basic properties of the inflationary background and perturbations, such as stability, power spectra and non-Gaussianity, crucially depend on the relative magnitudes of these terms. In the case that the problem involves widely separated scales (such as, for instance, the Hubble scale and the UV cutoff of the theory), it is not \textit{a priori} clear what the quantum-mechanically stable values of these Wilson coefficients\index{Wilson coefficients} are. In particular, treating them as arbitrary parameters can easily lead to results that require tuning, once quantum corrections are taken into account. 

%The easiest way to arrive at various hierarchies in quantum field theory is  using symmetries, exact or approximate. For example, chiral symmetry, although inexact, is the central reason behind the technical naturalness\index{technical naturalness} of fermion masses in the standard model of particle physics.
In the effective theory \eqref{EFTI-action-2}, one can use various symmetries to constrain the \acs{EFT} coefficients in different models of inflation. An obvious possible symmetry is provided by constant time translations, $t\to t+c$, which would constrain all coefficients (including the Hubble rate) in the \acs{EFT} to be time-independent, enforcing the background solution to be an exact de Sitter space. Another symmetry one can impose is given by arbitrary space independent reparametrizations of time, $t\to f(t)$ (see \textit{e.g.} \cite{Blas:2010hb}). In the exact symmetry limit, this would require $M^4$, $\hat{M}^3$ and $\bar{M}_3$ to vanish, leading to interesting theories in one corner of the inflationary theory space \cite{Creminelli:2012xb}. On the other hand, even if  time reparametrizations are not an exact symmetry, their \textit{weak} breaking (appropriately defined in what follows) would guarantee that any small values of these coefficients are technically natural\index{technical naturalness}. 

Building on the insights of the previous sections, we wish to investigate the consequences of yet another possible \textit{approximate} invariance of the action \eqref{EFTI-action-2}, namely that one induced by internal Galileon transformations\index{Galileon!transformations} and realized on the foliation scalar as in \eqref{g_inv}. If this scalar acquires a linear profile $\phi\propto t$ driving a de Sitter background, similarly to what we discussed before, the Galileon symmetry \eqref{g_inv} will have the following manifestation in the unitary gauge 
\begin{equation}
\label{gi0}
t\to t+\tilde b_\mu x^\mu~,
\end{equation}
where $\tilde b_\mu$ is a constant four-vector.
For more complicated $\phi$-profiles, the associated unitary gauge invariance will have a more complicated form. However, in practice it always makes life easier to restore the foliation scalar in the effective action \eqref{EFTI-action-2} in order to explore its approximate symmetries. We will take this route in what follows.

\subsection{Inflation with WBG symmetry}

\renewcommand{\arraystretch}{2.0}
\begin{table}[t]
\begin{adjustbox}{max width=\textwidth}
\begin{tabular}{|c||c|c|c|c|c|c|c|cl}
  \hline
$\mathcal{L}_I^{\rm WBG}$ &$\displaystyle {M^4}{\Lambda_2^{-4}} $ & $ \hat{M}^3 H  \Lambda_2^{-4} $ &$ \bar M ^{2} H^2  \Lambda_2^{-4}$ % &  $\tilde m^{2} H^2  \Lambda_2^{-4} $
\\
 \hline\hline
 $ I=2 $ & $4   X^2G_{2XX}$ &$\times$ &$\times$ % & $\times$
\\
  \hline
$ I=3 $ &$ \displaystyle -6  X Z \left( G_{3X}+2 X G_{3XX}\right)$ &$\displaystyle -2 {  X Z} G_{3X}$ &  $\times$ % & $\times$
\\
  \hline 
$ I=4 $ &  $ \displaystyle 24  X Z^2\left(3G_{4XX} + 2X  G_{4XXX}\right)$&{$ \displaystyle8 { X Z^2}\left( \frac{G_{4X}}{X}+2G_{4XX}\right)$}  & $\displaystyle -4 { Z^2} {G_{4X}}$ % & $\bar M^2 H^2  \Lambda_2^{-4} $
\\
 \hline
$ I=5 $ & {$ \displaystyle 2 X Z^3 \left(3\frac{G_{5X}}{X}+12 G_{5XX}+4X G_{5XXX}\right)$}&{$ \displaystyle 2 { XZ^3} \left(3 \frac{G_{5X}}{X}+2G_{5XX}\right)$} & $\displaystyle -2 { Z^3} {G_{5X}}$ % &  $\bar M^2 H^2  \Lambda_2^{-4} $
\\
  \hline
\end{tabular}\\
\end{adjustbox}
\caption{Contributions from the Lagrangian terms $\mathcal{L}^{\rm WBG}_I$ to various unitary-gauge operators, defined in Eq. \eqref{EFTI-action-2}. We have renamed $M^4\equiv M_2^4$ and $\hat{M}^3\equiv\hat{M}^3_1$, neglecting higher order operators. The relation $\bar{M}_1^2=-\bar{M}_2^2=2\bar{M}_3^2\equiv\bar{M}^2$ holds. The subscript $X$ means differentiation with respect to $X$ and all derivatives are evaluated on the background. We have assumed that all background quantities obey ``slow-roll'', \textit{i.e.} ${\rm d}/{\rm d}t\ll H$, so that terms involving derivatives of $X$ and $Z$ have been neglected. According to our assumption about the form of $G_4$ discussed in Sec.~\ref{WBG-secthree}, this is at least of order $X^2$ for small $X$, so that its contribution can be neglected. An analogous piece coming from $\mathcal{L}^{\rm WBG}_{5}$ is suppressed by a further factor of the slow-roll parameter.}
\label{tab1}
\end{table}

In \eqref{ordeftparest}, we have estimated the parameters of the \acs{EFTI} in the case of ``ordinary'' effective theories of the type \eqref{eftdbi}, proving that everything is determined in terms of the single scale $\Lambda$.
Now, we consider the theories characterized by \acs{WBG} invariance \eqref{full} and compute analogously the values of the various coefficients in Eq. \eqref{EFTI-action-2}, by means of reformulating the discussion of the previous section into the effective theory language \cite{Gleyzes:2013ooa}. The resulting contributions from each Lagrangian $\mathcal{L}^{\rm WBG}_I$ to different \acs{EFTI} operators are reported in Tab. \ref{tab1}. 

Interestingly, as a generic property of Horndeski theories, \textit{only one} combination of the operators $\delta K^2$, $\delta K_{ij}\delta K^{ij}$ and ${^{(3)}\!R}\delta N$, given by the following expression
\begin{equation}
\label{comb}
-  \delta K^2 + \delta K^{ij} \delta K_{ij} +  \, {}^{(3)}\!R  \, \delta N \;,
\end{equation}
appears in the \acs{EFT} action \cite{Gleyzes:2014rba}.
It follows from the discussion of Sec. \ref{WBG-sectwo} that this ``tuning'' is perfectly stable under loop corrections, as far as $X \lesssim 1$ and $Z \lesssim 1$, because of the (weakly broken) Galileon invariance. 
Moreover, the unique combination \eqref{comb} is in fact a redundant operator and can be removed by a redefinition of the metric \cite{Gleyzes:2014rba,Gleyzes:2015pma}\footnote{This redefinition corresponds to transforming into the frame where the tensor modes propagate at the unit speed of sound \cite{Creminelli:2014wna}; see also the discussion in Sec. \ref{sub-chap1-qf}.}. As a consequence, \textit{the single operator} $\delta N \delta K$\textit{,} \textit{associated with the cubic Galileon, is responsible for all the differences with respect to the more familiar \acs{DBI}/ghost inflation models, as far as stability and power spectra are concerned}\index{inflation!ghost}\index{inflation!DBI}. 

With this in mind, the quadratic action for the comoving curvature perturbation $\zeta$ defined in \eqref{ug-zeta} can be derived using the procedure outlined in Sec. \ref{sec-cswmwg}. Solving Eqs. \eqref{constraint-N}-\eqref{constraint-Ni} at the linear order in perturbation theory and plugging the solutions for $N$ and $N^i$ back into the theory \eqref{EFTI-action-2} with $\bar{M}^2_i=0$, one finds the action \eqref{dS-action} with the values \eqref{mg-N}-\eqref{mg-cs} in the case $\hat{M}^3\equiv\hat{M}^3_1$ and $M^4\equiv M_2^4$.
%
%\begin{equation}
%S_\zeta = \int d^4 x~a^3 \bigg [ A(t) \dot \zeta^2-B(t)\frac{(\vec \nabla\zeta)^2}{a^2}     \bigg ]~,%\end{equation}
%where
%\begin{align}
%\label{A}
%A &= \frac{M_{\rm Pl}^2 (-4M_{\rm Pl}^4 \dot H+3 \hat{M}^6+2 M_{\rm Pl}^2 M^4)}{(\hat{M}^3-2M_{\rm Pl}^2 H)^2}, \\
%\label{B}
%B &= \frac{M_{\rm Pl}^2 (-4M_{\rm Pl}^4 \dot H+2 M_{\rm Pl}^2 H\hat{M}^3 - \hat{M}^6+2 M_{\rm Pl}^2 \partial_t \hat{M}^3)}{(\hat{M}^3-2M_{\rm Pl}^2 H)^2}~,
%\end{align}
%and the speed of sound for short-wavelength perturbations is simply $c_s^2=B/A$.
%
Depending on the values of the parameters $X$ and $Z$, in a theory of inflation with \acs{WBG} invariance there are two phenomenologically distinct regimes, \textit{i.e.} a potentially\index{potentially dominated evolution} ($X\sim\sqrt{\varepsilon}$, $Z\sim 1$) and a kinetically\index{kinetically dominated evolution} ($X\sim Z\sim 1$) driven phase, which will be reviewed in details in the next chapters. These two cases give rise to phenomenologically different \acs{EFT}s, providing different contributions to the coefficients of the Lagrangian operators. Before discussing those situations, we conclude with some remarks about the consequences a \acs{WBG} theory can induce on the late-time evolution of the Universe, in the context of dark energy\index{dark energy}.

\section{Late-time universe in WBG theories}
\label{WBG-secsix}
\subsection{Dark energy}

The theories characterized by \acs{WBG} symmetry can also be applied in the context of the late-time cosmic acceleration\index{dark energy}\index{dark energy}. 
For simplicity, we will neglect the scalar potential $V(\phi)$ here, contrarily to the case discussed in the previous section. Moreover, we reabsorb the canonical scalar kinetic term in the definition of the function $G_2$, so that the action of interest reads as follows
\begin{equation}
\label{lt_full}
S=\int d^4x \sqrt{- g}  \left[ \frac{1}{2} M_{\rm Pl}^2 R +  \sum_{I=2}^5  \mathcal{L}^{\rm WBG}_I \right] ~.
\end{equation}
The homogeneous equations on a flat \acs{FRW} background obtained by varying this action are similar to those, discussed in Sec.~\ref{WBG-secfive}. However, in this section we choose to rewrite them in a slightly different form and notation. 

In terms of the two variables $X$ and $Z$ defined in Eqs.~\eqref{X_def} and \eqref{Z_def1}, the homogeneous scalar evolution equation reads \cite{Kobayashi:2010cm}
\begin{equation}
\frac{\D}{\D t} \left[ a^3\dot\phi F_{\rm DE}(X,Z) \right] =0 \;, \label{EOM_bg}
\end{equation}
where $ F_{\rm DE}(X,Y)$ 
is defined as
\begin{multline}
F_{\rm DE}(X,Z) \equiv G_{2X} -3   Z G_{3X} + 6 Z^2 \left( \frac{G_{4X}}{X} + 2 G_{4XX} \right)
\\
	+ Z^3 \left(3 \frac{G_{5X}}{X} + 2  G_{5XX} \right)  \;.
\end{multline}

In analogy with what we did for inflation, in order to study the dark energy\index{dark energy!EFT of} an \acs{EFT} approach can be adopted as well \cite{Gubitosi:2012hu,Gleyzes:2013ooa,Gleyzes:2014rba,Piazza:2013coa}. Nevertheless some further clarifications are necessary. It is worth emphasizing that the main difference of the late-time Universe with respect to inflation concerns the presence of several matter species. This yields deviations from the effective action \eqref{EFTI-action-2}. Indeed, assuming that the metric $g_{\mu\nu}$ is minimally coupled to matter fields, in principle a general free function of time $f(t)$ in front of the Ricci scalar has to be taken into account. In fact, with respect to inflation, one is not allowed to perform any redefinition of the metric, if a Jordan frame with a minimally-coupled matter has to be retained. Therefore, the first line of Eq. \eqref{EFTI-action-2} is modified as follows:
\begin{equation}
\label{s_pi}
S =   \int \ {\rm d}^4x \, N \sqrt{ \gamma } \, \left[ 
	\frac{1}{2}M_{\rm Pl}^2 f(t)R - 2\dot{f}(t)\frac{K}{N} + \frac{c(t)}{N^2} - \Lambda(t)
	+ \ldots \right] \, .
\end{equation}
The operators shown in \eqref{s_pi} alone contribute to the background evolution. In the case of inflation we had $f=1$ and
\begin{equation}
c = -M_{\rm Pl}^2 \dot H \;, \qquad \Lambda = M_{\rm Pl}^2 (3 H^2+\dot H)  \; .
\end{equation}
The Friedmann equations\index{Friedmann-Robertson-Walker (\ac{FRW})!equations} can be written as
\begin{align}
3 M_*^2 H^2 &=  \rho_{\rm m} + \rho_{\rm DE} \;, \\
- 2 M_*^2 \dot H & =  \rho_{\rm m} + p_{\rm m} + \rho_{\rm DE} + p_{\rm DE} \;,
\end{align}
where $\rho_{\rm m}$ and $p_{\rm m}$ respectively denote the energy density and pressure of matter. The effective Planck mass squared, $M_*^2$, is generically time-dependent due to the non-minimal couplings of the scalar in $\mathcal{L}^{\rm WBG}_4$ and $\mathcal{L}^{\rm WBG}_5$, and is explicitly given by the following expression
\begin{equation}
M_*^2 \equiv M_{\rm Pl}^2 \left(1 + 2 G_4 - 4 X G_{4X} - 2  ZX G_{5X}  \right)\;.
\end{equation}
In terms of the \acs{EFT} parameters of Eq.~\eqref{s_pi}, $M^2_* = M_{\rm Pl}^2 f + \bar M^2$.
Finally, the unperturbed  effective energy density and pressure read
\begin{multline}
\rho_{\rm DE} =  \ \Lambda_2^4 X\left[ - \frac{G_2}{X} + 2  G_{2X} - 6 Z  G_{3X} \right.
\\
 \left. + 12 Z^2 \left(\frac{G_{4X}}{X} +  2  G_{4XX} \right) + 4 Z^3 \left(\frac{G_{5X}}{X} +   G_{5XX} \right) \right] \;,
\label{rhoDE}
\end{multline}
\begin{multline}
p_{\rm DE} +\rho_{\rm DE}= \ 2 \Lambda_2^4 X F_{\rm DE} + 2 M_{\rm Pl}  \ddot \phi X 
	\left[  G_{3X}  - 4 Z \left( \frac{G_{4X}}{X} + 2  G_{4XX} \right) \right.
\\	
	\left. - Z^2 \left(3 \frac{G_{5X}}{X} + 2  G_{5XX} \right)  \right] \;.
\label{pDE}
\end{multline}

For Horndeski theories, the phenomenological deviations from the $\Lambda$CDM model at the level of linear perturbations can be conveniently parametrized in terms of four time-dependent parameters \cite{Gleyzes:2013ooa}. It is convenient to use the  dimensionless functions introduced in \cite{Bellini:2014fua}, which
can be expressed in terms of the \acs{EFT} coefficients of Eq.~\eqref{s_pi} as \cite{Gleyzes:2014rba}
\begin{equation}
\alpha_K =  \frac{2 c +  M^4}{M_*^2 H^2} \; ,   \quad \alpha_B  = \frac{M_{\rm Pl}^2 \dot f - \hat{M}^3}{ 2 M_*^2 H}   \;,   \quad
\alpha_M  =  \frac{M_{\rm Pl}^2 \dot f +  \dot{ \bar M}^2}{M_*^2H }  \;,  \quad
\alpha_T  = - \frac{  \bar M^2}{M_*^2 }  \;, \label{alphaEFT}
\end{equation}
where the function $c$ is fixed by the background Friedmann equations derived from action \eqref{s_pi},
\begin{equation}
2 c = p_{\rm DE} +\rho_{\rm DE} +  M_{\rm Pl}^2 (H \dot f - \ddot f) \;.
\end{equation}
For the theory of interest \eqref{lt_full}, their explicit expressions  are reported in
%App.~\ref{appC}. The contributions from the Lagrangians ${\cal L}_I^{\rm WBG}$ to each of these parameters is given in
Tab.~\ref{tab2}.
\renewcommand{\arraystretch}{2.2}
\begin{table}[t]
\begin{adjustbox}{max width=\textwidth}
\begin{tabular}{|c||c|c|c|c|c|c|c|cl}
  \hline
$\mathcal{L}_I^{\rm WBG}$&${M_*^2 H^2}{\Lambda_2^{-4}} \alpha_K $ & $ M_*^2 H^2 \Lambda_2^{-4} \alpha_B $ &$ M_*^2 H^2 \Lambda_2^{-4} \alpha_M$&  $M_*^2 H^2 \Lambda_2^{-4} \alpha_T $ \\
 \hline\hline
\rule{0pt}{3ex} 
$ I=2 $ & $2X ( G_{2X} + 2 X G_{2XX})$ &$\times$ &$\times$&  $\times$ \\
  \hline
$ I=3$ &$ \displaystyle -  12 X  Z (G_{3X} +  X G_{3XX})$ &$\displaystyle X Z  G_{3X}$ &  $\times$&  $\times$\\
  \hline 
$I=4 $ &  $ \displaystyle 12 X Z^2  \left(\frac{G_{4X}}{X} +8  G_{4XX} + 4X G_{4XXX} \right)$&{$ \displaystyle -4X Z^2 \left( \frac{G_{4X}}{X}+2G_{4XX}\right)$}  & $\displaystyle - \frac{ \ddot \phi  }{\Lambda_3^3}  \alpha_B $&  $\displaystyle 4 Z^2 {G_{4X}}$ \\
 \hline
$ I=5 $ & {$ \displaystyle 4 X Z^3  \left(3 \frac{G_{5X}}{X} + 7 G_{5XX} + 2 X G_{5XXX} \right)$}&{$ \displaystyle - X Z^3  \left(3 \frac{G_{5X}}{X}+2G_{5XX}\right)$} & $\displaystyle  2\frac{\dot H Z^3}{H^2 }  {G_{5X}} - 2\frac{ \ddot \phi  }{\Lambda_3^3}  \alpha_B $&  $\displaystyle 2 Z^2 \bigg( Z  - \frac{  \ddot \phi }{\Lambda_3^3 }  \bigg) {G_{5X}}$ \\
  \hline
\end{tabular}\\
\end{adjustbox}
\caption{Contribution from the Lagrangian terms $\mathcal{L}^{\rm WBG}_I$ to the various effective  $\alpha$-parameters defined in Eq.~\eqref{alphaEFT},  denoting the deviations from the $\Lambda$CDM model in linear perturbation theory.
%Their complete expressions are reported in App.~\ref{appC}, see Eqs.~\eqref{alphaK}--\eqref{alphaT}.
}
\label{tab2}
\end{table}

Let us first briefly comment on the physical meaning of the $\alpha$-parameters and their observational consequences (see, \textit{e.g.} \cite{Bellini:2014fua,Gleyzes:2014rba} for a more detailed discussion).
\begin{itemize}
\item The first of these functions, $\alpha_K$, parametrizes the kinetic energy of  scalar perturbations, induced by the four Lagrangian terms $\mathcal{L}^{\rm WBG}_I$. In terms of the \acs{EFT} for perturbations \eqref{s_pi}, the contributions to $\alpha_K$ arise from the operators $\propto \delta N^2$, described by the coefficients $c$ and  $M^4$. 
This parameter is enough to describe linear perturbations in the minimally coupled quintessence and $k$-essence models. In the minimal case, dark energy\index{dark energy} fluctuations behave as those of a perfect fluid with the speed of propagation $ c_s^2= 3 (1+w_{\rm DE} ) \Omega_{\rm DE} /\alpha_K$, where $\Omega_{\rm DE}$ is the energy density of $\phi$ in the units of the critical one.   
A well-studied example corresponds to the limit $\alpha_K \gg 1$, which leads to  zero sound speed  \cite{Creminelli:2008wc,Creminelli:2009mu}. 

\item The second function, $\alpha_B$, parametrizes the mixing between metric and scalar field fluctuations \cite{Deffayet:2010qz}. It can be induced by the Lagrangians $\mathcal{L}^{\rm WBG}_I$ with $ 3\le I \le 5$, or by  the operator containing $\delta N \delta K$ in the \acs{EFT} for perturbations, which describes kinetic mixing  with gravity. This operator induces a typical scale (approximately given by $k_B/a \simeq 3  H \sqrt{\Omega_{\rm m}  / 2} $ for $\alpha_M  = \alpha_T=0$) \cite{Bellini:2014fua},  below which dark energy\index{dark energy} can cluster with energy density fluctuations $\delta \rho_{\rm DE} \simeq 2 \delta \rho_{\rm m} \alpha_B^2/  \left( (\alpha_K^2 + 6 \alpha_B^2) c_s^2 \right)$. Above this scale, dark energy\index{dark energy} fluctuations simply behave as  those of a perfect fluid.

\item The function $\alpha_M$ is defined as  $\alpha_M \equiv {\rm d} \ln M_*^2 /{\rm d}\ln a$, and is thus related to the time variation of the effective Planck mass $M^2_*$ induced by the Lagrangians $\mathcal{L}^{\rm WBG}_4$ and $\mathcal{L}^{\rm WBG}_5$. It parametrizes the non-minimal coupling in scalar-tensor theories such as Brans-Dicke \cite{Brans:1961sx} and, as such, induces a slip between the gravitational potentials.
In terms of the \acs{EFT} operators, $\alpha_M$ can be generated by time variation of the parameters $f$ and $\bar M ^2$. In particular, in $f(R)$ theories, $\alpha_M = - \alpha_B$. 

\item Finally, the last function, $\alpha_T$, parametrizes the deviation of the speed of sound of tensor perturbations from unity in the presence of the Lagrangian $\mathcal{L}^{\rm WBG}_4$ and $\mathcal{L}^{\rm WBG}_5$ \cite{Kobayashi:2010cm}, and also induces a slip between the gravitational potentials. It is generated by the \acs{EFT} operator, multiplied by the coefficient $\bar M^2$.  
\end{itemize}

A particularly interesting case corresponds to a constant $X$, where the scalar field grows linearly with time, $\phi_0 \propto t$. In this case, the functions $G_I$ and all their derivatives are time-independent when evaluated on the background, while $Z$ changes proportionally to the Hubble rate. One can see from the form of the scalar equation, Eq. \eqref{EOM_bg}, that the $\phi_0 \propto t$ profile can only be a solution if the following relations are satisfied
\begin{equation}
G_{2X} = 0 \;, \qquad G_{3X} = 0 \;, \qquad G_{4X} + 2 X G_{4XX} = 0 \;, \qquad 3 G_{5X} + 2 X G_{5XX} = 0\;. \label{cond1}
\end{equation}
It then follows from Eqs.~\eqref{rhoDE} and \eqref{pDE} that the effective energy density on such backgrounds is given by $\rho_{\rm DE} = - \Lambda_2^4 (G_2 + 2 Z^3 G_{5X})$ and the equation of state is that of the cosmological constant, with $ w_{\rm DE} \equiv p_{\rm DE} / \rho_{\rm DE} =-1$. 
Moreover, while the dimensionless parameters $\alpha_K$, $\alpha_M$ and $\alpha_T$ in Tab.~\ref{tab2} are generally non-zero and time dependent, Eq.~\eqref{cond1} implies that $\alpha_B =0 $. In this case, in order to avoid ghost instabilities one must require that $\alpha_K \ge 0$ \cite{Gleyzes:2013ooa,Bellini:2014fua}.
Note that since $Z \propto H$, $\rho_{\rm DE}$ is time dependent, even though $w_{\rm DE} =-1$. This is because the effective Planck mass varies with time and the dark energy\index{dark energy} does not follow the usual conservation equation. Demanding that the energy density remains finite at early times requires it to be constant, \textit{i.e.} $G_{5X}=0$ and thus that also $\alpha_M=0$, see Tab.~\ref{tab2}. In this case, at the background level this model exactly behaves as a cosmological constant. At the level of linear perturbations, the dimensionless parameters $\alpha_K $ and  $\alpha_T$ do not vanish and contribute to the sound speed of scalar and tensor fluctuations, respectively as $c_s^2 = - 2 \alpha_T/\alpha_K$ and $c_T^2 = 1 + \alpha_T$ (see \cite{Bellini:2014fua,Gleyzes:2014rba}). One can avoid gradient instabilities\index{gradient instability} by requiring the positivity of both these speed of fluctuations squared, which implies $-1 \le \alpha_T \le 0$.

This is the simplest application of \acs{WBG} symmetry to the late-time acceleration. The assumption that  the background profile of the field is linear with time, $\phi_0 \propto t$, and that the background expansion history is the same as in $\Lambda$CDM considerably restricts the values that the parameters $\alpha_a$ can consistently  take. 
These solutions are not possible for the covariant Galileon \cite{Deffayet:2009wt}. In this case one usually assumes a tracker  solution with $Z =$ const.~\cite{DeFelice:2010pv}; this imposes a particular expansion history, which has been shown to lead to observations that are disfavoured with respect to $\Lambda$CDM (see \cite{Barreira:2014jha}  for a recent analysis and \textit{e.g.} \cite{Appleby:2011aa,Neveu:2013mfa} for analysis that do not assume the tracker solution). More sophisticated examples can be constructed using the Lagrangians \eqref{hor1}-\eqref{hor4}, by allowing the background solution for $\phi$ to be different from a linear profile in time. In this case, from Tab.~\ref{tab2} one generically expects that the parameters $\alpha_a$ may assume any value smaller than unity, $\alpha_a \sim X \lesssim 1$.

\subsection{Static sources and the Vainshtein mechanism}

The attempts to modify gravity\index{modified gravity} have a long story. One of the main physical motivations lies on the observation of the current cosmic acceleration\index{cosmic acceleration}.
%as inferred for instance from the type Ia supernovae\index{type Ia supernovae} and the baryon acoustic oscillations\index{baryon acoustic oscillations}.
Such an accelerated expansion can be well due to a cosmological constant\index{cosmological constant}, but it is known to suffer from a serious fine-tuning problem\index{fine-tuning}. Indeed, from \acs{CMB} measurements\index{Cosmic Microwave Background (\ac{CMB})} the vacuum energy density\index{vacuum energy density} is estimated to be $\rho_\text{vac}^{1/4}\sim 10^{-3}\text{ eV}$, which is ‘‘unnaturally'' small compared with the typical energy scales of the known interactions in the Standard Model of particle physics, \textit{e.g.} $\Lambda_\text{QCD}\sim 200\text{ MeV}$, $M_W\sim 80\text{ GeV}$ and $M_\text{Pl}\sim 10^{18}\text{ GeV}$. In other words, according to the standard renormalization procedure in quantum field theory, one should tune the scale in order to have a huge cancellation with the bare quantity, leading to the so-called cosmological constant problem\index{cosmological constant!problem}. This motivated the attempts to modify gravity in the infra-red, trying to justify the current acceleration without any cosmological constant, while the standard Solar System tests force such alternative theories to be degenerate with General Relativity in their predictions at small distances. Some examples of these attempts are provided by massive gravity\footnote{See, for instance, the reviews \cite{Hinterbichler:2011mg,Babichev:2013vm,Rubakov:2008mg} and the references therein.}\index{massive gravity}, the \acs{DGP} model\index{DGP model} \cite{Dvali:2000hr} and the Galileon \cite{Nicolis:2008in}. One common feature is the presence of extra-propagating degrees of freedom with respect to the standard two graviton helicities, that generate non-trivial, large-distance modifications of gravity. On the other hand, non-linear terms in the extra-field can provide screening effects, which allow to fulfil the Solar System tests.

In particular, in the context of theories with \acs{WBG} invariance, one can imagine to couple  $\phi$ to the trace of the matter stress tensor in the following way
\begin{equation}
\mathcal{L}_{\text{matter}}\sim \frac{1}{M_{\rm Pl} }\phi T_{\rm m}~.
\end{equation}
It is then necessary to make sure that the exchange of $\phi$ by realistic sources does not modify the Newtonian potential by any detectable amount. 
There are a number of ways to suppress the contribution from $\phi$ to the gravitational potential of astrophysical objects (stars, planets, etc.) in modified gravity. In the case of the theories with approximate Galileon invariance we are interested in, what guarantees that $\phi$ is screened beyond the observable values is the Vainshtein mechanism\index{Vainshtein!mechanism} \cite{Vainshtein:1972aa,Deffayet:2001uk}, crucially relying on non-linear contributions, as stated above. Indeed, in the decoupling limit\index{decoupling limit} \eqref{declim}, the theory reduces to the Galileon, which is known to naturally incorporate the screening below the Vainshtein radius\index{Vainshtein!radius} \cite{Nicolis:2008in}, which for a source of mass $M_{\rm source}$ reads
\begin{equation}
r_V=\left(\frac{M_{\rm source}}{M_{\rm Pl} \Lambda^3_3}\right)^{1/3}~.
\end{equation}
The decoupling limit, on the other hand, does capture all relevant astrophysical scales and all results obtained in that limit can be fully trusted. To see this directly, one can evaluate the quantity $X$ on a pure Galileon solution of Ref. \cite{Nicolis:2008in} and check that it is extremely small everywhere in space\footnote{In the case of the cubic Galileon, $X$ becomes of order one only at the Schwarzschild radius\index{Schwarzschild radius} of the source $r_S$, while, in the most general case, it is a constant within the Vainshtein radius\index{Vainshtein!radius}, of order of the tiny ratio $(r_S H_0)^{2/3}$, where $H_0$ is the current Hubble rate.}, so that all the results of Ref. \cite{Nicolis:2008in} apply to our case without any modification. In particular, for Galileons to be cosmologically relevant, one needs \cite{Nicolis:2008in} $\Lambda_3\sim (M_\text{Pl}H_0^2)^{1/3}\sim 10^{-3}\text{ Km}^{-1}$, where $H_0$ can be read in Tab. \ref{planck2015parinf}. For the solar mass, $M_{\odot}\sim 10^{57}\text{ GeV}$, one estimates\footnote{See also \cite{Andrews:2013} for a more quantitative analysis.} $r_V\sim 10^{16}\text{ Km}$, well beyond the Pluto's orbit radius.

%% file: chapters/2-03-chapter.tex
\chapter{Slow-roll inflation with WBG symmetry: the potentially driven phase}
\label{WBG-pot}

\begin{flushright}{\slshape    
    The next question is, what can we make out of laws which are nearly symmetrical?} \\ \medskip
    --- R. Feynman, R. B. Leighton and M. Sands, \textit{The Feynman Lectures on Physics}, Volume I.
%    --- \defcitealias{bentley:1999}{BBBBB}\citetalias{bentley:1999} \citep{bentley:1999}
\end{flushright}

% For an example of a full page figure, see Fig.~\ref{fig:myFullPageFigure}.

As already briefly discussed in Sec. \ref{N-Gsec}, it is a common feature of slow-roll models of inflation that the magnitude of non-Gaussianity\index{non-Gaussianity} is suppressed \cite{Maldacena:2002vr}, well below the observable level for any foreseeable future (we refer, for instance, to \cite{Alvarez:2014vva}).

Inspired by the features of \acs{WBG} theories\index{inflation!WBG}, we propose a novel slow-roll scenario where scalar perturbations propagate at a subluminal speed, leading to sizeable equilateral non-Gaussianity, $f_{\rm NL}\propto 1/c_s^4$, largely insensitive to the ultraviolet physics.

\section{Preliminary remarks}

In Part I, we have seen that in a generic low-energy inflationary \ac{EFT} additional higher-derivative corrections to the canonical action can not impact non-Gaussianity in any significant way without spoiling the consistency of the theory.
Only the presence of symmetries can justify an amplitude $f_{\rm NL}\gtrsim 1$.
As pointed out in Sec. \ref{subsec-stucktrick-dc}, a clear example of this fact is provided by \acs{DBI} inflation\index{inflation!DBI} with action \eqref{4_0-action2}.
%As pointed out in Sec. \ref{subsec-stucktrick-dc}, a clear example of this fact is provided by \acs{DBI} inflation\index{inflation!DBI} with action \eqref{4_0-action2}, which for a small speed of sound of the scalar perturbations, $c_s^2\ll 1$, predicts equilateral non-Gaussianity \cite{Creminelli:2005hu} with the amplitude  $f_{\rm NL}\sim 1/c_s^2$ \cite{Alishahiha:2004eh,Chen:2006nt}. Without the higher-dimensional space-time symmetry, non-linearly realized\index{symmetry!non-linearly realized} on $\phi$, allowing a re-summation of the series \eqref{eftdbi} and protecting the coefficients of the leading derivative operators from large quantum corrections, given the estimation \eqref{L3-hdnG} \cite{Creminelli:2003iq}, the prediction $f_{\rm NL}\gtrsim 1$ would have been unreliable, because the effective theory is tenable only for $(\partial \phi)^2\ll \Lambda^4$.

Now we study in details the potentially driven inflationary scenario introduced in the previous chapter. The energy density of the early Universe is dominated by the potential\index{potentially dominated evolution} of a slowly-rolling\index{inflation!slow-roll} scalar field, similarly to ordinary slow-roll theories. Yet, a definite set of higher-derivative interactions of the inflaton become relevant, leading to observably large non-Gaussianity. Nevertheless, the theory is predictive, since all the rest of the operators in the derivative expansion remain \textit{naturally} small in the full quantum theory. 
All these properties are consequences of the underlying \acs{WBG} invariance\index{Weakly Broken Galileon (\acs{WBG})!invariance}.

Just like in \acs{DBI} inflation\index{inflation!DBI}, enhanced scalar non-Gaussianity is generically associated with a reduced speed of sound of perturbations in our model. However, the enhancement is much stronger compared to the \acs{DBI} case, the amplitude of equilateral non-Gaussianity growing as $f_{\rm NL}\propto 1/c_s^4$ for small $c_s^2$.

\section{The slow-roll background}

We have shown in Chap. \ref{chap-WBG} that the properties of the theories with \acs{WBG} symmetry imply the possibility of a moderately coupled, yet predictive, regime characterized by
\begin{equation}
\label{stability}
X=\frac{\dot\phi_0^2}{\Lambda_2^4}\lesssim 1 ~, 
\qquad
Z\equiv \frac{H \dot\phi_0}{\Lambda_3^3}\lesssim 1~,
\end{equation}
for a homogeneous $\phi$-profile on a \acs{FRW} background with the Hubble rate $H$. From now on, $X$ will be understood as evaluated on the background solution. 
Despite the moderate coupling, quantum corrections are under control when the scalar background profile satisfies \eqref{stability}, even in the case that these inequalities are saturated, and the predictions of the classical theory can be trusted. 

As noted above, we are interested in the potentially-dominated models of inflation, characterized by \acs{WBG} invariance. These are governed by the Lagrangian \eqref{full}, where, since we have extracted the  canonical scalar and graviton kinetic terms, $G_2$ is assumed to start at least quadratic in $X$, while $G_3$ can have a linear piece. From now on we will set $G_4 = G_5 =0$ for the sake of simplicity;  generalization to the case of non-zero $G_4$ and $G_5$ is straightforward and will be commented on where appropriate. The ellipses in \eqref{full} denote an infinite number of other operators, present in the low-energy effective theory. 
We will assume that  the potential  $V(\phi)$ satisfies the ordinary slow-roll conditions, $\varepsilon_V \ll 1$ and $|\eta_V | \ll 1$, where the (potential-based) slow-roll parameters $\varepsilon_V$ and $\eta_V$ are defined as\index{slow-roll conditions}
\begin{equation}
\varepsilon_V \equiv  \frac{M_{\rm Pl}^2}{2} \left( \frac{V'}{V} \right)^2 \;,
\qquad
\eta_V \equiv M_{\rm Pl}^2 \frac{V''}{V}\;.
\label{srp}
\end{equation}
The analysis of quantum loops of Chap. \ref{chap-WBG}, leading to the non renormalization theorem\index{non-renormalization theorem} summarized above, has concentrated on the case with a vanishing potential. 
It is straightforward to show that the same results remain intact also in the presence of a non-zero, but sufficiently flat, $V(\phi)$ satisfying \eqref{srp}: this is shown in App. \ref{Appendix-WBG}, Sec. \ref{App-inf-pot}.

For the flat \acs{FRW} ansatz, $\D s^2 = -\D t^2 + a^2(t) \D \vec x^2$, the two Friedmann equations\index{Friedmann-Robertson-Walker (\ac{FRW})!equations} that follow from \eqref{full} are given by Eqs. \eqref{fried1}-\eqref{fried2}, which in the case $G_4 = G_5 =0$ reduce to
\begin{align}
\label{fried1-pot}
3M_{\rm Pl}^2H^2&=V - \Lambda_2^4 X\left[  \frac{1}{2}+ \frac{G_2}{X}-2 F(X,Z) \right] \, , \\
\label{fried2-pot}
M_{\rm Pl}^2\dot H& =-\Lambda_2^4 X F(X,Z) +M_{\rm Pl} X G_{3X} \ddot \phi_0~,
\end{align}
where
\begin{equation}
F(X,Z)= \frac{1}{2}+G_{2X}-3 Z G_{3X} \, .
\label{F-pot}
\end{equation}
Moreover, in the slow-roll regime ($\dot F\ll H F$, $\ddot\phi \ll H \dot\phi$) the homogeneous equation of motion of $\phi$ reduces to
\begin{equation}
\label{phiequation}
6 H \dot\phi_0 F(X,Z) \simeq -V'(\phi_0)~.
\end{equation}

We are interested in a regime where higher-derivative operators in \eqref{full} become important, while the quantum corrections are still under control. To this end, we assume  $Z\sim 1$, which also fixes the magnitude of the parameter $X$. Indeed, from the definition of $X$ and $Z$, Eq.~\eqref{stability}, it follows that
\begin{equation}
\sqrt{X}=\frac{\Lambda_2^2 Z}{M_{\rm Pl}H} \, .
\label{zepx}
\end{equation}
Making use of Eq. \eqref{fried2-pot}, one immediately obtains
\begin{equation}
X\sim \sqrt{\varepsilon} \, .
\label{xep}
\end{equation}
Note the order-unity slowly varying function of time $F(X,Z)$ in Eqs. \eqref{fried1-pot}-\eqref{phiequation}, which is strictly equal to $1/2$ in canonical slow-roll inflation\index{inflation!slow-roll}. Apart from this minor modification, all the equations that describe the background solution are similar to those of ordinary slow-roll models (up to corrections of higher order in $\varepsilon_V$ and $\eta_V$). In particular, unlike \textit{e.g.} the \acs{DBI} case, the usual flatness conditions $\varepsilon_V \ll 1$ and $|\eta_V | \ll 1$ need to hold for sustaining the quasi-de Sitter phase in our model. It is precisely for this reason that we refer to it as ``slow-roll''\footnote{Inflationary models based on particular subsets of the Lagrangian terms in \eqref{hor1}-\eqref{hor4}, \textit{G-inflation}\index{inflation!G-} and \textit{Galileon inflation}\index{inflation!Galileon}, have been studied in Refs. \cite{Kobayashi:2010cm, Burrage:2010cu} respectively. However, these references have focused on kinetically-driven inflation, corresponding to $\Lambda_2^4 \sim M_{\rm Pl}^2 H^2$ and $X \sim Z \sim 1$ in our notation (see also Tab. \ref{tabsummodels}).}.
At the level of perturbations, our scenario is of course very different from the canonical slow-roll inflation\index{inflation!slow-roll}; for example, unlike the latter, the scalar perturbations become strongly coupled at an energy scale parametrically smaller than $M_{\rm Pl}$, something we discuss in greater detail below.
 
It follows from the Friedmann equations that the contributions from the derivative operators in \eqref{full} to the inflationary energy density and pressure are proportional to $X\sim \sqrt{\varepsilon}$. One may wonder therefore whether loop corrections can outweigh these contributions for small values of $\varepsilon$. For $Z\sim 1$, the leading quantum corrections to the background stress tensor scale as $\sim \Lambda_3^4$. This should be much smaller than $\Lambda_2^4 \sqrt{\varepsilon}$, which implies a lower bound on the slow-roll parameter, $\varepsilon \gg(H/M_{\rm Pl})^2$. This is  the same bound on $\varepsilon$ as the one that arises from requiring quantum fluctuations of the inflaton to be small \cite{Linde:1986fd,Goncharov:1987ir}.

\section{Non-Gaussianity}

We now adopt the \acs{EFT} framework with Lagrangian \eqref{EFTI-action-2} with only the effective coefficients\index{inflation!EFT of} $M_2^4$, $M_3^4$, $\hat{M}_1^3$ and $\hat{M}_2^3$. 
In terms of the functions $G_2$ and $G_3$, these are 
\begin{align}
M_2^4 &= -2\Lambda_2^4 X\left[ 3Z G_{3X} +6 Z X G_{3XX}-2 X G_{2XX}-Y G_{3X}\right] \;,
\label{M24} \\
M_3^4 &=-2\Lambda_2^4 X \bigg[3 X G_{2XX}+\frac{2}{3} X^2 G_{2XXX} \nonumber \\
&\quad -Z \left(4 G_{3X}+11 XG_{3XX}+2 X^2G_{3XXX}\right)+Y G_{3X}\bigg] \;, 
\label{M34}\\
\hat M_1^3 &= -2\frac{\Lambda_2^4 X}{H} Z G_{3X} \, ,
\quad
\hat M_2^3 =-2\frac{\Lambda_2^4 X}{H}Z\left(2 G_{3X}+X G_{3XX}\right) \;,
\label{hM12}
\end{align}
where $Y = \ddot \phi_0/\Lambda_3^3$.
One can see from \eqref{M24}-\eqref{hM12} that approximate invariance\index{Weakly Broken Galileon (\acs{WBG})!invariance} under Galileon transformations imposes the following (radiatively stable) hierarchy among the various \acs{EFT} coefficients:
\begin{equation}
\label{eftmagnitudes}
M_2^4\sim M_3^4\sim M_{\rm Pl}^2 \dot H \, ,
\qquad
\hat M_1^3\sim \hat M_2^3\sim \frac{M_{\rm Pl}^2 \dot H}{H}~.
\end{equation} 
This is in stark contrast to what happens \textit{e.g.} in solely shift-symmetric theories, where the coefficients that stem from higher-derivative operators, such as $\hat M_1^3$ and $\hat M_2^3$, are much stronger suppressed. The latter hierarchy motivates to define the dimensionless, order-one coefficients\footnote{In order to deal with order-one coefficients, in this chapter we prefer the definitions \eqref{alphas} with respect to the previously introduced \eqref{coeffs}-\eqref{coeffs-bis}.}
\begin{equation}
\label{alphas}
\alpha_{1,3} \equiv - \frac{M_{2,3}^4}{2 M_{\rm Pl}^2 \dot H}\;,
\quad
\alpha_{2,4} \equiv - \frac{\hat M_{1,2}^3 H }{2  M_{\rm Pl}^2 \dot H} \;,
\end{equation}
convenient for describing the parameter space of the theories at hand. 

At sufficiently high energies, the dynamics of scalar perturbations is fully dominated by the dynamics of the adiabatic mode $\pi$, defined in Sec. \ref{sec-efti}. The scalar action \eqref{EFTI-action-2} in the decoupling limit\index{decoupling limit} reads (see also Sec. \ref{sec-stucktrick-dc})
\begin{multline}
\label{dlaction}
S_\pi = \int \D^4 x~a^3\left(-M_{\rm Pl}^2 \dot H\right)\bigg[ (1+\alpha_1)\left(\dot\pi^2-c_s^2 \frac{(\partial\pi)^2}{a^2}\right)    \\
+ \left(  {\alpha_2}-\alpha_1 \right) \dot\pi\frac{(\partial\pi)^2}{a^2} - 2 (  \alpha_1 + \alpha_3) \dot\pi^3 \\
+ 2 \frac{ \alpha_2 - \alpha_4}{H}\dot\pi^2 \frac{\partial^2\pi}{a^2} + \frac{\alpha_2}{H} \frac{(\partial\pi)^2\partial^2\pi}{a^4}  \bigg]~,
\end{multline}
where the speed of sound is
\begin{equation}
\label{cssq-pot}
c_s^2 \equiv  \frac{1 + \alpha_2}{1 + \alpha_1}  \;.
\end{equation}
It follows from Eq.~\eqref{cssq-pot} that if, for whatever reason, the parameter $\alpha_2$ happens to be close to $-1$, one can have strongly subluminal scalar perturbations. Most importantly, the approximate Galileon invariance guarantees that such an ``accidental'' arrangement of the parameters is respected by loop corrections. 
This is qualitatively different from how the small $c_s^2$ arises in models such as \acs{DBI}\index{inflation!DBI} inflation\footnote{In the limit $X\rightarrow 0$, using Eq. \eqref{M24}-\eqref{M34} with $G_4 = G_5=0$ in Eq. \eqref{alphas}, one finds $\alpha_1= 3\alpha_2$. This gives a negative kinetic term to $\pi$ for $\alpha_2 \simeq -1$. The parameter $X$ needs not to be very small, however; it is of order $X \sim \sqrt{\varepsilon}\simeq \text{a few}\times 0.1$  in \textit{e.g.} slow-roll models with monomial potentials. Moreover, the relation  $\alpha_1 = 3\alpha_2$ no longer holds for $G_4 \neq 0$ or $G_5 \neq 0$.}.

It is worth stressing at this point that the operators in the last line of \eqref{dlaction} can be  rewritten in terms of those in the second line via a perturbative field redefinition \cite{Creminelli:2010qf}. 
This simply amounts to use the linear equation of motion in \eqref{dlaction}. After straightforward manipulations, one finds\footnote{Explicitly, the coefficients $\gamma_1$ and $\gamma_2$ read
\begin{align}
\gamma_1 &\equiv  (c_s^2 - 1) \left(1+\frac{2}{c_s^2}\right) + (1+c_s^2)\alpha_1 \;, \nonumber\\
\gamma_2  &\equiv 2 \left(1- \frac1{c_s^{2}} \right) \left(2+ \frac1{c_s^{2}}\right) + \frac{2}{c_s^2} (  \alpha_1 - 2 \alpha_4) + 2 \alpha_1 - 2 \alpha_3 
\;.\nonumber
\end{align}} 
\begin{equation}
\label{dbiops}
S^{(3)}_{\pi} = \int \D^4 x ~a^3\left(- M_{\rm Pl}^2 \dot H\right)\left[\gamma_1 \dot\pi\frac{(\partial\pi)^2}{a^2} + \gamma_2  \dot\pi^3\right] \;.
\end{equation}
The two operators in \eqref{dbiops} are precisely those appearing in the decoupling limit\index{decoupling limit} of \acs{DBI} theories and the bispectrum they produce is close to the equilateral shape \cite{Creminelli:2005hu}. The genuine difference arises once the \textit{magnitude} of non-Gaussianity is concerned: instead of the $f^{\rm equil}_{\rm NL}\sim 1/c_s^2$ behaviour -- see Eq. \eqref{fnldbifull} -- characteristic of \acs{DBI} inflation\index{inflation!DBI}, in theories with \acs{WBG} symmetry non-Gaussianity scales as $f^{\rm equil}_{\rm NL}\sim 1/c_s^4$ in the small-$c_s^2$ limit. The latter scaling is due to the last operator in \eqref{dlaction}, whose precise contribution to the three-point function\index{bispectrum} of the curvature perturbation $\zeta$ reads\footnote{For notations we refer to Sec. \ref{N-Gsec}. In order to match the conventions the relation $\Phi=\frac{3}{5}\zeta$ has to be used.} \cite{Burrage:2010cu,Mizuno:2010ag}
\begin{multline}
\label{bzeta}
B_\zeta(k_1,k_2,k_3) =-\frac{1}{16}\frac{H^8}{A^2}\frac{\alpha_2}{1+\alpha_1}\frac{1}{c_s^{10}} \frac{k_1^2(k_1^2-k_2^2-k_3^2)}{k_t (k_1k_2k_3)^3} \\
\times \left(1+\frac{ \sum_{i>j}k_ik_j}{k_t^2}+3\frac{k_1k_2k_3}{k_t^3}\right)+2~\text{perms}~,
\end{multline}
where $k_t\equiv k_1+k_2+k_3$.
Here, $A$ denotes the normalization of the $\pi$-kinetic term in the decoupling limit, $A\equiv (-M_{\rm Pl}^2 \dot H)(1+\alpha_1)$. 
The amplitude of non-Gaussianity can be directly read off from Eq. \eqref{bzeta},
\begin{equation}
\label{fnl-pot}
f^{\rm equil}_{\rm NL}=\frac{5}{18}\frac{B_\zeta(k_*,k_*,k_*)}{P^2_{\zeta} (k_*)}=\frac{65}{162}\frac{\alpha_2}{1+\alpha_1}\frac{1}{c_s^4}~,
\end{equation}
where $P_{\zeta} (k) = H^4/(4Ak^3c_s^3)$ is the power spectrum\index{power spectrum} \eqref{dS-ps}, evaluated at a fiducial momentum scale $k_*$. A significantly reduced speed of sound, $\alpha_2 \simeq -1$ (see Eq. \eqref{cssq-pot}), implies a \textit{negative} $f_{\rm NL}$ . However, due to the strong dependence of $f_{\rm NL}$ on $c_s$, even slightly subluminal perturbations can produce a sizeable amount of non-Gaussianity; for example, a $ 10\%$ tuning of the $\alpha_2$ parameter can give rise to $f^{\rm equil}_{\rm NL}\simeq -70$ for $\alpha_1=1$ and $\alpha_2=-0.9$. We stress that such a tuning is not ``unnatural'' as a result of the non-renormalization theorem outlined in Chap. \ref{chap-WBG}.

The relative magnitudes of the \acs{EFT} coefficients, expressed in terms of the dimensionless quantities \eqref{coeffs}-\eqref{coeffs-bis}, in various single-field models of inflation are summarized in Tab. \ref{tabsummodels} of Chap. \ref{chap-WBG-constraints}.
Note that, the amplitude of non-Gaussianity in Galileon inflation\index{inflation!Galileon} \cite{Burrage:2010cu} scales as $f_{\rm NL}\sim 1/c_s^2$ \cite{Kobayashi:2011pc}, in the small $c_s^2$ limit. This is no longer true once $\mathcal{L}_4^{\text{WBG}}$ and $\mathcal{L}_5^{\text{WBG}}$ are included, in which case $f_{\rm NL} \propto c_s^{-4}$ also in Galileon\index{inflation!Galileon} inflation\footnote{Refs. \cite{Kamada:2010qe,Ohashi:2012wf,Gao:2011qe} have considered theories somewhat related to ours. However, these works have concentrated on the opposite regime, $Z\gg 1$. As a result, the $f^\text{equil}_\text{NL}\propto 1/c_s^4$ behaviour for $G_4=G_5=0$ has not been noticed in those papers as well.}.

If the theory \eqref{dlaction} is to be predictive, it is crucial that $\pi$ is weakly coupled at energies of the order of the inflationary Hubble rate, $\Lambda_\star\gg H$, where $\Lambda_\star$ is the strong coupling scale defined in Chap. \ref{chap-EFTinflation}, at which perturbative unitarity\index{unitarity bound} is violated in the $2\rightarrow 2$ scattering of $\pi$. In the $c_s^2\ll 1$ limit, $\Lambda_\star$ is set by the last interaction term in the action \eqref{dlaction}, and is estimated as $\Lambda_\star\sim \Lambda_3  c_s^{11/6}$. For $\alpha_1 \simeq 1$ and $\alpha_2 \simeq -1$,  using the experimental value $\Delta_\zeta^2 \simeq 2.5 \times 10^{-9}$,  one finds
\begin{equation}
\Lambda_\star^3 \sim \frac{\mathcal{O}(50)}{\big| {f_{\rm NL}^{\rm equil}} \big| }(8 H)^3 \;.
\end{equation}
Even for the largest $f_{\rm NL}^{\rm equil}$ compatible with the current observational bounds \cite{planckXVII:2015}, the strong coupling scale\index{strong coupling scale} is fairly above $H$, but well below the symmetry breaking scale\index{symmetry!breaking scale},
\begin{equation}
\Lambda_\text{b}^3 \sim {\mathcal{O}}(5) \big| {f_{\rm NL}^{\rm equil}} \big| \Lambda^3_\star  \, .
\label{wbg-sbs}
\end{equation}

%\begin{table}[t]
%\begin{center}
%\renewcommand{\arraystretch}{1.1}
%\begin{tabular}{|c||c|c|c|}
%  \hline
%{\textit {Inflationary model}}&${ |M_{\rm Pl}^2 \dot H|}$ & $ | M_2^4|$& $ |H\hat M_1^3|$ \\
% \hline\hline
% Canonical slow-roll\index{inflation!slow-roll}  & $1$ &$\sim 0$&$\sim 0$\\
% \hline
% \acs{DBI}\index{inflation!DBI} \cite{Silverstein:2003hf,Alishahiha:2004eh}  & $1$ &$\gg 1$&$\sim 0$\\
%  \hline
% Ghost\index{inflation!ghost} \cite{ArkaniHamed:2003uz,Senatore:2004rj} & $\sim 0$ &$ 1$&$\sim 0$\\
%  \hline
% Galileon \cite{Kobayashi:2010cm, Burrage:2010cu} & $\sim 0$ &$ 1$&$\sim 1$\\
%  \hline
% \acs{WBG}\index{inflation!WBG}  & $1$ &$\lesssim 1$&$\sim 1$\\
 % \hline
%\end{tabular}
%\caption{Relative magnitudes of the most relevant quadratic operators in \eqref{EFTI-action-2} in various inflationary models. 
%}
%\label{tab}
%\end{center}
%\end{table}

We close this section with a remark concerning the regime of validity of the decoupling limit\index{decoupling limit}. For the small values of the speed of sound we are interested in, one should be  careful with mixing terms that involve spatial derivatives. For example, consider the $\delta N\delta K$ operator in Eq. \eqref{EFTI-action-2}.  The most important mixing of scalar modes with gravity that arises from this operator is suppressed by a factor of $\varepsilon/c_s^2$ compared to the $\pi$-kinetic term at horizon crossing. Therefore, the validity of the decoupling limit analysis requires that $ c_s^2 \gtrsim \varepsilon$ hold, which puts an upper bound ($f_{\rm NL}\lesssim 1/\varepsilon^2$) on the amplitude of non-Gaussianity attainable within the decoupling limit.

%% file: chapters/2-04-chapter.tex
\chapter{The kinetically driven phase}
\label{WBG-kin}

\begin{flushright}{\slshape    
    SALVIATI: It is true that the Copernican system creates disturbances in the Aristotelian Universe, but we are dealing with our own real and actual Universe.} \\ \medskip
    --- Galileo Galilei, \textit{Dialogue Concerning the Two Chief World Systems}.
%    --- \defcitealias{bentley:1999}{BBBBB}\citetalias{bentley:1999} \citep{bentley:1999}
\end{flushright}

% For an example of a full page figure, see Fig.~\ref{fig:myFullPageFigure}.

In the previous chapter, we have studied a potentially dominated inflationary evolution with an underlying weakly broken Galileon symmetry. On the theoretical side, \acs{WBG} invariance has provided a notable extension of the non-renormalization theorem, which turns out to be only ``slightly violated'' if a suitable coupling to gravity is chosen, selecting a very special sub-class of the Horndeski theories.
%increasing the reliability of a particular sub-class of the Horndeski theories.
On the phenomenological side, it yielded an unnoticed enhancement of the amplitude of non-Gaussianity, with respect to the \acs{DBI} model\index{inflation!DBI} and the standard Galileon inflation\index{inflation!Galileon}.

In this chapter, we consider a class of \acs{WBG} theories\index{inflation!WBG} with negligible potential in the evolutionary equation \eqref{fried1}. In this case, the dynamics is driven by the derivative operators.
We will prove that in this regime the decoupling limit does not apply: one can not disentangle the scalar mode from the gravity sector, which turns out to be order-one important at all energy scales. This has in principle dramatic phenomenological consequences, dominating non-Gaussian effects and giving rise to a strong enhancement,
\begin{equation}
\label{largeng}
f_{\rm NL}\propto \frac{1}{c_s^6}~,
\end{equation}
in the limit of small speed of sound.
Such an abrupt growth of $f_{\rm NL}$ would have never been seen from a decoupling limit perspective, which gives $f_{\rm NL}\propto 1/c_s^2$ for the same region of the parameter space\footnote{See Refs. \cite{Burrage:2010cu,Mizuno:2010ag} for the treatment in the decoupling limit.}.
We anticipate that an amplitude of the form \eqref{largeng} is already ruled out by the existing experimental bounds on primordial gravitational waves. However, we find of some theoretical interests to briefly present the derivation.
%Nevertheless, giving up the excluded region of the parameter space where the enhancement \eqref{largeng} arises, we will show in the next chapter that our kinetically driven phase can fit the experimentally allowed region, providing a possible explanation of the inflationary stage.

\section{Decoupling limit unreliability}

The strongly coupled backgrounds -- \textit{i.e.} the ones for which all terms 
\eqref{hor1}--\eqref{hor4} are of the same order -- correspond to the function $F(X_0,Z_0)$ vanishing at the leading order, as already noted in Chap. \ref{WBG-secfour}. For $X$ and $Z$ strictly constant, one has an exact de Sitter space, which can be made quasi-de Sitter by turning on a small potential for $\phi$. One can see from Eq.~\eqref{f} that for $F$ to vanish requires at least $Z_0$ to be order-one. Although less evident, it is easy to show that also $X_0$ has to be order-one, if the potential is to provide a sub-leading contribution to the energy density. Indeed, it follows from Eq.~\eqref{fried1} that the energy density scales as $M_{\rm Pl}^2 H^2\sim \Lambda_2^4 X$ in this case. Together with the following equality, 
\begin{equation}
\label{zandx}
Z=\sqrt{X}~\frac{M_{\rm Pl} H}{\Lambda_2^2}~,
\end{equation}
which simply follows from the very definitions of $X$ and $Z$,  this implies that $Z\sim X$.
One thus concludes that whenever the theory approaches strong coupling on a (quasi) de Sitter background, $Z\sim1$, all functions $G_I$ should be generically re-summed. 

For large backgrounds that nearly saturate the bound \eqref{stability}, one consequence of the approximate symmetry under \eqref{g_inv} is the following \textit{radiatively stable} hierarchy 
\begin{equation}
\label{hierarchy 1}
M_{\rm Pl}^2 H^2 \sim M^4\sim  \hat{M}^3 H \sim ~ \bar M^2 H^2  \;,
\end{equation}
as can be seen from Tab.~\ref{tab1}.
Therefore, the strongly-coupled case at hand corresponds to the largest possible values of the coefficients $M^4_i$, $\hat{M}^3_i$ and $\bar M^2_i$ in the effective action \eqref{EFTI-action-2}, \textit{i.e.} $M_{\rm Pl}^2 H^2$, $M_{\rm Pl}^2 H$ and $M_{\rm Pl}^2$ respectively. 
Remarkably enough, for a sufficiently large hierarchy between $\Lambda_2$ and $\Lambda_3$, or equivalently between $M_{\rm Pl}$ and $\Lambda_3$, the strong coupling of our backgrounds does not necessarily imply the breakdown of the classical description, as follows from the estimates of Eq.~\eqref{estimates}.
%For example, in terms of the \acs{EFT} language, the operator $(\Box\phi)^2/\Lambda_3^2$ leads to the $\delta K^2$ term of order $\delta\bar M^2\sim \Lambda_2^4/\Lambda_3^2\sim M_{\rm Pl}^2 (H/M_{\rm Pl})^{2/3}$, which is suppressed by a tiny factor compared to the leading contribution\footnote{Note that quantum corrections generically introduce higher time derivatives (\textit{e.g.} $\delta \dot N^2$) at the same scales as the higher space derivatives. These lead to loss of unitarity associated with ghost excitations with masses above the \acs{EFT} cutoff (cured by whatever completes the theory in the UV). The scales associated with the higher space- and time-derivative operators can only be disentangled when the speed of sound is small, $c_s\ll1$, in which case the latter scale becomes enhanced by powers of $c_s^{-1}$.}. 

As anticipated, there is another important characteristic to the ``large'' backgrounds $X\sim 1$, $Z\sim 1$ -- or, equivalently, to large values of the \acs{EFT} coefficients given in Eq. \eqref{hierarchy 1}. Assuming $M_{\rm Pl}^2 \dot H \lesssim M^4 $, as required by being close to de Sitter space, there is no short-distance limit in which the metric can be decoupled from the inflaton perturbations. Therefore, \textit{mixing with gravity is order-one important at all scales}. To see this, let us restore the Goldstone boson\index{Goldstone field} $\pi$, non-linearly realizing the spontaneously broken time diffeomorphisms\index{time-diffeomorphisms!breaking of}: $$g^{00} \to g^{00}=-N^{-2}=-1-2\dot\pi-\dot\pi^2+\frac{(\partial\pi)^2}{a^2}, \qquad \delta K\to \delta K-\frac{\partial^2\pi}{a^2}+\mathcal{O}(\pi^2) ~.$$ 
The relevant part of the $\pi$ action then reads
\begin{equation}
\label{piaction}
-M_{\rm Pl}^2 \dot H\frac{1}{N^2}+ \frac{1}{2} M^4 (\delta N)^2 -\hat{M}^3 \delta N\delta K\to \dot\pi_c^2+\frac{\hat{M}^3}{M_{\rm Pl} f^2_\pi}\delta N_c\frac{\partial^2 \pi_c}{a^2} + \dots \;,
\end{equation}
where we have defined the ``decay constant'' $f^2_\pi \equiv ( M^4/2-M_{\rm Pl}^2\dot H )^{1/2}$ and canonically normalized the Goldstone and the lapse variable as $\pi=\pi_c/f_\pi^2$ and $\delta N=\delta N_c/M_{\rm Pl}$, respectively. One can see from the last term in this expression, that the mixing with gravity is indeed important for the Goldstone dynamics at all distance scales if the \acs{EFT} coefficients saturate the strong coupling bound (\textit{i.e.} if $\hat{M}^3 \sim M_{\rm Pl} M^2$). This means that, in this case, one has to perform the full analysis including dynamical gravity in order to extract reliable predictions from the theory. The calculation of the cubic action and non-Gaussianity in this regime, much more involved due to the non-decoupling of gravitational effects, has been reported in Sec. \ref{sec-cswmwg}.

In terms of the definitions \eqref{coeffs}-\eqref{coeffs-bis}, the kinetically driven inflation with weakly broken Galileon symmetry (\acs{KWBG}) occurs whenever
\begin{equation}
\alpha , \, \beta , \,  \gamma , \,  \delta \sim 1 \, .
\label{defKWBG}
\end{equation}
This hierarchy has important physical consequences, as anticipated at the end of Chap. \ref{chap-EFTinflation}. Indeed, it means that not only the underlying weakly broken symmetry allows to explore a wider region of the parameter space than that one defined by $\hat{M}^3H\ll M^4$, which one would have inferred requiring naively the validity of the derivative series in the effective Lagrangian, but also a larger set of values for the effective parameters, compared for instance to slow-roll models and \acs{DBI} inflation, is under control. The comparison with the recent experimental bounds will be presented in the next chapter.

\section{KWBG inflation}

We now give a detailed account of the kinetically driven inflation with weakly broken Galileon symmetry, characterized by the hierarchy \eqref{defKWBG}. Well within the regime of validity of the low energy \acs{EFT}, the fast growth of non-Gaussianity in Eq. \eqref{largeng} arises as a result of subtle effects associated with the mixing of the adiabatic perturbations with the metric degrees of freedom, partially explaining why it has gone unnoticed in the literature. As we will discuss in the next chapter, such an amplitude is ruled out by experimental constraints and the \acs{KWBG} regime \eqref{defKWBG} has to be slightly modified to be consistent with observations. Nevertheless, we present here the derivation of the result \eqref{largeng}, which is an important consequence from the theoretical point of view of the weak breaking of the underlying symmetry and the relevance of the gravity sector at all energies. The uninterested reader can move directly to the discussion about the experimental constraints on the \acs{EFTI} of the next chapter. 

It is useful to re-write the coefficients \eqref{mg-N}-\eqref{mg-cs} in terms of \eqref{coeffs}-\eqref{coeffs-bis}. Neglecting $\partial_t\hat{M}_1^3$,
\begin{align}
\mathcal{N} &= M_\text{Pl}^2 ~\frac{3\alpha(\alpha-2)+\beta+\varepsilon}{(\alpha-1)^2}~,
\label{k-cssq1}\\
c_s^2 &= \frac{\alpha (1-\alpha)+\varepsilon}{3\alpha^2-6\alpha+\beta+\varepsilon}~.
\label{k-cssq2}
\end{align}

For our purposes, it will be sufficient to set $\varepsilon=0$ to avoid complication of the expressions. 
Moreover, it will prove convenient to introduce also the combination
\begin{equation}
x\equiv \frac{1-\alpha}{c_s^2}~,
\end{equation}
so that $x$ is generically an order-one constant. 
The normalization factor for the curvature perturbation then becomes
\begin{equation}
\mathcal{N} =M_\text{Pl}^2 \frac{1-x c_s^2}{x c_s^4}~.
\end{equation}
The latter quantity has a strong dependence on the speed of sound: for small $c_s^2$, it grows like $\mathcal{N}\propto 1/c_s^4$, and this appears to make the scalar perturbations weakly coupled, suppressing the self-interactions of the canonically normalized $\zeta$, and therefore suppressing non-Gaussianity. This observation is decisive, however: in order to make a conclusive statement regarding non-Gaussianity, one has to study the $c_s$-dependence of the cubic $\zeta$ interactions in the theory at hand. In fact, we will find that the cubic interactions grow as fast as $1/c_s^{10}$, eventually leading to non-Gaussianity of order $f_{\rm NL}\propto 1/c_s^6$. 
The action for the comoving curvature perturbation, up to the cubic order in non-linearity, is given in Eq. \eqref{our-14}. In the remainder of this section, we will confine ourselves to the leading order in the $1/c^2_s$-expansion, which allows to extract the fastest-growing effects in the deep subluminal region of the parameter space of interest. There are seven operators that contribute in this limit: assuming $\alpha\simeq 1$, the relevant $\zeta$-action up to cubic order becomes
\begin{multline}
\label{reducedaction}
S_\zeta =M_\text{Pl}^2 \int \D^4 x ~a^3 ~ \bigg\{ \frac{1}{x c_s^4} \bigg[ \dot\zeta^2 -c_s^2 \frac{(\partial_i\zeta)^2}{a^2}  \bigg]  +\frac{1}{x^3 c_s^{10}}\bigg[\frac{1}{H}\dot\zeta^3 - 3\zeta\dot\zeta^2  
\\
+ c_s^2 \zeta \frac{(\partial_i\zeta)^2}{a^2} 
-\frac{3}{2} \dot\zeta\partial_i\zeta\partial_i\partial^{-2}\dot\zeta   
-\frac{3}{4}\partial^2\zeta \partial_i\partial^{-2}\dot\zeta \partial_i\partial^{-2}\dot\zeta  
\\
 +2\frac{c_s^2}{H}\frac{\partial^2\zeta}{a^2}\partial_i\zeta\partial_i\partial^{-2}\dot\zeta -\frac{c_s^4}{H^2} \frac{\partial^2\zeta}{a^2} \frac{(\partial_i\zeta)^2}{a^2}  \bigg] \bigg\}~.
\end{multline}
Various terms in \eqref{reducedaction} appear to be of different order in $1/c_s^2$; however, one should keep in mind that higher spatial derivatives lead to additional factors of $1/c_s$ in the amplitude of non-Gaussianity, so that \textit{e.g.} the two cubic operators $\dot\zeta^3/c_s^{10}$ and $\zeta (\partial_i\zeta)^2/ c_s^8$ contribute comparably to $f_{\rm NL}$ in the small $c_s^2$ limit. 

\begin{figure}
\centering
\includegraphics[width=.75\textwidth]{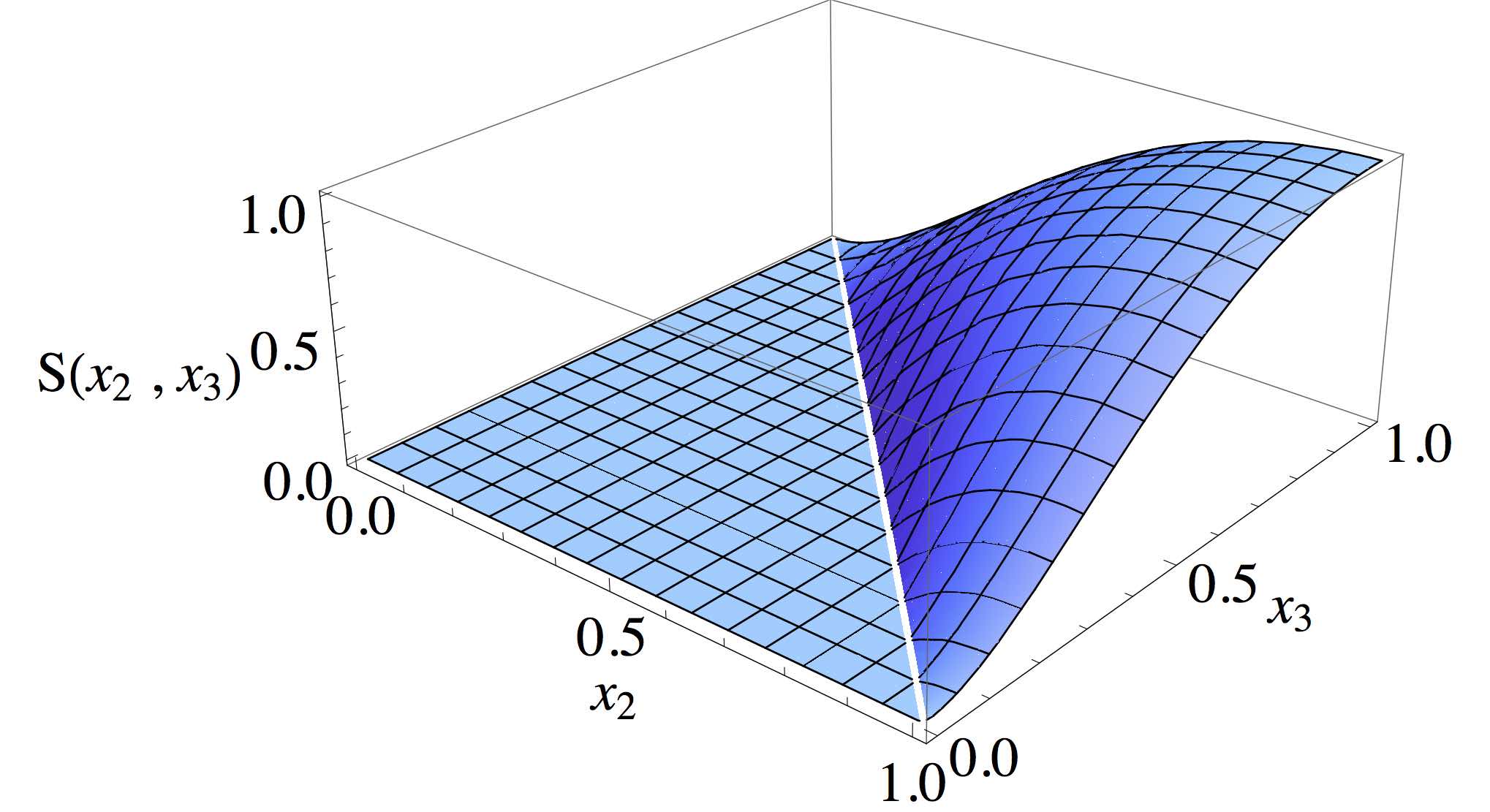} 
\caption{The shape of non-Gaussianity, corresponding to Eq. \eqref{limitnongauss}. We use the standard notation: $x_2 \equiv k_2/k_1$ and $x_3 \equiv k_3/k_1$.}
\label{fig:shape}
\end{figure}

Summing up all the contributions to non-Gaussianity described in Sec. \ref{sec-cswmwg} yields a simple expression at the leading order in the $1/c_s^2$ expansion
\begin{multline}
\label{limitnongauss}
B_\Phi(k_1,k_2,k_3)=-12 \left(\frac{3}{20}\right)^3\frac{M_\text{Pl}^2 H^{4}}{(\mathcal{N}x)^3 c_s^{16}}\frac{1}{k_t^3(k_1k_2k_3)^2} 
\\
\times\left[ \sum_i k_i^3 -\sum_{i\neq j} k_i^2k_j+2k_1k_2k_3\right] ~.
\end{multline}
The shape of the bispectrum is close to the equilateral one, see Fig. \ref{fig:shape}. We note, that the second and the third operators above lead to squeezed non-Gaussianity, which can not characterize a derivatively coupled theory of the sort we are considering. Indeed, all non-equilateral contributions cancel out in the full bispectrum, representing a nice consistency check of our results. This leaves us with the simple expression in Eq. \eqref{limitnongauss}. 

The fact that all of these terms are equally important at typical frequency scales of order of the inflationary Hubble rate can also be seen from the analysis of the $2\to2$ scattering of $\zeta$. 
As discussed in Chap. \ref{chap-EFTinflation}, it is a fairly generic fact that the strong coupling scale of a theory shrinks to zero (or, more precisely, becomes dominated by higher-order effects in the dispersion relation) as the speed of sound is taken to vanish, namely in the limit we are interested in here. In particular, in order to be able to trust our low-energy effective theory, as done in Chap. \ref{WBG-pot}, we should make sure that the strong coupling scale\index{strong coupling scale} of that theory is parametrically greater than the Hubble rate $H$, the typical energy/frequency scale fixed by the measurements of the inflationary observables.
Using the definitions of Chap. \ref{chap-EFTinflation}, one can estimate for instance the strong coupling scale associated with the operator $\dot{\zeta}^3$:
\begin{equation}
\Lambda^2_{\star,\dot{\zeta}^3} = x^{3/2} c_s^{11/2} M_\text{Pl}  H~.
\end{equation}
Demanding this scale to be larger than the inflationary Hubble rate yields the following bound
\begin{equation}
\label{hubblecond}
 x^{3/2} c_s^{11/2} M_\text{Pl} > H~.
\end{equation}
What about the rest of the cubic operators in \eqref{reducedaction}? We have seen, that they all contribute by an equal order of magnitude to non-Gaussianity, so that there should exist a well-defined sense in which they are all ``equally strongly coupled'' around the Hubble frequencies. 
Repeating the above analysis, it is easy to see that the six remaining cubic interactions in \eqref{reducedaction} provide, in general, different strong coupling scales with respect to $\Lambda_{\star,\dot{\zeta}^3}$. For example, the operator $\zeta \dot\zeta^2$ starts violating perturbative unitarity\index{unitarity bound} in a $2\to 2$ scattering of $\zeta$ around the frequency scale of order
\begin{equation}
\Lambda_{\star,\zeta \dot\zeta^2} = x^{3/2} c_s^{11/2} M_\text{Pl}  ~.
\end{equation}
Nevertheless, being different from $\Lambda_{\star,\dot{\zeta}^3}$, the expression for $\Lambda_{\star,\zeta \dot\zeta^2}$ implies the exact same condition \eqref{hubblecond} if the scattering at Hubble frequencies is to be unitary. One can straightforwardly check that the same conclusion applies to all operators in \eqref{hubblecond}, fixing the sense in which all of these operators are equally important for the physics at the horizon. 

We conclude noticing another interesting effect, in addition to the behaviour \eqref{largeng}. It is the inverse-proportional growth (for non-zero $\gamma$ and $\delta$) of the amplitude of non-Gaussianity with a small tensor-to-scalar ratio $r$:
\begin{equation}
\label{1overr}
f_{\rm NL} = \gamma\frac{c_s}{r}\frac{80}{81} \frac{\alpha^2-3\alpha+2-\varepsilon+3(\alpha-1)^2c_s^2}{ (\alpha-1)^4}
-\delta \frac{c_s^3}{r} \frac{80}{81}\frac{1}{ (\alpha-1)^3} + \dots
\end{equation}
This formula, supplemented by the present limits on the primordial gravitational waves, will play an important role in constraining the theories under consideration.

%% file: chapters/2-05-chapter.tex
\chapter{Constraints on single-field inflation}
\label{chap-WBG-constraints}

\begin{flushright}{\slshape    
	VLADIMIR: What are you insinuating? That we've come to the wrong place?} \\ \medskip
	--- Samuel Beckett, \textit{Waiting for Godot}, Act I.
%	--- \defcitealias{bentley:1999}{BBBBB}\citetalias{bentley:1999} \citep{bentley:1999}
\end{flushright}

% For an example of a full page figure, see Fig.~\ref{fig:myFullPageFigure}.

Given the action \eqref{EFTI-action-2} and the parametrization \eqref{coeffs}-\eqref{coeffs-bis}, we have proved that an underlying weakly broken Galileon symmetry allows to explore the parameter space defined by
\begin{equation}
0\lesssim \big \{~|\alpha|, ~|\beta|, ~|\gamma|,~ |\delta| ~\big \} \lesssim \mathcal{O}(1)~,
\label{def-ps}
\end{equation}
being populated by the models we have analysed in the previous chapters.

Here, we want to figure out how the experimental bounds on the physical observables, that have been outlined in Part I, constrain the parameters in the effective action \eqref{EFTI-action-2}. This is of crucial importance because it allows, in principle, to get some indications about the fundamental theory underlying the inflationary phase, among the plethora of models that the \acs{EFTI} is able to capture in a single framework.

However, before proceeding, we find instructive to summarise the defining regimes of some known inflationary models, in order to appreciate and highlight the main differences and novelties of our \acs{WBG} regime. This is the content of the next section.

\section{WBG inflation and comparison with other models}
\label{sec-comparison}

Without the conceit of providing a full description, we recall here the features of just a couple of models, that are interesting for our purposes. Moreover, we report in Tab. \ref{tabsummodels} the main differences with respect to the \acs{WBG} scenarios.

{\renewcommand{\arraystretch}{1.8}
		\begin{table}[t]
		\centering
%		\resizebox{\columnwidth}{!}{	% Serve per fissare la larghezza come la colonna;
									% Il simbolo "!" per far scalare automaticamente l'altezza
		\begin{tabular}{cccc}
  		\hline
  		Inflationary models
			& Parameter hierarchy
			& $c_s^2$
			& $f_{\rm NL}$
 		\\ \hline\hline
		Canonical slow-roll\index{inflation!slow-roll}
			& $\varepsilon\ll1$; \, $\alpha,\beta,\gamma,\delta=0$
			& $1$
			& $\sim\varepsilon$
 		\\ \hline
 		\acs{DBI}-like theories\index{inflation!DBI}
 			& $\alpha, \gamma=0$; \,  $\varepsilon\ll\beta\ll\delta$ 
 			& $\frac{\varepsilon}{\beta}$
 			& $\sim\frac{1}{c_s^2}$
		\\ \hline
 		Galileon inflation\index{inflation!Galileon}
 			& $\varepsilon\ll\alpha,\beta,\gamma,\delta$ 
 			& $\frac{\alpha(1-\alpha)}{\beta-3\alpha(2-\alpha)}$
 			& $\sim\frac{1}{c_s^2}$
		\\ \hline 
		\acs{SRWBG} inflation\index{inflation!WBG}
 			& $\varepsilon\sim\alpha,\beta,\gamma,\delta$
 			& $\frac{\varepsilon+\alpha}{\varepsilon+\beta-6\alpha}$
 			& $\sim\frac{1}{c_s^4}$
		\\ \hline
		\acs{KWBG} inflation\index{inflation!WBG}
 			& $\varepsilon\ll\alpha,\beta,\gamma,\delta\sim 1$
 			& $\frac{\alpha(1-\alpha)}{\beta-3\alpha(2-\alpha)}$
 			& $\sim\frac{1}{c_s^6}$
		\\ \hline
		\end{tabular}
%		}
		\caption{Phenomenological predictions of the analysed models. We neglect higher derivative operators, restricting the discussion to the operators in \eqref{EFTI-action-2} proportional to $M^4_i$ and $\hat{M}^3_i$.}
\label{tabsummodels}
		\end{table}
}

\subsection{DBI, and related models}
\label{dbipar}
 
In the unitary gauge, these models give rise to the following relations between the dimensionless parameters\footnote{See Sec. \ref{subsec-stucktrick-dc} and \cite{Alishahiha:2004eh, Chen:2006nt}.}
\begin{equation}
\big \{ \alpha, ~\gamma \big \} \sim 0~, \qquad  \big\{ ~|\beta|, ~|\delta|~ \big \}\gtrsim \varepsilon~,
\end{equation}
implying the following behaviour of the amplitude of non-Gaussianity for strongly subluminal perturbations
\begin{equation}
\label{fnldbi}
f_{\rm NL}\propto \frac{1}{c_s^2}~.
\end{equation}
In fact, the two non-zero coefficients $\beta$ and $\delta$ parametrically exceed the slow-roll parameter $\varepsilon$ in the small $c_s^2$ limit:
$\beta \propto \varepsilon/c_s^2$, and $\delta \propto \varepsilon/c_s^4~$.
However, current experimental bounds imply $c_s^2\gtrsim \varepsilon$ for the \acs{DBI} model\index{inflation!DBI} \cite{planckXX:2015}, so perhaps the optimal values for these parameters to keep in mind are
\begin{equation}
| \beta | \lesssim 1, \qquad |\delta|\sim \frac{\beta}{c_s^2}~.
\end{equation}

\subsection{G-inflation/Galileon inflation}
\label{ginfpar}
 
In this category, we collect inflationary theories characterized by the dimensionless parameters of \eqref{EFTI-action-2} satisfying the following conditions
\begin{equation}
\label{ginfpar1}
\big \{~ |\beta| ~, |\gamma|, ~|\delta|~\big \} \sim 1~, \qquad |\alpha |< 1 ~,
\end{equation}
so that the subluminal limit corresponds to
\begin{equation}
\label{ginf1}
c_s^2\sim \frac{\alpha}{\beta} < 1~.
\end{equation} 
This can be the case in G-inflation\index{inflation!G-} \cite{Kobayashi:2010cm} and Galileon inflation\index{inflation!Galileon} \cite{Burrage:2010cu}.
For a suppressed speed of sound, the amplitude of non-Gaussianity grows similarly to \acs{DBI} models\index{inflation!DBI}
\begin{equation}
f_{\rm NL}\propto \frac{1}{c_s^2}~.
\label{fnldbi1}
\end{equation}
It has been noticed by Burrage et al. \cite{Burrage:2010cu}, however, that in the unitary gauge the most general theory of Galileon inflation\index{inflation!Galileon} introduces extra cubic operators, such as $\sqrt{-g}~\delta N(\delta K^i_j\delta K^j_i-\delta K^2)$ (with an order-one coefficient in Planck units), and these can result in a faster growth 
\begin{equation}
f_{\rm NL}\propto \frac{1}{c_s^4}~.
\label{nggi}
\end{equation}
We have not included such operators in our analysis. The major reason is that, as we will see in the next section, constraints on various inflationary models described by the \acs{EFT} \eqref{EFTI-action-2} are already quite strong regardless of the cubic operators; including the latter can loosen the constraints due to non-Gaussianity, but only at an expense of some cancellations. 

\subsection{Slow-roll inflation with WBG symmetry}
\label{srwbgpar}

This class of theories, which we will refer to as \acs{SRWBG} from now on, has been introduced in Chap. \ref{WBG-pot} and gives rise to the values of the dimensionless parameters of the \acs{EFT}  \eqref{EFTI-action-2} suppressed by slow-roll 
\begin{equation}
\label{srwbg}
\big \{~|\alpha|,~ |\beta|,~|\gamma|,~|\delta|~\big \} \sim \varepsilon ~.
\end{equation}
The speed of sound of scalar perturbations is generically order-one, but not necessarily strictly one, in contrast to canonical slow-roll inflation\index{inflation!slow-roll}. The background evolution, along with all background characteristics (spectral tilt, number of e-folds from freeze-out of the \acs{CMB} modes until the end of inflation, spectrum of gravitational waves, etc.) are all parametrically similar to garden-variety slow-roll models. What is different, though, is that the scalar perturbations can be more strongly coupled than in canonical slow-roll inflation, in a way that, nevertheless, allows to keep control over the derivative expansion. This leads to an amplitude of non-Gaussianity that grows like
\begin{equation}
f_{\rm NL} \propto \frac{1}{c_s^4}
\label{fnlsrwbg}
\end{equation}
in the subluminal limit. Note that, while this behaviour is similar to \eqref{nggi}, the underlying models are very different: the \acs{SRWBG} model is a minimal deformation of slow-roll inflation\index{inflation!slow-roll}, unlike Galileon inflation\index{inflation!Galileon} which describes a \textit{kinetically-driven} background. Moreover, as already mentioned above, the version of Galileon inflation\index{inflation!Galileon} described by the \acs{EFT} \eqref{EFTI-action-2} in fact yields a \acs{DBI}-like growth, Eq. \eqref{fnldbi}, while the slow-roll theories with \acs{WBG} invariance lead to \eqref{fnlsrwbg} already within the realm of the \acs{EFT} \eqref{EFTI-action-2}. The strong dependence of non-Gaussianity on the speed of sound allows to generate appreciable values for $f_{\rm NL}$ even for mildly subluminal perturbations, arising for
\begin{equation}
\label{srwbgngcond}
\alpha \simeq - \varepsilon 
\end{equation}
in the \acs{SRWBG} model (see Eq. \eqref{k-cssq2}). 

\subsection{Kinetically driven inflation with WBG symmetry}
\label{kwbgpar}

This model, discussed in Chap. \ref{WBG-kin}, is characterized by the hierarchy
\begin{equation}
\label{kwbg-h}
\big \{~|\alpha|,~ |\beta|,~|\gamma|,~|\delta|~\big \} \sim 1 ~,
\end{equation}
closing the range of values \eqref{def-ps}, admitted by the \acs{WBG} symmetry. Such a configuration is also captured by the models in Sec. \ref{ginfpar}, but we have preferred to dedicate a separate paragraph to the regime $\alpha\simeq 1$ yielding the strong enhancement \eqref{largeng}, which was unnoticed in the previous works. Moreover, this has provided the simplified $\zeta$-action \eqref{reducedaction} and the bispectrum \eqref{limitnongauss}.

\begin{figure}
\centering
\includegraphics[width=.75\textwidth]{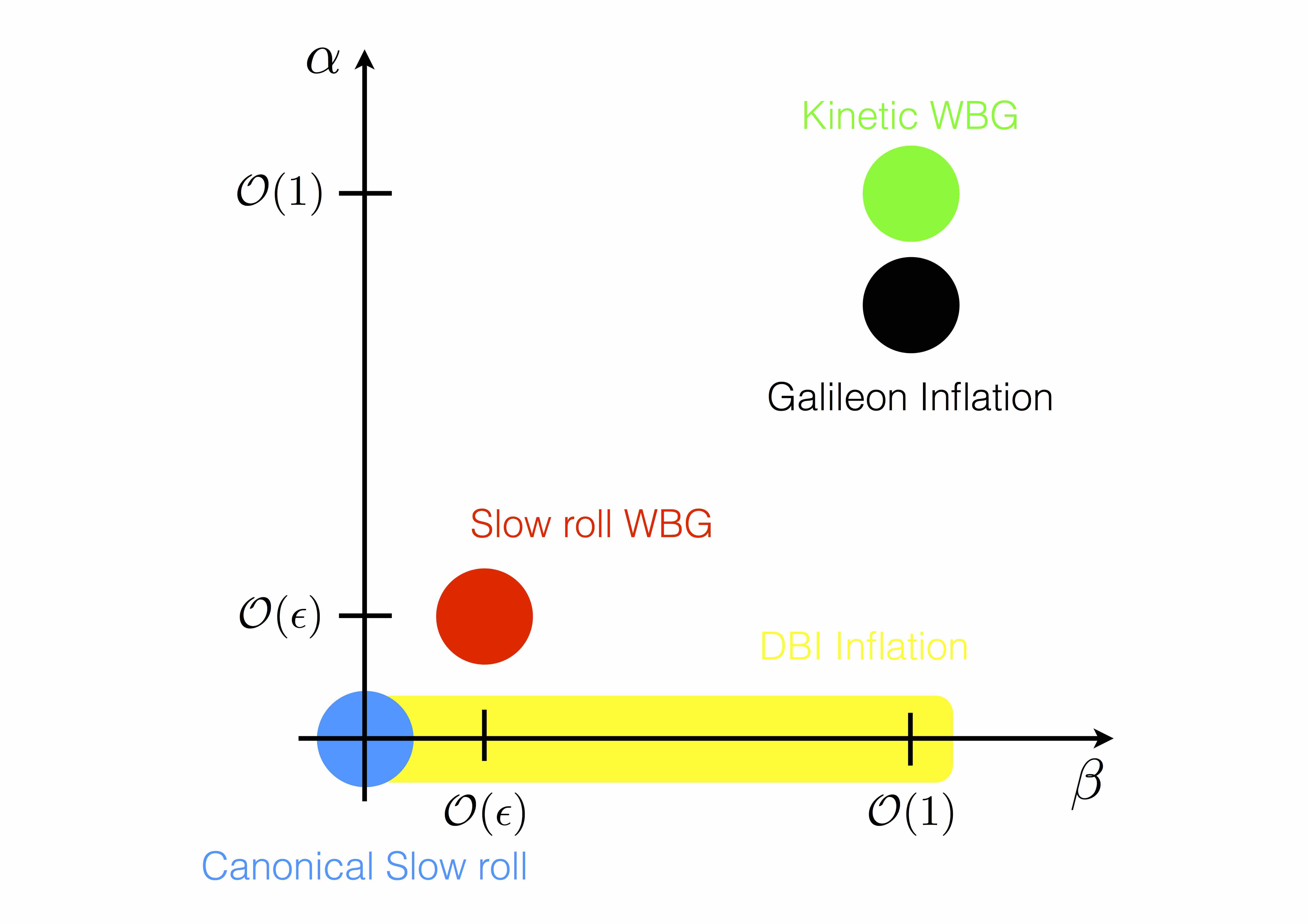} 
\caption{Various single-field models of inflation in the $\alpha$-$\beta$ plane.}
\label{fig:models}
\end{figure}

In Fig. \ref{fig:models}\index{inflation!WBG}\index{inflation!single-field}, we identify the defining regimes of all these models.
The purpose of the rest of this chapter is to add the experimental constraints and draw the excluded regions on the sketch of Fig. \ref{fig:models}.

\section{Preliminaries}

Concerning the phenomenology, we find that the constraints on the effective theory \eqref{EFTI-action-2} are rather robust: despite the apparent multitude of the \acs{EFT} coefficients, basic theoretical considerations (such as the absence of instabilities and of superluminal scalar modes) and current limits on the primordial gravitational waves and non-Gaussianity already limit most of the parameter space. These constraints operate in a coordinated way, ruling out complementary regions. Moreover, we find that the theoretical and experimental viability of a given region of the parameter space is to a great extent determined by the set of just three numbers characterizing the operators in \eqref{EFTI-action-2}: $\alpha$, $\beta$ and $\varepsilon$. Indeed, at the quadratic level, the action  \eqref{EFTI-action-2} captures all single-field models with scalar perturbations obeying the usual, phonon-like dispersion relation
\begin{equation}
\label{phdr}
\omega = c_s k~,
\end{equation} 
at energy scales of order $H$, and our results apply to any theory with the latter property. 
Actually, there are other two additional operators one can add to the Lagrangian \eqref{EFTI-action-2} to be consistent with Eq. \eqref{phdr}. These are $\sqrt{-g}~(\delta K^i_j\delta K^j_i-\delta K^2)$ and $\sqrt{-g} ~\delta N \upleft{3}{R}$~. However, both these operators are redundant and can be removed by a perturbative field redefinition \cite{Creminelli:2014wna,Gleyzes:2015egt}, so at least the quadratic piece of our action \eqref{EFTI-action-2} is very generic. Since, as remarked above, the majority of the constraints on the \acs{EFT} \eqref{EFTI-action-2} stem precisely from the \textit{quadratic} Lagrangian, our analysis should (at least qualitatively) capture the phenomenology of any model satisfying \eqref{phdr}. 

Current data still allow for an appreciable range of parameters for the \acs{EFTI} \eqref{EFTI-action-2}\index{inflation!EFT of}, leaving room for detectable non-Gaussianity. Reducing the existing upper bound on the tensor-to-scalar ratio\index{tensor-to-scalar ratio} by less than an order of magnitude,
\begin{equation}
\label{smallr}
r< 10^{-2}~, 
\end{equation}
would however put the theories that predict the tensor and the scalar tilts of the same order ($ |n_T| \sim |n_s-1|\sim \varepsilon$) in a serious tension with experiment. In this case, the slow-roll inflation with \acs{WBG} symmetry (\acs{SRWBG}), introduced in the previous chapters, has a slightly better chance of surviving the bound \eqref{smallr}. Unlike the canonical slow-roll models with plateau-like potentials famously in agreement with \eqref{smallr}, this model can be consistent with the latter constraint even if driven by a convex potential, as well as give rise to somewhat strongly coupled and highly non-Gaussian ($|f_{\rm NL}|\sim 1\div20$) scalar perturbations.

\section{Constraints}
\label{sec3}

There is a number of constraints, both theoretical and experimental, that the models \ref{dbipar} through \ref{kwbgpar} discussed in the previous section are subject to. 
Above all, there is a constraint expressing the absence of negative norm states (or, alternatively, boundedness from below of the Hamiltonian) and of gradient instability\index{gradient instability}. Moreover, motivated by the discussion in Chap. \ref{chap-EFTinflation}, we demand the absence of superluminal scalar perturbations.
%While it is not fully clear whether superluminality within a low-energy \acs{EFT} is unconditionally unacceptable, there are good reasons to believe that at the very least it is inconsistent with the standard properties (Lorentz-invariance\index{symmetry!Lorentz}, locality, analyticity, etc.) of a hypothetical UV completion \cite{Adams:2006sv}. To be on the safe side, we will thus demand that the scalar excitations are subluminal as well. 
The above considerations then summarize into the following conditions on the parameters of the theory
\begin{equation}
\mathcal{N}>0~, \qquad 0<c_s^2\leq 1~.
\end{equation}

Furthermore, an important role for our analysis will be played by the current limits on the amplitude of primordial gravitational waves.  The tensor-to-scalar ratio \eqref{te-to-s} can be readily read off the quadratic $\zeta$ action and Eqs. \eqref{k-cssq1}-\eqref{k-cssq2}:
\begin{equation}
\label{stratio}
r = 16\frac{\mathcal{N} c_s^3}{M_\text{Pl}^2}=16 \frac{\varepsilon+\alpha-\alpha^2}{(\alpha-1)^2}\sqrt{\frac{\varepsilon+\alpha-\alpha^2}{3\alpha^2-6\alpha+\beta+\varepsilon}}~.
\end{equation}
In the slow-roll limit\index{inflation!slow-roll}, $\alpha=\beta=0$, this reduces to the familiar expression $r_\text{sr}=16 \varepsilon$, while for \acs{DBI} inflation\index{inflation!DBI} ($\alpha=0$) $r_\text{DBI} = 16\varepsilon c_s$~.

Finally, we will impose the experimental limits concerning primordial non-Gaussianity. The full calculation of the scalar bispectrum for the theory \eqref{EFTI-action-2} has been presented in Sec. \ref{sec-cswmwg-sub2}. 

The precise expression for $f_{\rm NL}$ in terms of the dimensionless parameters of the action \eqref{EFTI-action-2} (including the slow-roll parameter $\varepsilon$) is not particularly illuminating. As discussed previously, in different regions of the parameter space corresponding to significantly subluminal scalar perturbations, $f_{\rm NL}$ acquires a simple leading behaviour of the type $f_{\rm NL}\propto 1/c_s^{2p}$, with $p=1,2$ or $3$. Moreover, for non-zero $\gamma$ and $\delta$, there is a "$f_{\rm NL}\propto 1/r$ effect", mentioned in Eq. \eqref{1overr}.

\begin{figure}
\centering
\includegraphics[width=.45\textwidth]{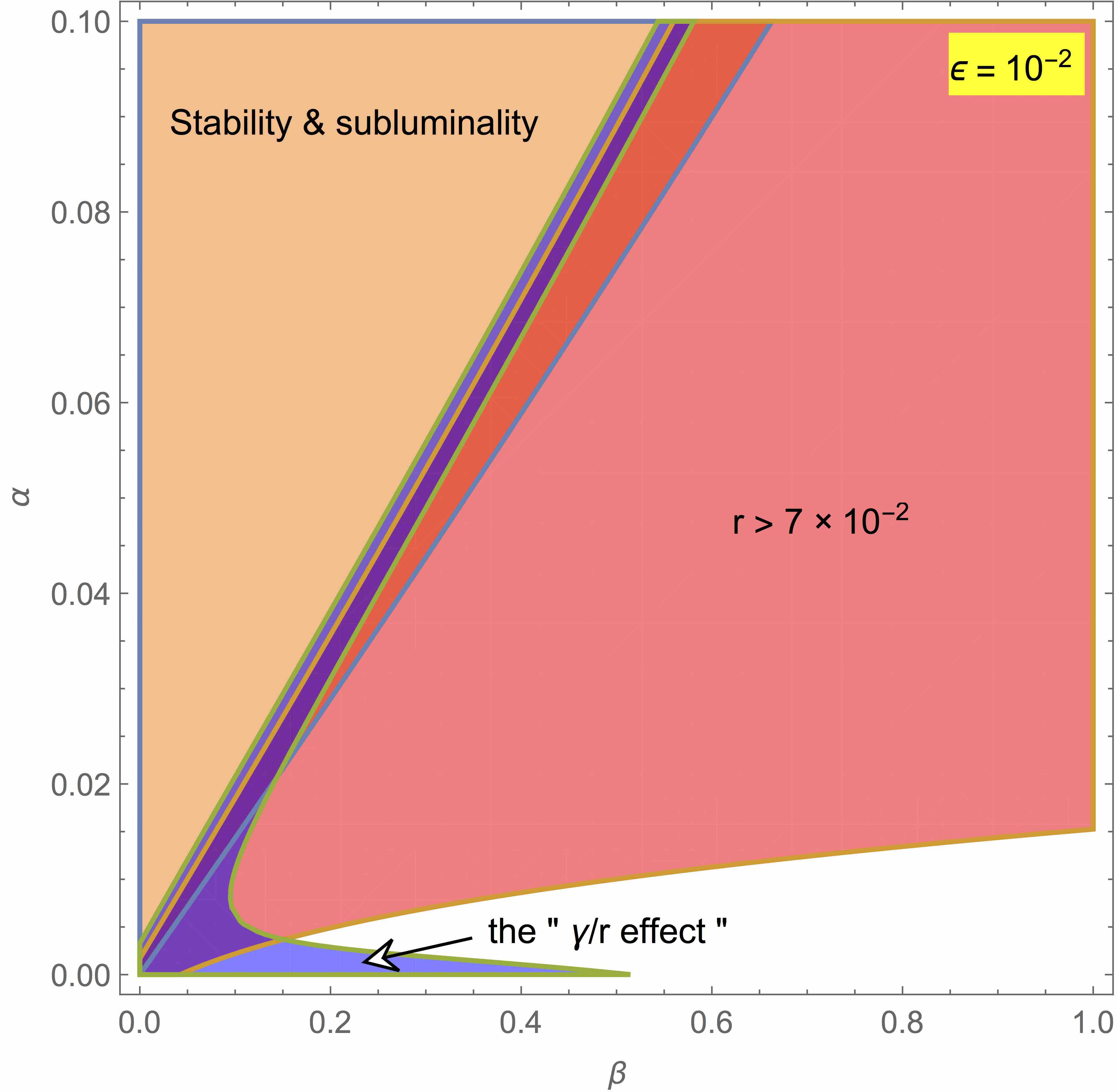} \quad
\includegraphics[width=.45\textwidth]{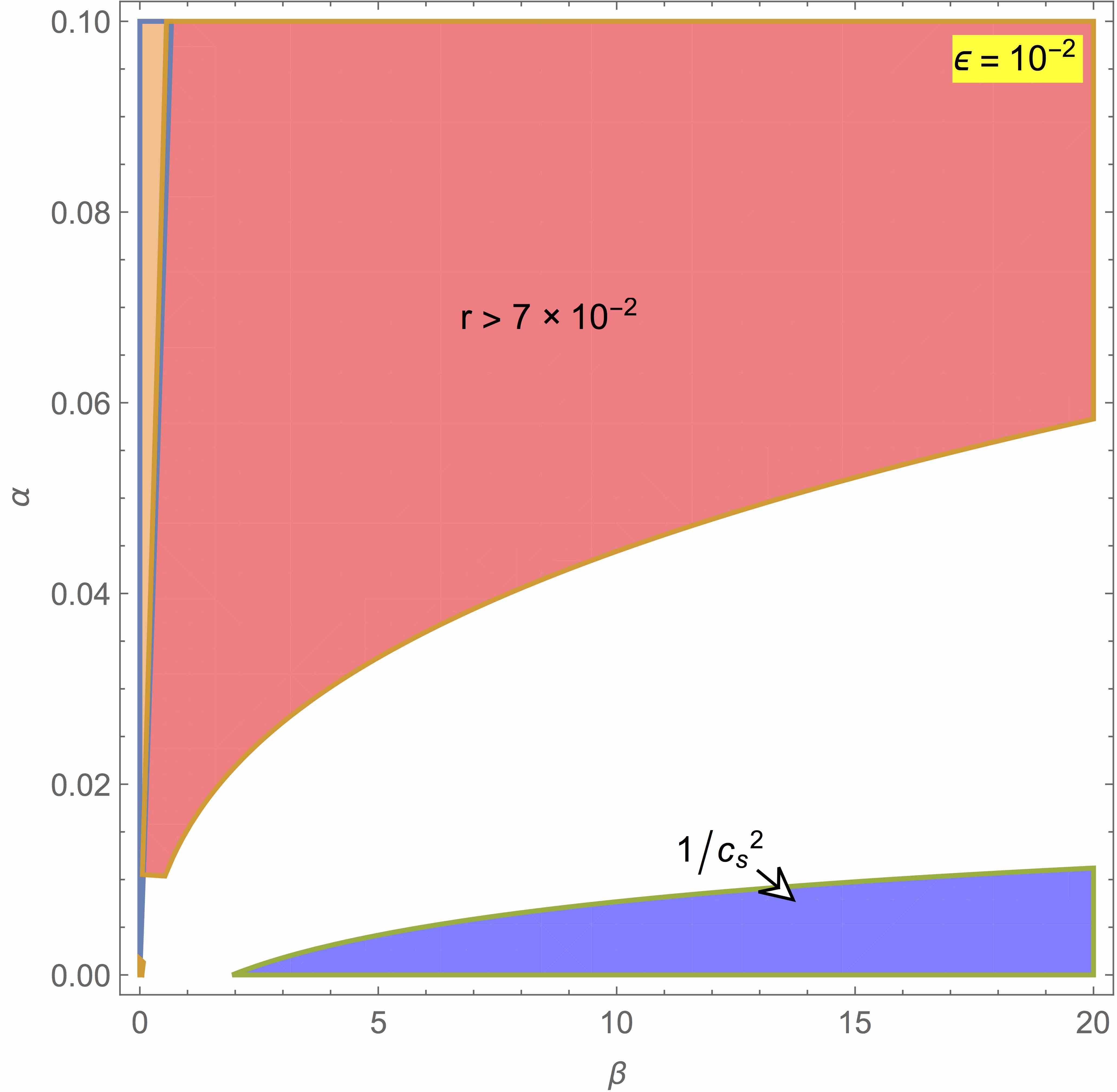} \\
\includegraphics[width=.45\textwidth]{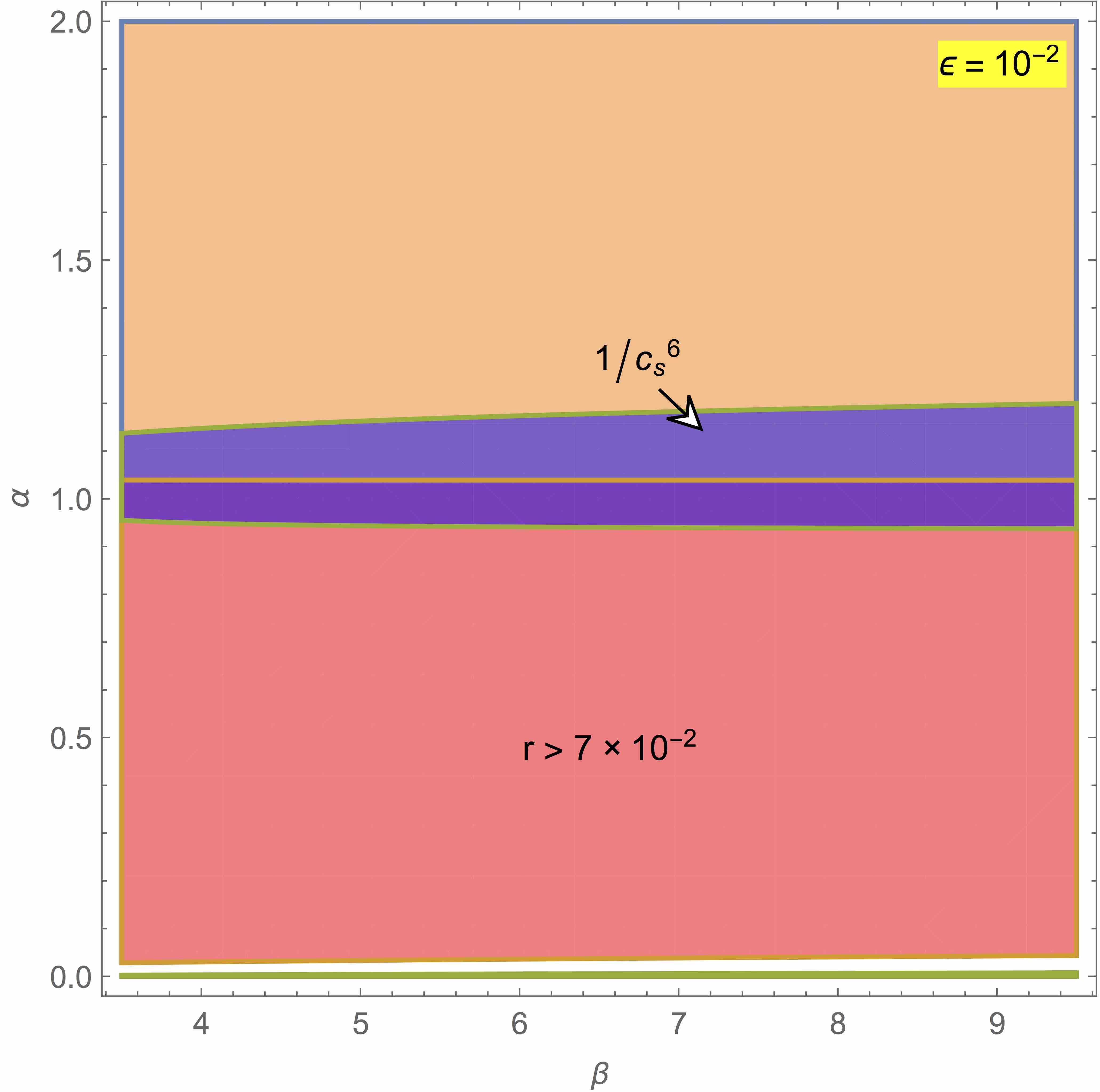} \quad
\includegraphics[width=.45\textwidth]{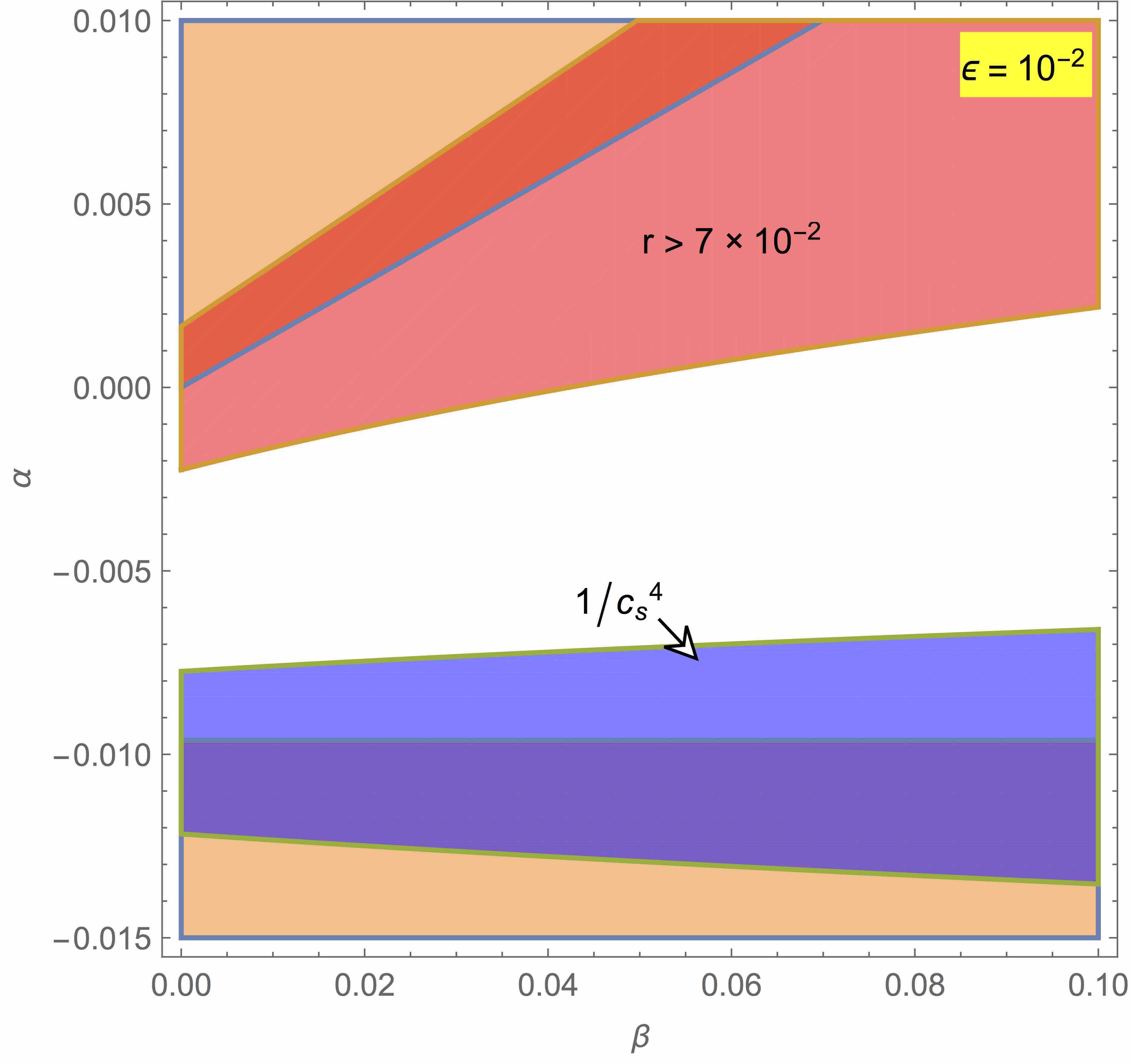}
\caption{Various regions of the parameter space, excluded by the requirements of stability \& subluminality (orange), current limits on the amplitude of primordial gravitational waves (red), and on non-Gaussianity (blue). The slow-roll parameter has been fixed to $\varepsilon=10^{-2}$, and the red and blue bands correspond to regions excluded respectively by the bounds $r<0.07$ and $-50<f_{\rm NL}<50$. The parameters $\gamma$ and $\delta$ have been chosen to vanish everywhere except the upper left panel, where they have been set to $\gamma=\delta=5$. }
\label{fig:1}
\end{figure}
\begin{figure}
\centering
\includegraphics[width=.45\textwidth]{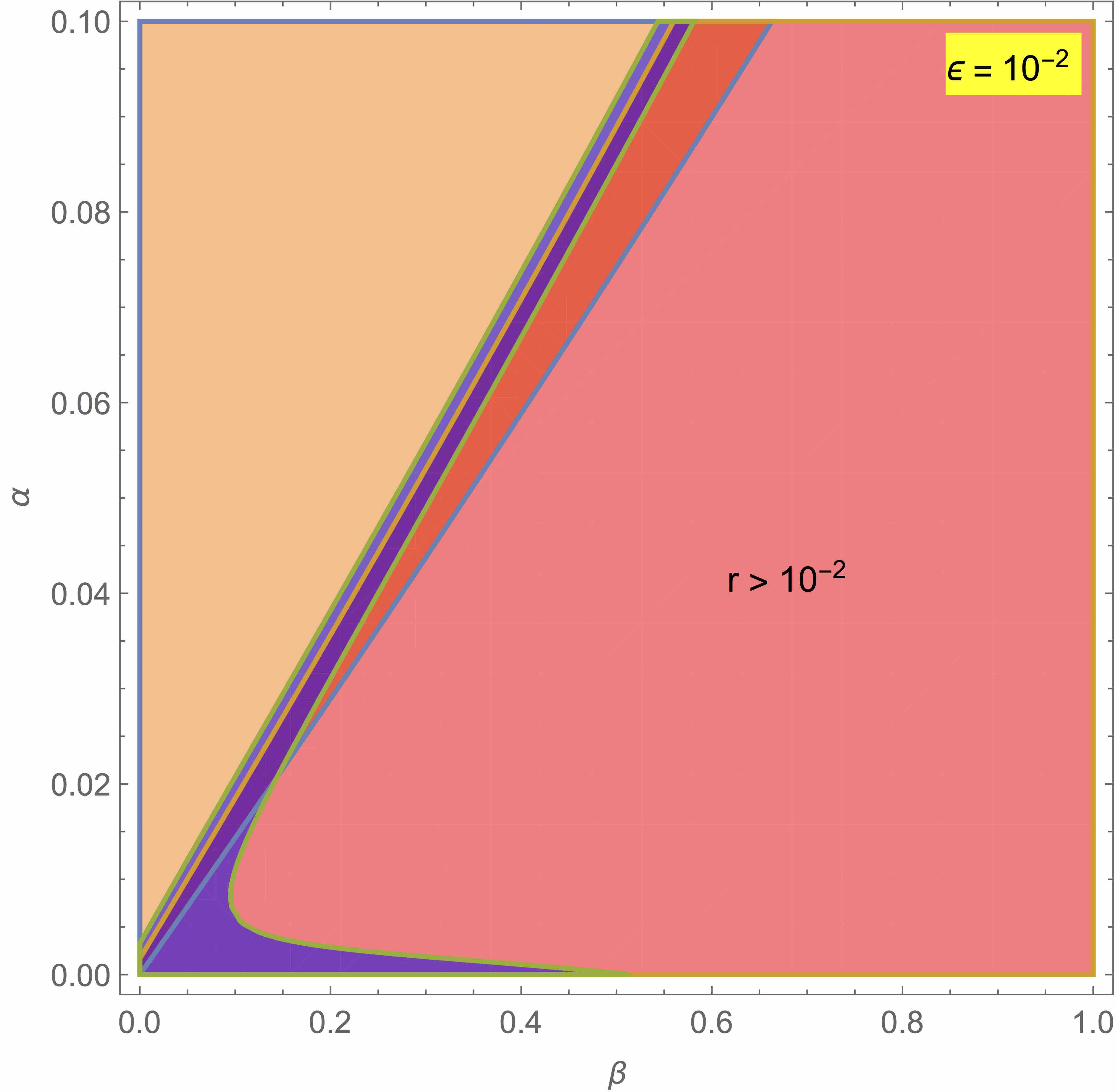} \quad
\includegraphics[width=.45\textwidth]{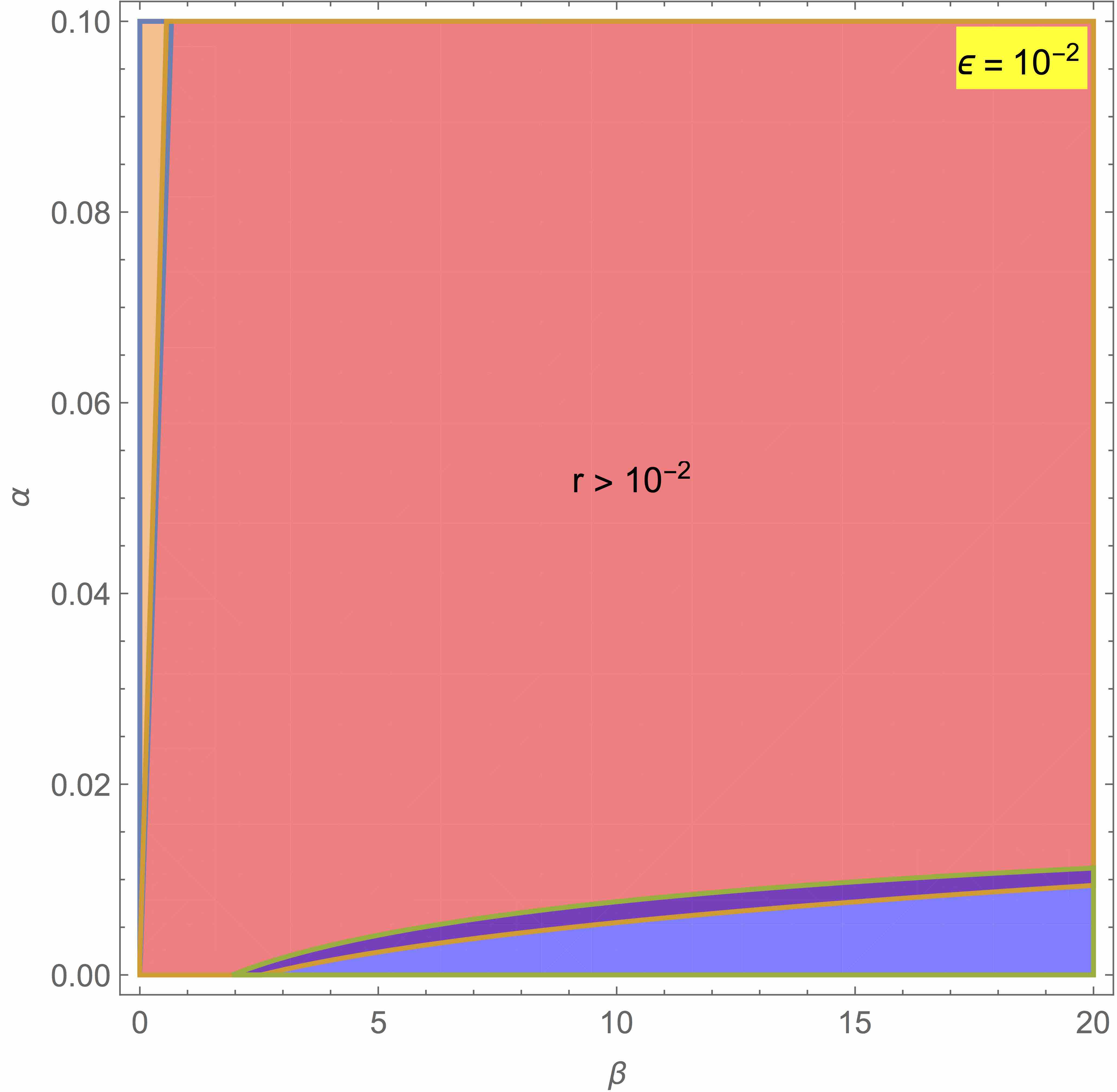} \\
\includegraphics[width=.45\textwidth]{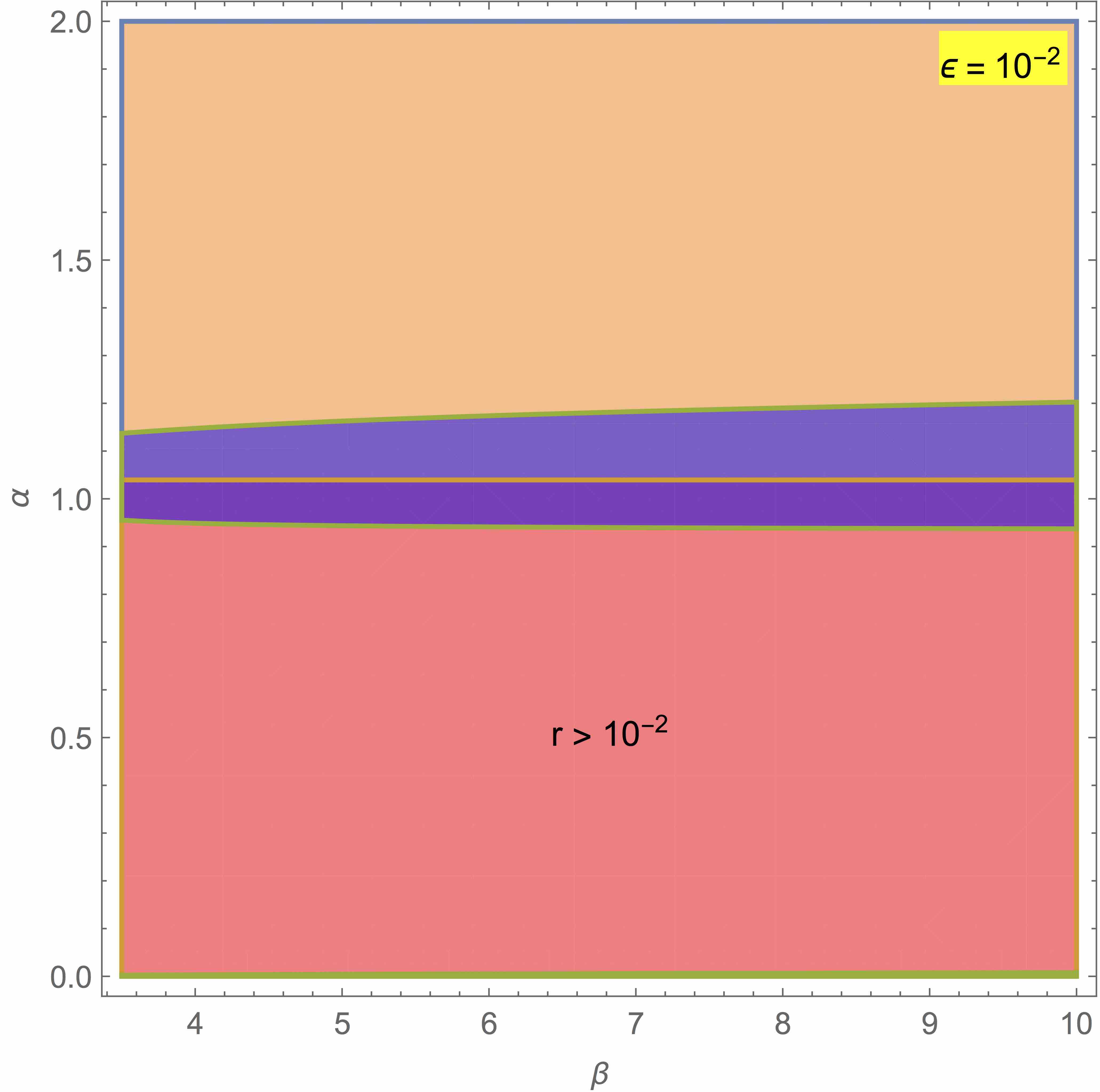} \quad
\includegraphics[width=.45\textwidth]{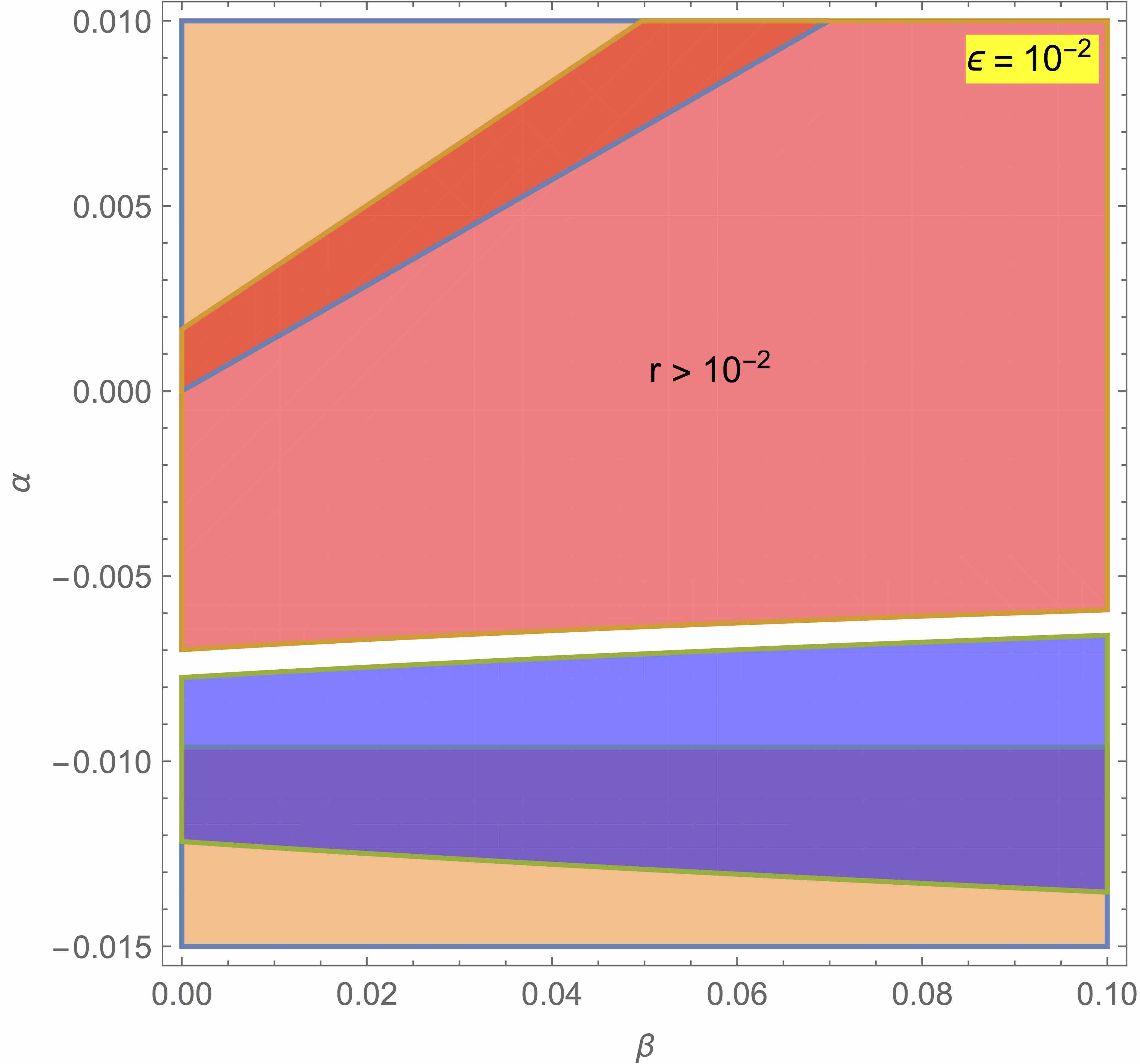}
\caption{Here we illustrate the exact same situation as in Fig. \ref{fig:1}, with the only difference in the exclusion from the tensor-to-scalar ratio: the red band now corresponds to the regions where $r>10^{-2}$. }
\label{fig:2}
\end{figure}

We will ultimately project the parameter space onto the $\alpha$-$\beta$ plane, so some input regarding the magnitude of the slow-roll parameter $\varepsilon$ is needed. The measured tilt of the primordial scalar spectrum $n_s$ suggests that $\varepsilon\lesssim |n_s-1|\sim 10^{-2}$. The latter bound is saturated for many inflationary models, \textit{e.g.} those with convex potentials. On the other hand, there are models characterized by plateau-like potentials such as Starobinsky $R^2$ inflation\index{Starobinsky model} \cite{Starobinsky:1980te}, or the so-called IR \acs{DBI} inflation\index{inflation!DBI} \cite{Chen:2005fe}, where $\varepsilon$ can be much smaller than the scalar tilt. In order to capture both classes of models, we will assume two values for the slow-roll parameter in our analysis: $\varepsilon = 10^{-2}$ and $\varepsilon\sim 0$ (the latter precisely defined below).  

The three constraints discussed above lead to an interesting interplay, in many cases excluding complementary regions of the parameter space. Consider, for example, a \acs{DBI}-like model with a generic power-law potential, so that $\varepsilon \sim 10^{-2}$.
The tensor-to-scalar ratio\index{inflation!DBI},  $r_{\rm DBI} = 16\varepsilon c_s$, has to be below $\sim 0.1$ according to the current experimental limits\index{experiments!Planck} \cite{planckXX:2015}, requiring a somewhat suppressed speed of sound. On the other hand, significantly suppressing $c_s$, one runs into tension with the current limits on non-Gaussianity, in accord with Eq. \eqref{fnldbi}. The constraint for \acs{DBI} models is roughly $c_s\gtrsim 0.1$ \cite{planckXVII:2015} (see also Sec. \ref{subsec-stucktrick-dc}). This means that measuring $r\sim10^{-2}$ would rule out the given class of theories. In contrast, \acs{DBI} theories\index{inflation!DBI} driven by plateau-like potentials, like the IR model of Ref. \cite{Chen:2005fe}, are characterized by $\varepsilon \ll 10^{-2}$ and therefore have a better chance of being consistent with a small tensor-to-scalar ratio. 
 
\begin{figure}
\centering
\includegraphics[width=.45\textwidth]{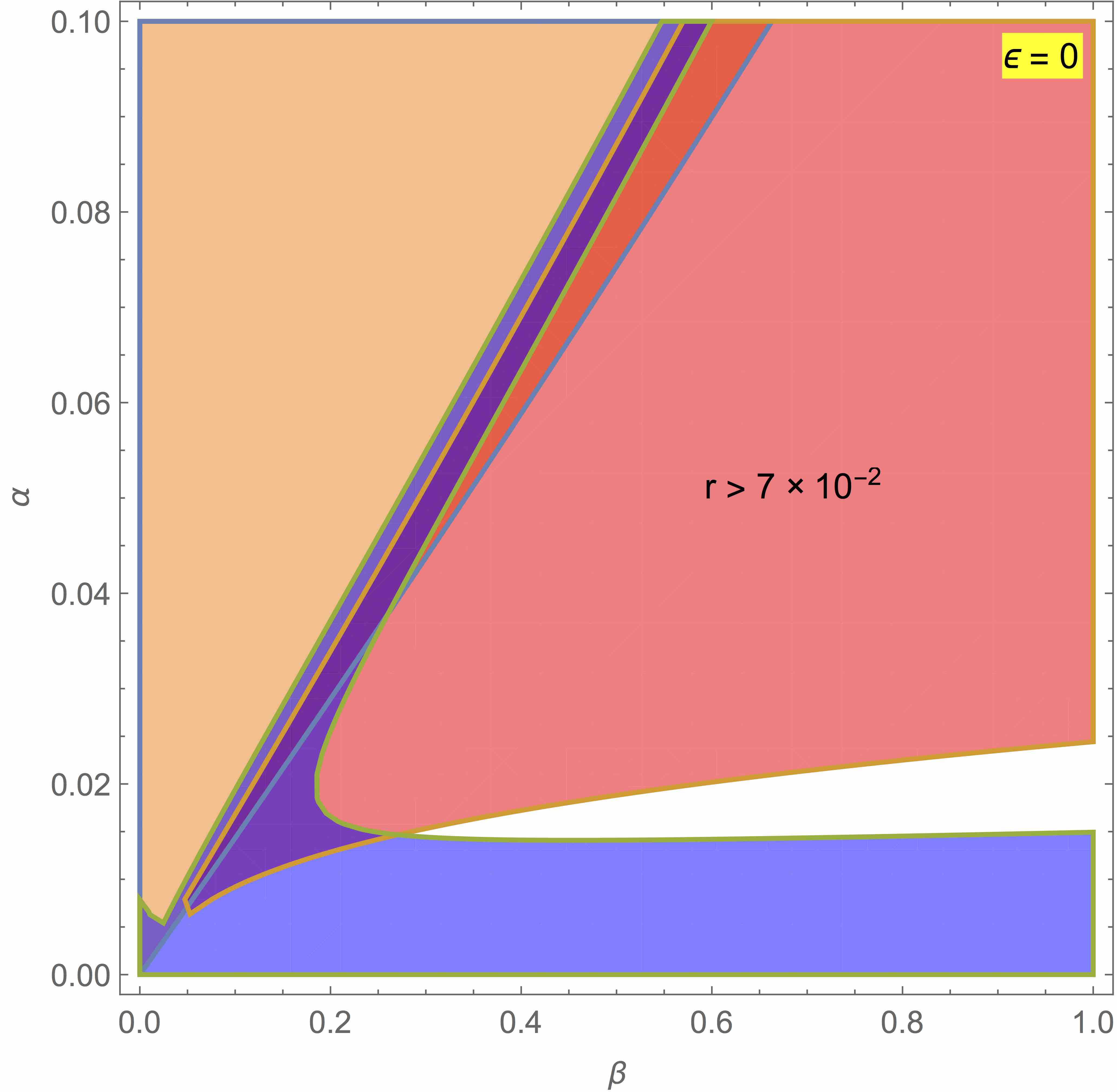} \quad
\includegraphics[width=.45\textwidth]{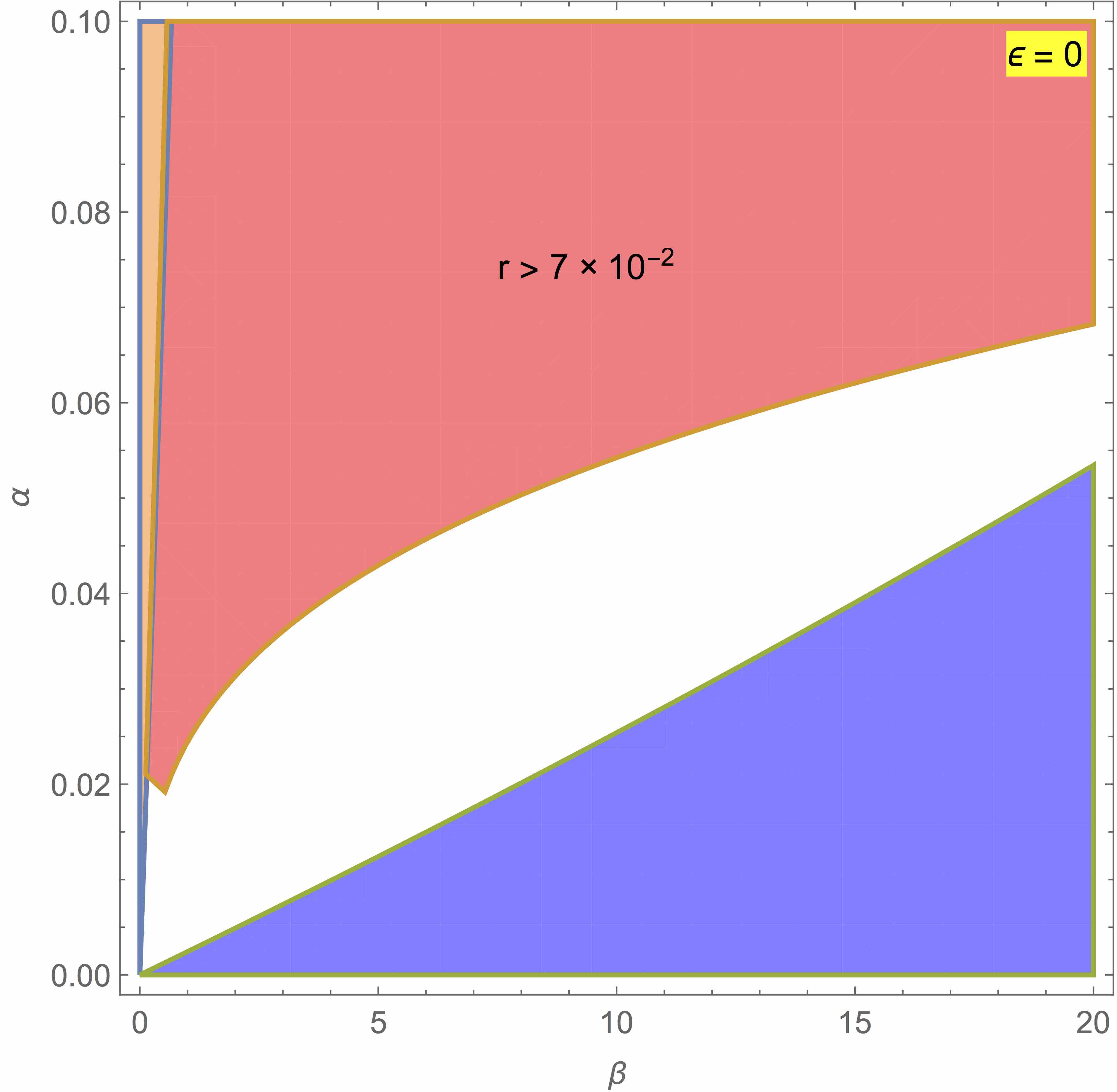} \\
\includegraphics[width=.45\textwidth]{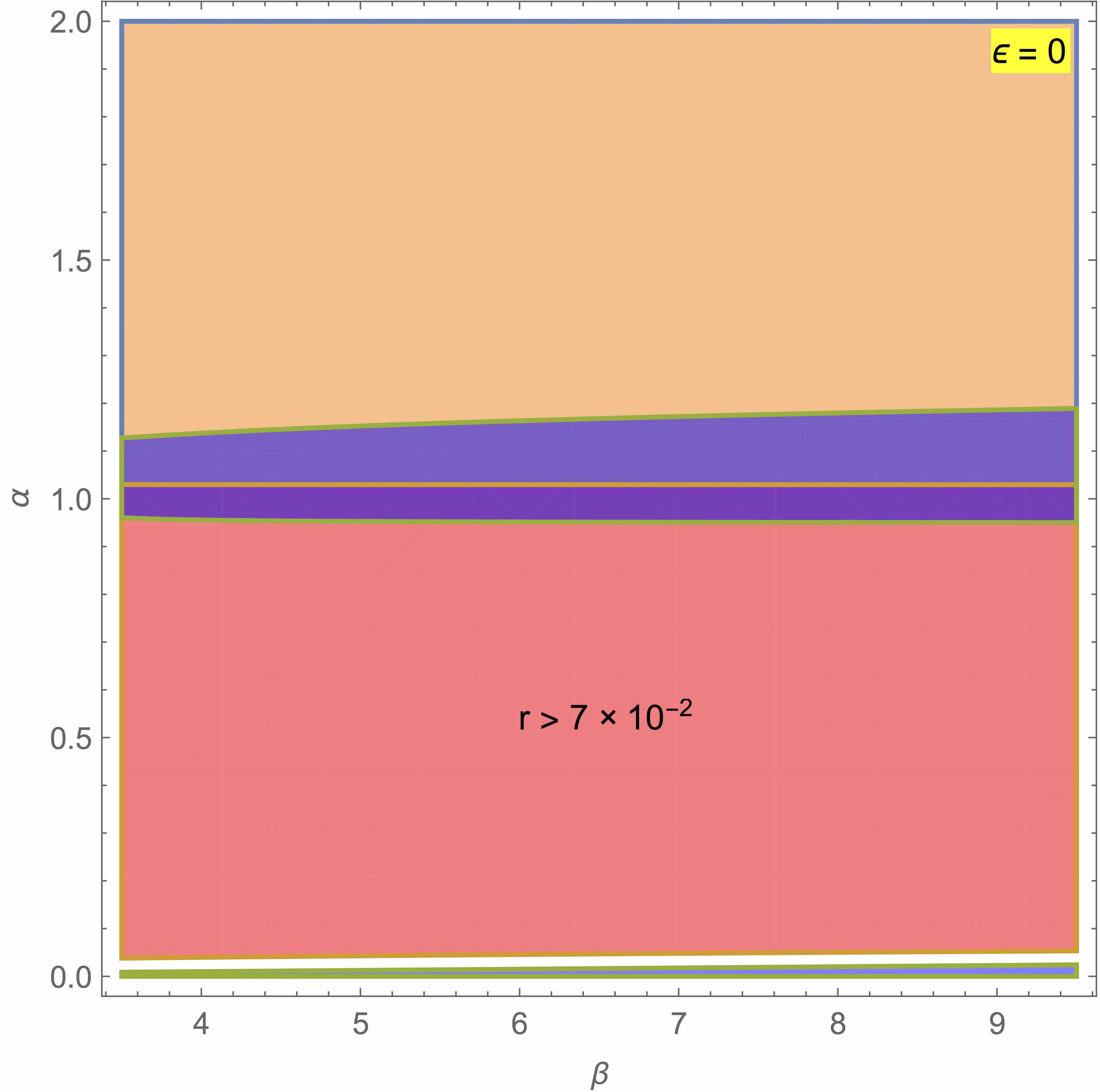} \quad
\includegraphics[width=.45\textwidth]{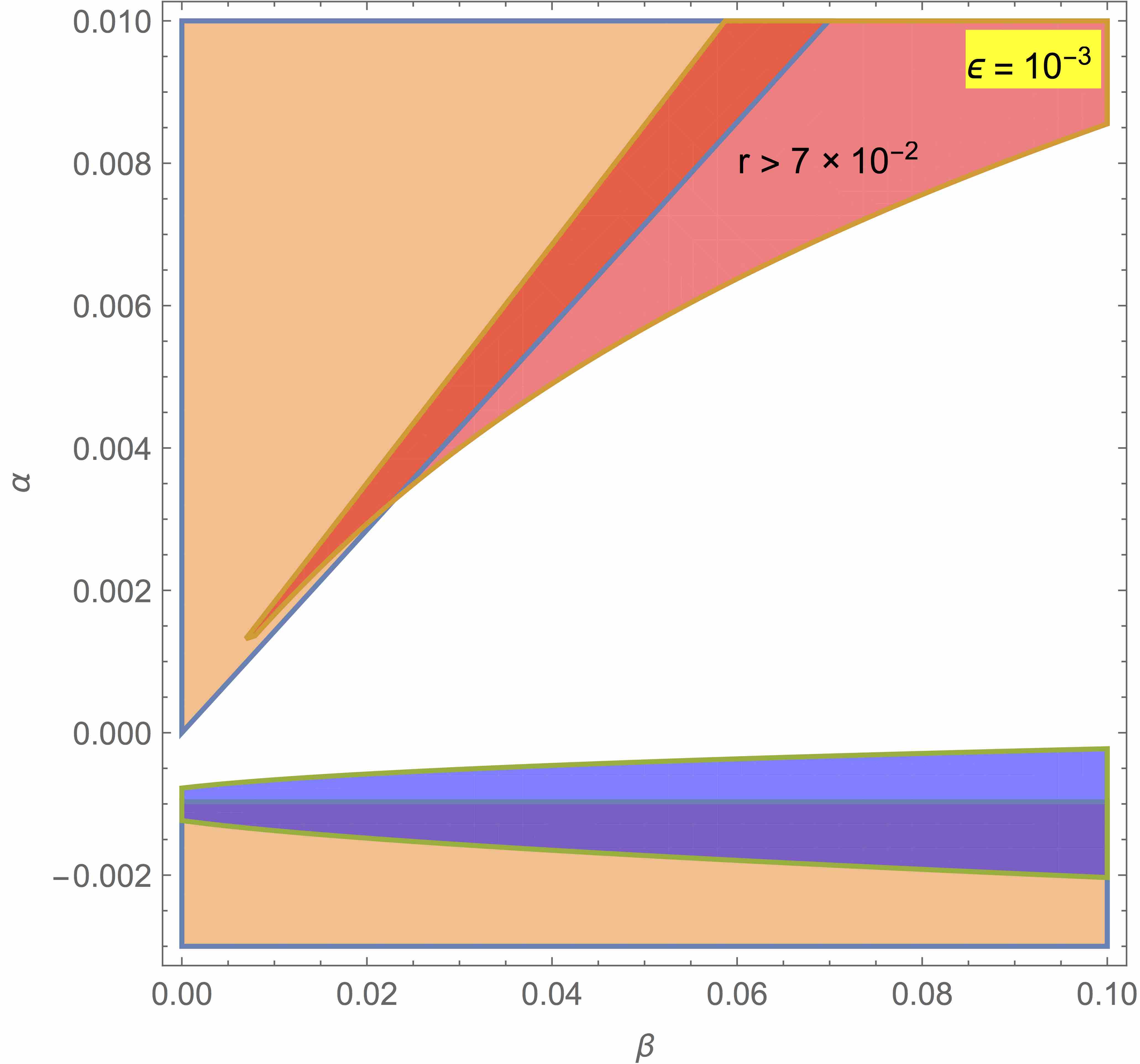}
\caption{Various regions of the parameter space corresponding to models discussed in Sec. \ref{sec-comparison}. The red band shows the regions excluded by requiring $r<7\times 10^{-2}$, while $\varepsilon = 0$ in all panels but the lower right one, which has $\varepsilon = 10^{-3}$. }
\label{fig:3}
\end{figure}

\begin{figure}
\centering
\includegraphics[width=.45\textwidth]{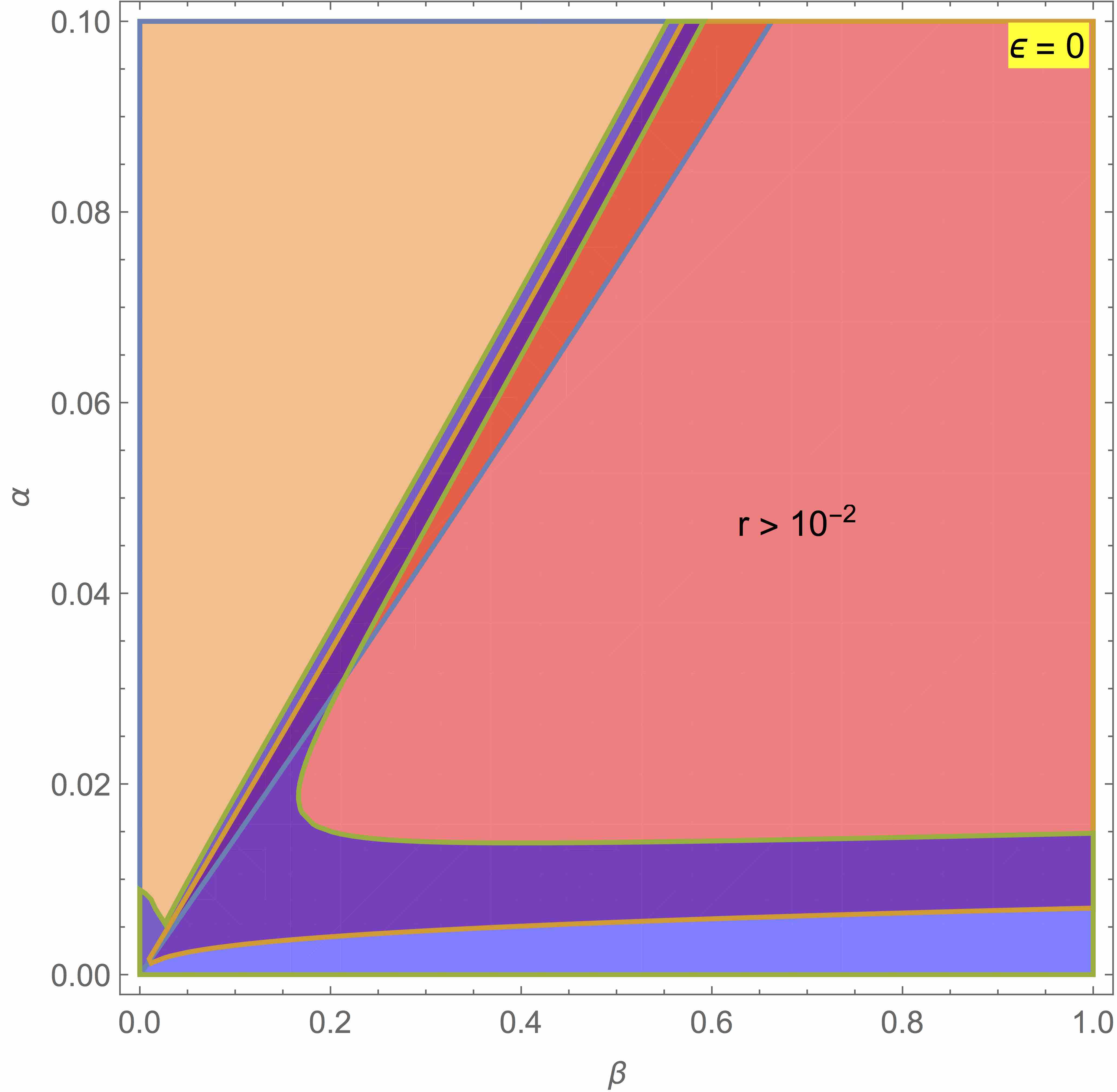} \quad
\includegraphics[width=.45\textwidth]{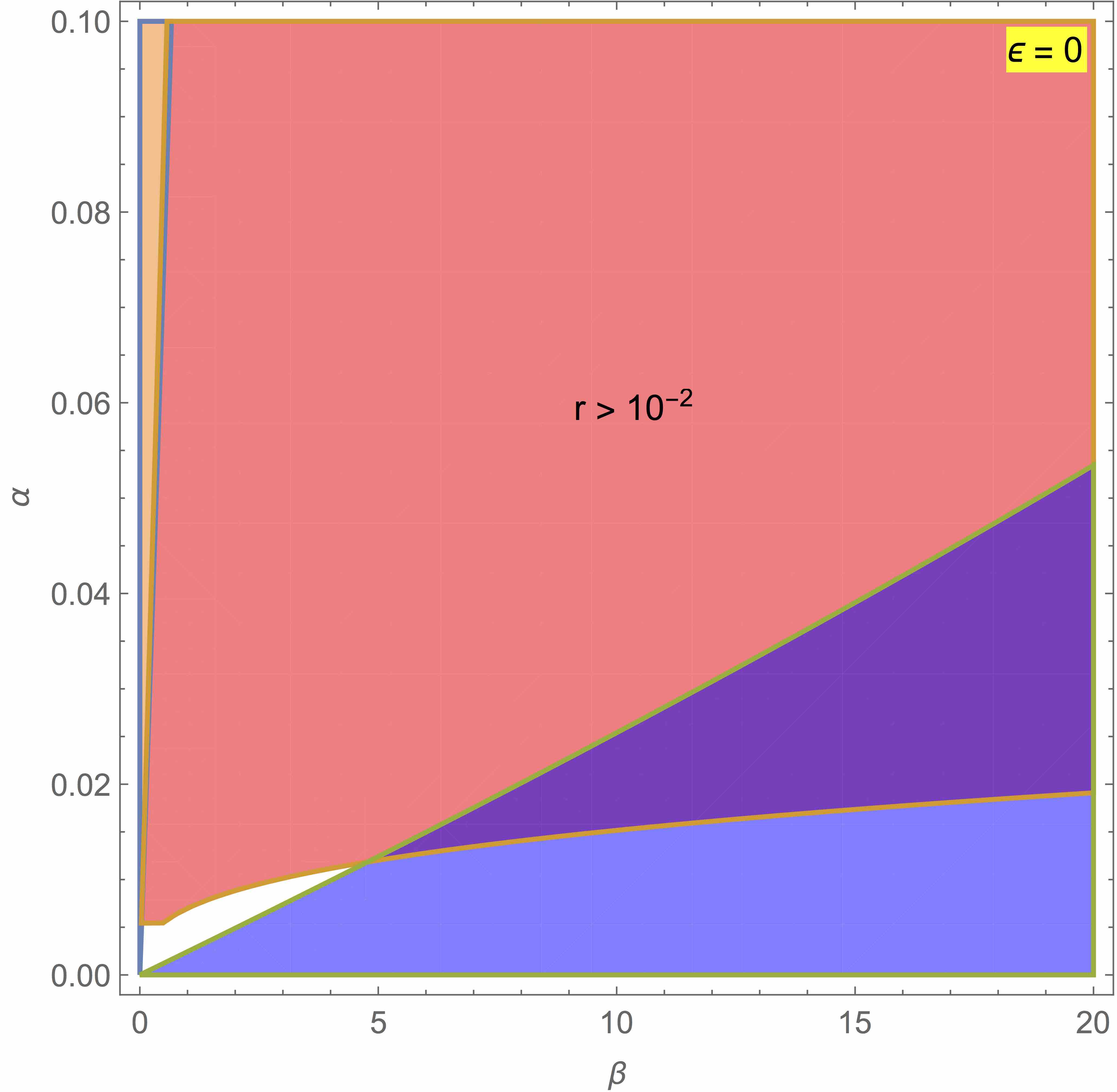} \\
\includegraphics[width=.45\textwidth]{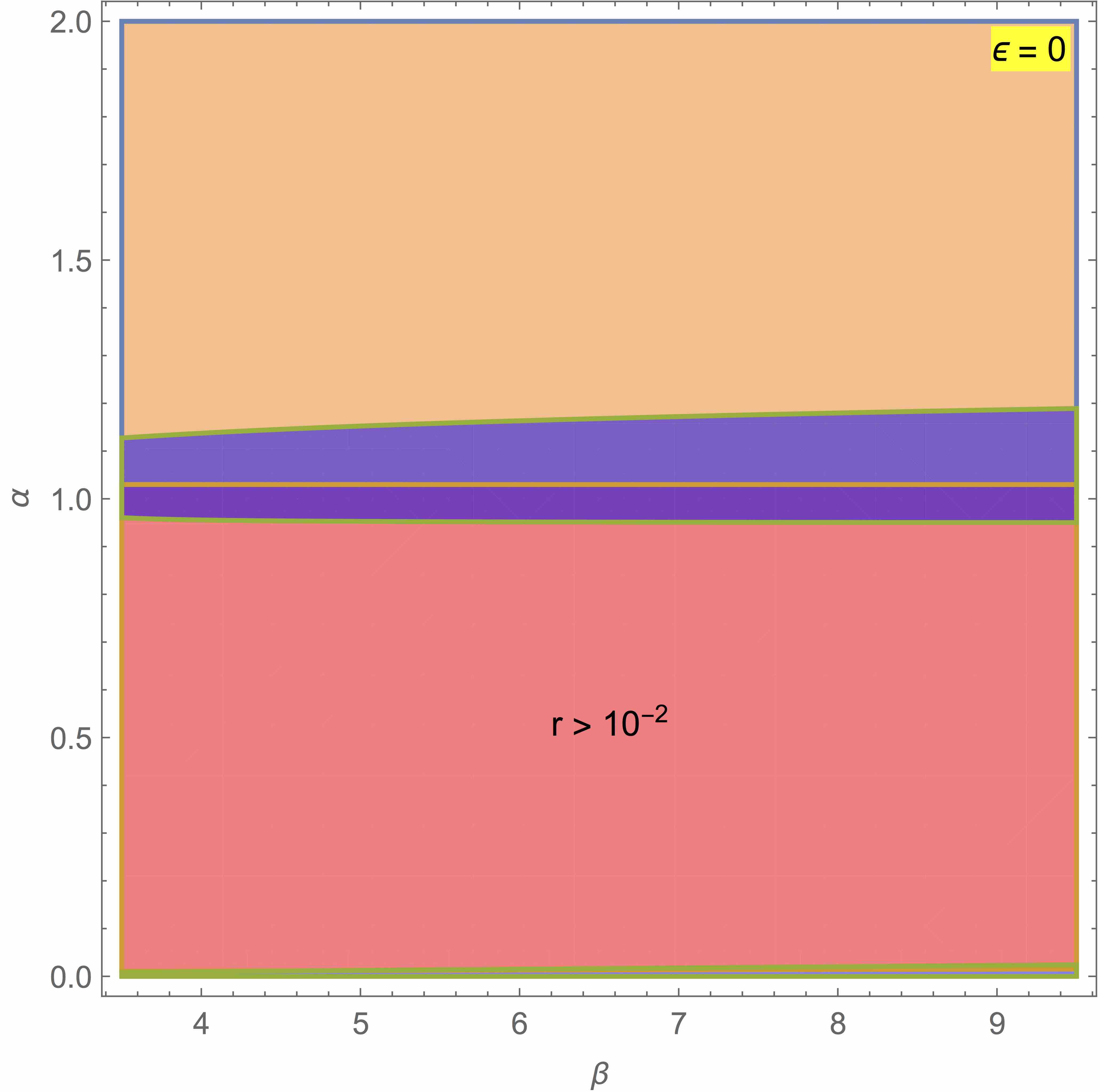} \quad
\includegraphics[width=.45\textwidth]{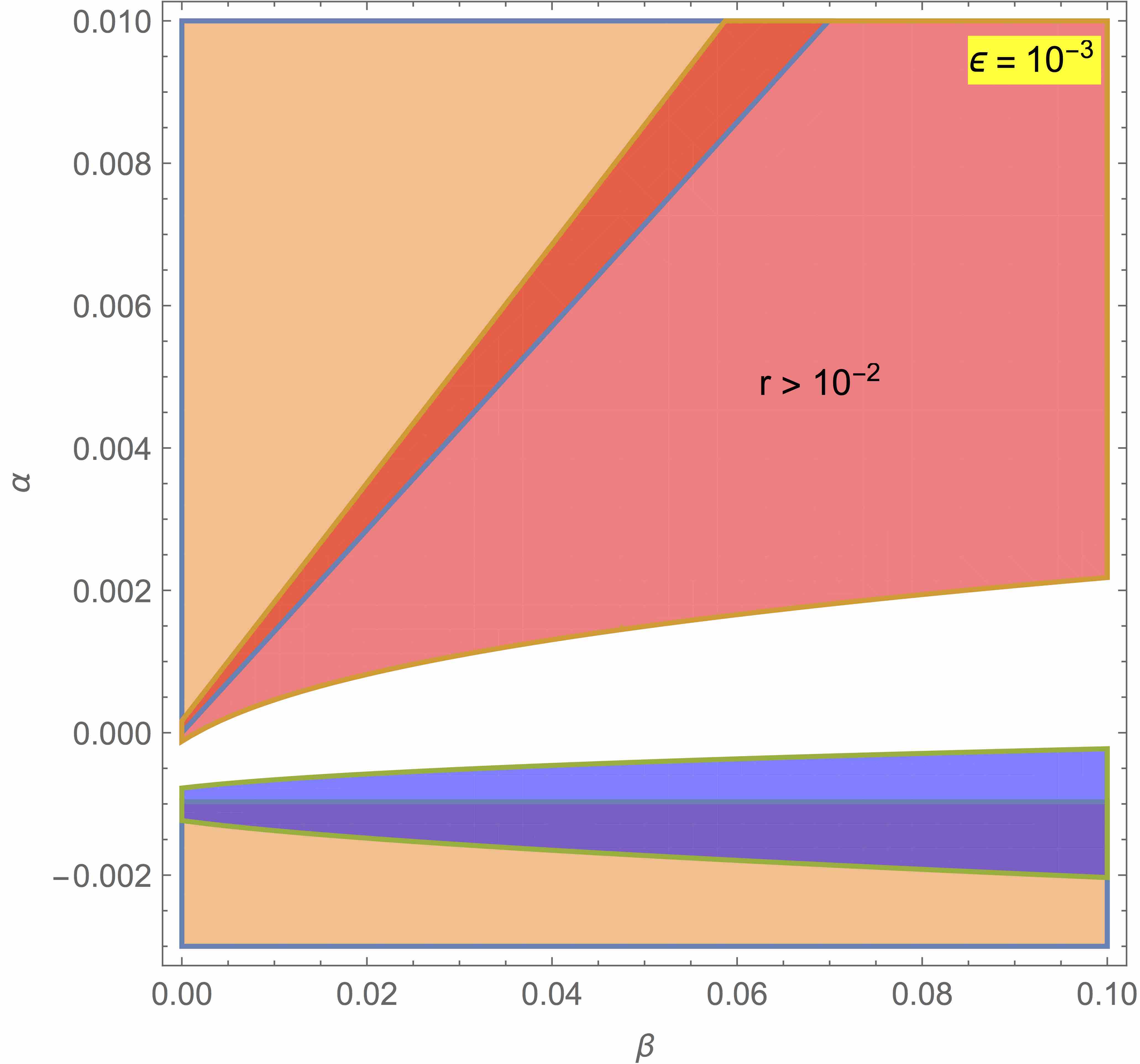}
\caption{Here we illustrate the exact same situation as in Fig. \ref{fig:3}, with the only difference in the exclusion from the tensor-to-scalar ratio: the red band now corresponds to the regions where $r>10^{-2}$. }
\label{fig:4}
\end{figure}

The examples of exclusion plots for the general parameter space of interest are shown on Figs. \ref{fig:1}-\ref{fig:4}. On the first two of these figures, we assume $\varepsilon = 10^{-2}$, while the last two correspond to $\varepsilon\sim 0$. Moreover, we require $f_{\rm NL}$ to be in the range $-50<f_{\rm NL}<50$, motivated by the current limits on equilateral non-Gaussianity\index{experiments!Planck} \cite{planckXVII:2015}, displayed in Tab. \ref{planck2015fNL}. The orange, red and blue regions depict parts of the parameter space excluded by instabilities and/or superluminality, limits on the tensor-to-scalar ratio, and on non-Gaussianity respectively. 

\subsection{Models with $\varepsilon \sim |n_s-1|$}

In Fig. \ref{fig:1}, the red regions are the exclusion bands due to the present $95$\% C.L. bound on the amplitude of the primordial gravitational waves, $r_{0.05}<0.07$ \cite{Bicepnew}\index{experiments!KeckArray}\index{experiments!BICEP}. One can see, that the data prefer a significantly suppressed $\alpha$, effectively ruling out inflationary theories with $\alpha\gg 10^{-2}$: this is a rather general result, true for all cases that we consider below\footnote{The allowed region can in fact reach out to $\alpha\simeq 0.1$, but this only happens for $\beta \sim 50$, casting shadow on the quantum stability of the corresponding theories.}.

For $\alpha$ much smaller than $\varepsilon$, the boundary between the \acs{DBI} and G-/Galileon inflation\index{inflation!Galileon}\index{inflation!G-}\index{inflation!DBI} is blurred. However, an important discriminant that remains is the fact that $\gamma$ can be much larger in the latter class of models. For this reason, we have chosen $\gamma=\delta=5$ in the upper left panel, which results in an additional exclusion region in G-/Galileon inflation\index{inflation!Galileon}\index{inflation!G-} due to large non-Gaussianity stemming from the "$1/r$" effect of Eq. \eqref{1overr} (from the two terms in this equation, only the one proportional to $\gamma$ contributes significantly). Had we chosen $|\gamma| \ll1$ as in \acs{DBI} inflation\index{inflation!DBI}, this band would have completely disappeared from the plot. 
The regions of the $\alpha$-$\beta$ plane explored in the rest of the panels in Fig. \ref{fig:1} are not affected by $\gamma$ and $\delta$ for reasonable values of these parameters\footnote{For the upper right and the lower left panels, even setting $\gamma\sim\delta\sim 10$ has little effect on the exclusion regions. The lower right panel corresponds to the model \ref{srwbgpar} (\acs{SRWBG}), where both of these parameters are naturally of order $\varepsilon$ and lead to negligible effects.} so we have set them to zero everywhere except in the upper left one. 

The upper right panel of Fig. \ref{fig:1} shows the blue exclusion region due to non-Gaussianity in the $f_{\rm NL}\propto 1/c_s^2$ regime, characteristic of the models \ref{dbipar} and \ref{ginfpar}. The blue band shown here appears for larger values of $\beta$, where the speed of sound becomes small enough -- see Eq. \eqref{k-cssq2} --
\begin{equation}
c_s^2 \sim \frac{\mathcal{O}(\alpha, \varepsilon)}{\beta}~,
\end{equation}
so as to trigger the growth of $f_{\rm NL}$ according to Eqs. \eqref{fnldbi} and \eqref{fnldbi1}. 

Another part of the parameter space, corresponding to the model \ref{kwbgpar} (\acs{KWBG}) is shown on the lower left panel. One can see, that the region excluded by non-Gaussianity due to the abrupt growth $f_{\rm NL}\propto 1/c_s^6$ is concentrated around $\alpha\simeq 1$. Unfortunately, this model is already ruled out by the limits on the primordial gravitational waves, combined with the theoretical requirements of stability and subluminality.

Finally, on the lower right panel, we zoom onto the parameter space corresponding to the \acs{SRWBG} model of Chap. \ref{WBG-pot}. One can see the exclusion band from non-Gaussianity around $\alpha\simeq -\varepsilon$, corresponding to $f_{\rm NL}$ growing like $\sim 1/c_s^4$. Just like in \acs{DBI}/G-/Galileon inflation\index{inflation!DBI}\index{inflation!G-}\index{inflation!Galileon}, there remains an appreciable portion of the parameter space still allowed by our constraints with $r\lesssim 0.07$, including regions of the $\alpha$-$\beta$ plane characterized by detectable non-Gaussianity. We stress again, however, that the \acs{SRWBG} model, being in a well-defined sense a minimal deformation of canonical slow-roll inflation\index{inflation!slow-roll}, is very different from the rest of the models considered in Sec. \ref{sec-comparison}. 

While the current data still leave some room for most of the models with $\varepsilon\sim |n_s-1|$, the situation can change dramatically if the upper limit on $r$ decreases to $r\lesssim 10^{-2}$ (which is less than an order of magnitude improvement in current precision). The plots, corresponding to this case, are shown in Fig. \ref{fig:2}. One can see that the regions that were previously allowed are now fully covered by the exclusion bands from gravitational waves. The only region that still remains is a narrow band in the slow-roll model with weakly broken Galileon invariance, shown on the lower right panel of Fig. \ref{fig:2}. 

The canonical models of slow-roll inflation\index{inflation!slow-roll} sit at the origin of the $\alpha$--$\beta$ plane and are of course not visible on our plots. Measuring $r\lesssim 10^{-2}$ would rule out most of these, with an exception of models with plateau-like potentials, such as Starobinsky $R^2$ inflation\index{Starobinsky model} \cite{Starobinsky:1980te}. In these models, the tilt of the scalar spectrum is mostly determined by the second slow-roll parameter, $\eta_V\equiv M_\text{Pl}^2 V''/V$, so that $\varepsilon$ can be much smaller than $|n_s-1|$ to suppress the tensor-to-scalar ratio. Needless to say, falling into the category of canonical slow-roll theories, $R^2$ inflation predicts undetectable non-Gaussianity, $f_{\rm NL}\sim 10^{-2}$ \cite{Maldacena:2002vr}. In contrast, the (non-canonical) slow-roll inflation\index{inflation!slow-roll} with \acs{WBG} symmetry, even if driven by the simplest convex potentials (with $\varepsilon\sim |n_s-1|$), does possess a parameter space consistent with tensor-to-scalar ratios as small as $r \lesssim 10^{-2}$, as seen from the lower right panel of Fig. \ref{fig:2}. Moreover, close to the blue non-Gaussianity exclusion band, this model can generate detectable (equilateral) non-Gaussianity, $|f_{\rm NL}|\lesssim 50$. 


\subsection{Models with $\varepsilon \ll |n_s-1|$}

It is well-known that the tensor-to-scalar ratio $r$ can be significantly suppressed (and therefore the bounds coming from this observable ameliorated) in models where the variation of the inflationary Hubble rate does not significantly contribute to the scalar tilt. A famous example are theories (slow-roll or not) driven by plateau-like potentials. To capture this class of models, we repeat the analysis of the previous subsection setting now $\varepsilon = 0$ for all plots\footnote{Ideally speaking, \acs{DBI} inflation corresponds to a vanishing coefficient $\alpha$. Setting $\varepsilon = 0$ then sends the speed of sound to zero, or equivalently, $f_{\rm NL}$ to infinity, ruling out the \acs{DBI} models with vanishing $\varepsilon$. This can be seen \textit{e.g.} on the upper right panel of Fig. \ref{fig:3}, where the non-Gaussianity exclusion band covers the whole $\alpha=0$ axis. Of course, in a more realistic situation with a small but non-zero $\varepsilon$, an allowed region with $\alpha=0$ and a small enough $\beta$ opens up.} but that corresponding to slow-roll \acs{WBG} inflation (in which the regime with large non-Gaussianity crucially depends on the presence of a non-zero slow-roll parameter $\varepsilon$ -- see Eq. \eqref{srwbgngcond}). In the latter case we set $\varepsilon = 10^{-3}$.
Apart from these modifications, the Figs. \ref{fig:3} and \ref{fig:4} correspond to the exact same choices of parameters as in Figs. \ref{fig:1} and \ref{fig:2} respectively. 

The situation for $\varepsilon \ll |n_s-1|$ is qualitatively similar to the previous case ($\varepsilon\sim |n_s-1|$). The current data still allows a parameter space consistent with the existing bounds on $r$ and corresponding to measurable non-Gaussianity. However, improving the limits on the amplitude of primordial gravitational waves could still induce dramatic changes. One novelty compared to the case of the previous subsection is that for $\varepsilon \sim 0$ there would still remain a small allowed parameter space for \acs{DBI}/G-/Galileon inflation\index{inflation!DBI}\index{inflation!G-}\index{inflation!Galileon} even if $r < 10^{-2}$ (see the upper right panel of Fig. \ref{fig:4}). Moreover, a much larger fraction of the allowed parameter space would survive the $r < 10^{-2}$ bound in the \acs{SRWBG} model, as seen from the lower right panels of Figs. \ref{fig:3} and \ref{fig:4}.

\section{Conclusive remarks}

In this chapter, we have outlined qualitatively the experimental bounds on the parameters of the \acs{EFTI} introduced in Part I, comparing some inflationary scenarios and focusing on theories with an underlying \acs{WBG} invariance. We have argued above that, at the quadratic order, the effective action \eqref{EFTI-action-2} only up to the operator $\delta N \delta K$ is the most general one that captures all the theories characterized by scalar perturbations with usual, phonon-like dispersion relation \eqref{phdr}\footnote{We want to remark on the fact that our qualitative analysis of Chap. \ref{chap-WBG-constraints} is rather general but it is not without loopholes. 
Indeed, there exist models such as ghost inflation\index{inflation!ghost} \cite{ArkaniHamed:2003uz,Senatore:2004rj,Ivanov:2014yla} where $\alpha\sim0$ and the background describes a perfect de Sitter space (\textit{i.e.} $\varepsilon=0$), so that the scalar speed of sound vanishes at the zero-th order, consistently with our general expression for $c_s^2$, Eq. \eqref{k-cssq2}. In such a case, one ought to consider effects of higher-derivative operators in \eqref{EFTI-action-2}, \textit{e.g.} $\delta K^2$, that will dominate the gradient energy of the scalar modes at horizon crossing, namely at characteristic frequencies of order $\omega \sim H$. This results in a rather different $\omega = k^2/M$ dispersion relation with some cutoff scale $M$, which would modify our analysis.}.
From an \acs{EFT} point of view, the main consequence of the \acs{WBG} symmetry is that the higher derivative operator $\delta N \delta K$ can be as relevant as $\delta N^2$, without invalidating the effective description. This effect is reflected in possibly enhanced non-Gaussianity in the scalar spectrum. The amplitude can be significantly larger with respect to the predictions of other models of inflation involving the same operators in the effective formulation. Probably, the phenomenologically most interesting \acs{WBG} inflationary model consists in a slow-roll evolution admitting large non-Gaussianity, as discussed in Chap. \ref{WBG-pot}. Furthermore, as shown in Chap. \ref{chap-WBG}, the \acs{WBG} invariance determines the existence of non-renormalization properties, that guarantee the quantum stability of the result and the validity of the effective expansion. This concept has been the main reason that motivated the present work. In Part II, it has been discussed mostly in the context of the standard inflationary paradigm. In the next part, it will be employed in a different class of cosmic evolutions.

%% file: chapters/3-01-chapter.tex
\chapter{Inflation from Minkowski space}
\label{inf-mink}

\begin{flushright}{\slshape    
    Naturally, we were all there - old Qfwfq said - where else could we have been? Nobody knew then that there could be space. Or time either: what use did we have for time, packed in there like sardines?
    I say ``packed like sardines", using a literary image: in reality there wasn't even space to pack us into. Every point of each of us coincided with every point of each of the others in a single point, which was where we all were. In fact, we didn't even bother one another, except for personality differences, because when space doesn't exist, having somebody unpleasant like Mr. $\text{Pber}^\text{t}$ $\text{Pber}^\text{d}$ underfoot all the time is the most irritating thing. } \\ \medskip
    --- Italo Calvino, \textit{Le Cosmicomiche}.
%    --- \defcitealias{bentley:1999}{BBBBB}\citetalias{bentley:1999} \citep{bentley:1999}
\end{flushright}

% For an example of a full page figure, see Fig.~\ref{fig:myFullPageFigure}.

\section{Towards non-standard cosmologies}
\label{inf-mink-0}

The Big Bang\index{Big Bang model} cosmology is essentially inspired by the observational fact that the Universe is expanding. Indeed, this suggests that, going backward in time, the Universe shrinks and becomes smaller.
On the other hand, the time derivative of \eqref{Fe1} yields
\begin{equation}
\dot{\rho} = -3H(\rho+p) \, .
\label{ims-nec}
\end{equation}
In other words, in a standard Universe satisfying the \acf{NEC}\index{Null Energy Condition (\acs{NEC})} $\rho+p\geq 0$ the energy density can never increase. This means that not only the Universe should have been very compact in the past but also extremely high energies should have been involved. This provides a connection between cosmology and high energy physics, which could shed light on the extreme initial phases. However, at the Planck scale General Relativity is expected to break down and some completion\index{UV completion} is required. These ingredients define the standard Big Bang paradigm.

As discussed in Chap. \ref{introduction}, such a picture is not devoid of defects. Indeed, an inflationary phase has to be invoked in order to solve some issues. The prize we have to pay is the violation of the \acf{SEC}\index{Strong Energy Condition (\acs{SEC})}, $\rho+3p\geq 0$.

At this point, a natural and fascinating question arises: could the \acs{NEC} be \textit{consistently} violated during the cosmic evolution?\index{non-standard cosmologies} If this were possible, as it is clear from \eqref{ims-nec}, one could get rid of the initial extreme regimes, disregarding any short-distance completion of gravitational interactions in the past, which is unavoidable in any \acs{NEC}-satisfying cosmology. In other words, the Big Bang singularity could be smoothed down and the Universe would emerge from a flat space-time\footnote{Using the Friedmann equations \eqref{Fe1}-\eqref{Fe2}, the violation of the \acs{NEC} is equivalent to $\dot{H}=-4\pi G(\rho+p)\geq 0$, which justifies the introduction in this chapter of a different definition for the $\varepsilon$-parameter, with respect to Eq. \eqref{eps}. Indeed, we will define $\varepsilon\equiv \dot{H}/H^2$, in order to make it positive definite.}. As we will see, this is not only of academic interest, because it automatically solves for instance the problems of the standard Big Bang model.
%for instance the flatness\index{flatness problem} and horizon problems\index{horizon problem}. Such a possibility was originally suggested in \cite{Creminelli:2010ba}.

It is fair to say that violating the \acs{NEC} is usually synonymous with instabilities, at least for a system consisting of an arbitrary number of scalar fields with up to one derivative per field in the action \cite{Dubovsky:2005xd,Hsu:2004vr}. The simplest example is provided by a ghost field with the ``wrong sign'' in the kinetic term: indeed, from \eqref{T00}-\eqref{Tii} one immediately infers that $\rho+p<0$. However, a ghost field is generically associated with a catastrophic vacuum instability.
More in general, in \cite{Dubovsky:2005xd} Dubovsky et al. proved that if a solution violates the \acs{NEC} either it is unstable or, if it is stable, both its stress-energy tensor is anisotropic and some superluminal perturbations propagate \cite{Nicolis:2009qm}.

Nevertheless, the statement is not without consistent loopholes. One possibility of evasion is provided by the ghost condensate\index{ghost condensate} \cite{ArkaniHamed:2003uy}, that crucially relies on (spontaneously) breaking Lorentz invariance\index{symmetry!Lorentz} in a way that gives rise to a non-standard $\omega \sim k^2$ infrared dispersion relation for the scalar field driving the \acs{NEC} violation. Indeed, it was argued in \cite{Creminelli:2006xe} that ghost condensation can lead to consistent alternative cosmologies with a weak ($\dot H\ll H^2$) violation of the \acs{NEC}. Another loophole has historically emerged with the discovery of the Galileon \cite{Nicolis:2008in}, which is ghost free, as we have seen. It has immediately been realized that (conformal) Galileons can be implemented in building a \acs{NEC}-violating alternative scenario to the standard Big Bang paradigm, referred to as \ac{GG} \cite{Creminelli:2010ba}. In this class of models, conformal transformations, or sometimes just the dilatations\index{dilations} \cite{Creminelli:2012my}, are assumed to be a symmetry of the flat-space theory, non-linearly realized\index{symmetry!non-linearly realized} by a scalar field $\pi$, while couplings of $\pi$ to gravity are assumed to weakly break that symmetry.
A crucial difference from the standard Big Bang model is that gravity is largely irrelevant for the early Universe described by \ac{GG}: the cosmological phase of interest, during which the perturbations (relevant for the \acs{CMB}\index{Cosmic Microwave Background (\ac{CMB})}) are produced, effectively takes place on a quasi-Minkowski space-time, while scale-invariant density perturbations are naturally produced due to the unbroken dilatation\index{dilations} invariance of the time-dependent scalar background\footnote{Similar ideas lie behind other constructions, such as that of a complex scalar rolling down a negative quartic potential \cite{Rubakov:2009np,Osipov:2010ee}, the \textit{pseudo-conformal Universe} \cite{Hinterbichler:2011qk} and \acs{DBI} genesis \cite{Hinterbichler:2012fr,Hinterbichler:2012yn}. The corresponding \acs{NEC}-violating backgrounds are characterized by the same symmetry-breaking pattern, albeit technically realized in different ways.}.
As mentioned before, flatness\index{flatness problem}, homogeneity and horizon problems\index{horizon problem} are automatically solved due to the quasi-Minkowski nature of the background space-time and the gradual shrinking of the comoving Hubble horizon $(a H)^{-1}$. It is thus fair to say that, as far as the standard problems of the Big Bang cosmology as well as density perturbations are concerned, Galilean Genesis\index{Galilean Genesis (\acs{GG})} is degenerate in its predictions with inflation. 

The differences come with the inclusion of tensor modes: irrelevance of gravity in genesis cosmologies results in a strongly blue-tilted and completely unobservable (at least as far as the \acs{CMB} experiments are concerned) spectrum of tensor perturbations\index{tensor modes} \cite{Creminelli:2010ba}. For that reason, it is commonly believed that any possible detection of primordial gravitational waves would strongly disfavour genesis models, as well as their many variations. 
Indeed, a detectable, scale-invariant\index{scale-invariance} tensor spectrum requires the background space-time to be (quasi-) de Sitter at the time of freeze-out of the relevant set of modes (for instance, we refer to \cite{Creminelli:2014wna} for a recent discussion). In the case of a detection of primordial tensor modes, this would mean that any scenario that aims to describe the early Universe should allow for a sufficiently extended period of de Sitter evolution.
This apparently singles out the standard inflation as the preferred paradigm to provide our flat and homogeneous Universe.

\begin{figure}
\includegraphics[width=0.75\textwidth]{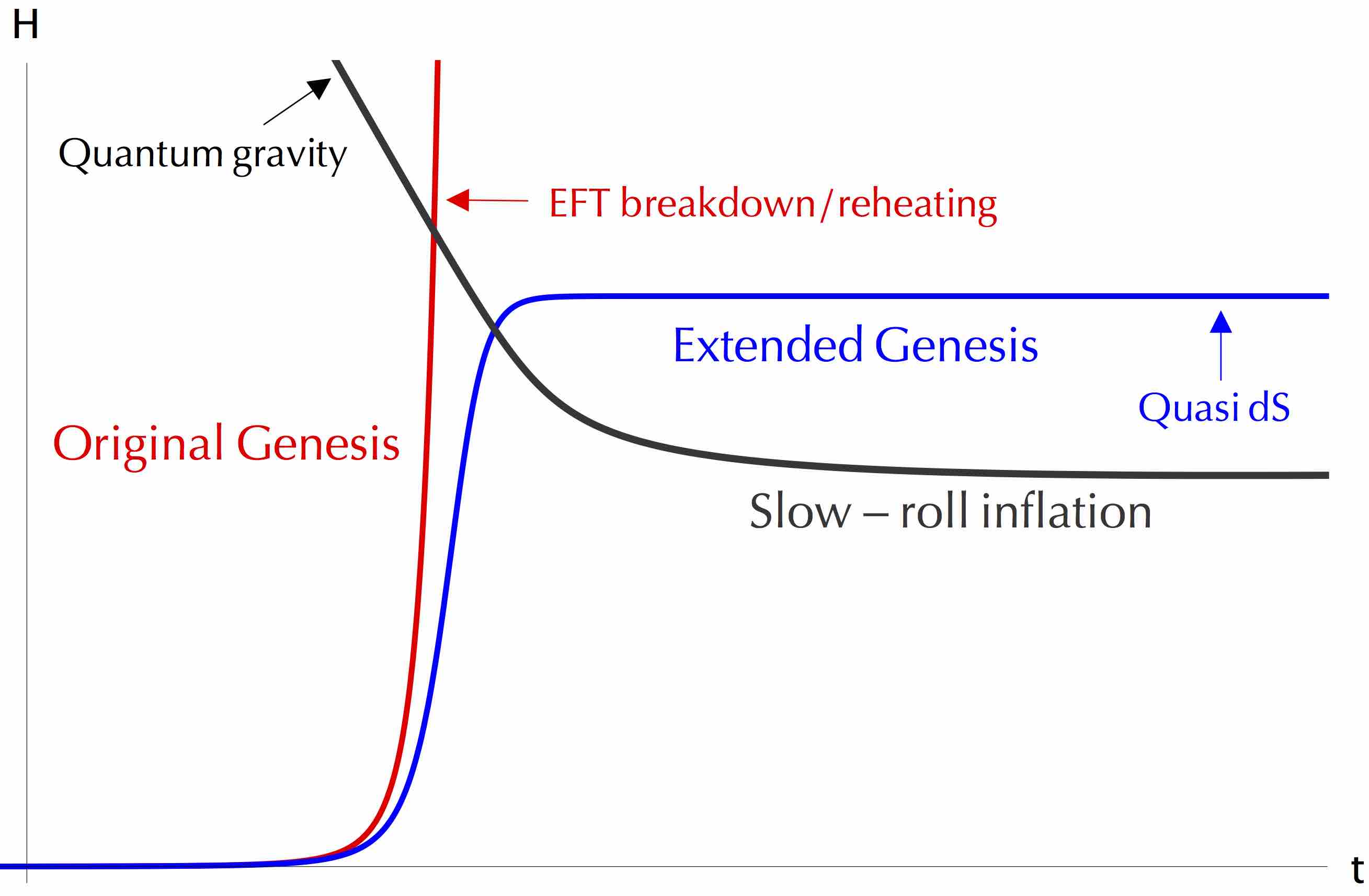}\centering
\caption{A sketch of the early Universe expansion rate as a function of time for the standard slow-roll inflation (black), as well as original (red) and extended (blue) genesis scenarios.}
\label{3fig:1}
\end{figure}

One motivation of this third part is to re-assess the latter observation, with a focus on Galilean Genesis as an alternative to inflation. We will broadly define the genesis as a phase of the Universe with a strongly \acs{NEC}-violating ($\varepsilon\equiv \dot H/H^2 \geq 1$) expansion that starts out in a low-curvature, maximally symmetric (essentially Minkowski or de Sitter\index{de Sitter!space}) space-time. Can such initial conditions result in a scale-invariant\index{scale-invariance} and unsuppressed tensor spectrum in a sufficiently broad range of physical scales? As noted above, at least for scalar-tensor theories we will be discussing below, generating scale-invariant tensor modes requires the geometry to be close to de Sitter for a certain period of time during the system evolution. The question therefore reduces to that of the possibility for the Universe to consistently evolve from a low/zero-curvature background in the far past to a much higher curvature inflationary de Sitter space-time capable of generating observable tensor spectrum at intermediate stages of its history. Because the system has to pass through a quasi de Sitter regime, one should be able to keep good theoretical control over the dynamics beyond the point when gravity starts playing a non-negligible role. Indeed, in the original \acs{GG}, the moment of time $t_0$ at which gravity becomes order-one important is roughly the moment of the \acs{EFT}\index{Effective Field Theory (\acs{EFT})} breakdown and not too long after that the Universe is assumed to reheat, while all relevant cosmological perturbations are generated at times $t\ll t_0$ (we will assume time to flow from $t=-\infty$ towards $t=0$ throughout). This situation is sketched by the red curve on Fig. \ref{3fig:1}. In terms of the model parameters, 
\begin{equation}
\label{tzero}
t_0\sim -\frac{f}{M_{\rm Pl}} \frac{1}{H_0}\, ,
\end{equation}
where $f$ is the decay constant of $\pi$, while $H_0\ll f$ is a free parameter, setting the scale for the expansion rate around $t \sim t_0$ (the natural value for the decay constant is $f\sim M_{\rm Pl}$, which we will assume for definiteness in what follows). The slow-roll parameter $\varepsilon$, starting out formally infinite at $t=-\infty$, decreases with time and is naively estimated to be of order unity at $t_0$. This means that the geometry can not be approximated by de Sitter space at any time during the genesis phase. 

While most of the qualitative features of \acs{GG} directly follow from the scale-invariance\index{scale-invariance} of the (flat-space) $\pi$-Lagrangian, the latter symmetry is badly broken by gravity around $t=t_0$. The background field value can be estimated at that time as
\begin{equation}
\phi\equiv \e^\pi \simeq \mathcal{O}(1)~,
\end{equation} 
whereas throughout the genesis phase $\phi\ll 1$. Let us imagine that the scale-invariant\index{scale-invariance} \acs{GG} Lagrangian is supplemented by dimension-five or higher, dilatation\index{dilations} non-invariant operators. One is then led to conclude that the symmetry-breaking terms in the effective action for $\pi$ itself can start influencing the dynamics for $t\sim t_0$ -- even in the extreme case that these are down by the Planck scale. Indeed, the canonically normalized field $\pi_c$ becomes of order $\pi_c(t_0) \sim f $, making \textit{e.g.} the Planck-suppressed operator  $\pi_c (\partial\pi_c)^2$ of the same order as the kinetic term -- at least in case that the decay constant $f$ is not far from $M_{\rm Pl}$. 
These estimates motivate the extension of the $\pi$ action with dilatation-breaking operators\index{dilations} that, while irrelevant throughout the genesis phase, could in principle strongly influence the dynamics around the time when gravity becomes order-one important. 

We will show below that, at least for a well-defined subclass of the resulting extensions, cosmological solutions resembling Galilean Genesis at early times, but smoothly extending beyond the time $t=t_0$ as illustrated by the blue curve on Fig. \ref{3fig:1}, do exist. These solutions, starting from some time $t_{i}$, asymptotically reach an inflationary (quasi) de Sitter space on which both the scalar and the tensor modes are generated with scale-invariant spectrum, just like in inflation. Nevertheless, the scenario at hand -- referred to as \textit{Extended Genesis} below -- crucially differs from inflation in the Universe evolution at early times ($t\ll t_i$). Most importantly, \acs{NEC} violation provides the possibility to avoid the singularity in the past, with the Universe gradually relaxing to a low- (or even zero-) curvature space as it is run backwards in time. Due to the latter property, Extended Genesis can be alternatively viewed as an early-time complete realization of inflation\index{inflation}.

In the cases we consider below, the late-time dynamics of the Extended Genesis will be described by \acs{NEC}-violating versions of G-/Galileon inflation\index{inflation!Galileon}\index{inflation!G-}, equipped with the weakly broken Galileon symmetry and all its phenomenologically attractive properties. First, it can produce a large tensor-to-scalar ratio without trans-Planckian field excursions, unlike the standard slow-roll inflation\index{inflation!slow-roll} \cite{Lyth:1996im}. Second, a \acs{WBG} phase can lead to a sizeable equilateral non-Gaussianity, with distinctive signatures with respect to ghost and \acs{DBI} models, and the standard Galileon inflation\index{inflation!Galileon}, as we have widely discussed in the previous part. Finally, since $\pi$ itself acquires a scale-invariant\index{scale-invariance} spectrum, it is in principle unnecessary to invoke spectator fields, required in many alternative models to inflation, to generate the observed density perturbations.

\section{Generalities}
\label{inf-mink-1}

Before diving into a more detailed discussion, we briefly highlight the major properties of theories allowing for the genesis-de Sitter transition. We expect these properties to be the defining ingredients of any other construction capable of achieving our goals. Most importantly, the theories of interest enjoy an enhanced symmetry both for small and for large values of the ``sigma model'' field $\phi$. In both limits $\e^\pi\ll 1$ and $\e^\pi\gg 1$, the (flat-space) $\pi$-Lagrangian will acquire invariance either under dilatations\index{dilations},
\begin{equation}
\label{dil}
\pi(x)\to \pi(\e^\lambda x)+\lambda~, 
\end{equation}
describing the scale-invariant\index{scale-invariance} (and, in special cases, conformal) Galileon \cite{Nicolis:2008in}, 
or under constant shifts,
\begin{equation}
\label{shift}
\pi(x)\to \pi(x)+\lambda~,
\end{equation}
describing for instance $P(X)$ theories\footnote{By ‘‘$P(X)$'' we mean an arbitrary polynomial in $X\equiv-(\partial\pi)^2$.} or the ghost condensation \cite{ArkaniHamed:2003uy,ArkaniHamed:2003uz}. Apart from the two asymptotically exact symmetries, for $\e^\pi\gtrsim 1$ it is useful to equip the Lagrangians under consideration with a weakly broken invariance under internal Galileon transformations\index{Galileon!transformations}, 
\begin{equation}
\label{galinv}
\pi\to\pi+b_\mu x^\mu~,
\end{equation}
leading to a theory of the type \eqref{hor1}-\eqref{hor4}, because of the important consequences we have demonstrated in Part II.
Galileon invariance becomes more and more pronounced as $\pi\to 0$.
As we will see in Sec. \ref{analytic}, in certain cases the small-field regime will itself consist of two qualitatively different stages -- the system gradually evolving from ghost condensate\index{ghost condensate} (described by an effectively shift-symmetric theory) in the asymptotic past into Galilean Genesis with an enhanced scale-invariance\index{scale-invariance} \eqref{dil} -- all while $\phi\ll 1$.
%\begin{comment}
\begin{figure}
\centering
\includegraphics[width=.7\textwidth]{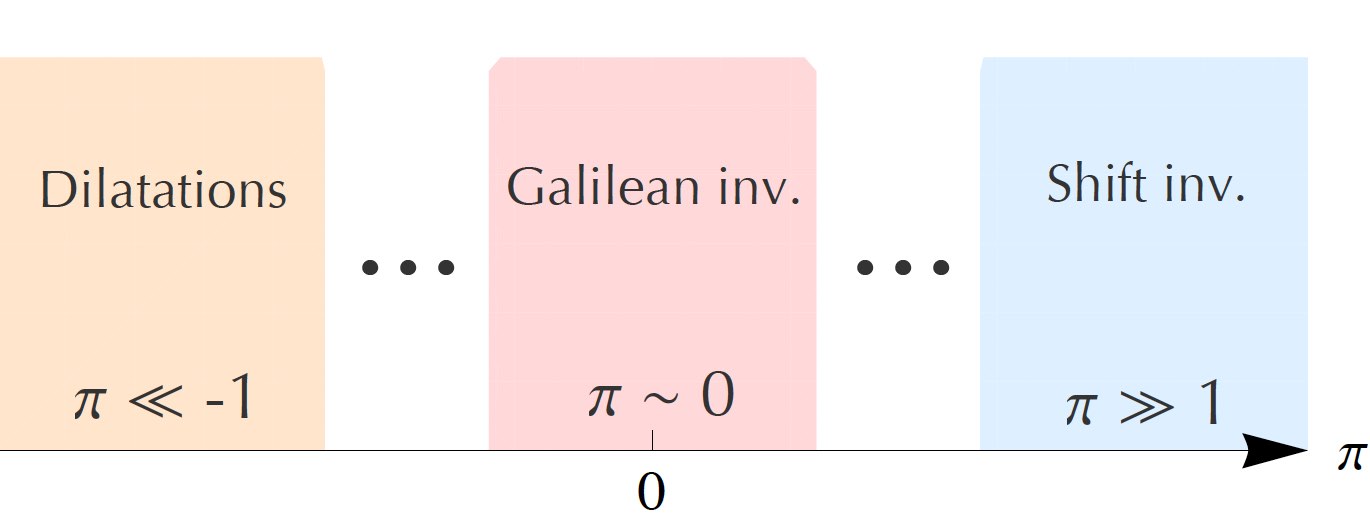} 
\caption{Illustration of emergent symmetries on backgrounds with various expectation values of the scalar $\pi$.}
\label{symmetries}
\end{figure}
%\end{comment}

The asymptotically emergent symmetries are precisely what makes the existence of \acs{NEC}-violating cosmologies, interpolating between Minkowski and de Sitter space-times possible. Let us consider for instance the genesis-de Sitter transition of Fig. \ref{3fig:1}. The enhanced conformal invariance at early times/small field values\footnote{To avoid confusion, we note again that ``small field values'' refers to the expectation value of the sigma model field $\e^\pi$, while the Goldstone\index{Goldstone field} $\pi$ is characterized by large negative values in the given regime.} generically gives rise to a \acs{GG}-like evolution of the Universe, whereby the conformal invariance $SO(4,2)$\index{conformal!$SO(4,2)$ group} gets broken down to the maximal de Sitter subgroup $SO(4,1)$\index{de Sitter!$SO(4,1)$ group} by a time-dependent background\footnote{We stress that while de Sitter group is the (linearly realized) symmetry group of the scalar action, the geometry throughout the Galilean Genesis phase remains close to flat.}, the Hubble rate and the sigma model field $\e^\pi$ growing as time flows from $t=-\infty$ towards $t=0$. On the other hand, whenever $\e^\pi$ starts exceeding unity, the emergent shift symmetry naturally leads to an attractor solution with de Sitter \textit{geometry} on which the scalar acquires a linear profile, $\pi\propto t$ \cite{ArkaniHamed:2003uz, Kobayashi:2010cm}. This qualitatively explains the gradual transformation between \acs{GG} and de Sitter phases, as illustrated by the blue curve in Fig. \ref{3fig:1}. 

Last but not least, the enhanced symmetries for large and small field values lead to the quantum robustness of the whole qualitative picture. Indeed, both symmetries \eqref{dil} and \eqref{shift} are broken at order one when $\e^\pi\sim 1$, making it hard to argue in favour of quantum stability of the detailed intermediate-time behaviour of our solutions. Nevertheless, the presence of asympotic exact (or even only weakly broken) symmetries determines the radiative stability of both the early- and the late-time dynamics. Backgrounds exhibiting the genesis-de Sitter transition can thus be expected to exist \textit{generically}, since both of the asymptotic solutions arise solely from symmetry considerations. A similar discussion of quantum robustness has been given in Ref. \cite{Elder:2013gya} in the context of flat-space constructions interpolating between \acs{NEC}-satisfying and \acs{NEC}-violating vacua.

For completeness, in the rest of the section we give a relatively detailed overview of the two asymptotic regimes of the solutions we wish to study.

\subsection{Galilean Genesis}

The conformal symmetry\index{symmetry!conformal}, $SO(4,2)$\index{conformal!$SO(4,2)$ group}, can be generically broken down to its maximal, de Sitter subgroup $SO(4,1)$\index{de Sitter!$SO(4,1)$ group} by a time-dependent scalar profile \cite{Nicolis:2008in,Nicolis:2009qm,Fubini:1976jm}. One way to achieve such breaking is via the (simplest non-trivial) \textit{conformal Galileon} Lagrangian\index{Galilean Genesis (\acs{GG})}
\begin{equation}
\label{gg}
S_{\text{1}}=\int \D^4 x \sqrt{-g}~\bigg[ f^2 \e^{2\pi}(\partial\pi)^2+\frac{f^3}{\Lambda^3}(\partial\pi)^2\Box\pi+\frac{f^3}{2\Lambda^3} (\partial\pi)^4\bigg ]~.
\end{equation}
It can be straightforwardly checked that the theory possesses an exact rolling solution on flat space-time \cite{Creminelli:2010ba,Nicolis:2009qm}
\begin{equation}
\label{ggsol}
\e^\pi=-\frac{1}{H_0 t}, \qquad H_0^2=\frac{2\Lambda^3}{3f}~,
\end{equation}
leading precisely to the $SO(4,2) \rightarrow SO(4,1)$ breaking pattern. The dilatation\index{dilations} invariance, left unbroken by the background, leads to a vanishing energy density\footnote{This immediately follows from scale-invariance ($\rho\propto \frac{1}{t^4}$) plus the energy conservation ($\dot\rho=0$).}, $\rho=0$, while the pressure $p=-2f^2/(H_0^2t^4)$ is negative -- implying a strongly \acs{NEC}-violating ($\dot H\gg H^2$) expansion \cite{Creminelli:2010ba}. The Universe described by \acs{GG} starts out in flat space-time, the Hubble rate growing according to the second Friedmann equation\index{Friedmann-Robertson-Walker (\ac{FRW})!equations}
$2M_{\rm Pl}^2 \dot H=-(\rho+p)$, which upon integration yields
\begin{equation}
\label{ggsol1}
H\simeq-\frac{1}{3}\frac{f^2}{M_{\rm Pl}^2}\frac{1}{H^2_0t^3}~.
\end{equation}
The time $t_0$ at which gravity starts playing non-negligible role ($H\sim \dot\pi$), can be estimated as in \eqref{tzero}. It roughly coincides with the time of \acs{EFT} breakdown/start of reheating. Scalar perturbations, relevant for the \acs{CMB} are instead produced at earlier times $t\lesssim t_0$, via minimally coupling an additional, scaling dimension-$0$ field $\varphi$ to the ``fake de Sitter'' metric $g^{\text{dS}}_{\mu\nu}=\e^{2\pi}\eta_{\mu\nu}$. This leads to a scale-invariant spectrum for the spectator $\varphi$ (despite the background metric being practically flat), that can be later imprinted on the physical curvature perturbation $\zeta$ through one of the standard mechanisms \cite{Enqvist:2001zp,Lyth:2001nq,Dvali:2003em}. The near-to-flat geometry on the other hand implies a strongly blue-tilted tensor spectrum $P_T(k)\sim k^2$, largely irrelevant for \acs{CMB} observations \cite{Creminelli:2010ba}. 

\subsection{Inflation with WBG symmetry}

Inflation with a \acs{WBG} symmetry\index{inflation!WBG} has been the central topic of Part II. However, it is worth dedicating here another paragraph in order to review it with a more useful notation for the purposes of this third part. We collect here the main results and specialise to a particular case.
Describing the late-time de Sitter asymptotic phase, we will be interested in the following \acs{WBG} action
\begin{equation}
\label{ginf}
S_{\text{2}}=\int \D^4 x \sqrt{-g} ~\bigg[f^2 (\partial\pi)^2+\gamma_3\frac{f^3}{\Lambda^3}(\partial\pi)^2\Box\pi+\gamma_4\frac{f^3}{2\Lambda^3} (\partial\pi)^4\bigg ]~,
\end{equation}
where $\gamma_{3,4}$ are constant parameters. The above action represents a particular case of the theory \eqref{full}, possessing all the nice quantum properties and the physical consequences extensively studied in Part II.

The form of the action \eqref{ginf} is dictated by the early-time asymptotic genesis. Indeed, both the interactions in \eqref{ginf} are also present in \eqref{gg}, the only difference between the two theories being that the former lacks scale-invariance. Moreover, the Galileon term is crucial for the speed of sound in the inflationary regime to be strictly positive. 

In the notations of \eqref{ginf}, the Friedmann equation and the equation of motion for $\pi$ take on the following form on spatially flat \acs{FRW} backgrounds 
\begin{gather}
H^2=\frac{f^2}{3 M_{\rm Pl}^2 H_0^2} \left( \gamma_4 \dot\pi^4+4\gamma_3 H\dot\pi^3-H_0^2\dot\pi^2    \right)  ~, \\
\left( 4 \gamma_4\dot\pi^2 +8\gamma_3 H\dot\pi -2 H_0^2\right)\ddot\pi +4 \gamma_4 H\dot\pi^3+4\gamma_3\left(3 H^2+\dot H\right)\dot\pi^2-6 H_0^2 H\dot\pi=0 ~,
\end{gather}
making the existence of de Sitter vacua ($H=\text{const.}$) with a linear $\pi\propto t$ profile explicit -- a direct consequence of shift-invariance \eqref{shift} of the $\pi$-Lagrangian. %\footnote{The shift invariance of the cubic galileon gets promoted to \textit{galilean invariance} $\pi\to\pi+b_\mu x^\mu$ on flat spacetime, which is however broken by the quartic term.}. 
Furthermore, the expansion rate and the scalar profile can be estimated as
\begin{equation}
\label{sol1}
H^2\sim \frac{f^2}{M_{\rm Pl}^2} H_0^2~,\qquad \dot \pi\sim H_0~.
\end{equation}

As we have learnt in Part I, the simplest and the most straightforward way of studying the spectrum of scalar perturbations is based on the \acs{EFTI}\index{inflation!EFT of} (see Chap. \ref{chap-EFTinflation}). However, for matter of convenience, we prefer the notations of App. \ref{Appendix-Eij}. Therefore, the two operators in the effective theory \eqref{EFTI-action-Eij} that lead to non-trivial dynamics at high energies are the $\delta N^2$ and $\delta N\delta E^i_{~i}$ terms.
As inferred from Tab. \ref{tab1} by simply setting $X\sim 1$, $Z\sim f/M_{\rm Pl}$ and $\Lambda_2^4=f\Lambda^3$, the coefficients of these terms are of order
\begin{equation}
\label{sol2}
m^4_2\sim f^2 H_0^2, \qquad \hat{m}^3_1\sim f^2 H_0~.
\end{equation}
Alternatively, this can be inferred also from Eq. \eqref{M's}, related to the generalized theory of Sec. \ref{ggal}, where the particular choice $\mathcal{F}_2=1$ corresponds exactly to the present case.

We can now estimate the scalar power spectrum and the size of the tensor modes. In doing so, we use the definition \eqref{dS-ps}-\eqref{dS-ps-delta}, where the normalization factor $\mathcal{N}$ is given by the computation \eqref{mg-N} and is estimated to be $\mathcal{N}\sim M_\text{Pl}^2$, because of the result \eqref{sol2}. Moreover, from Eq. \eqref{mg-cs} one finds $c_s^2\sim f/M_\text{Pl}$.
On the other hand, the tensor spectrum\index{tensor modes} is given by the universal formula $\Delta^2_T\sim H^2/M_{\rm Pl}^2$. Using Eqs. \eqref{sol1}, one finds the following expressions for the dimensionless power spectra\index{tensor-to-scalar ratio}\index{power spectrum}
\begin{equation}
\Delta^2_\zeta \sim \frac{f^{1/2} H_0^2}{M_{\rm Pl}^{5/2}},\qquad
\Delta^2_T \sim \frac{f^2 H_0^2}{M_{\rm Pl}^4}, \qquad
r=\frac{\Delta^2_T}{\Delta^2_\zeta } \sim \left(\frac{f}{M_{\rm Pl}}\right)^{3/2}~.
\end{equation}
Moreover, one can see from the above expression that the tensor-to-scalar ratio can easily be made large enough to be detectable if $f$ is sufficiently close to $M_{\rm Pl}$ -- all within the regime of validity of the underlying effective field theory. 

The amount of non-Gaussianity\index{non-Gaussianity} in these kinds of inflationary evolutions has already been estimated in Part II and summarized in Tab. \ref{tabsummodels}, therefore we do not insist on this further. 

\begin{figure}
\centering
\includegraphics[width=.4\textwidth]{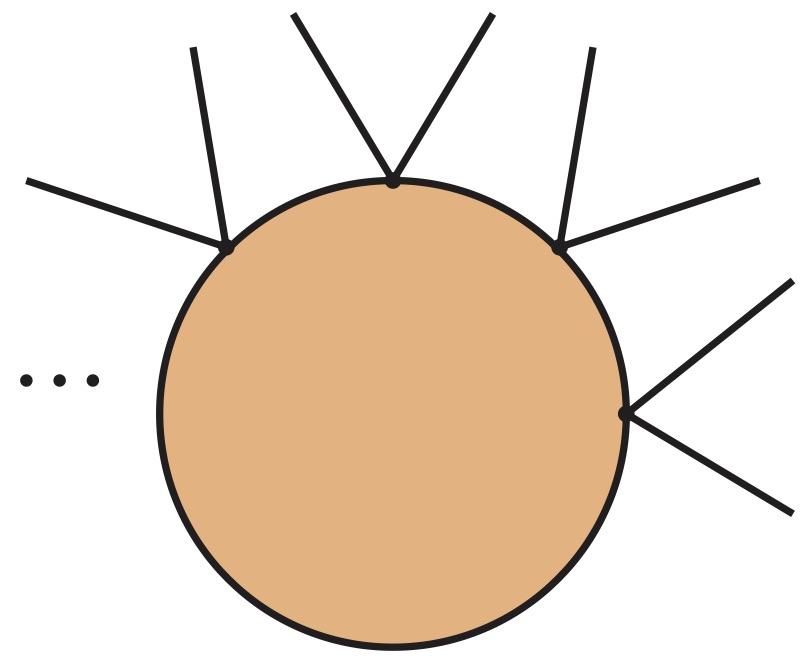} 
\caption{The diagram, responsible for the dominant quantum correction to the background solution in a \acs{WBG} theory of inflation.}
\label{diagram}
\end{figure}

However, we recall that, belonging to the \acs{WBG} class, at the quantum level the theory \eqref{ginf} can only generate effective vertices of the form in Fig. \ref{diagram}, where each pair of fields $(\partial\pi)^{2}$ is suppressed by at least one factor of $f$. In other words, the loop-generated self-interactions of the scalar field, estimated as in Chap. \ref{chap-WBG}, in the current notations are
\begin{equation}
\mathcal{L}_{\text{loop}}= \frac{(\partial\pi)^{2n}}{f^n \Lambda^{3k}}, \qquad k=n-\frac{4}{3}~.
\end{equation}
Evaluating $\mathcal{L}_{\text{loop}}$ on the classical de Sitter background gives
\begin{equation}
\mathcal{L}_{\text{loop}}=f^{4/3} H_0^{8/3}\ll \rho_{\text{dS}}~,
\end{equation}
independently of $n$ and where $\rho_{\text{dS}}\equiv f^2H_0^2$. As we already know, this means that quantum corrections of the form $(\partial\pi)^{2n}$ do not modify the background obtained from the Lagrangian (\ref{ginf}), establishing a criterion of consistency of the theory at quantum level. 

Finally, recall that the $\pi\propto t$ solution describes a perfect de Sitter space, leading to exactly scale-invariant perturbations; adding a small potential (or deforming the form of the action otherwise), both the scalar and the tensor modes can be produced with slightly tilted spectra -- just as they are in the canonical inflationary case. In addition, to complete the picture one of course has to specify a mechanism for exiting the de Sitter phase. There are known ways of achieving this, and we refer the interested reader to works dealing with similar issues in various contexts \cite{ArkaniHamed:2003uz,Senatore:2004rj,Ivanov:2014yla,Osipov:2010ee}.

\subsection{An explicit example}

As a simple example of a theory with the above-described asymptotic features, one can consider the deformed Galilean Genesis Lagrangian
\begin{equation}
S=\int \D^4x~ \sqrt{-g} ~\bigg[\frac{1}{2}M_{\rm Pl}^2 R +f^2~ \frac{\e^{2\pi}}{1+\beta \e^{2\pi}} ~(\partial\pi)^2+\frac{f^3}{\Lambda^3}(\partial\pi)^2\Box\pi +\frac{f^3}{2\Lambda^3}(\partial\pi)^4  \bigg] 
\label{simplemodel}
\end{equation}
with $\beta$ an arbitrary constant. For $\beta=0$ the theory is just the conformal Galileon: the Universe starts growing in the \acs{GG} phase, then the expansion rate diverges and the background exits the regime of validity of the \acs{EFT} at some finite time (see the red curve in Fig. \ref{3fig:1}) -- the scalar profile growing as $\e^\pi\sim 1/t$ throughout. For a non-zero $\beta$ on the other hand, the dynamics of the system is completely altered as soon as $\beta \e^{2\pi}$ becomes of order, or greater than one: the theory becomes effectively described by a \acs{WBG} Lagrangian with a cubic Galileon self-interaction, resulting in transition into an inflationary de Sitter phase. 
The corresponding solutions are studied in App. \ref{num}, where the existence of Extended Genesis cosmologies is illustrated via numerical analysis: the system clearly exhibits transition from genesis into a quasi-de Sitter regime precisely around the time $t_0$ given in \eqref{tzero}, see Fig. \ref{figuretwo}. Perhaps the only downside of this simple model is the short temporal region with gradient instability\index{gradient instability} at intermediate times: while completely free from ghosts, the squared speed of sound of the scalar perturbation on the given background becomes slightly negative around $t\sim t_0$ for a period of roughly a Hubble time, as shown in Fig. \ref{figuretwo}. However, while being certainly a problem in the classical theory, we argue in Sec. \ref{hdim}  (see also App. \ref{num}) that higher-order effects in the effective theory for perturbations can easily take care of this issue, rendering the entire cosmological evolution free from instabilities.

We refer the reader to App. \ref{num} for a detailed discussion of the numerical solutions to the illustrative model \eqref{simplemodel}, and turn to a systematic construction of theories leading to early Universe cosmology with the genesis -- de Sitter transition in the next section.

\section{Generalized Galileons}
\label{ggal}

In the present and the next section we will take on the task of obtaining (analytic) cosmological solutions exhibiting Extended Genesis. Rather than constructing solutions to a particular theory obeying the asymptotic scale and shift symmetries described in Sec. \ref{inf-mink-1}, we will employ the trick used in Ref. \cite{Elder:2013gya}, where the appropriate theory itself is inversely engineered, being based on a postulated ansatz for the desired cosmological solution. The asymptotic symmetries, as we will see, then follow automatically from the construction that we describe in what follows.

Consider the following (generally dilatation-breaking) deformation of the \acs{GG} Lagrangian
\begin{equation}
\label{ggg}
\mathcal{S}_\pi=\int d^4 x ~\sqrt{-g} ~\bigg[f^2 \mathcal{F}_1(\pi) (\partial\pi)^2+\frac{f^3}{\Lambda^3} (\partial\pi)^2\Box\pi +\frac{f^3}{2\Lambda^3} \mathcal{F}_2(\pi) (\partial\pi)^4    \bigg ]
\end{equation} 
where $\mathcal{F}_{1,2}$ are \emph{a priori} arbitrary dimensionless functions of the Galileon field $\pi$. We will interchangeably use the two scales $\Lambda$ and $H_0$ -- as defined in \eqref{ggsol} -- throughout. The dynamics of the system is governed by the Einstein equations plus the scalar equation of motion. However, these are not independent: as a consequence of diffeomorphism invariance, the scalar equation can be traded for the conservation of its stress-energy tensor via
\begin{equation}
\label{s.eom}
\nabla_\mu T^\mu_{~\nu}=-\frac{\delta S}{\delta \pi} \partial_\nu\pi~.
\end{equation}
On homogeneous \acs{FRW} backgrounds, it is the energy conservation, $\dot \rho + 3H (\rho+p)=0$, that yields the $\pi$ equation of motion. On the other hand, energy conservation follows from the temporal and space components of the Einstein equations. Therefore we can choose the latter two to make up a complete system determining the background evolution. 
The stress-energy tensor, sourced by $\pi$ in \eqref{ggg} is  
\begin{multline}
T^\pi_{\mu\nu}
	= -f^2 \mathcal{F}_1(\pi)[2\partial_\mu\pi\partial_\nu\pi-g_{\mu\nu} (\partial\pi)^2]
\\
	-\frac{f^3}{\Lambda^3}[2\partial_\mu\pi\partial_\nu\pi\Box\pi-
\partial_\mu\pi\partial_\nu(\partial\pi)^2-\partial_\nu\pi\partial_\mu(\partial\pi)^2
+g_{\mu\nu}\partial_\lambda\pi\partial^\lambda(\partial\pi)^2]
\\
	-\frac{f^3}{2 \Lambda^3}\mathcal{F}_2(\pi) ~[4 (\partial\pi)^2\partial_\mu\pi\partial_\nu\pi-g_{\mu\nu} (\partial\pi)^4]~,
\end{multline}
leading to the following expressions for the energy density and pressure due to a homogeneous $\pi$-profile
\begin{align}
\label{rho}
\rho &=\frac{f^2}{H_0^2} ~\dot \pi^2 \big [ \mathcal{F}_2(\pi)\dot\pi^2+4 H\dot\pi -  H_0^2\mathcal{F}_1(\pi)\big ] ~,\\
\label{press}
p &=\frac{f^2}{3 H_0^2}~\dot\pi^2 \bigg[ \mathcal{F}_2(\pi) \dot\pi^2-4\ddot \pi-3H_0^2\mathcal{F}_1(\pi) \bigg]~.
\end{align}
The two functions $\mathcal{F}_{1,2}(\pi)$ can be solved with the help of the temporal and spatial components of Einstein equations, $3 M_{\rm Pl}^2 H^2=\rho$ and $M_{\rm Pl}^2 (3 H^2+2 \dot H)=-p$, which yield
\begin{align}
\label{f1}
\mathcal{F}_1&=\frac{6 M_{\rm Pl}^2 H_0^2 H^2+3M_{\rm Pl}^2 H_0^2\dot H-2 f^2 H\dot \pi^3-2f^2 \dot\pi^2\ddot\pi}{f^2 H_0^2\dot\pi^2} \, , \\
\label{f2}
\mathcal{F}_2&=\frac{9M_{\rm Pl}^2 H_0^2 H^2+3M_{\rm Pl}^2 H_0^2\dot H-6 f^2 H\dot \pi^3-2f^2 \dot\pi^2\ddot\pi}{f^2 \dot\pi^4}~.
\end{align}
Now, for any postulated homogeneous profile of the scalar and the Hubble rate, one can find the theory (\textit{i.e.} find $\mathcal{F}_{1,2}(\pi)$) such that the desired background solves its equations of motion. The recipe for constructing the relevant solutions is given as follows:
\begin{itemize}
\item Postulate background profiles $\pi_0(t)$ and $H(t)$;

\item For the chosen background solutions, find the time-dependent functions $\mathcal{F}_{1,2} (t)$  with the help of \eqref{f1} and \eqref{f2};

\item Invert the expression for $\pi_0(t)$ to find $t=t(\pi_0)$;

\item Using the previous steps, find $\mathcal{F}_{1,2}$ as functions of $\pi_0$:

$\mathcal{F}_{1,2}=\mathcal{F}_{1,2}\left(t(\pi_0)\right)$.

\end{itemize}
One can formally construct theories admitting arbitrary cosmological profiles for $\pi$ and $H$. Although such an \textit{ad hoc} construction might look uncomfortable, we will see that at least for the solutions we will be interested in, it will lead to theories that enjoy various types of asymptotic symmetry, making them highly non-generic in the sense discussed in Sec. \ref{inf-mink-1}. 

\subsection{Perturbations}

As a next step, we check now whether the cosmological solutions, obtained through the above procedure, are stable. This can be done with the help of the analysis spelled out in App. \ref{appA}. In the unitary gauge, $\pi(x,t)=\pi_0(t)$, the only scalar degree of freedom present in the theory is captured by the standard massless curvature perturbation of equal-density hypersurfaces $\zeta$, defined by \eqref{ug-zeta}.

Having the background quantities at hand, one can readily derive the quadratic $\zeta$ action following the standard procedure of \cite{Maldacena:2002vr}, that we have described in Sec. \ref{sec-cswmwg}:
\begin{equation}
\label{quadact}
S_\zeta^{(2)}=\int d^4x~ a^3~\bigg[A(t)~\dot\zeta^2-B(t)~\frac{1}{a^2}\left(\vec{\nabla}\zeta\right)^2-C(t)~\frac{1}{a^4}\left(\vec{\nabla}^2\zeta\right)^2   \bigg]~.
\end{equation}
For our action \eqref{ggg}, the kinetic coefficients $A$ and $B$ are readable in Eqs. \eqref{mg-N}-\eqref{mg-cs}. In the notations of App. \ref{Appendix-Eij} they are 
\begin{align}
\label{A}
A(t) &=\frac{M_{\rm Pl}^2 (-4 M_{\rm Pl}^4 \dot H-12M_{\rm Pl}^2 H \hat m^3_1+3\hat m^6_1+2M_{\rm Pl}^2 m_2^4)}{(2M_{\rm Pl}^2H-\hat m^3_1)^2}~,\\
\label{B}
B(t)&=\frac{M_{\rm Pl}^2 \left(-4 M_{\rm Pl}^4 \dot H+2M_{\rm Pl}^2 H \hat m^3_1-\hat m^6_1+2M_{\rm Pl}^2\partial_t\hat m^3_1\right)}{(2M_{\rm Pl}^2H-\hat m^3_1)^2}~,
\end{align}
while $C(t)\equiv0 $. It will be non-zero once we include higher-order terms in the effective theory in Sec. \ref{hdim}.
Explicit expressions for the time-dependent coefficients $\hat m^3_1$ and $m^4_2$ are given in Eq. \eqref{M's}. %\footnote{As is explicit in \eqref{A} and \eqref{B}, for  $2M_{\rm Pl}^2H = \hat M^3$ both $A$ and $B$ diverge, suggesting that the given point in the `parameter space' corresponds to a non-dynamical scalar perturbation. However, $H$ and $\hat M^3$ are not free parameters of the theory, but are fixed by the background solution at hand. Indeed, one can show that for this particular relation between $H$ and $\hat M^3$, the equation of motion for the perturbation of the standard lapse variable ($\delta N\equiv N-1$), instead of fixing it, implies $\dot \zeta=0$ -- leaving the two tensor modes as the only propagating degrees of freedom on the given background. Close to this point on the other hand, $\zeta$ can be thought of as dynamical, however very weakly coupled.} 
Apart from other background quantities, these explicitly depend on the function $\mathcal{F}_2(\pi_0)$. Using the expression \eqref{f2} for the latter, one finds
\begin{align}
\label{Anec}
A&=3 M_{\rm Pl}^2 (36 M_{\rm Pl}^4 H_0^4 H^2+9M_{\rm Pl}^4 H_0^4\dot H-18 M_{\rm Pl}^2 f^2 H_0^2H \dot\pi ^3 \nonumber \\
&\qquad
	-6M_{\rm Pl}^2 f^2 H_0^2 \dot\pi ^2\ddot\pi+4 f^4 \dot\pi^6)(3M_{\rm Pl}^2 H_0^2 H-2f^2 \dot\pi^3)^{-2} \, , \\
\label{Bnec}
B&=\frac{-9M_{\rm Pl}^6 H_0^4 \dot H+6M_{\rm Pl}^4 f^2H_0^2H\dot\pi^3+18M_{\rm Pl}^4 f^2  H_0^2 \dot \pi^2\ddot\pi  -4M_{\rm Pl}^2 f^4\dot\pi^6}{(3M_{\rm Pl}^2 H_0^2 H-2f^2 \dot\pi^3)^2} \, ,
\end{align}
while the speed of sound for short wavelength scalar perturbations is given by $c_s^2=A/B$.
Positive $A$ and $B$ throughout the entire course of cosmological evolution guarantee the absence of ghost and gradient instabilities\index{gradient instability} respectively. 

As a quick check, one can apply the above piece of formalism to Galilean Genesis \cite{Creminelli:2010ba}.
Plugging the scalar and Hubble profiles, Eqs. \eqref{ggsol} and \eqref{ggsol1}, into the expressions for the kinetic coefficients \eqref{Anec}-\eqref{Bnec} of the curvature perturbation, one obtains the following values at the leading order in $M_{\rm Pl}$:
\begin{equation}
A(t)=B(t)=\frac{9M_{\rm Pl}^4H_0^2}{f^2} ~t^2~.
\end{equation}
This precisely agrees with the expressions found in \cite{Creminelli:2010ba}.

\section{Extended Genesis: analytic solutions}
\label{analytic}

\index{Extended Genesis}While the recipe spelled out in the previous section formally allows to construct theories admitting essentially arbitrary cosmological solutions, most of these fail to be physically meaningful in one way or another. Such a generic solution will lead to either ghost or gradient instability\index{gradient instability} at the level of small perturbations. Moreover, most of the resulting theories will be free from symmetries -- even the asymptotic ones, casting shadow on quantum robustness of the whole picture. Nevertheless, we will show in this section that a class of theories, which admit completely stable cosmological solutions interpolating between a low/zero curvature maximally symmetric space-time in the far past and a larger curvature inflationary de Sitter space-time in the future with a strong/moderate violation of the null energy condition in between, exists. Importantly, we will see that asymptotically these theories enjoy symmetries of the kind described in Sec. \ref{inf-mink-1}.

Let us work in a coordinate system such that time runs from $t =-\infty$ towards $t=0$ over the cosmological phase of interest. At (or shortly after) $t=0$, the system is assumed to reheat, or exit the given phase otherwise. Inspired by the early-time Galilean Genesis asymptotic regime \eqref{ggsol} and \eqref{ggsol1}, we will adopt the following ansatz for the Hubble rate
\begin{equation}
\label{ansatz}
H=\lambda +\beta ~\frac{f^2}{M_{\rm Pl}^2 H_0^2} ~\dot\pi_0^3~,
\end{equation}
where $\lambda$ and $\beta$ are free parameters (of mass dimension one and zero respectively) of the theory, giving rise to the solution of interest.
For the scalar, we will assume an ansatz of the following form
\begin{equation}
\label{ansatz1}
\e^{\pi_0}=\frac{1}{H_0} \frac{1}{t_*-t}~,
\end{equation}
motivated again by the genesis solution
Here, $t_*$ is yet another free parameter with inverse mass dimension. While resembling \acs{GG} at early times (and for sufficiently small $\lambda$), Eqs. \eqref{ansatz} and \eqref{ansatz1} describe a cosmology regularized towards $t\to 0^-$, so that none of the invariants in the theory grow unbounded over the entire interval $t\in (-\infty,~0]$. 
Galilean Genesis is recovered at all times for the particular values of the parameters $\lambda=0$, $\beta=1/3$ and $t_*=0$; for $\lambda\neq 0$ on the other hand, there is a crucial difference: rather than the flat, Minkowski space-time, the system starts out evolving from de Sitter space with curvature set by the parameter $\lambda$. 

In order for the Universe to be described by an inflationary de Sitter geometry at $t\to 0^-$, the parameters of the theory should satisfy certain constraints.   
One of such constraints arises from requiring the Hubble rate not to vary considerably over a single e-fold at $|t|\ll t_*$. The necessary condition for that is: 
\begin{equation}
\label{dscond}
1\gg \varepsilon \equiv \frac{\dot H}{H^2}\bigg |_{t\to 0}\sim 
\begin{cases}
\frac{M_{\rm Pl}^2 H_0^2}{\beta f^2} ~t_*^2~, &\text{if} ~~\lambda \ll \beta~\frac{ f^2}{M_{\rm Pl}^2 H_0^2}~\frac{1}{ t_*^3} \\
\frac{\beta}{\lambda^2}~ \frac{f^2}{M_{\rm Pl}^2 H_0^2}~\frac{1}{ t_*^4}~, &\text{if} ~~ \lambda \gg \beta~\frac{ f^2}{M_{\rm Pl}^2 H_0^2}~\frac{1}{ t_*^3}~.
\end{cases}
\end{equation}
Not surprisingly, this condition is equivalent to the one constraining $\dot \pi$ to be quasi-constant at late times: 
\begin{equation}
\frac{1}{H}\frac{\D}{\D t} \ln \dot\pi_0\ll 1 ~.
\end{equation}
This shows that $\pi$ can indeed be approximated by a linear profile towards $t\to 0^-$, leading to the late-time inflation discussed in Sec. \ref{inf-mink-1}.

In the rest of the section we will study various interesting regions in the six-dimensional space spanned by the free parameters $\left( M_{\rm Pl}, f,H_0,\lambda,\beta, t_* \right)$ of the theory.

\subsection{$\lambda =0$} 

We begin with the case that, in the asymptotic past, the system starts out evolving from flat space-time. This happens for $\lambda=0$. As a quick consistency check, one can derive the conformally invariant \acs{GG} Lagrangian \eqref{gg} from our ansatz for the Extended Genesis cosmology, following the inverse construction of the previous section. Indeed, plugging \eqref{ansatz} and \eqref{ansatz1} (with $\beta=1/3$) into the expressions for the $\mathcal{F}$-functions, \eqref{f1} and \eqref{f2}, we find at the leading order in $1/M_{\rm Pl}^2$ (and at times $|t|\gg t_*$)
\begin{equation}
\mathcal{F}_1=\frac{1}{H_0^2 t^2}=\e^{2\pi}, \qquad \mathcal{F}_2=1~.
\end{equation}
This precisely corresponds to the conformal Galileon. For values of $\beta$ other than $1/3$, on the other hand, our ansatz describes subluminal versions of \acs{GG} \cite{Creminelli:2012my} at $|t|\gg t_*$.

Concentrating on the full solution, including times $|t|\leq t_*$, stability of the system requires that the kinetic coefficients in \eqref{quadact} are positive at all times. For $\lambda=0$, they are given as follows 
\begin{align}
\label{atilde}
\frac{3 (2M_{\rm Pl}^2 H-\hat m_1^3)^2 H_0^4}{4M_{\rm Pl}^2}A&=2(4+15 x +18x^2) f^4 \dot\pi^6+3(4+9x)M_{\rm Pl}^2 f^2 H_0^2\dot\pi^2\ddot\pi~,~~~~~\\
\label{btilde}
\frac{3 (2M_{\rm Pl}^2 H-\hat m_1^3)^2 H_0^4}{4M_{\rm Pl}^2}B&=\left(2 x f^2\dot\pi^4-9x M_{\rm Pl}^2 H_0^2\ddot\pi\right)f^2 \dot\pi^2~,
\end{align}
where we have defined $x=\beta-2/3$ for further convenience. As an immediate observation, we note that $A$ is manifestly positive for positive $x$ (both $\dot\pi$ and $\ddot\pi$ are positive at all times for our ansatz), while $B$ does not have a definite sign. For the special  case that the parameter $x$ is small however, $B$ can be made arbitrarily small, compared to $A$, implying a vanishing speed of sound for $\zeta$. This is similar to what happens in ghost condensation, where the absence of gradient instability\index{gradient instability} is determined by higher-order operators in the effective theory.

It is straightforward to see that $B$ cannot be positive over the entire temporal interval of interest -- at least for our ansatz \eqref{ansatz}. Indeed, we are interested in solutions that start in Galilean Genesis at $t\to -\infty$ and end up in the inflationary phase at $t\to 0^-$. 
As shown in the previous section, the latter phase requires $\dot\pi$ to be practically constant, meaning that the second term in the parentheses on the r.h.s. of \eqref{btilde} should be negligible compared to the first one at late times. Positivity of $B$ at late times then requires $x>0$. On the other hand, Galilean Genesis corresponds to the second term prevailing at sufficiently early times, since $\ddot \pi\sim 1/t^2$ decreases parametrically slower than $\dot\pi^4\sim 1/t^4$ at large and negative $t$. For $x>0$ however, this would lead to gradient instability\index{gradient instability} at early times. In contrast, in the opposite case of $x<0$, one would recover gradient instability at late times, while the early-time genesis phase would be completely stable. One is therefore led to conclude that gradient instability is unavoidable for the given choice of the ansatz \eqref{ansatz} in the $\lambda=0$ case -- at least at the leading order in derivative expansion.

Concentrating on negative $x$ (so that the genesis phase is stable), the time at which gradient instability\index{gradient instability} occurs (\textit{i.e.} when $B$ flips sign) is of order $|\tau|\sim f/(M_{\rm Pl} H_0)$. The slow-roll parameter at that time can be readily estimated, $\varepsilon\sim M_{\rm Pl}^2 H_0^2 \tau^2/f^2\sim 1$, see Eq. \eqref{dscond}. This means that the gradient instability for $\lambda=0$ solutions necessarily kicks in before the onset of the de Sitter regime, explaining the pattern we have found via numerical analysis in Sec. \ref{inf-mink-1} (see also App. \ref{num}).

We conclude the present subsection with a couple of consistency checks for our calculations. First, we note that for $x=-1/3$, corresponding to Galilean Genesis, one recovers an exactly luminal scalar mode, $c_s^2=\tilde B/\tilde A=1$ at early times. Moreover, as stressed several times above, the late-time de Sitter phase should correspond to an enhanced shift symmetry on $\pi$. That this is indeed the case is the result of the quasi-constancy of the $\mathcal{F}$ functions
\begin{equation}
\frac{1}{H} ~\frac{\D}{\D t}\ln \mathcal{F}_{1,2} \ll 1~,
\end{equation}
which, as can be straightforwardly verified, directly follows from (the $\lambda=0$ version of) Eq. \eqref{dscond} -- the condition for the Universe to be described by de Sitter geometry at $|t|\ll t_*$.

\subsection{$\lambda \neq 0$} 

We now turn to the case that in the asymptotic past the Universe starts out evolving from de Sitter space, rather than Minkowski, $\lambda\neq 0$. The curvature of the initial state is of order $R\sim \lambda^2$ and is a free parameter of the theory. If its value is strictly zero, we have seen that the resultant cosmological solution suffers from a gradient instability\index{gradient instability} before the onset of de Sitter regime for much of the parameter space, at least if one ignores higher-order operators in the effective theory. However, for non-zero $\lambda$, as we will now demonstrate, gradient instabilities\index{gradient instability} can be avoided even in the ``classical'' theory, that is without invoking higher-derivative terms in the \acs{EFT} for perturbations.  

The kinetic coefficients \eqref{Anec} and \eqref{Bnec}, evaluated on the given ansatz are:
\begin{align}
A&=\frac{M_{\rm Pl}^2}{3}\big\{ 36  M_{\rm Pl}^4 H_0^4 \lambda^2\tau^6+3 M_{\rm Pl}^2 f^2 H_0^2 [-(10+24 x)\lambda\tau +4+9x]\tau ^2 \nonumber
\\
&\qquad
	+f^4 (8+30x+36 x^2)\big\}(f^2 x-M_{\rm Pl}^2 H_0^2\lambda \tau^3)^{-2}~, \\
B&=\frac{M_{\rm Pl}^2}{3}~ \frac{M_{\rm Pl}^2 f^2 H_0^2 (-2\lambda\tau-9x)\tau^2 + 2 f^4 x}{(f^2 x-M_{\rm Pl}^2 H_0^2\lambda \tau^3)^2}~ ,
\end{align}
where we have defined $\tau\equiv t-t_* \leq -t_*$ . An important observation that we will use in what follows is that for positive $x$, and for $\bar \varepsilon \equiv \lambda t_*>9 x/2$, both $A$ and $B$ are manifestly positive (and finite) \textit{at all times}, as can be readily verified by inspecting the above expressions. 

Given that a strictly vanishing $\lambda$ is not allowed by stability, how small can it be? The smallness of the initial curvature can be conveniently characterized by 
\begin{equation}
\label{hsep}
\frac{H(t=0)}{H(t=-\infty)}=\left(\lambda\frac{M_{\rm Pl}^2 H_0^2 t_*^3}{\beta f^2}\right)^{-1}\sim \frac{1}{\varepsilon \bar\varepsilon}~.
\end{equation}
Note that, while $\varepsilon\ll 1$ is required by the late-time de Sitter space, $\bar \varepsilon$ is in principle an unconstrained parameter of the theory. 

To summarize, choosing $x< 2\bar \varepsilon/9$, one can arrange for a manifestly stable cosmological solution, interpolating between two de Sitter space-times with an arbitrary ratio of the corresponding asymptotic curvatures. Moreover, the larger is the separation between the asymptotic Hubble rates \eqref{hsep}, the smaller is the deviation of the late-time geometry from perfect de Sitter space. The speed of sound of the curvature perturbation at $t=0$ can be readily evaluated from the above expressions for the kinetic coefficients
\begin{equation}
\label{cssq1}
c_s^2(t=0)=\frac{x (2-9 \varepsilon)+2\varepsilon \bar\varepsilon}{8+30 x+36 x^2+\mathcal{O}(\varepsilon)}~.
\end{equation} 
Note that the asymptotic $c_s^2$ is finite. For $\varepsilon=0$, its magnitude is bounded from above by $c_s^2<0.031$, which can be found by maximizing the expression \eqref{cssq1} for the squared speed of sound\footnote{Cf. the analytic bound on the scalar speed of sound $c_s^2<0.031$ in Galileon inflation\index{inflation!Galileon}, quoted in \cite{Kobayashi:2010cm}.}.

Let us for simplicity set $x= 0$ from now on. One distinct property of our ansatz is that the coefficient $A$, having a contribution constant in time, becomes parametrically greater than $B$ at $|t|\gg t_0$, as $B\sim -1/t^3$ at large and negative $t$. This means that the speed of sound of the curvature perturbation tends to zero at early times. 

What is the theory describing the asymptotic past of the background solutions at hand? To answer this question, we evaluate the $\mathcal{F}$ functions from our deformed Galileon action \eqref{ggg}. At the leading order in $1/t$, one finds
\begin{align}
\mathcal{F}_1&=6\frac{M_{\rm Pl}^2 \lambda^2}{f^2 H_0^2} (H_0 t)^2=6\frac{M_{\rm Pl}^2 \lambda^2}{f^2 H_0^2}\e^{-2\pi}~, \\
\mathcal{F}_2&=9\frac{M_{\rm Pl}^2 \lambda^2}{f^2 H_0^2} (H_0 t)^4=9\frac{M_{\rm Pl}^2 \lambda^2}{f^2 H_0^2}\e^{-4\pi}~,
\end{align} 
which imply the following form of the scalar action 
\begin{multline}
\label{earlydslag}
\mathcal{S}^{\text{early}}_\pi=\int d^4 x ~\sqrt{-g} \bigg[ 6\frac{M_{\rm Pl}^2 \lambda^2}{ H_0^2}\e^{-2\pi} (\partial\pi)^2
\\
+\frac{2}{3}\frac{f^2}{H_0^2} (\partial\pi)^2\Box\pi
+3\frac{M_{\rm Pl}^2 \lambda^2}{H_0^4}\e^{-4\pi}  (\partial\pi)^4    \bigg ]~.
\end{multline} 
In the regime of interest, $\e^\pi\sim1/t$ and the first and the third terms in the parentheses are constant, while the second (the cubic Galileon) goes as $\sim 1/t^3$ and is thus completely irrelevant in the asymptotic past\footnote{The latter estimate comes from the $H \dot \pi^3 $ piece, coming from the expansion of the covariant derivative on a de Sitter background.}. 
\begin{figure}
\centering
\includegraphics[width=.46\textwidth]{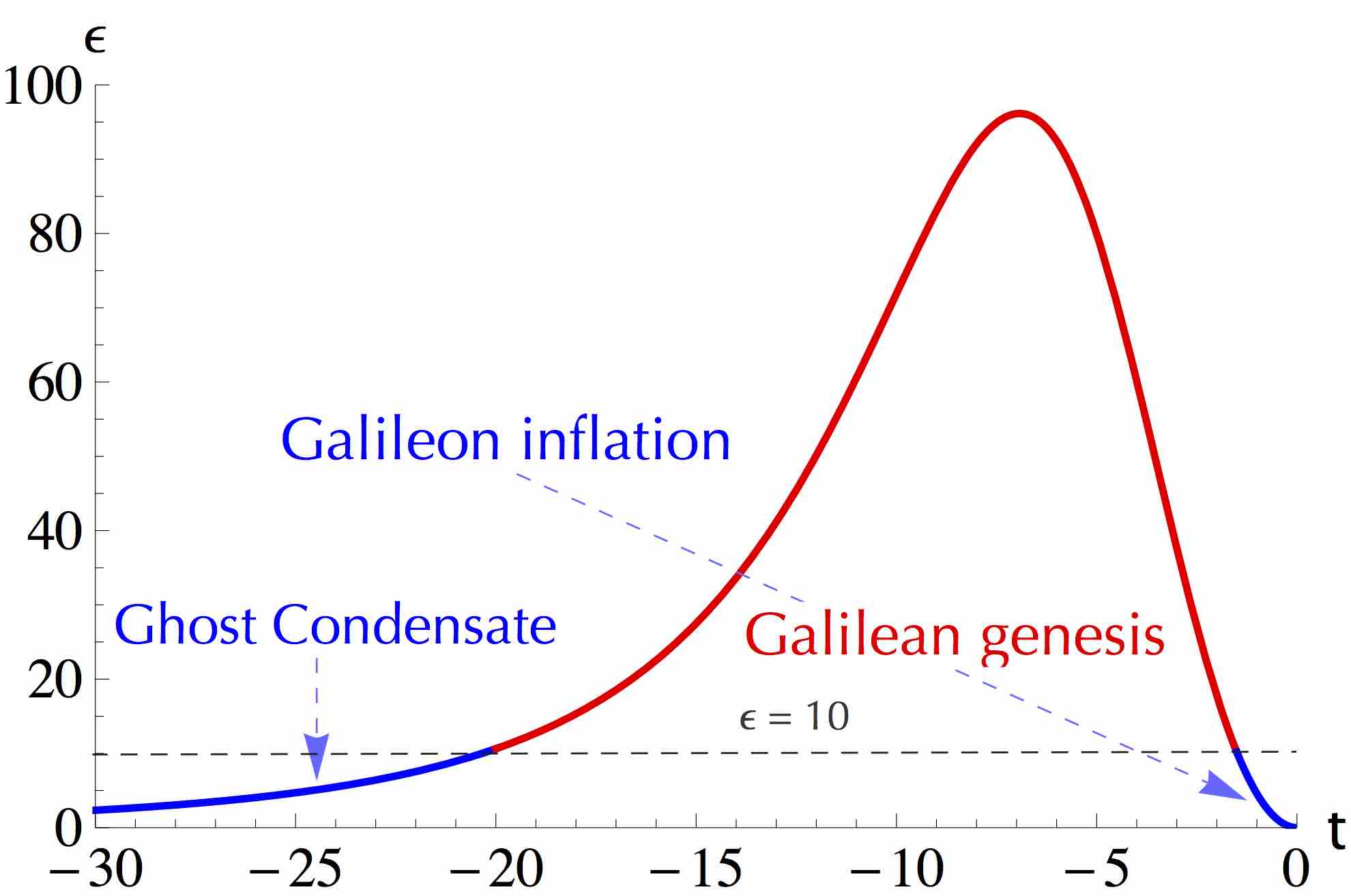} \quad
\includegraphics[width=.46\textwidth]{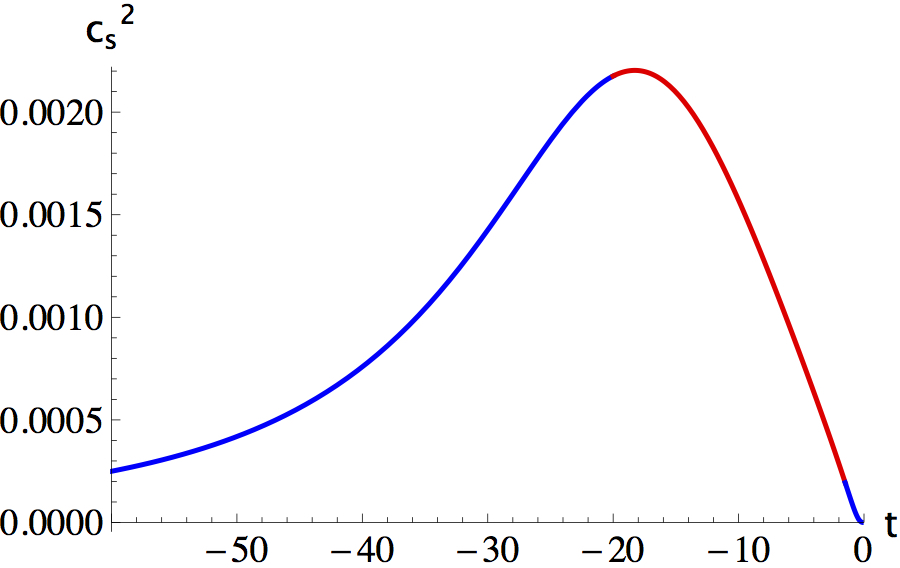}
\caption{The parameter $\varepsilon$ (left) and the speed of sound of the curvature perturbation $c_s^2$ (right) as functions of time on the solution \eqref{ansatz}-\eqref{ansatz1}. The scales $f$ and $M_{\rm Pl}$ have been assumed equal, while the rest of the parameters have been chosen to be: $H_0=1$, $\lambda=10^{-3}$, $t_*=10^{-2}$, $x=0$. The two colours correspond to $\varepsilon <10$ (blue) and $\varepsilon>10$ (red).}
\label{figurethree}
\end{figure}
However, once the cubic Galileon is neglected, the theory acquires a global symmetry. To see this, it is useful to define a new field $\chi=\e^{-\pi}$, in terms of which the two relevant operators are simply $(\partial\chi)^2$ and $(\partial\chi)^4$, and the new symmetry is immediately identified as invariance under constant shifts $\chi\to \chi+c$, while in terms of $\pi$ it looks more complicated: $\pi\to -\ln \left(\e^{-\pi}+c\right)$. This shows that the early-time $\lambda\neq 0$ cosmology is effectively described by a ghost condensate-type theory\index{ghost condensate}, albeit written in obscure variables (and hence with a vanishing speed of sound). Needless to say, the emergent global symmetry comes hand-in-hand with all the attractive properties, classical or quantum, characteristic of ghost condensation\footnote{That the given solution indeed describes ghost condensation can also be seen from the fact that $\chi$ acquires a linear profile, $\chi=-H_0 t$, just as the ghost field does on self-accelerated backgrounds.} (see \cite{ArkaniHamed:2003uz,ArkaniHamed:2003uy}). 

To get a more quantitative perspective on the above discussion, let us consider the solutions \eqref{ansatz} and \eqref{ansatz1} for a specific set of available parameters. As an immediate observation, we note that the Hubble rate does not depend on the magnitude of $f$ and $M_{\rm Pl}$ separately (as far as external matter or bare cosmological constant are not introduced into the system): physical quantities are only sensitive to the ratio of the two scales. As a result, one can arbitrarily set the physical units for any quantity at any instant of time. For example, the Hubble scale at time $t=0$ can be freely chosen to be $H(0)=10^{14}$ GeV in some putative system of units where $H_0\equiv 1$. With this in mind, we set $f=M_{\rm Pl}$, and consider the following values for the rest of the parameters: $H_0=1$, $\lambda=10^{-3}$, $t_*=10^{-2}$, $x=0$, satisfying (the first case of) the late-time de Sitter condition, Eq. \eqref{dscond}.

The time-dependence of the slow-roll parameter $\varepsilon=\dot H/H^2$ (left) and the speed of sound of the curvature perturbation $c_s^2$ (right) for the above choice of the theory parameters is shown in Fig. \ref{figurethree}. From how $\varepsilon$ depends on time, one can distinguish three stages of evolution, according to whether the system violates the \acs{NEC} strongly (red), or weakly (blue). The Universe starts out in de Sitter space ($\varepsilon \simeq 0$) with tiny curvature $\sim \lambda^2$, the Hubble rate as well as the slow-roll parameter $\varepsilon$ gradually increasing with time. When $\varepsilon \simeq 10$, it enters into the Galilean Genesis phase with strong violation of the null energy condition.  Peaking at $\varepsilon \sim 10^2$ at intermediate times, \acs{NEC}-violation weakens down back to $\varepsilon \simeq 10$ at $t\simeq -1.5$ (signalling the beginning of the third, Galileon inflation\index{inflation!Galileon} stage), $\varepsilon$ decreasing to sub per-cent values shortly afterwards (the final phase of the system corresponds to the blue ends of the curves near $t\to 0$ in Fig. \ref{figurethree}). 

While the concluding, inflationary de Sitter phase seems rather short in its extension in time, the large magnitude (in units of $H_0$) of the expansion rate at those times allows it to accommodate a large number of e-folds. Indeed, from $t=-0.1$ ($\varepsilon \simeq 5\cdot 10^{-2}$) up to $t=0$ ($\varepsilon \simeq 5\cdot 10^{-4}$), the number of times the scale factor doubles can be easily estimated
\begin{equation}
N_{\text{e}}=\int\limits_{-0.1}\limits^{0} H dt\simeq 3300~,
\end{equation}
showing that the de Sitter phase towards the end of the temporal interval of interest is in fact very extended. Furthermore, the Hubble parameter at $t=0$ is $H(0)\sim 10^6$, implying a huge ratio of de Sitter expansion rates in the asymptotic future and the asymptotic past
\begin{equation}
\frac{H(0)}{H(-\infty)}\sim 10^9~.
\end{equation}

The right panel of Fig. \ref{figurethree} shows the evolution of the scalar speed of sound. As remarked above, $c_s^2$ starts evolving from nearly zero value at early times, as required by ghost condensate-type cosmologies\index{ghost condensate}. Peaking at $c_s^2\simeq 2\cdot 10^{-3}$ during the genesis stage, it drops down again towards late-time Galileon inflation\index{inflation!Galileon}.

While ghost condensation, described by a $P(X)$-type theory implies vanishing speed of sound of the scalar perturbation at the leading order \cite{ArkaniHamed:2003uy}, Galileon inflation\index{inflation!Galileon} does not necessarily lead to $c_s^2=0$ although, as discussed before, there is an upper bound $c_s^2\leq 0.031$ in the latter class of models with a single cubic Galileon operator \cite{Kobayashi:2010cm}. %These results resonate with the small speed of sound we are finding throughout the entire course of expansion for the backgrounds under consideration. 
However, our solutions qualitatively and crucially differ from ``tilted'' ghost condensate\index{ghost condensate} with \acs{NEC} violation considered in \cite{Creminelli:2006xe} in the fact that the speed of sound, although small, is strictly \textit{positive} at all times. The latter is not true for pure $P(X)$ theories: violation of the null energy condition unambiguously implies gradient instabilities at the leading order in the ghost condensate \cite{Dubovsky:2005xd,Hsu:2004vr}.

At early times, the tiny speed of sound of the scalar mode suggests that higher-order operators in the effective theory for perturbations could be qualitatively affecting the dynamics of the system. Moreover, depending on the nature of the UV completion\index{UV completion}, higher-derivative terms could also play a role in the intermediate, Galilean Genesis phase. In order to estimate these effects, we turn to explore the structure of the next-to-leading-order action in the \acs{EFT} formalism in the following section.

\section{Beyond the leading order}
\label{hdim}

The tiny asymptotic scalar speed of sound found for the Extended Genesis solutions motivates to go beyond the leading order in the \acs{EFT} for perturbations to assess the role of higher-derivative operators in the stability of the system. The generic action for metric fluctuations on a \acs{FRW} background driven by a single ``clock'' is given by Eq. \eqref{EFTI-action-Eij}\index{inflation!EFT of} in App. \ref{Appendix-Eij}. In particular, we are interested in the following operators
\begin{equation}
\begin{split}
\label{blo-s_pi2}
S=&\int d^4x~\sqrt{\gamma}N\bigg[-M_{\rm Pl}^2 \dot H \frac{1}{N^2}-M_{\rm Pl}^2(3 H^2+\dot H) \\
&+\frac{m_2^4}{2} (\delta N)^2-\hat m_1^3\delta E\delta N
-\frac{\bar m_1^2}{2}\delta E^2-\frac{\bar m_2^2}{2}\delta E^{ij} \delta E_{ij}+\dots  \bigg] \, ,
\end{split}
\end{equation}
excluding the Einstein-Hilbert part.
The ``classical'' theory \eqref{ggg} generates only the first two terms in the second line of \eqref{blo-s_pi2}, and all the above analysis has assumed vanishing $\bar m_1^2$ and $\bar m_2^2$ (as well as higher-derivative operators, implied by the ellipses). In practice, the latter coefficients are expected to be present, although suppressed in derivative expansion. We refer to App. \ref{Appendix-Extended_Genesis} for a detailed discussion.

In what follows, we assume non-zero $\bar m_1^2$ and $\bar m_2^2$ in computing the $\zeta$ quadratic action on Extended Genesis backgrounds. The results, given in \eqref{A'}-\eqref{C'} of App. \ref{Appendix-Extended_Genesis}, are rather tedious and reluctant to simple analysis in their exact form. For simplicity, we will expand all relevant quantities to linear order in  $\bar m_1^2$ and $\bar m_2^2$, assuming these are small in the sense that higher order terms in the expansion give subleading corrections -- something we will justify \emph{a posteriori}. The procedure yields the following expressions for the kinetic coefficients\footnote{The signs are defined so that all kinetic coefficients have to be positive for the complete stability (at all wavelengths) of the corresponding background. Moreover, it is worth noticing that $C$ vanishes if $\bar m_1^2=-\bar m_2^2$, in accordance with what we have said in Part II. In particular, we recall that in Chap. \ref{chap-WBG-constraints}, having focused on the case of a linear dispersion relation of the form \eqref{phdr}, the analysis has been crucially based on the possibility of getting rid of the combination $\delta E^2-\delta E_{ij}\delta E^{ij}$.} $A$, $B$ and $C$ 
on backgrounds corresponding to the second case ($\lambda\neq 0$) of the previous section:
\begin{align}
A
&	=\frac{2}{3M_{\rm Pl}^2 H_0^4\lambda^2\tau^6}\left[18 M_{\rm Pl}^4 H_0^4\lambda^2 \tau^6-3M_{\rm Pl}^2f^2H_0^2(5\lambda \tau-2)\tau^2+4 f^4\right] \nonumber
\\
&\qquad
	+p_1\bar m_1^2+p_2\bar m_2^2 \, ,
\\
B
	&=-\frac{2}{3}\frac{f^2}{H_0^2\lambda}\frac{1}{\tau^3}+p_3\bar m_1^2+p_4\bar m_2^2+q_3\partial_t\bar m_1^2+q_4\partial_t\bar m_2^2 \, ,
\\
C
	&=\frac{\bar m_1^2+\bar m_2^2}{2\lambda^2} \, ,
\end{align}
where we have defined $\tau \equiv t-t_*<0$ and introduced auxiliary coefficients $p_i$ and $q_i$, given as follows
\begin{equation}
p_1=-\frac{\left[27 M_{\rm Pl}^4 H_0^4\lambda^2\tau^6-6M_{\rm Pl}^2 f^2 H_0^2(5\lambda\tau-2)\tau^2+8 f^4\right]^2}{18 M_{\rm Pl}^8 H_0^8\lambda^4\tau^{12}} \, ,
\end{equation}
\begin{multline}
p_2=-\frac{1}{18 M_{\rm Pl}^8 H_0^8\lambda^4\tau^{12}}\big(1107 M_{\rm Pl}^8 H_0^8\lambda^4\tau^{12}-1980M_{\rm Pl}^6 f^2H_0^6\lambda^3\tau^9
\\
+792M_{\rm Pl}^6 f^2H_0^6\lambda^2\tau^8 +1428M_{\rm Pl}^4 f^4 H_0^4\lambda^2\tau^6-720 M_{\rm Pl}^4 f^4 H_0^4\lambda \tau^5
\\
+144M_{\rm Pl}^4 f^4 H_0^4\tau^4-480M_{\rm Pl}^2 f^6 H_0^2\lambda\tau^3+192 M_{\rm Pl}^2 f^6 H_0^2 \tau^2+64 f^8\big) \, ,
\end{multline}
\begin{multline}
p_3=-\frac{1}{18M_{\rm Pl}^6 H_0^6 \lambda^3\tau^9}\big[81M_{\rm Pl}^6 H_0^6\lambda^3 \tau^9-18 M_{\rm Pl}^4 f^2 H_0^4 (8\lambda^2\tau^2-17\lambda\tau+8)\tau^4
\\
+84 M_{\rm Pl}^2 f^4 H_0^2 (\lambda \tau-2)\tau^2-16 f^6 \big] \, ,
\end{multline}
\begin{multline}
p_4=-\frac{1}{18M_{\rm Pl}^6 H_0^6 \lambda^3\tau^9}\big[99M_{\rm Pl}^6 H_0^6\lambda^3 \tau^9-6M_{\rm Pl}^4 f^2 H_0^4 (26\lambda^2\tau^2-51\lambda\tau+24)\tau^4
\\
+84 M_{\rm Pl}^2 f^4 H_0^2 (\lambda \tau-2)\tau^2-16 f^6 \big] \, ,
\end{multline}
\begin{equation}
q_3=-\frac{27M_{\rm Pl}^4 H_0^4 \lambda^2\tau^6-6M_{\rm Pl}^2 f^2 H_0^2(5\lambda \tau-2)\tau^2+8 f^4}{6 M_{\rm Pl}^4 H_0^4\lambda^3\tau^{6}} \, ,
\end{equation}
\begin{equation}
q_4=-\frac{33M_{\rm Pl}^4 H_0^4 \lambda^2\tau^6-6M_{\rm Pl}^2 f^2 H_0^2(5\lambda \tau-2)\tau^2+8 f^4}{6 M_{\rm Pl}^4 H_0^4\lambda^3\tau^{6}} ~.
\end{equation}
An immediate and important observation is that all of the coefficients $p_i$ and $q_i$ are sign-definite (negative) \emph{at all times}. Moreover, since different linear combinations of $\bar m_1^2$ and $\bar m_2^2$ enter into $B$ and $C$, nothing prevents us from choosing the former pair of \acs{EFT} coefficients such that both $B$ and $C$ are positive, thus avoiding any instability over the entire cosmological period of interest.

Furthermore, we have found in the previous section that the speed of sound of the scalar mode goes to zero ($c_s^2\to 0^+$) in the asymptotic past for the backgrounds corresponding to the Extended Genesis. This suggests that $t\to -\infty$ is precisely the regime where higher-order corrections in the \acs{EFT} for perturbations could play an important role. In fact, in the case that $\bar m_1^2$ and $\bar m_2^2$ fall off slower than $1/t^3$ at early times, higher-order effects give contributions that dominate over the leading-order piece in the coefficient $B$ at early times\footnote{This seemingly casts shadow on the meaning of our expansion in small $\bar m_1^2$ and $\bar m_2^2$; fortunately, a closer inspection of \eqref{A'}-\eqref{C'} shows that the expansion parameters at $t\to -\infty$ are in fact $\bar m_1^2/M_{\rm Pl}^2$ and $\bar m_2^2/M_{\rm Pl}^2$, meaning that next-order corrections in the series indeed give subleading effects.}. Focusing on the $t\to-\infty$ ghost condensate\index{ghost condensate} regime and neglecting time derivatives of $\bar m_1^2$ and $\bar m_2^2$ for simplicity, we find
\begin{align}
\label{App}
A&=12 M_{\rm Pl}^2  +\mathcal{O}\left(\bar{m}^2_1,\bar{m}^2_2\right)~, \\
\label{Bpp}
B&=-\frac{2}{3}\frac{f^2}{H_0^2\lambda}\frac{1}{t^3}-\frac{9}{2} \bar{m}^2_1-\frac{11}{2} \bar{m}^2_2+\mathcal{O}\left(\frac{\bar{m}_1^4}{M_{\rm Pl}^2},\frac{\bar{m}_2^4}{M_{\rm Pl}^2}\right)~, \\
\label{Cpp}
C&=\frac{\bar{m}^2_1+\bar{m}^2_2}{\lambda^2}+\mathcal{O}\left(\frac{\bar{m}^4_1}{M_{\rm Pl}^2\lambda^2},\frac{\bar{m}^4_2}{M_{\rm Pl}^2\lambda^2}\right) ~.
\end{align}
Again, since $B$ and $C$ involve different linear combinations of $\bar{m}^2_1$ and $\bar{m}^2_2$, one can freely choose the values for the latter two coefficients, such that the theory is free from any instability\footnote{Note that there is in fact even more freedom: one could always make the coefficient $C$ positive by adding a term of the form $N\sqrt{\gamma}\upleft{3}{R}^2$ to \eqref{blo-s_pi2}; see the discussion in App. \ref{Appendix-Extended_Genesis}.}. 

Moreover, in the case that $\bar{m}_1^2$ and $\bar{m}_2^2$ drop off slower than $1/t^3$ for large and negative times, the asymptotic speed of sound of the scalar perturbation is set by the ratio 
\begin{equation}
c_s^2 = \frac{|9 \bar{m}_1^2+11\bar{m}_2^2|}{24 M_{\rm Pl}^2}~,
\label{mroba}
\end{equation}
and is not necessarily infinitesimally close to zero if at least one of the two \acs{EFT} coefficients $\bar{m}_1^2$ and $\bar{m}_2^2$ tends to a constant at early times\footnote{Note, that the squared speed of sound also sets the magnitude for the expansion parameter in \eqref{App}-\eqref{Cpp}.}.

To summarize, we have found that the beyond-the-leading-order structure of the effective theory for perturbations allows to cure possible classical gradient instability\index{gradient instability}, such as the one found for a subclass of the backgrounds studied above. Moreover, it does so naturally for the $\lambda=0$ solutions of Sec. \ref{analytic} and the numerical example of Sec. \ref{inf-mink-1}, for which the leading-order instability is weak  (\textit{i.e.} $c_s^2$ goes negative, but very small in magnitude).
Whenever the speed of sound of the scalar mode vanishes at the leading order, on the other hand, higher-order effects can push the corresponding solution into a completely stable direction.

\section{Conclusive remarks}
\label{concl}

We would like to conclude this third part emphasising again that, despite the standard Big Bang paradigm provides us an extremely compelling picture of the early Universe, it is not the only possible one. The question of how far alternative scenarios can go in adequately describing the observed Universe has been a strong motivation for expanding the theory space in non-standard directions. Perhaps the most dramatic departure from the inflationary paradigm corresponds to theories that violate the \acs{NEC}, thereby allowing for a qualitatively different evolution of the early Universe that, among other interesting features, is capable of smoothing out the Big Bang singularity. The fact that this can happen without instabilities for a Universe starting out from the flat, Minkowski space-time has been shown in Refs. \cite{Creminelli:2010ba,Creminelli:2006xe,Hinterbichler:2012fr}. 

A common feature of alternatives to inflation based on \acs{NEC} violation is the usual prediction of a strongly blue-tilted and unobservable (at least in the \acs{CMB} experiments) spectrum of tensor perturbations. The ultimate reason for this lies in the fact that the phenomenologically interesting phase of cosmological evolution happens on quasi-flat backgrounds. Would then a detection of primordial $B$ modes in \acs{CMB} polarization conclusively rule out these theories? 

In this part, we have argued that the answer to this question is negative.   
We have constructed explicit theories that lead to an early Universe cosmology interpolating between a small/zero curvature maximally symmetric (de Sitter or Minkowski) space-time in the far past and an inflationary de Sitter space-time, capable of generating a scale-invariant tensor spectrum of significant amplitude, in the asymptotic future. This is possible because, at intermediate times, the system can \textit{strongly} violate the \acs{NEC} ($\dot H\gg H^2$) as it happens in genesis models, all without developing any kind of instability. The corresponding backgrounds can be viewed as a regularized extension of Galilean Genesis: now, none of the physical quantities grows beyond the cutoff scale. Alternatively, one can view them as a certain early-time complete realization of inflation, that leads to a (almost) flat pre-inflationary Universe.

In the game, symmetries are the key ingredients.
Being deformations of the conformal Galileon, the theories constructed above enjoy non-linearly realized\index{symmetry!non-linearly realized} emergent symmetries at both the early- and the late-time asymptotic stages. Indeed, it is precisely the nature of these regimes that determines the qualitative picture of the cosmological solutions of interest: these are described by quantum-mechanically stable, robust theories based solely on symmetry principles.

In particular, the analysis of Part II has served as an input for the construction of the late-time stage: fitting into a class of more general theories with an underlying weakly broken symmetry, the Galileon inflationary\index{inflation!Galileon} phase enjoys important non-renormalization properties, guaranteeing the quantum consistency of the classical solution.

%% file: chapters/appendix-ADM.tex
\chapter{ADM formulation of General Relativity}
\label{Appendix-ADM}

This appendix is devoted to the introduction of notations and variables of the \acs{ADM} formalism\index{Arnowitt-Deser-Misner (\acs{ADM}) formalism} \cite{Arnowitt:1962hi}, which turns out to be the most convenient approach for what we discuss in the main text of the work.

The starting point is the introduction of a foliation of a space-time manifold $\mathcal{M}$ in terms of space-like three-dimensional sub-manifolds. Namely, we consider a generic hypersurface $\Sigma(t)$, defined by the condition $t(x^\mu)=\text{constant}$ and parametrized by the values of the ``time'' $t$. The orthogonal direction on each hypersuface is identified by the unit normal
\begin{equation}
n_\mu \equiv \frac{\partial_\mu t}{\sqrt{-\partial^\alpha t\partial_\alpha t}} \, ,
\label{adm-n}
\end{equation}
satisfying $n^\mu n_\mu=-1$.

If $y^i$ denotes some spatial coordinates lying on each $\Sigma$, a suitable four-dimensional coordinate system $(t,y^i)$ can be introduced. By definition, the projector $e^\mu_i\equiv\partial x^\mu/\partial y^i$ satisfies the condition $n_\mu e^\mu_i=0$. Decomposing the vector $t^\mu\equiv\partial x^\mu/\partial t$ in terms of normal and tangent components as $t^\mu = Nn^\mu+N^ie^\mu_i$ for some values $N$ and $N^i$, called \textit{lapse} and \textit{shift} respectively, from the coordinate transformation $x^\mu = x^\mu(t,y^i)$ the following relation holds
\begin{equation}
\D x^\mu = t^\mu \D t + e^\mu_i\D y^i = (N\D t)n^\mu + (\D y^i+N^i\D t)e^\mu_i \, .
\label{adm-dx}
\end{equation}
The space-time metric is obtained as
\begin{equation}
\D s^2 = g_{\mu \nu}\D x^\mu\D x^\nu = - N^2\D t^2 + \gamma_{ij}(\D y^i+N^i\D t)(\D y^j+N^j\D t) \, ,
\label{adm-met}
\end{equation}
where $\gamma_{ij}$ is the $3$-metric induced on $\Sigma(t)$ which can be uniquely determined by using the two conditions
\begin{equation}
\gamma_{\mu\nu}n^\mu = 0 \, ,
\qquad
\gamma_{\mu\nu}e^\mu_i = g_{\mu\nu}e^\mu_i \, ,
\label{condh}
\end{equation}
leading to
\begin{equation}
\gamma_{\mu\nu} = g_{\mu\nu} + n_\mu n_\nu \, .
\label{h-g}
\end{equation}

In the new coordinate system $(t,y^i)$, we have $n_i=0$ and $n_0=(-g^{00})^{-1/2}$. Moreover the following relations hold:
\begin{align}
g_{00} & = N^iN_i-N^{2} \, , & g_{0i} & = N_i \, , & g_{ij} & = \gamma_{ij} \, ,
\label{rm-1}\\
g^{00} & = -N^{-2} \, , & g^{0i} & = N^{-2}N^i \, , & g^{ij} & = \gamma^{ij}-N^{-2}N^iN^j \, .
\label{rm-2}
\end{align}

Denoting with $\nabla_\mu$ the covariant derivative on the manifold $\mathcal{M}$, the corresponding differential operator on the three-dimensional hypersurface $\Sigma$, acting on a generic tensor $T^{\mu_1\ldots\mu_n}_{\nu_1\ldots\nu_k}$, can be written as
\begin{equation}
D_\alpha T^{\mu_1\ldots\mu_n}_{\nu_1\ldots\nu_k} \equiv \gamma_\alpha^\beta \gamma^{\mu_1}_{\rho_1}\ldots \gamma^{\mu_n}_{\rho_n}\gamma^{\sigma_1}_{\nu_1}\ldots \gamma^{\sigma_k}_{\nu_k} \nabla_\beta T^{\rho_1\ldots\rho_n}_{\sigma_1\ldots\sigma_k} \, ,
\label{cder-h}
\end{equation}
which satisfies $D_\alpha \gamma_{\mu\nu}=0$.

The quantity which measures how much the surface $\Sigma$ is curved in its embedding in the manifold $\mathcal{M}$ is called \textit{extrinsic curvature}\index{extrinsic curvature} and is defined as
\begin{equation}
K_{\mu\nu} \equiv D_\mu n_\nu \, .
\label{ex-curv}
\end{equation}
Moreover, $K_{\mu\nu}$ is symmetric and has only purely spatial components of the form
\begin{equation}
K_{ij} = \frac{1}{2N}\left(\partial_t \gamma_{ij} - D_iN_j-D_jN_i\right) \, .
\label{ex-curvij}
\end{equation}

In terms of these foliated quantities, up to surface contributions, the Einstein-Hilbert action\index{Einstein-Hilbert action} for General Relativity is
\begin{equation}
S_{\text{EH}} = \frac{M_{\rm Pl}^2}{2}\int\D^4x \, N\sqrt{\gamma}
	\left[\upleft{3}{R} + K_{\mu\nu}K^{\mu\nu} - K^2 \right] \, ,
\label{EH-action}
\end{equation}
where $\upleft{3}{R}$ is the scalar curvature of the three-dimensional space and $K\equiv K^\mu_\mu$ is the trace of the extrinsic curvature.

%% file: chapters/appendix-bispectrum.tex
\chapter{The bispectrum in the EFT of inflation}
\label{Appendix-bispectrum}

In this appendix, we show the explicit form of each contribution $S_i(k_1,k_2,k_3)$, entering the bispectrum \eqref{mgbispr} and related to the different operators in the cubic action \eqref{our-14}. The results are expressed in terms of the dimensionless parameters \eqref{coeffs}-\eqref{coeffs-bis}.
Introducing the sum $k_t\equiv k_1+k_2+k_3$, the explicit computation yields:

\paragraph{1. Contribution from $\dot \zeta^3$}
\begin{gather}
S_1(k_1,k_2,k_3) =\frac{4}{k_t^3 k_1k_2k_3} \, , \\
\begin{split}
c_1=-\frac{M_\text{Pl}^2}{c_s^4}
&	(\alpha-1)^{-4}\big[
	\alpha^2-\alpha-\varepsilon+ c_s^2(3\alpha-\gamma+1)(\alpha^2-\alpha-\varepsilon)
\\
&	+c_s^4 (\alpha-1)(6\alpha^2-3\alpha\gamma+3\gamma+\delta-2\varepsilon)\big] \, ,
\end{split}
\end{gather}

\paragraph{2. Contribution from $\zeta \dot \zeta^2$}
\begin{gather}
S_2(k_1,k_2,k_3) = 2~\frac{k_t+k_1}{k_t^2 k_1^3 k_2k_3} \, , \\
c_2= 3~\frac{M_\text{Pl}^2}{c_s^4}~ \frac{(\alpha^2-\alpha-\varepsilon)\left[1-(\alpha-1)^2 c_s^2\right]}{(\alpha-1)^4} \, ,
\end{gather}
\paragraph{3. Contribution from $\zeta (\partial_i \zeta)^2$}
\begin{gather}
\begin{split}
S_3(k_1,k_2,k_3) = (k_1^2-k_2^2-k_3^2)\bigg[
	-\frac{k_t}{(k_1k_2k_3)^3}&+\frac{\sum_{i>j}k_ik_j}{k_t (k_1k_2k_3)^3}
\\
&	+\frac{1}{k_t^2(k_1k_2k_3)^2}\bigg] \, ,
\end{split}
\\
c_3 = -\frac{c_2}{3}  \, ,
\end{gather}
\paragraph{4. Contribution from $\dot\zeta\partial_i\zeta\partial_i\partial^{-2}\dot\zeta$}
\begin{gather}
 S_4(k_1,k_2,k_3) =\frac{k_1^2-k_2^2-k_3^2}{2 k_t k_1(k_2k_3)^3}~\left(2+\frac{k_2+k_3}{k_t}\right) \, , \\
 c_4 = -\frac{M_\text{Pl}^2}{c_s^4}~\frac{(\alpha^2-5\alpha-\varepsilon+4)(\alpha^2-\alpha-\varepsilon)^2}{2(\alpha-1)^6} \, ,
\end{gather}
\paragraph{5. Contribution from $\partial^2\zeta \partial_i\partial^{-2}\dot\zeta \partial_i\partial^{-2}\dot\zeta$}
\begin{gather}
  S_5(k_1,k_2,k_3) = \frac{k_1^2-k_2^2-k_3^2}{k_t k_1(k_2k_3)^3}~\left(1+\frac{k_1}{k_t}\right)\, , \\
c_5 = \frac{M_\text{Pl}^2}{c_s^4}~\frac{(\alpha^2-\alpha-\varepsilon)^2(3\alpha^2-3\alpha+\varepsilon)}{4(\alpha-1)^6} \, ,
\end{gather}
\paragraph{6. Contribution from $\partial^2\zeta\partial_i\zeta\partial_i\partial^{-2}\dot\zeta$}
\begin{gather}
 S_6(k_1,k_2,k_3) = \frac{k_1^2-k_2^2-k_3^2}{2 k_t k_1(k_2k_3)^3}~\left(2+\frac{2 k_1+k_2+k_3}{k_t}+\frac{2k_1(k_2+k_3)}{k_t^2}\right)  \, , \\
 c_8 = -\frac{M_\text{Pl}^2}{c_s^4} ~\frac{2\alpha (\alpha^2-\alpha-\varepsilon)}{(\alpha-1)^4} \, ,
\end{gather}
\paragraph{7. Contribution from $\partial^2\zeta(\partial\zeta)^2$}
\begin{gather}
S_7(k_1,k_2,k_3) =2~ \frac{k_1^2-k_2^2-k_3^2}{k_t k_1(k_2k_3)^3}~\left(1+\frac{\sum_{i>j}k_ik_j}{k^2_t}+\frac{3k_1k_2k_3}{k_t^3}\right) \, , \\
c_7 = \frac{M_\text{Pl}^2}{c_s^4}~\frac{\alpha}{(\alpha-1)^3} \, ,
\end{gather}
\paragraph{8. Contribution from $\partial^2\zeta\dot\zeta^2$}
\begin{gather}
S_8(k_1,k_2,k_3) =4~ \frac{k_t+3k_1}{k_t^4k_1k_2k_3} \, , \\
c_8 = \frac{M_\text{Pl}^2}{c_s^2}~\frac{4\alpha-\gamma}{(\alpha-1)^3} \, .
\end{gather}

%% file: chapters/appendix-WBG.tex
\chapter{Quantum corrections in WBG theories}
\label{Appendix-WBG}

In this appendix, we extend the discussion of Sec.~\ref{WBG-sectwo} (Part II) to show that the vertices  with three solid lines and one graviton, as well as five solid lines and two gravitons (see Fig.~\ref{fig2}) can be removed in a suitable curved-space extension of the Galileon terms.
Moreover, the presence of a potential in the Lagrangian is taken into account.

\section{Technical analysis}

We have considered the case of the cubic Galileon in the main text. One has to work a little more to understand the case of quartic minimally coupled Galileon. The contribution to the second vertex from the lower line of Fig.~\ref{fig2} can be extracted by picking up a factor of $\partial h$ from the covariant derivative acting on one of the scalars. The relevant term reads,
\begin{equation}
\label{cd}
\nabla_\mu\nabla_\nu\phi \sim -\frac{1}{2}\left(\partial_\mu h_{\rho\nu}+\partial_\nu h_{\rho\mu}-\partial_\rho h_{\mu\nu}\right) \partial_\rho\phi~.
\end{equation}
Here and in the rest of the present section, by ``$\sim$'' we mean ``equals up to a total derivative and up to terms with fewer factors of $\partial\phi$'', and we do not distinguish between upper and lower indices for simplicity (everything is contracted with the flat-space metric). Using \eqref{cd}, we find after a little bit of algebra 
\begin{equation}
\label{contr1}
-4 \mathcal{L}_4^{\text{min}}\sim - 4 (\partial\phi)^2 h_{\mu\nu} \partial_\rho\phi \left(\partial_\mu\partial_\nu-\Box \eta_{\mu\nu}\right)\partial_\rho\phi~,
\end{equation}
where $\mathcal{L}_4^{\text{min}}$ denotes the Lagrangian term, obtained by minimally covariantizing \eqref{gal2}. Inserting the latter vertex into a generic \acs{1PI} loop diagram would induce Galileon symmetry-breaking operators at the scale $M_{\rm Pl}^{1/6} \Lambda_3^{5/6}$, parametrically exceeding $\Lambda_2$, as discussed in the main text. Fortunately, it turns out possible to raise it by adding non-minimal couplings, capable of eliminating the vertices with three solid lines coming from the minimally coupled theory. For the quartic Galileon, the right coupling is $\sqrt{-g}(\partial\phi)^4R$. Indeed, expanding the Ricci scalar to the linear order in $h$ and integrating by parts, we obtain
\begin{equation}
\label{contr2}
\sqrt{-g}(\partial\phi)^4 R \sim  4 (\partial\phi)^2 h_{\mu\nu} \partial_\rho\phi \left(\partial_\mu\partial_\nu-\Box \eta_{\mu\nu}\right)\partial_\rho\phi+\mathcal{O}(h^2)~,
\end{equation}
implying that the two contributions to the possible vertex with one graviton and three solid lines exactly cancel for our resulting generalized theory. 

The way we have chosen to couple the Galileon to gravity has been dictated by our desire to raise the scale suppressing loop-generated symmetry-breaking operators. However, there is something more to it: the resulting non-minimal theory can be recognized as that of Horndeski type, leading to second-order equations both for the metric and the scalar. In retrospect, this is not surprising: the ``three solid line'' vertices, if present, are the only ones that would introduce higher-order equations for the metric in the theory of minimally coupled Galileon, as can be easily seen from \textit{e.g.} Eq.~\eqref{contr1}. Our non-minimal term exactly cancels this contribution. We thus observe something similar to what happens in massive gravity \cite{deRham:2010ik}: there, eliminating terms that lead to higher-order equations (or, equivalently, to the Boulware-Deser ghost) automatically raises the quantum cutoff of the theory; in our case, the cutoff ($\Lambda_3$) remains unaltered, but a similar procedure raises the scale of breaking for the Galileon symmetry. 

The same result can be straightforwardly extended to the case of the quintic Galileon. A brute-force expansion of the minimally-coupled term, using \eqref{cd}, yields
\begin{multline}
\label{contr3}
\frac{2}{3}\mathcal{L}_5^{\text{min}}
	\sim (\partial\phi)^2\partial_\rho\phi \Big\{ \partial_\rho h\left[(\Box\phi)^2-(\partial_\mu\partial_\nu\phi)^2\right]
\\
	-2\partial_\rho h_{\mu\nu}\left(\partial_\mu\partial_\nu\phi \Box\phi-\partial_\mu\partial_\alpha\phi \partial_\nu\partial_\alpha\phi\right)
	- 2 \partial_\mu h_{\rho\mu}\left[(\Box\phi)^2-(\partial_\alpha\partial_\beta\phi)^2\right]
\\
	+4 \partial_\mu h_{\rho\nu}(\partial_\mu\partial_\nu\phi \Box\phi-\partial_\mu\partial_\alpha\phi \partial_\nu\partial_\alpha\phi)\Big\}~.
\end{multline}
It is impossible to manipulate the last expression by partial integration or otherwise, so as to be left with less than three factors of $\partial\phi$. This would lead to the last vertex with three solid lines in Fig.~\ref{fig2}, and therefore higher-order equations for the metric. One can however show that this term can be eliminated by adding a non-minimal piece with an exact coefficient corresponding to the Horndeski theory. 
To this end, we note that to the linear order in the metric perturbation, the following relation holds
\begin{equation}
\sqrt{-g}(\partial\phi)^4 G_{\mu\nu}\nabla^\mu\nabla^\nu\phi = -\frac{1}{2}(\partial\phi)^4\epsilon_{\mu\alpha\rho\lambda} \epsilon_{\nu\beta\sigma\lambda}\partial_\alpha\partial_\beta h_{\rho\sigma}\partial_\mu\partial_\nu\phi+\mathcal{O}(h^2)~,
\end{equation}
where $\epsilon$ denotes the totally antisymmetric symbol with 
$\epsilon_{1234}=1$. Expanding the antisymmetric product and integrating by parts, we obtain
\begin{multline}
\label{contr4}
(\partial\phi)^4 G_{\mu\nu}\nabla^\mu\nabla^\nu\phi \sim (\partial\phi)^2 \partial_\rho\phi \Big\{\partial_\rho h\left[(\partial_\mu\partial_\nu\phi)^2-(\Box\phi)^2\right]
\\
+2 \partial_\rho h_{\mu\nu}\left(\partial_\mu\partial_\nu\phi \Box\phi-\partial_\mu\partial_\alpha\phi \partial_\nu\partial_\alpha\phi\right)\Big\} \, .
\end{multline}
This exactly cancels the first line of the minimal term contribution \eqref{contr3}. Partially integrating the second line of \eqref{contr3} on the other hand gives
\begin{equation}
 -2(\partial\phi)^2\partial_\rho\phi \partial_\nu h_{\rho\mu}\big[\partial_\mu\partial_\alpha\phi\partial_\nu\partial_\alpha\phi-\partial_\mu
 \partial_\nu\phi\Box\phi
+\partial_\mu\Box\phi\partial_\nu\phi-\partial_\mu\partial_\nu\partial_\alpha
\phi\partial_\alpha\phi  \big]~.
\end{equation}
One can now check that whatever stands to the right of $(\partial\phi)^2\partial_\rho\phi$ in the last expression is a total derivative ($\partial_\nu$ of a local operator). This means that one final partial integration can get rid of one solid line in the corresponding vertex, reducing the number of solid lines to two. This proves our statement regarding the elimination of vertices with one graviton and more than two solid lines.

One last statement we wish to prove regards the absence of vertices with two gravitons and more than four solid lines. The only possible obstruction comes from the non-minimal coupling obtained above, $(\partial\phi)^4 G_{\mu\nu}\nabla^\mu\nabla^\nu\phi$, in which we take the Einstein tensor linear in the metric perturbation, and pick up another factor of $h$ from the covariant derivative\footnote{A vertex with more than four solid lines coming from the minimally coupled quintic Galileon would involve at least three gravitons.}. One can show, however, that the corresponding term can always be put into the form in which there are no more than four scalars with a single derivative acting on them. Indeed, using the expression for $G_{\mu\nu}$ in terms of the antisymmetric symbol, we have
\begin{equation}
(\partial\phi)^4 G_{\mu\nu}\nabla^\mu\nabla^\nu\phi \sim \frac{1}{4}(\partial\phi)^4 \epsilon_{\mu\alpha\rho\lambda} \epsilon_{\nu\beta\sigma\lambda}\partial_\alpha\partial_\beta h_{\rho\sigma} (2\partial_\mu h_{\gamma\nu}-\partial_\gamma h_{\mu\nu})\partial_\gamma\phi~.
\end{equation}
This form makes it straightforward to convince oneself that the factors involving two gravitons in the above expression collect into either a total derivative or a total derivative up to corrections involving at most four factors of $\partial\phi$:
\begin{equation}
\begin{split}
\epsilon_{\mu\alpha\rho\lambda} \epsilon_{\nu\beta\sigma\lambda}\partial_\alpha\partial_\beta h_{\rho\sigma} \partial_\mu h_{\gamma\nu}&=\partial_\mu\left(\epsilon_{\mu\alpha\rho\lambda} \epsilon_{\nu\beta\sigma\lambda}\partial_\alpha\partial_\beta h_{\rho\sigma} h_{\gamma\nu}\right) \\
(\partial\phi)^4 \partial_\gamma\phi~\epsilon_{\mu\alpha\rho\lambda} \epsilon_{\nu\beta\sigma\lambda}\partial_\alpha\partial_\beta h_{\rho\sigma}\partial_\gamma h_{\mu\nu}&\sim \frac{1}{2} (\partial\phi)^4 \partial_\gamma\phi~\partial_\gamma \left(\epsilon_{\mu\alpha\rho\lambda} \epsilon_{\nu\beta\sigma\lambda}\partial_\alpha\partial_\beta h_{\rho\sigma} h_{\mu\nu}\right)~.
\end{split}
\end{equation}
At this point, all that remains is to do a single partial integration to complete the proof of the statement, made in the beginning of the present section.

\section{Quantum corrections in presence of the inflaton potential}
\label{App-inf-pot}

The previous analysis has provided the notion of \acs{WBG} invariance relying on the absence of any potential in the theory. However, the study of slow-roll, potentially dominated models\index{potentially dominated evolution} of inflation requires to understand whether the results get spoiled by the introduction of such a potential $V(\phi)$ in the Lagrangian. This might appear dangerous regarding the quantum stability, essentially because super-Planckian excursions of the scalar field, well beyond the cutoff of the theory, $\Lambda_3$, are expected in a slow-roll evolution.

The aim of this section is to show that the previous results are still valid in the presence of a potential $V(\phi)$, which is therefore compatible with the notion of \acs{WBG} invariance.

\begin{figure}
\includegraphics[width=0.5\textwidth, natwidth=610,natheight=600]{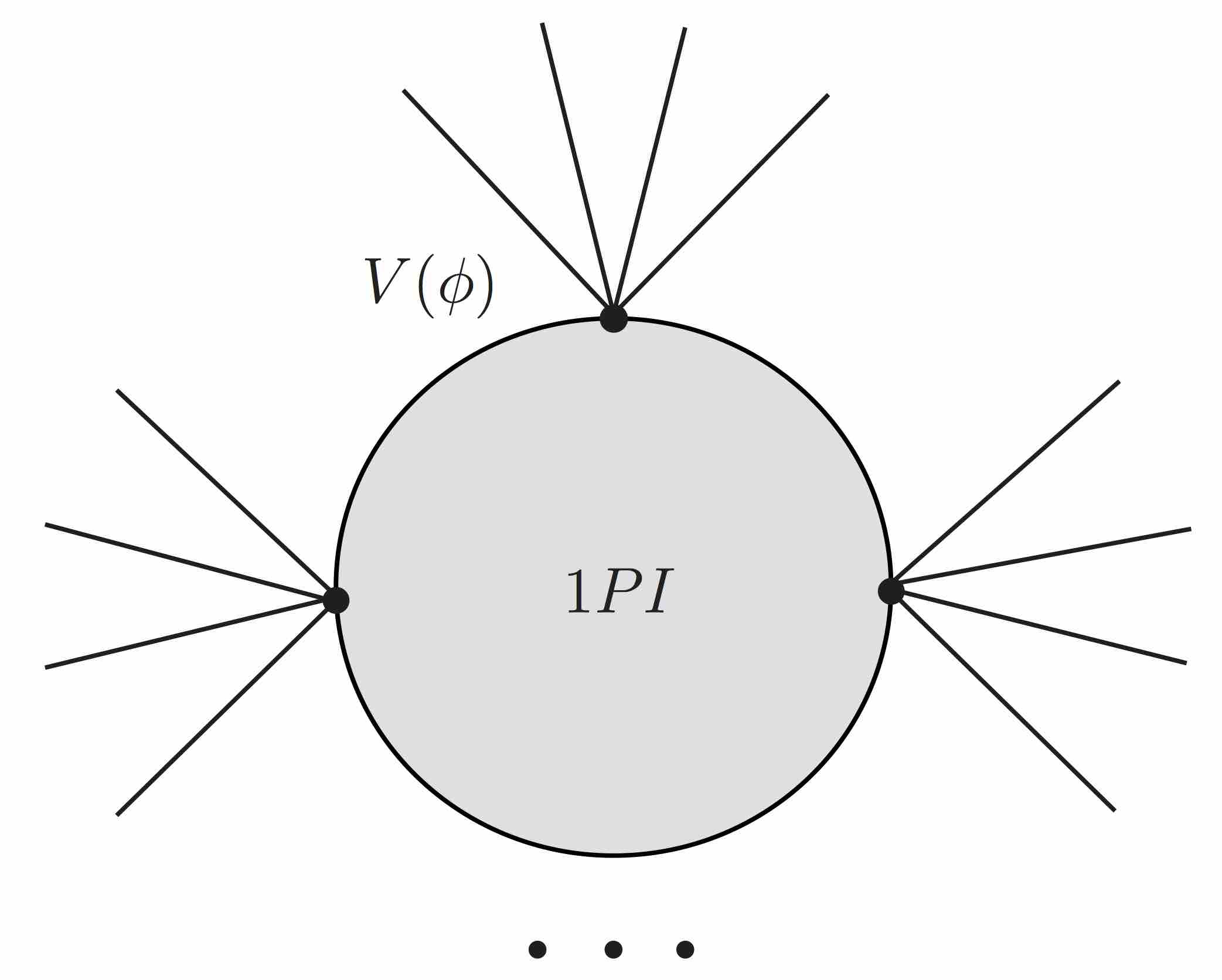}\centering
\caption{A \acs{1PI} loop diagram that contributes to the quantum corrections for the potential $V(\phi)\sim\phi^m$.}
\label{fig-Vloop}
\end{figure}

Previously we have proved that \acs{1PI} diagrams that arise from anywhere except $V(\phi)$ in \eqref{full} do not renormalize the action up to corrections at least of order $\Lambda_3/M_{\rm Pl}$. We now consider the insertions from the potential: it is straightforward to realize that this remains true. A generic Lagrangian term, proportional to $V(\phi)$, is of the following schematic form
\begin{equation}
V(\phi)\frac{h^n_c}{M_{\rm Pl}^n} \, ,
\label{qcVp}
\end{equation}
where $h_c$ stands for the canonically normalized metric perturbation and $n$ is an integer. On can easily get convinced that any loop correction to the Lagrangian terms \eqref{qcVp}, as shown in Fig. \ref{fig-Vloop}, is suppressed by a tiny factor. Given that during inflation $\phi\gg M_{\rm Pl}$, possible dangerous contributions could arise when the largest number of $\phi$ is allocated to external legs. The presence of internal graviton lines in the loop yields suppressions with powers of $M_{\rm Pl}^{-1}$, providing quantum corrections to $V(\phi)$ that are proportional to the ratio $(\Lambda_3/M_{\rm Pl})^n$. Moreover, one can consider the case in which the scalar field is employed as internal line. For instance, assuming $V(\phi)\sim\phi^m$ for some integer $m$, the employment of one graviton and one $\phi$ in the loop provides the vertex with an effective coefficient of order
\begin{equation}
\sim \frac{V'\Lambda_3^2}{VM_{\rm Pl}} \sim \sqrt{\varepsilon}\left(\frac{\Lambda_3}{M_{\rm Pl}}\right)^2 \, ,
\label{qcVpeffc1}
\end{equation}
which is even more suppressed, as expected. Involving more scalar fields in the loop induces higher derivatives of the potential.

One can similarly consider all the other possible loop corrections to the potential, generated by the interacting terms in the action \eqref{full}, and realize that the presence of $V(\phi)$ does not spoil the quantum properties of the theories under consideration.

%% file: chapters/appendix-Eij.tex
\chapter{Notations for the effective field theory of inflation}
\label{Appendix-Eij}

This brief appendix is devoted to some comments on notations in the \acs{EFT} of inflation\index{inflation!EFT of}, in order to avoid ambiguities. Instead of the extrinsic curvature\index{extrinsic curvature} $K_{\mu\nu}$ defined as in Eq. \eqref{ex-curvij}, it will be useful to write the effective action for inflation in terms of
\begin{equation}
E_{ij}\equiv NK_{ij} = \frac{1}{2}\left(\partial_t \gamma_{ij} - D_iN_j-D_jN_i\right) \, .
\label{ex-curv-Eij}
\end{equation}
In analogy with Eq. \eqref{EFTI-action-2}, we will write the action as
\begin{equation}
\begin{split}
S = & \int\D^4x \, N\sqrt{\gamma} \bigg[
	\frac{1}{2}M_{\rm Pl}^2\left(\upleft{3}{R} + \frac{E_{\mu\nu}E^{\mu\nu} - E^2}{N^2} \right)
	- \frac{M_{\rm Pl}^2\dot{H}}{N^2} - M_{\rm Pl}^2(3H^2+\dot{H}) 
\\
&	+ \frac{1}{2}m_2(t)^4\delta N^2 + m_3(t)^4\delta N^3 + \ldots
\\
&	- \hat{m}_1(t)^3\delta N\delta E
	+ \hat{m}_2(t)^3\delta N^2\delta E + \ldots
\\
&	- \frac{1}{2}\bar{m}_1(t)^2\delta E^2
	- \frac{1}{2}\bar{m}_2(t)^2\delta {E^\mu}_\nu\delta {E^\nu}_\mu 
	+ \bar{m}_3(t)^2 \upleft{3}{R} \delta N + \ldots
\bigg] \, ,
\label{EFTI-action-Eij}
\end{split}
\end{equation}
where the relations between the effective coefficients can be trivially found to be
\begin{align}
m_2^4 & = M_2^4 + 6H\hat{M}_1^3 - 3H^2\left(3\bar{M}_1^2+\bar{M}_2^2\right) \, ,
\label{massr-1}\\
m_3^4 & = M_3^4 - 3H\left(\hat{M}_1^3+\hat{M}_2^3\right) \, ,
\label{massr-2}\\
\hat{m}_1^3 & = \hat{M}_1^3 - H\left(3\bar{M}_1^2+\bar{M}_2^2\right) \, ,
\label{massr-3}\\
\hat{m}_2^3 & = \hat{M}_1^3+\hat{M}_2^3 \, ,
\label{massr-4}\\
\bar{m}_1^2 & = \bar{M}_1^2 \, ,
\qquad
\bar{m}_2^2 = \bar{M}_2^2 \, ,
\qquad
\bar{m}_3^2 = \bar{M}_3^2 \, ,
\label{massr-5}
\end{align}
neglecting the contributions from higher order operators.

The constraint equations\index{Hamiltonian constraints} \eqref{constraint-N}-\eqref{constraint-Ni} become
\begin{multline}
\frac{M_P^2}{2} \left[ \upleft{3}{R} - \frac{1}{N^2}\left(E^{ij}E_{ij}-E^2\right)
		+\frac{2}{N^2}\dot{H}
		- 2(3H^2+\dot{H})\right] 
\\
	+ m_2^4\delta N 
	-\hat{m}_1^3\delta E 
	+ \bar{m}_3^2\upleft{3}{R} = 0 
\label{constraint-N-Eij}
\end{multline}
and
\begin{align}
\hat{\nabla}_i \left[ \frac{M_P^2}{N}(E^i_j - E \delta^i_j)
	- \hat{m}^3_1 \delta^i_j \delta N
	- \bar{m}_1^2 \delta^i_j \delta E
	- \bar{m}_2^2 \delta E^i_j \right] = 0 \, ,
\label{constraint-Ni-Eij}
\end{align}
respectively.

%% file: chapters/appendix-Extended_Genesis.tex
\chapter{Analysis of the Extended Genesis}
\label{Appendix-Extended_Genesis}

\section{The EFT for cosmological perturbations}
\label{appA}

\index{Extended Genesis}In this appendix we summarize some of the technical details on computing the two point function of adiabatic perturbations on \acs{NEC}-violating cosmological backgrounds. We closely follow the discussion of \cite{Creminelli:2010ba,Creminelli:2006xe}, generalizing the relevant expressions found in those references whenever appropriate. Furthermore, we present some numerical results and discuss the effect of higher order contributions.

\subsection{Galileons in ADM variables}

It will prove convenient to work in the $(3+1)$ form of our generalized Galileon action \eqref{ggg}, with the metric 
\begin{equation}
\D s^2=-N^2 \D t^2+g_{ij}(N^i \D t+\D x^i)(N^j \D t+\D x^j)
\end{equation}
in \acs{ADM} variables. 
The necessary expressions have been derived in \cite{Creminelli:2010ba}, and we just summarize their results, with a minimal amount of adjustment relevant to our case. 
In the unitary gauge defined by the absence of $\pi$-perturbations, $\pi(x,t)=\pi_0(t)$, the full action \eqref{ggg} can be written in terms of these variables in the following way (see \cite{Creminelli:2010ba} for derivation in the case of Galilean Genesis, $\mathcal{F}_1=e^{2\pi_0}$ and $\mathcal{F}_2=1$)
\begin{align}
\label{adm}
S&=S_g+S_\pi \, ,\\
S_g&=\frac{M_{\rm Pl}^2}{2} \int \D^4x~N\sqrt{\gamma}\left[\upleft{3}{R}+(K^{ij}K_{ij}-K^2) \right] \, ,\\
S_{\pi}&=f^2 \int \D^4 x ~N\sqrt{\gamma} \left[ -\mathcal{F}_1\left(\pi_0\right)\frac{\dot \pi_0^2}{N^2} + \frac{4 \dot \pi_0^3}{9 H_0^2} \frac{1}{N^3}K^i_{~i} + \mathcal{F}_2\left(\pi_0\right)\frac{\dot \pi_0^4}{3 H_0^2}\frac{1}{N^4}\right] \, .
\end{align}

\subsection{Effective field theory}

To study the scalar spectrum of \eqref{adm}, one can readily employ the standard \acs{EFT}\index{inflation!EFT of} of inflation formalism (see Chap. \ref{chap-EFTinflation} and App. \ref{Appendix-Eij}). The generic action for matter perturbations can then be written as in \eqref{EFTI-action-Eij}. The first line gives the only terms that start linearly in metric perturbations, therefore their coefficients are completely fixed by the background equations. On the other hand, the coefficients of the other operators are \textit{a priori} unconstrained. While only two are generated at the ``classical'' level by the action \eqref{ggg},
\begin{equation}
\label{M's}
m^4_2=\frac{4}{3} \frac{f^2}{H_0^2} \left(2 \mathcal{F}_2\left(\pi_0\right) \dot\pi_0^4+\dot \pi_0^2\ddot\pi_0+9 H\dot\pi_0^3\right), \quad \hat m_1^3=\frac{4}{3}\frac{f^2}{H_0^2}\dot\pi_0^3~, 
\end{equation}
the rest of the operators are expected to be present ($\bar{m}_i \neq 0$), although suppressed by whatever the quantum expansion parameter of the theory is.

%\subsection{Generalized Galileons} 

%Starting with the case $\bar{m}_i=0$ and using the parametrization \eqref{ug-zeta}, the solutions of the constraints \eqref{constraint-N-Eij}-\eqref{constraint-Ni-Eij} are
%\begin{align}
%\label{lapse}
%\delta N&=\frac{2M_{\rm Pl}^2}{2M_{\rm Pl}^2 H-\hat m^3_1}\dot \zeta \, \\
%\nabla_iN^i &=-\frac{2M_{\rm Pl}^2}{2M_{\rm Pl}^2 H-\hat m_1^3}\frac{\nabla^2\zeta}{a^2} +\frac{-4M_{\rm Pl}^4\dot H-12M_{\rm Pl}^2 H\hat m_1^3+3 \hat m_1^6+2M_{\rm Pl}^2 m_2^4}{(2M_{\rm Pl}^2 H-\hat m_1^3)^2}\dot\zeta \, .
%\end{align}
%Having obtained the lapse and shift, one can plug their expressions back into \eqref{EFTI-action-Eij}, that after a few integrations by parts finally yields the quadratic action for $\zeta$
%\begin{equation}
%S_\zeta=\int \D^4x~ a^3\left[ A(t)\dot\zeta^2-B(t)\frac{(\vec{\nabla}\zeta)^2  }{a^2} \right] ~,
%\end{equation}
%where the kinetic coefficients are given by the following expressions
%\begin{align}
%A(t) &=\frac{M_{\rm Pl}^2 (-4 M_{\rm Pl}^4 \dot H-12M_{\rm Pl}^2 H \hat m_1^3+3\hat m_1^6+2M_{\rm Pl}^2 m_2^4)}{(2M_{\rm Pl}^2H-\hat m_1^3)^2} \, , \\
%B(t)&=\frac{M_{\rm Pl}^2 (-4 M_{\rm Pl}^4 \dot H+2M_{\rm Pl}^2 H \hat m_1^3-\hat m_1^6+2M_{\rm Pl}^2\partial_t\hat m_1^3)}{(2M_{\rm Pl}^2H-\hat m_1^3)^2}~.
%\end{align}
%The speed of sound of the (short-wavelength) curvature perturbation $\zeta$ is 
%\begin{equation}
%c_s^2(t)=\frac{B(t)}{A(t)}~.
%\end{equation}
%These expressions have been used in Sec. \ref{analytic} and \ref{num} in testing for stability of solutions with the genesis - de Sitter transition. 

\subsection{Effects of higher (spatial) derivative operators} 
 
The case $\bar{m}_i=0$ has been already discussed in Chap. \ref{chap-EFTinflation}. The solutions of the Hamiltonian constraints, leading to the result \eqref{quadact} once plugged back into the action with the coefficients \eqref{A}-\eqref{B} and $C=0$, can be directly read in \eqref{chap3-ce-1}-\eqref{chap3-ce-2} with the help of the relations \eqref{massr-1}-\eqref{massr-5}. Such expressions have been used in Sec. \ref{analytic} and \ref{num} in testing for stability of solutions with the genesis - de Sitter transition. 

In order to assess the role of higher-order operators, we generalize the calculation of Sec. \ref{sec-cswmwg} (in the notations of App. \ref{Appendix-Eij}) to the case of non-zero $\bar{m}_1$ and $\bar{m}_2$, still keeping $\bar{m}_3=0$.
Linearising and solving \eqref{constraint-N-Eij} and \eqref{constraint-Ni-Eij}, and substituting back into \eqref{EFTI-action-Eij} yields the following quadratic action for the curvature perturbation 
\begin{equation}
S_\zeta=\int \D^4x~ a^3\bigg[A(t)\dot\zeta^2-B(t)\frac{1}{a^2}\left(\vec{\nabla}\zeta\right)^2-C(t)\frac{1}{a^4}\left(\vec{\nabla}^2\zeta\right)^2 \bigg]~,
\end{equation}
with the kinetic coefficients 
\begin{align}
\label{A'}
A &=(M_{\rm Pl}^2-\bar{m}_2^2)\cdot X\\
\label{B'}
B&=-M_{\rm Pl}^2-\frac{1}{a}~\partial_t Y ~, \\
\label{C'}
C&=\frac{2 M_{\rm Pl}^4 (\bar{m}_1^2+\bar{m}_2^2)}{Z} ~,
\end{align}
where the three auxiliary functions $X$, $Y$ and $Z$ have been defined as follows
\begin{multline}
X=[(2M_{\rm Pl}^2+3\bar m_1^2+\bar m_2^2) (m_2^4-2M_{\rm Pl}^2 \dot H)-6M_{\rm Pl}^2 H^2 (3\bar m_1^2+\bar m_2^2)
\\
+3 \hat m_1^3 (\hat m_1^3-4 M_{\rm Pl}^2 H)]
[2M_{\rm Pl}^2 H^2 (2 M_{\rm Pl}^2-3\bar m_1^2-3\bar m_2^2)
\\
+(\bar m_1^2+\bar m_2^2) (m_2^4-2M_{\rm Pl}^2\dot H)+\hat m_1^3 (\hat m_1^3-4M_{\rm Pl}^2 H)]^{-1} ~,
\end{multline}
\begin{multline}
Y=a[2 M_{\rm Pl}^2 (M_{\rm Pl}^2-\bar m_2^2) (\hat m_1^3-2M_{\rm Pl}^2 H)]
[2M_{\rm Pl}^2 H^2 (2 M_{\rm Pl}^2-3\bar m_1^2-3\bar m_2^2)
\\
+(\bar m_1^2+\bar m_2^2) (m_2^4-2M_{\rm Pl}^2\dot H)+\hat m_1^3 (\hat m_1^3-4M_{\rm Pl}^2 H)]^{-1} \, ,
\end{multline}
\begin{multline}
Z=2M_{\rm Pl}^2 H^2 (2 M_{\rm Pl}^2-3\bar m_1^2-3\bar m_2^2)
\\
+(\bar m_1^2+\bar m_2^2) (m_2^4-2M_{\rm Pl}^2\dot H)+\hat m_1^3 (\hat m_1^3-4M_{\rm Pl}^2 H) \, .
\end{multline}

\section{Extended Genesis: numerical study}
\label{num}

In this section we study an explicit illustrative model possessing cosmological solutions with the genesis - de Sitter transition. In principle, any theory described by $S_{1,2}$ of Sec. \ref{inf-mink-1} for $\e^\pi\ll 1$ and $\e^\pi\gg 1$ is expected to reproduce the Extended Genesis cosmologies. Perhaps the simplest example is provided by \eqref{simplemodel}. For $\beta=0$ the theory is just the conformal Galileon and when starting out in the \acs{GG} phase, the expansion rate of the Universe  diverges and the background exits the regime of validity of the \acs{EFT} at some finite time, the scalar profile gradually growing as $\e^\pi\sim 1/t$. 
For a non-zero $\beta$ however, the dynamics of the system is completely altered as soon as $\beta \e^{2\pi}$ becomes of order, or greater than, unity: the theory becomes effectively described by a $P(X)$-type Lagrangian with a cubic Galileon self-interaction, resulting, as we will show shortly, in transition into an inflationary de Sitter phase. 

\begin{figure}[t]
\centering
\includegraphics[width=.38\textwidth]{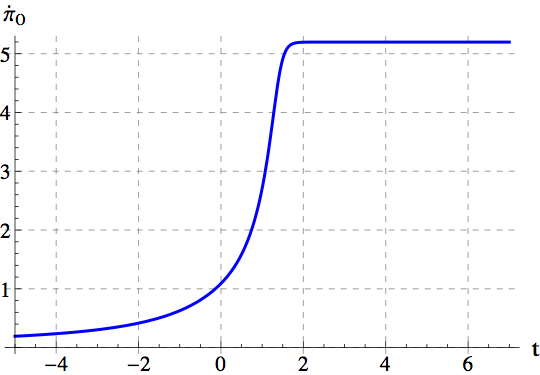} \quad
\includegraphics[width=.38\textwidth]{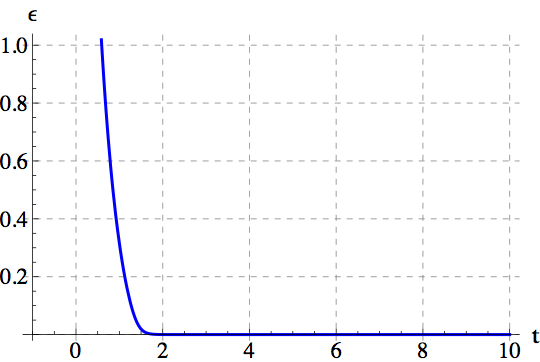}\\
\includegraphics[width=.38\textwidth]{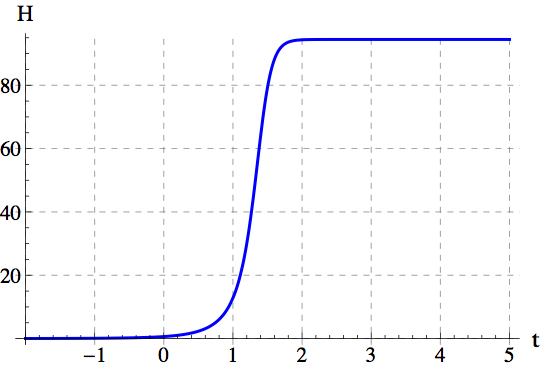}\quad
\includegraphics[width=.41\textwidth]{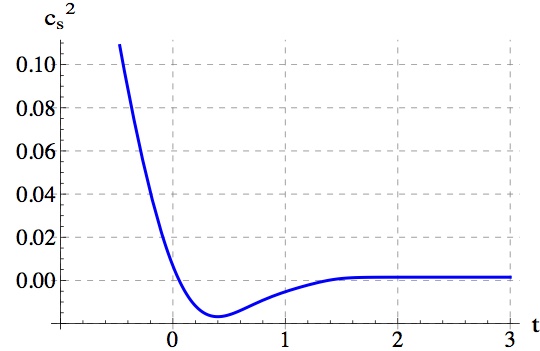}
\caption{Numerical solutions to the theory \eqref{simplemodel}, exhibiting the genesis - de Sitter transition. The time evolutions of $\dot\pi_0$, $\varepsilon=\dot H/H^2$, $H$ and the squared speed of sound $c_s^2$ of the curvature perturbation $\zeta$ (for early enough times not displayed in the plots, $c_s^2$ goes asymptotically to $1$ as required by Galilean Genesis \cite{Creminelli:2010ba}) are shown.}
\label{figuretwo}
\end{figure}

Fig. \ref{figuretwo} illustrates a typical solution from our numerical study, obtained by integrating expressions for $\dot H$ and $\ddot \pi$ with the initial conditions, relevant for Galilean Genesis. We have assumed $\beta=0.001$ and $M_{\rm Pl}=f=10^6  H_0$, setting $H_0$ (related to $\Lambda$ as in \eqref{ggsol}) as the unit mass scale.
The (time dependent) background quantities $\dot\pi_0$, $\varepsilon=\dot H/H^2$, $H$ and the squared speed of sound $c_s^2$ of the curvature perturbation $\zeta$ are shown. For early enough times, not displayed in the plots, $c_s^2$ goes asymptotically to $1$, due to the emergent conformal symmetry. The graphs for the Hubble rate and the time derivative of $\pi_0$ clearly show the genesis - de Sitter transition, the scalar field acquiring a linear $\pi\propto t$ profile at late times. 

The main steps of the explicit computation of the quadratic $\zeta$ action for the theory \eqref{simplemodel} have been retraced in the previous section.
While complete stability and (sub-) luminality of the given backgrounds can be readily checked analytically in both asymptotic regimes, the short transition region between 
the two phases displays gradient instability, at least for the values of the parameters that we have been able to cover in numerical studies (we have checked explicitly that for all considered solutions, the flip of sign of the $c_s^2$ quantity stems from the gradient energy becoming negative, not the kinetic one, that would lead to a more severe ghost instability). 
For all solutions displaying the genesis - de Sitter transition, the squared speed of sound varies from unity at early times (as required by Galilean Genesis \cite{Creminelli:2010ba}) to a small value $c_s^2 \lesssim 0.03$ in the asymptotic future\footnote{It can be shown analytically \cite{Kobayashi:2010cm} that for the most general Galileon theory of the form \eqref{ginf}, the speed of sound for the scalar perturbation on a de Sitter background is bounded from above, $c_s^2\leq 0.031$, precisely what we are finding numerically for the Extended Genesis future asymptotic stage.}, via a slight dip below zero in between that lasts from a few Hubble times to a fraction thereof, depending on a solution. 
In principle, gradient instability has a characteristic time scale of order at least the quantum cutoff of the theory, therefore a background with this feature can not be considered fully legitimate.  

We present theories admitting fully stable cosmologies with the genesis - de Sitter transition in Sec. \ref{analytic}, but note that even for the present simple theory the small (order per-cent) negative squared speed of sound corresponding to the gradient instability of Fig. \ref{figuretwo} suggests that it can be naturally cured by incorporating higher-order corrections in the effective theory for perturbations. While we carry out a systematic study of higher-order effects in Sec. \ref{hdim} (see also Sec. \ref{appA}), we give a quick argument here. The interplay between higher (spatial) derivative operators, contributing $\sim k^4$ to the IR dispersion relation for the scalar perturbation, and the presence of the cosmological horizon can stabilize the system against potential gradient instability (see \textit{e.g.} Ref. \cite{Creminelli:2006xe}).  This can be seen as follows.
At the level of four derivatives, one can add to the effective theory for perturbations on our background solution of Fig. \ref{figuretwo} the following term\footnote{There are other and more relevant terms beyond the leading order in the \acs{EFT} (see Sec. \ref{hdim} for a systematic study). However, for the illustrative purposes we are after, we neglect them here.}  
\begin{equation}
\Delta S\sim \int \D^4x~\sqrt{-g} ~\kappa(t) \upleft{3}{R}^2~,
\end{equation} 
where $\upleft{3}{R}$ is the scalar curvature of the three-dimensional metric, induced on equal time hypersurfaces and $\kappa$ is an arbitrary dimensionless time dependent coupling. This term adds a higher-spatial derivative contribution to  the (unitary gauge) quadratic action for the curvature perturbation
\begin{equation}
\Delta S_\zeta \sim -\int \D^4 x ~ \frac{\kappa}{a}(\vec{\nabla}^2\zeta)^2~.
\end{equation} 
Since $\kappa$ is an arbitrary function, it can always be chosen in such a way to render the instability scale for the background solution of Fig. \ref{figuretwo} smaller than the relevant instantaneous Hubble rate. Indeed, at frequencies larger than Hubble, the canonically-normalized curvature perturbation is described by the following action
\begin{equation}
S_\zeta=\int \D^4x~ a^3\left[\dot\zeta_c^2-c_s^2~\frac{(\vec{\nabla}\zeta_c)^2}{a^2} -\frac{\kappa}{A}\frac{(\vec{\nabla}^2\zeta_c)^2}{a^4}\right] \, ,
\end{equation}
where $c_s^2$ is negative in the region with gradient instability. At large enough (physical) momenta, $  k^2 \gtrsim |c_s^2| A/\kappa$, the system is stabilized by higher-order effects. Requiring the corresponding frequency to be less than the instantaneous Hubble rate then yields the condition on $\kappa$ for a completely stable background solution,
\begin{equation}
\kappa(t) \gtrsim \frac{c_s(t)^4 A(t)}{H(t)^2}~.
\end{equation}
Note the strong dependence ($\propto c_s^4$) on the scalar speed of sound of the lower bound on the coefficient $\kappa$. In particular, for small $c_s^2$, one can expect higher-order effects to easily cure the leading-order gradient instability.

%% file: FrontBackmatter/Bibliography.tex
%********************************************************************
% Bibliography
%*******************************************************
% work-around to have small caps also here in the headline
\manualmark
\markboth{\spacedlowsmallcaps{\bibname}}{\spacedlowsmallcaps{\bibname}} % work-around to have small caps also
%\phantomsection 
\refstepcounter{dummy}
\addtocontents{toc}{\protect\vspace{\beforebibskip}} % to have the bib a bit from the rest in the toc
\addcontentsline{toc}{chapter}{\tocEntry{\bibname}}
\bibliographystyle{unsrt} 
\label{app:bibliography} 
\bibliography{Bibliography}

%% file: FrontBackmatter/Colophon.tex
\pagestyle{empty}

\hfill

\vfill

\pdfbookmark[0]{Colophon}{colophon}
\section*{Colophon}
This document was typeset using the typographical \texttt{classicthesis} developed by Andr\'e Miede. 
%The style was inspired by Robert Bringhurst's seminal book on typography ``\emph{The Elements of Typographic Style}''. 
%\texttt{classicthesis} is available for both \LaTeX\ and \mLyX: 
%\begin{center}\url{http://code.google.com/p/classicthesis/}\end{center}
%Happy users of \texttt{classicthesis} usually send a real postcard to the author, a collection of postcards received so far is featured here: 
%\begin{center}\url{http://postcards.miede.de/}\end{center}
 
\bigskip

\noindent\finalVersionString

%% file: FrontBackmatter/Index.tex
%********************************************************************
% Index
%********************************************************************

\manualmark
\markboth{\spacedlowsmallcaps{\indexname}}{\spacedlowsmallcaps{\indexname}}
\refstepcounter{dummy}
\pagestyle{scrheadings}
\addcontentsline{toc}{chapter}{\tocEntry{\indexname}}
\printindex